\documentclass[a4paper]{osudissert96}

\usepackage{subfigure}
\usepackage{amsfonts}
\usepackage{color}
\usepackage{amsmath}
\usepackage{graphicx}  %
\usepackage{bm}  %
\usepackage{amssymb}
\usepackage{dcolumn}
\usepackage{multirow}
\usepackage{simplewick} 
\usepackage{xr} 

\def\BibTeX{{\rm B\kern-.05em{\sc i\kern-.025em b}\kern-.08em
    T\kern-.1667em\lower.7ex\hbox{E}\kern-.125emX}}

\graphicspath{{./figures/}}

\newcommand{\beqn}{\begin{equation}}
\newcommand{\eeqn}{\end {equation}}
\newcommand{\bea}{\begin{eqnarray}}
\newcommand{\eea}{\end{eqnarray}}
\newcommand{\ben}{\begin{eqnarray*}}
\newcommand{\een}{\end{eqnarray*}}
\newcommand{\be}{\begin{enumerate}}
\newcommand{\ee}{\end{enumerate}}
\newcommand{\bi}{\begin{itemize}}
\newcommand{\ei}{\end{itemize}}
\newcommand{\bfig}{\begin{figure}}
\newcommand{\efig}{\end{figure}}
\newcommand{\bc}{\begin{center}}
\newcommand{\ec}{\end{center}}
\newcommand{\et}{\end{center} \end{table}}

\newcommand{\np}{n_{p}}
\newcommand{\nq}{n_{q}}
\newcommand{\npp}{n'_{p}}
\newcommand{\nqp}{n'_{q}}

\newcommand{\vlowk}{V_{{\rm low}\,k}}

\newcommand{\Trel}{T_{\rm rel}}

\newcommand{\nmax}{N_{\rm max}\,}

\renewcommand\typesetChapterTitle[1]{\uppercase{#1}}

\newcommand{\bra}[1]{\ensuremath{\langle#1|}}
\newcommand{\ket}[1]{\ensuremath{|#1\rangle}}

\newcommand{\braket}[2]{\ensuremath{\langle#1|#2\rangle}}

\newcommand{\chieft}{\ensuremath{\chi}EFT}
\newcommand{\comments}[1]{}

\newcommand{\oneSzero}{$^1$S$_0$}
\newcommand{\threeSone}{$^3$S$_1$}
\newcommand{\onePone}{$^1$P$_1$}
\newcommand{\infm}{\,\mbox{fm}^{-1}}
\newcommand{\avpot}{Argonne $v_{18}$}
\newcommand{\Lambdaho}{\Lambda_{HO}}
\newcommand{\Lho}{L_{HO}}

\begin{document}
%
%
%
%
%
\author{Brian Dainton}
\title{Toward Universality in Similarity Renormalization Group Evolved Few-body Potential Matrix Elements}
\authordegrees{B.S., M.S.}  
\unit{Physics}
\advisorname{Prof. R. J. Perry}
\member{Prof. R. J. Furnstahl}
\member{Prof. T. Humanic}
\member{Prof. E. Braaten}

%
%

\maketitle


\disscopyright

%
\begin{abstract}
  As calculations of many-body nuclear systems become more and more precise, the question has shifted from ``can we calculate observables?'' to ``how precise can our calculations be?''.
Improvements in many-body methods have identified the need for better few-body potentials as input.
There are a wide variety of input potentials, which all accurately describe few-body systems, and all must be transformed with similarity renormalization group (SRG) transformations before they are computationally efficient enough to be used in certain many-body calculations.
It was realized that modern realistic 2-body potentials with very different matrix elements evolve under SRG transformations to the same universal low-energy shape.
Understanding the requirements for universality in evolved potential matrix elements can aid in the construction of better potentials for the many-body problem.
Furthermore, understanding the precision of few-body evolution is a key step to setting error bars on theoretical predictions.

We first examine how the universality of two-nucleon interactions evolved using similarity renormalization
group (SRG) transformations correlates with T-matrix equivalence, with the ultimate goal
of gaining insight into universality for three-nucleon forces. 
Because potentials are fit to low-energy data, they are (approximately) phase-shift equivalent only
up to a certain energy, and we find universality in evolved potentials up to the corresponding momentum.
More generally, we find universality in local energy regions, reflecting a local decoupling by
the SRG. 
The further requirements for universality in evolved potential matrix elements are explored
using two simple alternative potentials. 
We see evidence that in addition to predicting the same observables, common long-range behavior (i.e., explicit pion physics) is required for universality.
In agreement with observations made previously for $\vlowk$ evolution, regions of universal potential matrix elements are restricted to where half-on-shell T-matrix equivalence holds.

To continue the study in the 3-body sector, we create a simple 1-D spinless boson ``theoretical laboratory'' for a dramatic improvement in computational efficiency.
We introduce a basis-transformation, harmonic oscillator (HO) basis, which is used for current many-body calculations and discuss the imposed truncations.
We confirm that evolution to universal low-energy 2-body potential matrix elements is the same for 1-D bosons as 3-D fermions, and show that a further simplification of using positive-valued eigenvalues rather than phase shifts is valid.
When SRG evolving in a HO-basis, we show that the evolved matrix elements, once transformed back into momentum-representation, differ from those when evolving with momentum representation.
This is because the generator in each basis is not exactly the same due to the truncation.
In the 2-body sector, this can be avoided by increasing the basis size, but it remains unclear whether this is possible in the 3-body sector, as computational power required is greatly increased for three-body evolution.

In our efforts to study universal matrix elements in the 3-body sector by observing momentum representation matrix elements, we observe oscillations much like those appearing from truncation errors in the 2-body sector.
We can identify that the spectator particle adds strength in far-off diagonal potential matrix elements of the embedded 2-body potential, thus truncation errors appear in 3-body HO-basis evolution at much higher values of the decoupling scale than expected from 2-body calculations.
Observing matrix elements of a part of the SRG evolution that should be the same in 2- and 3-body sectors shows that they are different, which indicates the difference is an error in the HO-basis 3-body evolution.
With a better understanding of the source of errors in 3-body input potentials, we hope to gain further insight into how to make progress towards precision nuclear many-body calculations.

\end{abstract}
%
%
%
\dedication{For Laura}
%
%
\begin{acknowledgments}

First, I would like to extend my sincere gratitude to my research advisor, Dr. Robert Perry.
With the five undergraduate courses taught by him and as my graduate advisor, he has had the greatest impact upon my education.
His enthusiasm and dedication serve as an inspiration to all of his students, and he has a unique way of making physics exciting.
It has been a privilege to learn from him for so long.

I must also thank Dr. Richard Furnstahl, who could be credited as a secondary advisor (and actually \emph{was} my undergraduate faculty advisor).
His door is nearly always open and he has been an invaluable source of physics, computational, and editorial knowledge.

Thanks to the other members of my committee, Dr. Thomas Humanic and Dr. Eric Braatan, for their time and helpful comments.
Thank you Kyle Wendt, with whom I shared my office for most of my career, for sharing several python scripts and assisting as I learned how to force a computer to bow to my whims.
I learned much in discussing each of our research topics in the day-to-day.
Thanks to Heiko Hergert, who could get faulty code to properly run simply by entering the room, thank you for your shared, near-omnipotence on all matters physics and computers, and for much needed youtube distractions and holiday baked-goods.
Thanks also to Sushant More, Sebastian K\"{o}nig, Eric Anderson, Kai Hebeler, and everyone else who I have interacted with as we strive to solve the mysteries of low-energy nuclear physics.

Thanks to my family and friends, who have been a constant source of love and support.  
Thank you Laura, my wife, who has been ever patient and understanding; encouraging me when facing a stubborn problem and sharing in the elation of the achievements.
Thanks to Axel, who certainly would have liked more time to play with his daddy.
Thanks to my mother, who always remained positive during countless moody phone calls on my drives home.

And thanks to everyone else who have enriched my life and otherwise contributed to this effort.

\end{acknowledgments}

\begin{vita}

\dateitem{April 6, 1986}{Born - Middleburg Heights, OH, U.S.A.}
\dateitem{June 2008}{B.S. Physics, The Ohio State University,
Columbus, Ohio}
\dateitem{July 2013}{M.S. Physics, The Ohio State University,
Columbus, Ohio}
\dateitem{September 2008 - present}{Department of Physics, The
Ohio State University, Columbus, Ohio} 

\begin{publist}
\researchpubs
\pubitem{{``Universality in Similarity Renormalization Group Evolved Potential Matrix Elements and T-matrix Equivalence"}, 
B. Dainton, R.J.~Furnstahl, R.J.~Perry, 
Phys.\ Rev.\ {\bf C} 78, 014001 (2014). arXiv:1310.6690 [nucl-th]}
\end{publist}

\begin{fieldsstudy} 
\majorfield{Physics} 
\end{fieldsstudy}

\end{vita}

%
\tableofcontents
\listoftables
\listoffigures
%
\chapter{Introduction}
\label{chapt:introduction}










The nucleus is made up of positively charged protons and neutral neutrons, and thus the force binding it together cannot be electromagnetic in origin.
The strong interaction that is responsible for this binding has been the focus of study of many physicists for over 80 years, since the discovery of the neutron in 1932.
Despite many triumphs, however, there is still much work to be done to achieve accuracy goals for calculations of many-nucleon systems.

Initially, much of the theory of low-energy nuclear physics was driven by experiment, including the discovery of new particles.
Yukawa proposed that the force between nucleons is caused by exchanging a particle, later called a pion.
This interaction was much like the exchange of massive photons, and with the discovery of the pion in 1947, meson-exchange potentials became more sophisticated.
In the 50's and 60's more mesons were discovered and the interaction between nucleons could be modeled by exchange of single mesons.
In the 70's more advanced models including multiple meson exchange (such as 2-pion exchange) were created.
At this time, however, it was not possible to perform accurate few- and many-body calculations with the potentials due to insufficient computing power, which slowed progress.
By the end of the 70's quantum chromodynamics (QCD) was largely established as the fundamental theory of the strong interaction, which called into question the use of meson-exchange potentials, as mesons are not fundamental particles in QCD.
At the time, QCD was not solvable in the low-energy regime in which nuclei exist, becoming perturbative only at much higher energies due to asymptotic freedom.
Because of this, meson-exchange potentials were still used, but deemed phenomenological.

Phenomenological potentials became sophisticated enough to accurately reproduce 2-body observables and required certain considerations in their construction.
Nuclear saturation implies that the potential must be short-ranged and strongly attractive at separations of a few fm.
The deuteron (a nucleus made of one proton and one neutron) has an electric quadrupole moment, which means that it is not spherically symmetric, and therefore the wavefunction is not purely S-wave, but a mixture of partial waves.
This requires a non-central force, which has a tensor character.
Scattering experiments imply that the interaction turns repulsive at higher energies, which led to a ``hard core'' being put into the interaction~\cite{Kanada:1967uf}.
One such phenomenological potential we use in this study is \avpot~\cite{av18orig,av18}.
These potentials are very accurate in the 2-body sector, but the hard core is prohibitive in the many-body regime for most solution methods, because the necessary single-particle basis sizes are too large for even today's computing capabilities.
Some of the phenomenological potentials lead to triumphs in calculations.
First, they can very accurately reproduce 2-nucleon observables, and with them irrefutable evidence for the necessity of 3-body forces was found.
Secondly, they are local in particle separation, which is a requirement for quantum Monte Carlo calculations. This is a particular class of many-body methods which can accurately describe energy spectra and shapes up to $^{12}$C.

Around 1990, another breed of potentials began to be developed.
Weinberg applied effective field theory methods to multi-nucleon systems by constructing the most general Lagrangian consistent with low-energy QCD \cite{Weinberg:1990rz}.
The new theory was called chiral effective field theory (\chieft), and the degrees of freedom were once again nucleons and pions.
\chieft\ potentials have beneficial qualities; they have the relevant degrees of freedom for low-energy nuclear physics, are consistent with QCD symmetries, and are much softer than the hard-core potentials.
Both phenomenological and \chieft\ potentials are used in modern calculations and continue to be improved upon.
The most beneficial quality of \chieft\ potentials is that they are derived in a model-independent systematic expansion, which makes theoretical error estimates possible.
Accurate \chieft\ potentials are much softer than the hard-core potentials, but they still are ``too hard'' for convergence in large many-nucleon systems.

To address this problem, Lee-Suzuki transformations were applied in free space to integrate out high-energy degrees of freedom and soften an initial realistic potential, 
generating phase-shift-equivalent ``low-momentum'' or ``$\vlowk$'' potentials~\cite{vlowkuniv,bognerfurnstahlschwenk}. 
This can be done in one step or incrementally using a renormalization group (RG) equation for the
potential~\cite{Bogner:2001jn}. 
Bogner and collaborators observed that a wide variety of realistic potentials have very similar low-momentum matrix elements after softening, which they termed the \textit{model independence} of $\vlowk$ potentials~\cite{Bogner:2001gq,vlowkuniv}.  
The diagonal $\vlowk$ potential matrix elements were found to match in regions of phase-shift equivalence of the realistic potentials while the off-diagonal matrix elements matched in regions of half-on-shell (HOS) T-matrix equivalence~\cite{vlowkuniv}.  
They suggested that differences in the HOS T-matrix and thus the off-diagonal $\vlowk$ potential matrix elements occur because of different treatments of pion physics~\cite{vlowkuniv}.

Subsequently, similarity renormalization group (SRG) unitary transformations have been used 
to soften nuclear potentials while preserving 
observables~\cite{GlazekWilson,Wegner,Bogner:2006pc,bognerfurnstahlschwenk,Roth:2011vt,Furnstahl:2013oba}.
The Lee-Suzuki transformations are derived using the $T$-matrix, and extension into the 3-body sector is difficult, whereas SRG transformations in the 3-body sector are conceptually straight-forward.
Like $\vlowk$ transformations, the SRG decouples high-energy from low-energy physics, 
allowing one to truncate the matrices above some decoupling 
scale~\cite{decouplingSRG,SRGgeneral,bognerfurnstahlschwenk}.  
Further, the low-energy matrix elements of initial realistic potentials also
flow to the same form, but differ in detail from those found using $\vlowk$ transformations. 
There is preliminary evidence that the SRG flow to common matrix elements extends to 
three-body forces~\cite{Hebeler:2012pr,Wendt:2013bla}, which are important ingredients for consistent treatments of nuclei 
with RG methods~\cite{Hammer:2012id,Furnstahl:2013oba}.

In analogy to the behavior of other Hamiltonians under RG transformations, this model independence is naturally interpreted as a flow to universality of the evolved potential matrix elements. 
Operators can be classified according to how their dimensionless coupling changes under RG transformations as relevant (coupling increase with decreasing resolution scale), marginal (coupling remains of the same order), and  irrelevant (coupling decreases)~\cite{Wilson:1974mb}.
The universal ability of different potentials to describe the same behavior is a consequence of the potentials possessing the same relevant and marginal operators, and the flow toward universal potential matrix elements naturally occurs as irrelevant operators diminish.
This form of universality can have powerful consequences if it can be understood and exploited.
It suggests that for low-energy problems, a broad class of starting potentials that fit data will be nearly equivalent after evolution~\cite{Timoteo:2011tt,Arriola:2013nja}. 
If realized for many-body forces, it may be possible to more easily construct accurate potentials (choosing operators based solely on their ease of use, then fitting constants to data), if they flow to a universal form after running the SRG.

In the effort to improve the nucleon-nucleon interaction, much attention is given to improving the potentials from phenomenological considerations or including more terms and degrees of freedom in \chieft.
Because the SRG is such an important tool to increase computational efficiency, and because the realistic potentials flow toward a universal form, in this study we focus on better understanding criteria for universal SRG evolved potential matrix elements and identifying errors in potential matrix elements after few-body harmonic oscillator (HO)-basis SRG evolution.
Many of the results of previous SRG studies have used 2-body evolution in partial-wave momentum representation \cite{GlazekPerry,BlockDiag,bognerfurnstahlschwenk,SRGgeneral,decouplingSRG,Wendt:2011qj}, but in practice, because of the importance of the 3-body forces, many-body problems require 3-body potentials and benefit from evolution in HO-basis as input (3-body momentum representation evolution has recently been done, but is not widely applied yet \cite{Hebeler:2012pr}).
Errors in few-body HO-basis evolved Hamiltonians cannot always be distinguished in 3-body observables because even the approximate transformations are unitary.
These errors can lead to lower precision in many-body calculations.
Because these errors in the 3-body Hamiltonian do not affect 3-body observables but can effect many-body observables, we can naturally interpret them as the introduction of spurious 4-body and further forces.
As unevolved potentials are increasingly more difficult to generate, it is important to utilize short-cuts gained from universality in evolved potential matrix elements.
We also must understand the precision of few-body SRG evolution, which in turn will lead to better input Hamiltonians and a better understanding of propagation of uncertainties for many-body methods.

\section{Motivation to Improve the Few-Body Hamiltonian}

\begin{figure}
\includegraphics[width=.9 \textwidth]{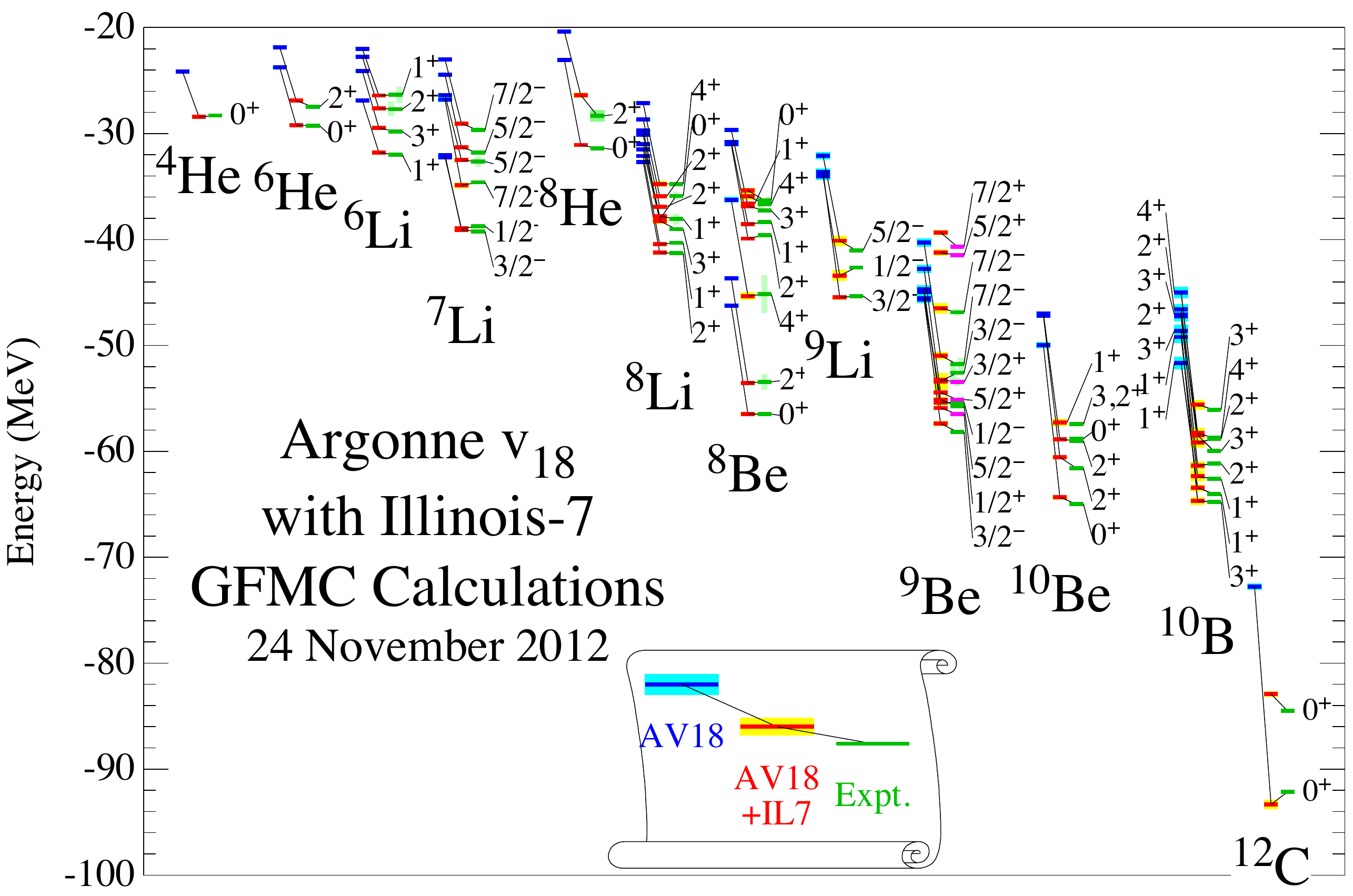}
	\caption{  Binding energies of various nuclei calculated using a quantum Monte Carlo many-body method with only 2-body input potential (AV18), with 2- and 3-body input potential (AV18+IL7), and experimental values.  Including the 3-body shifts the energies by a fraction of the 2-body results, increasing the accuracy systematically for all energy levels.  Accuracy generally decreases in heavier nuclei.  Figure from Ref. [25]}. 
	\label{fig:many_body_needs_3b_pot}
\end{figure}

As mentioned earlier, phenomenological potentials were able to reproduce 2-body observables with high accuracy, but attempts at few-body nuclear calculations could not reproduce experimental values using these ``high-precision'' 2-body potentials alone.
Few-body calculations had enough precision to confirm what many had been resisting because of the prohibitive leap in computational difficulty; that interactions depending on all three particles are not negligible and must be included.
Once developed, phenomenological 3-body potentials vastly improved the accuracy of bound-state calculations for light nuclei.
Figure~\ref{fig:many_body_needs_3b_pot} shows calculated binding energies of light nuclei with only a 2-body potential, and with the inclusion of two different 3-body potentials~\cite{3b_evidence}.
This picture enforces the idea of a many-body hierarchy; that 2-body forces are the greatest contribution to observables and 3-body forces are non-negligible and must be included for improved accuracy.
Higher-body forces can be added, each with decreasing contribution.
The importance of the 4-body force is still an open area of study.

\begin{figure}
\includegraphics[width=.9 \textwidth]{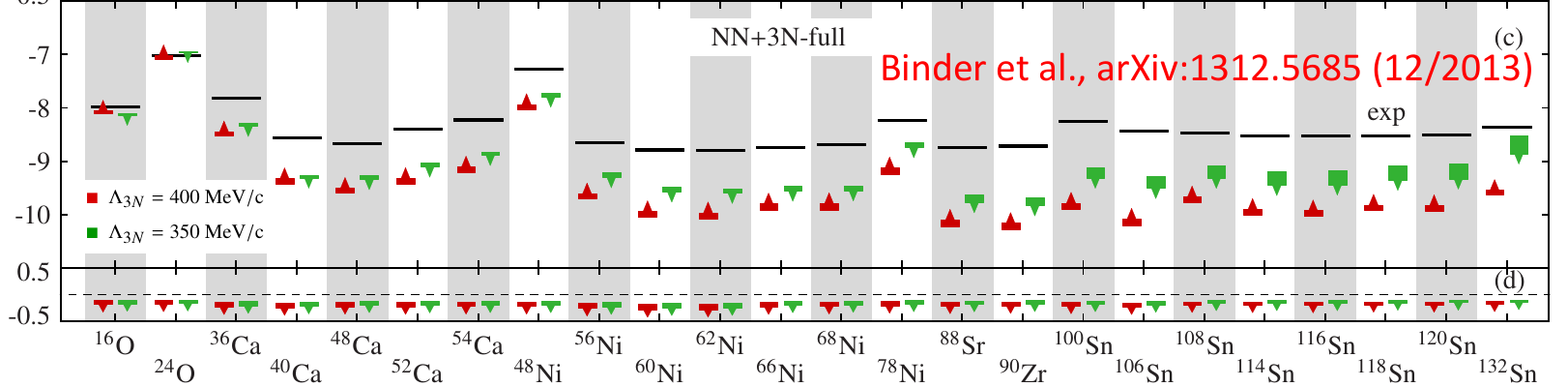}
	\caption{  Binding energy per particle in MeV of various nuclei calculated with a modern many-body method with input 3-body \chieft\ potential.  Precision estimates are made using an upper and lower bound generated from two different regularization cutoffs.  Accuracy and precision generally decrease in heavier nuclei. Figure from Ref. [12]} 
	\label{fig:many_body_needs_better_few_body}
\end{figure}

Modern many-body techniques which use as input modern 2- and 3-body potentials have become very precise.
These potentials are able to reproduce light-nuclei bound-state energies, but heavier nuclei remain a challenge.
Figure~\ref{fig:many_body_needs_better_few_body} shows a recent calculation for binding energies and approximated errors of many nuclei using a modern many-body calculation~\cite{Binder:2013xaa}.
The precision of calculations is a new and developing field, and to estimate error bands, a calculation is run with several different realistic potentials, such that a maximum and minimum value in observables gives some idea of the precision.
In Fig.~\ref{fig:many_body_needs_better_few_body}, we see that within the estimated precision for the heavier nuclei, the calculation does not accurately reproduce experimental values.
With confidence in the precision of the many-body methods, the implication is that this inaccuracy stems from the input Hamiltonian.
Furthermore, this method uses an HO-basis SRG-evolved Hamiltonian as input, and there is circumstantial evidence that the evolution itself is a major source of the inaccuracy ~\cite{Binder:2013xaa}.


\section{Discretization and Numerical Error}

Because we examine the precision of certain aspects of our calculation, a basic knowledge of numerically solving integrals and integral equations is critical.
As an example, we will briefly review numerical quadrature to introduce key terms and discuss numerical errors that will be critical in later chapters.
Computationally, we can approximate an integral as a finite sum of the product of a set of weights (often called mesh weights), $w_{i}$, multiplied by a set of functional values, $f_{i} \equiv f(x_{i})$, evaluated at each element in a set of values within the limits of integration (often called mesh points):
\beqn
	\int_{a}^{b} dx\ f(x) = \sum_{i=0}^{\nmax}( w_{i}\ f_{i} ) + \xi .  
\eeqn
$\xi$ is an error which can depend on mesh points, the function being integrated, integration limits $a$ and $b$ and the integration method.
Integration routines are a large topic by themselves, and we refer the interested reader to Ref.~\cite{computational}.
With a finite mesh size, one can neither have infinite integration limits nor infinitesimal spacing, thus the integration error, $\xi$, is created from two sources, which are analogous to infrared and ultraviolet errors.  
First, truncating infinite integrals at a finite value introduces a \emph{truncation error} which will be common in integrals over momentum in later sections.
Second, there is an error associated with the mesh spacing which would exist even over finite integration limits but typically will decrease as our mesh size increases.
Increasing the range of the mesh but keeping mesh size constant will generally improve cut off errors for indefinite integrals, but increase mesh width errors.
With these definitions, wavefunctions and operators are easily discretized when numerically solving integral equations.

One first must choose a basis to work in; then operators become finite matrices, and wavefunctions become vectors; all of which introduce numerical errors related to the integration errors previously mentioned.
For our calculations, when possible, we simply allow the range of the mesh \emph{and} mesh size to be sufficiently large that errors appearing in the 2-body sector from the discretization are negligible when compared to the errors we are studying in the 3-body sector.
Specifically, we choose mesh points and mesh weights from Gauss-Legendre quadrature, which is typically much more accurate than Newton-Cotes (constant mesh spacing) or a simple Riemann-summation (for more on these methods, see Ref.~\cite{computational}).


\section{A Word on Units}

Throughout this paper, units of the potential may look unfamiliar.
We choose for simplicity to use natural units, $\hbar = m = 1$.
Units depend on basis and dimensionality of the system.
With this definition, our units all must be powers of fm.
To be explicit, as a rule, position is always measured in fm, and with natural units this means momentum always in fm$^{-1}$.
The units of the representation of an operator in one basis are not necessarily the same as in another, and we will see that the potential has units which depend on the number of dimensions and representation.
To find the units of the potential in a particular basis, we examine the Schr\"{o}dinger eigenvalue equation in that basis and match units for each term.
For three dimensions (3-D), observing the Schr\"{o}dinger eigenvalue equation in coordinate representation,
\beqn
	( \frac{1}{2} \bm{\nabla}^{2} + V(\bm{r}) ) \Psi(\bm{p}) = E\ \Psi(\bm{p}),
\eeqn
we see that the potential, $V(r)$, and energy, E, both have units of fm$^{-2}$, in order to match the units of $\bm{\nabla}^{2}$.
The same is true for HO-basis.

In momentum representation in 3-D, the Schr\"{o}dinger eigenvalue equation is
\beqn
	\frac{p^{2}}{2} \Psi(\bm{p}) + \int d \bm{p}'\ V(\bm{p},\bm{p}') \Psi(\bm{p}') = E\ \Psi(\bm{p}),
\eeqn
and we see that the potential, $V(\bm{p},\bm{p}')$, must have units of fm, so that combined with the fm$^{-3}$ from $d \bm{p}'$, the integral has the same units as $E\ \Psi(\bm{p})$ and $\frac{p^{2}}{2} \Psi(\bm{p})$.

Similarly, in one dimension (1-D) in coordinate representation and HO-basis the potential has units of fm$^{-2}$.
In momentum representation, we again look to the Schr\"{o}dinger equation,
\beqn
	\frac{p^{2}}{2} \Psi(p) + \int dp'\ V(p,p') \Psi(p') = E\ \Psi(p),
\eeqn
and match units to see that the potential now has units of fm$^{-1}$.

\section{Organization of Thesis}

First, in chapter~\ref{chapt:SRG}, we review important aspects of similarity renormalization group transformations.
We will provide details on the different generators used in the following chapters and discuss the flow equations that dictate the transformation.
We also mention a strategy to evolve the 3-body potential while avoiding complications from ``spectator $\delta$-functions,'' which will enter into the discussion of alternate methods not used in this study.

In chapter~\ref{chapt:2BUniv}, we re-examine for the SRG the conclusions of Ref.~\cite{vlowkuniv} for $\vlowk$ potentials,
that the potentials must be phase-shift equivalent up to a certain resolution scale but also have consistent, explicit handling of the long-range pion physics~\cite{bognerfurnstahlschwenk}.  
We use an inverse scattering separable potential (ISSP) to test if universality in potential matrix elements emerges at high energies and without explicit pion-exchange terms.  
The ISSP can reproduce all observables in the two-nucleon problem, and we will see explicitly that this is not enough for all low-momentum matrix elements to flow towards a universal form at finite cutoff.  
Also, when creating the ISSP we are free to choose a binding energy independent of the phase shifts, thus we can see the effect of differences in the binding energy on evolved 
low-momentum matrix elements.
Furthermore, with the ISSP we are able to generate potentials with local phase-shift-equivalence and universal regions, which implies a local decoupling by the SRG.

To test the idea that the same explicit long-range treatment is required for flow to a universal form,
we introduce a second simple potential that is phase-shift equivalent at low energies and includes explicit 
one-pion exchange (OPE).  We use the model proposed by Navarro P\'erez \textit{et al}.~\cite{Arriola,Perez:2013jpa}, which
combines the OPE potential with a sum of $\delta$-shell potentials.  This potential replaces the 
short-range physics with simple terms to be fit to phase shifts, while preserving the long-range force.
At the close of chapter~\ref{chapt:2BUniv}, we will set up the 3-body problem with a necessary simplification of the phase-shift fitting procedure.

In chapter~\ref{chapt:HOBasis}, we provide details for the treatment of the 3-body problem.
We discuss the benefits and drawbacks of momentum representation and HO-basis.
We discuss HO-basis in great detail in the 2-body 3-D sector and for 1-D two- and three-body problems. 
We introduce the imposed truncation errors and implications for transformation between bases.
When evolving HO-basis potentials with SRG using relative kinetic energy as the generator, we identify evolution errors once the decoupling scale becomes too low.
Because of the singular nature of the kinetic energy, accurate transformation between bases is impossible, and different finite bases represent a slightly different operator due to the truncated basis space.
The differences in the generator lead to differences in SRG evolution between HO-basis and momentum representation.
Much of the formalism in setting up the HO-basis will be reviewed, however the implications of the basis truncation errors and especially the SRG errors will be of critical importance for the following chapter.

In chapter~\ref{chapt:3BUniv}, we create a simple ``test laboratory'' in 1-D to simplify the 3-body problem.
We test that our tools reproduce the results of previous papers \cite{Akerlund:2011cb,Jurgenson:2008jp} and then introduce a more ideal potential for our needs.
We confirm that our model follows the same rules for SRG evolution to universal, low-energy potential matrix elements.
Then we launch into the 3-body sector and discover that the evolution errors of 2-body potentials in HO-basis appear at much higher $\lambda$.
These errors prevent us from studying universality of matrix elements in the three-body sector.
We discuss the origin of the 3-body matrix elements that produce this evolution error.

In chapter~\ref{chapt:conclusion}, we review our findings and provide ideas for further study.
Specifically, a number of interesting developments are underway in the low-energy nuclear theory community that will be able to reconcile the differences in HO-basis and momentum representation calculations.

For a list of important abbreviations used throughout, see appendix~\ref{app:app_abbr}.
\chapter{Similarity Renormalization Group}
\label{chapt:SRG}

The similarity renormalization group~\cite{Bogner:2006pc,Szpigel:2000xj} is a continuous series of infinitesimal unitary transformations acting on the Hamiltonian.  The simplest SRG transformations can
be expressed in differential form as a flow equation:
\begin{equation}
	\frac{dH_{s}}{ds} = [\eta_{s},H_{s}] = [[G_{s},H_{s}],H_{s}] \;,
	\label{eq:srgdiffeq}
\end{equation}
where $s$ is a flow parameter~\cite{GlazekWilson,Wegner,bognerfurnstahlschwenk}.  
For most nuclear applications to date, the operator $G_{s}$ is chosen to be the kinetic energy operator, 
denoted $T$.  
(We will refer to $G_s$ in this work as the SRG ``generator''.)
The most commonly used diagonalizing generator for non-nuclear applications is known as the
Wegner generator~\cite{Kehrein:2006}.  
It uses the diagonal of the Hamiltonian defined in momentum representation, $H_{s}^{d}$ instead of $T$ for $G_{s}$.  
Flows using the Wegner generator are indistinguishable from $T$ for the range of evolution in the present study but can differ drastically if the SRG cutoff becomes very low~\cite{GlazekPerry} or if a large-cutoff chiral potential is used~\cite{Wendt:2011qj}.

The goal of the SRG is to decouple high-energy from low-energy degrees of freedom 
in the Hamiltonian by driving far off-diagonal matrix elements to zero.  
Instead of $s$, we usually refer to the decoupling scale, $\lambda = s^{-\frac{1}{4}}$ 
for $T$ and $H_{s}^{d}$, where $\lambda$ is chosen to have the same units as momentum.  
In the SRG flow with the $T$ generator, 
the dominant term of Eq.~\eqref{eq:srgdiffeq} for far off-diagonal matrix elements  
is the term linear in the potential, $[[T,V_{s}],T]$, where $V_{s} \equiv H_{s} - T$.
If we keep just this term, the flow equation is immediately solved for these
matrix elements, yielding (with mass $m=1$)
\begin{equation}\label{eq:srg_formfactor}
	V_{s}(k,k') \simeq V_{s=0}(k,k')\ e^{-(\frac{k^{2}-k'^{2}}{\lambda^{2}})^{2}}
	\;. 
\end{equation}
Thus $\lambda^{2}$ is roughly the maximum difference between
kinetic energies of nonzero matrix elements.
Once the Hamiltonian is sufficiently evolved to exhibit decoupling, low-energy observables 
can be obtained from a truncated Hamiltonian~\cite{decouplingSRG} or one finds naturally
that a smaller expansion basis is needed for a desired degree of convergence.  
%
\begin{figure}
\includegraphics[width=.9 \textwidth]{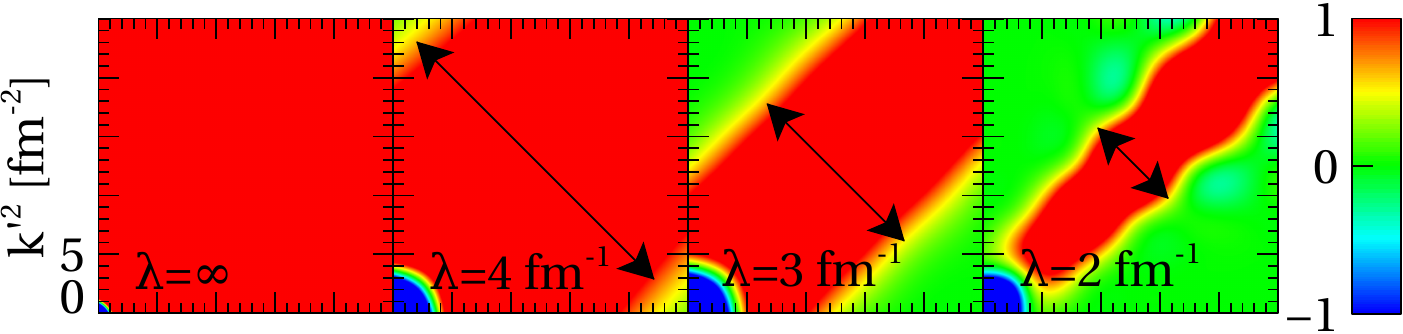}
	\caption{  Matrix elements $V(k,k')$ plotted against kinetic energy, $k^2$, of a realistic potential evolved with $G_{s}=\Trel$ to different $\lambda$ ($\infty$, 4 fm$^{-1}$, 3 fm$^{-1}$, and 2 fm$^{-1}$).   The matrix elements are scaled by the largest absolute value, then multiplied by 10 to emphasize the shape of the SRG-imposed form factor.  The colorbar range the first scaling, but not the multiplication by 10.  The width of the bands is roughly $\lambda^{2}$ as implied by Eq.~\eqref{eq:srg_formfactor}.  
	\label{fig:trel_evo_shape}}
\end{figure}
Fig.~\ref{fig:trel_evo_shape} shows potential matrix elements evolved to different $\lambda$.
In all contour plots following, we will always normalize the colors to the maximum absolute value of the matrix elements ($V_{ij}/\rm{max}[V_{ij}]$), and when noted we further scale the matrix elements by a constant for better visualization of shapes.
We can clearly see the decoupling of energies larger than the width of the red band.
When evolving to low enough $\lambda$, the flow equation term is quadratic in V and numerical inaccuracies cause some oscillation about the width of the SRG form factor of the dominant term.

A \emph{nondiagonalizing} alternative for $G(s)$ is the momentum-representation block generator~\cite{BlockDiag}, 
$H_{s}^{bd}$, defined as:
\bea
	H_{s}^{bd} &=& P H_{s} P + Q H_{s} Q , \\
	\bra{p} P \ket{p'} &=& \Theta(\Lambda - p)\ \delta(p-p') , \\
	Q &=& 1 - P,
\eea
where $\Theta$ denotes the Heaviside step function.
Versions of this generator with smoother cutoffs exist as well.
$H_{s}^{bd}$ matrix elements are the block diagonal elements of the evolved Hamiltonian $H_s$, 
separated at a fixed chosen cutoff parameter $\Lambda$. 
(That is, the generator $H_{s}^{bd}$ in a momentum basis 
is obtained from $H_{s}(k,k')$ by setting to zero the matrix elements where $k<\Lambda$
and $k'>\Lambda$ or $k>\Lambda$ and $k'<\Lambda$.) 
This yields the same basic pattern of decoupling achieved with $\vlowk$ Lee-Suziki 
transformations~\cite{bognerfurnstahlschwenk,Arriola,vlowkuniv}.  
In fact, the $\vlowk$ and SRG block diagonal transformations have been shown to result in very similar 
Hamiltonians for the lower energy block if the SRG transformation is run to
$\lambda \ll \Lambda$~\cite{BlockDiag}.  
For $H_{s}^{bd}$, $\lambda = s^{-\frac{1}{2}}$ and represents the maximum difference in
energy for coupling between the blocks above and below $\Lambda$.
%
\begin{figure}
\includegraphics[width=.9 \textwidth]{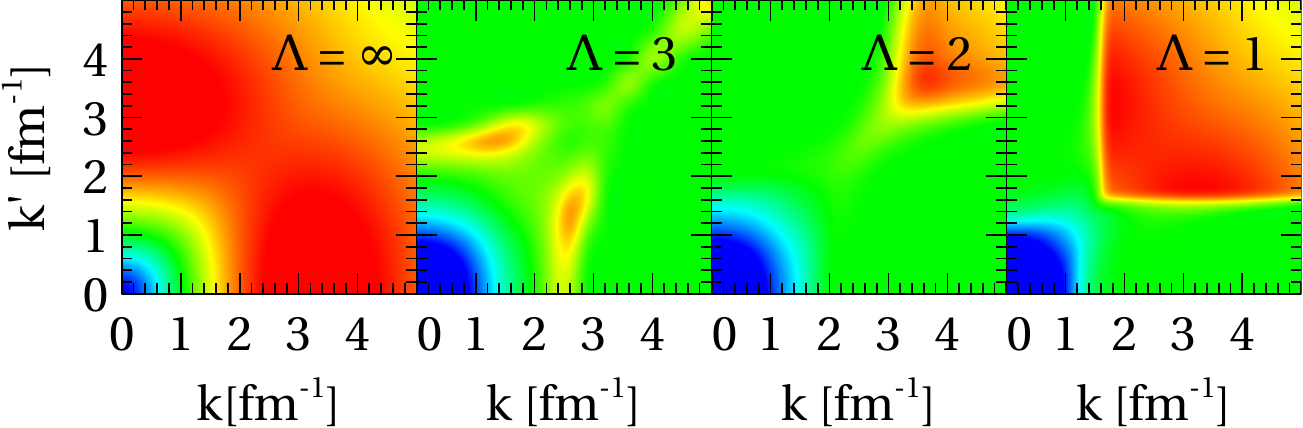}
	\caption{  Matrix elements $V(k,k')$ of a realistic potential evolved with $G_{s}=H_{s}^{bd}$ to different $\Lambda$ ($\infty$, 4 fm$^{-1}$, 3 fm$^{-1}$, and 2 fm$^{-1}$).  All $\lambda$ are 1.5 fm$^{-1}$, which allows for some non-zero matrix elements around the connected corners of the two blocks.}
	\label{fig:block_evo_shape}
\end{figure}
Fig.~\ref{fig:block_evo_shape} shows a realistic potential evolved to different $\lambda$.
Notice that there are two distinct blocks of nonzero matrix elements, at $k,k'$ both above $\Lambda$ or both below.

We will see in chapter~\ref{chapt:2BUniv} how different potentials evolve to a low-energy universal form.
It has been shown that the SRG with the $T$ generator~\cite{SRGgeneral} and $\vlowk$~\cite{bognerfurnstahlschwenk,Arriola,vlowkuniv} each drive realistic potentials to separate low-energy universal forms, and we will show that $H_{s}^{bd}$ also drives potential matrix elements to a different universal form in the 2-body sector.  

When evolving in a many-body sector, one can simply plug in the many-body $\Trel$ and $H_{s}$ into the flow equation, Eq.~\eqref{eq:srgdiffeq}.
In a momentum representation, $\delta$-functions over spectator particles make this strategy numerically impossible, but evolution in a discrete basis, such as HO-basis, has no singular $\delta$-functions, thus it apparently solves this problem.
This is historically the evolution strategy for 3-nucleon potentials and will be the strategy we use for our simple system of 1-D spinless bosons in chapter~\ref{chapt:3BUniv}.
Large numerical inaccuracies with this method occur, however, as we will see in chapter~\ref{chapt:HOBasis} such that evolution in different finite bases can be different.
It is possible that the inaccuracies of the Hamiltonian observed in heavy-nuclei problems~\cite{Roth:2011vt} is partially due to SRG evolution errors in HO-basis.

Recent work with SRG evolution in momentum representation takes a different approach, explicitly subtracting the singular parts from the flow equation at each step~\cite{3bsrgsimple,Hebeler:2012pr,Bogner:2006pc}.
We can define our 2- and 3-body potential and kinetic energy operators as:
\bea
	H &=& T_{12} + T_{3} + V_{12} + V_{13} + V_{23} + V_{123}, \\
	T_{12} + T_{3} &=& T_{13} + T_{2} = T_{23} + T_{1},
\eea
where ${12}$ denotes the relative kinetic energy or 2-body potential between particles 1 and 2, $T_{3}$ is the kinetic energy of the third particle relative to the center of mass of the pair, and $V_{123}$ is the 3-body potential.
The flow equation then becomes,
\beqn
	\frac{dV_{s}}{ds} = \frac{dV_{12}}{ds} + \frac{dV_{13}}{ds} + \frac{dV_{23}}{ds} + \frac{dV_{123}}{ds} = \left[ [T_{rel},V_{s}],H_{s} \right] .
\eeqn
With some algebra, we can subtract out the 2-body flow equation for each pair,
\beqn
	\frac{dV_{ij}}{ds} = \left[ [T_{ij},V_{ij}],T_{ij}+V_{ij} \right].
\eeqn
This subtracts away the ``disconnected diagrams'' and thus the dangerous spectator deltas~\cite{3bsrgsimple,Bogner:2006pc}, and we are left with:
\bea
	\frac{dV_{123}}{ds} =  &&\left[ [T_{12},V_{12}],V_{13}+V_{23}+V_{123} \right] \\
							&&+ 	\left[ [T_{13},V_{13}],V_{12}+V_{23}+V_{123} \right] \\
							&&+ 	\left[ [T_{23},V_{23}],V_{12}+V_{13}+V_{123} \right] \\
							&&+ 	\left[ [T_{rel},V_{123}],H_{s} \right].
\eea
This form of the flow equation is not nearly as convenient to implement, as it requires precalculating the 2-body evolved potential with the same grid and weights and inserting it at every step of solving the 3-body differential equation.
The benefit, however is that the 2-body potential in the 2-body space doesn't have pathologies involved with embedding it into a 3-body space (such as spectator $\delta$-functions).
A very useful feature of this form of the flow equation is that we can see explicitly that the 3-body evolved potential has terms that depend only on 2-body potentials.
Thus, in a 3-body evolution without an initial 3-body potential, one will be induced.


\chapter{Realistic 2-body Universality}
\label{chapt:2BUniv}

We discuss universality in matrix elements of modern realistic potentials in 
Section~\ref{sec:realistic}.  
In Section~\ref{sec:ISSP}, we provide a working description of the ISSP formalism, examine universality in ISSP\rq{}s, and discuss the resulting insight into the prerequisites for universality.  
Section~\ref{sec:OPE} gives a description of the $\delta$-shell plus OPE potential
and examines the SRG flow of this potential to a universal form. 
We also comment on the SRG flow of the JISP16 potential. 
Section~\ref{sec:ps_eig_equiv} details the relationship between phase-shift and eigenvalue equivalence, which will simplify the $\delta$-shell fitting procedure for different bases and more bodies.
Finally, we conclude in Section~\ref{sec:conclusion} with a summary and the outlook
for the three-body problem.


\section{Modern Realistic Potentials} \label{sec:realistic}
	
\begin{figure*}[tbh!]
	\includegraphics[width=.3 \textwidth]{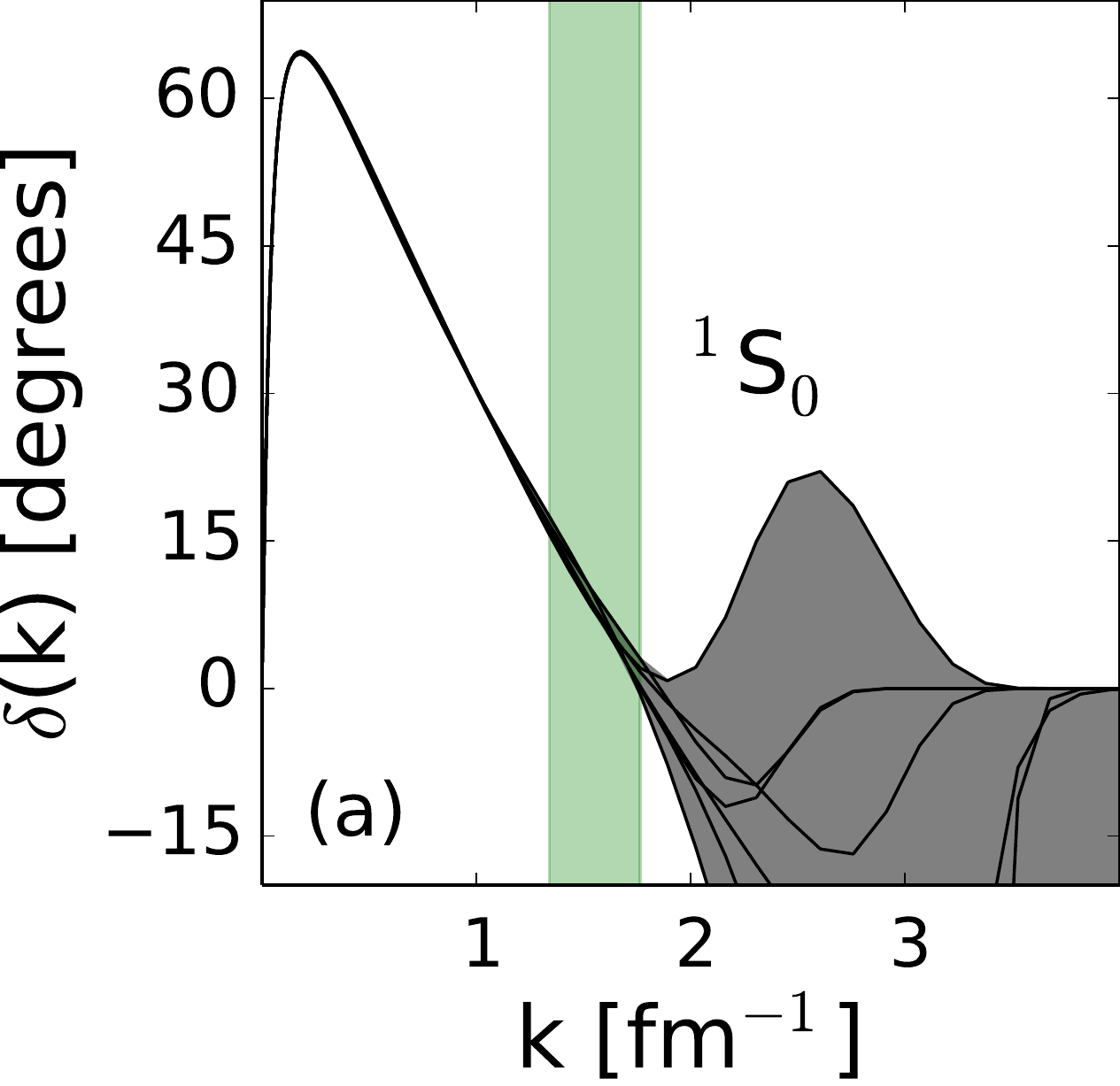}~~~%
	\includegraphics[width=.3 \textwidth]{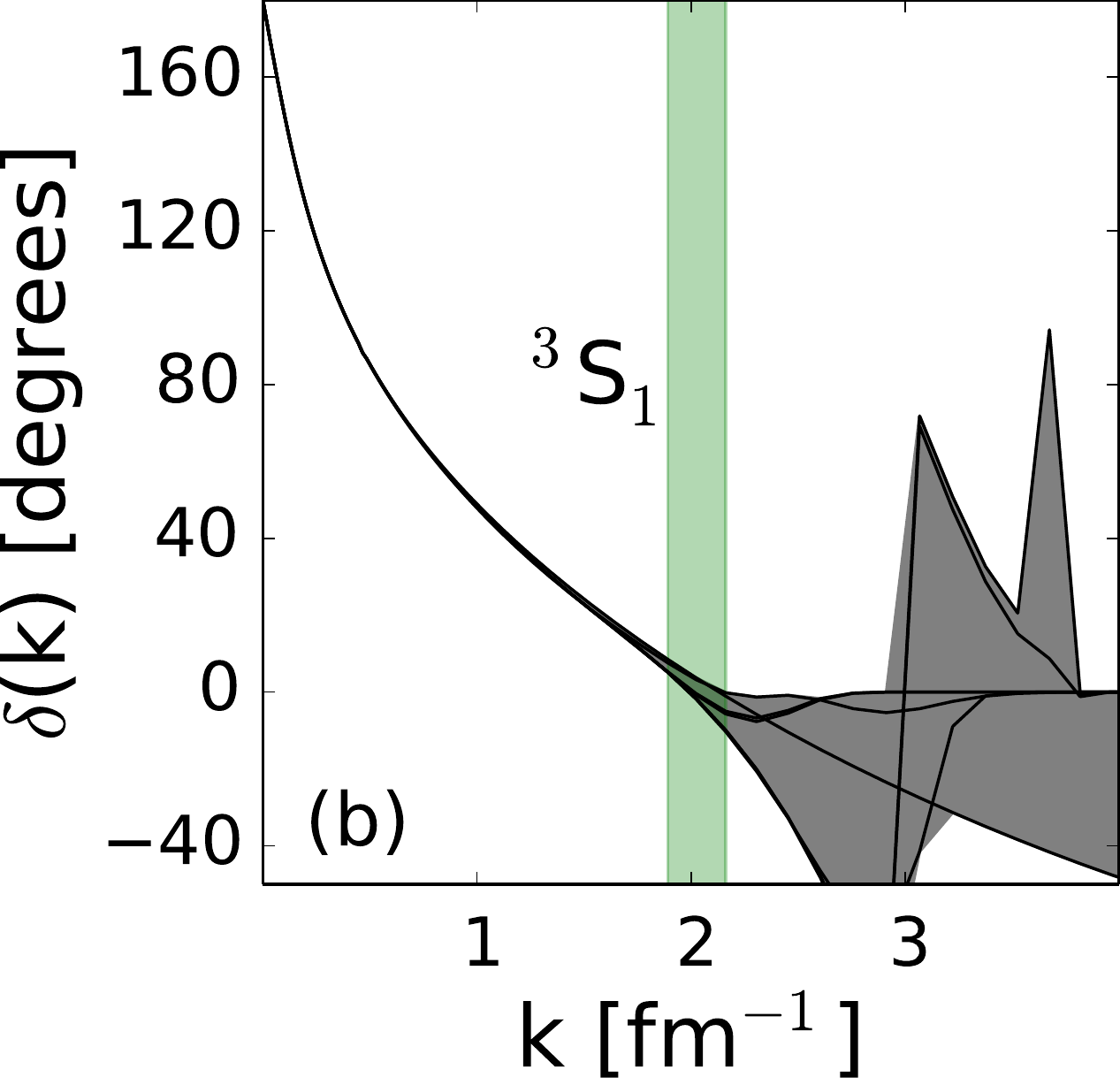}~~~%
	\includegraphics[width=.3 \textwidth]{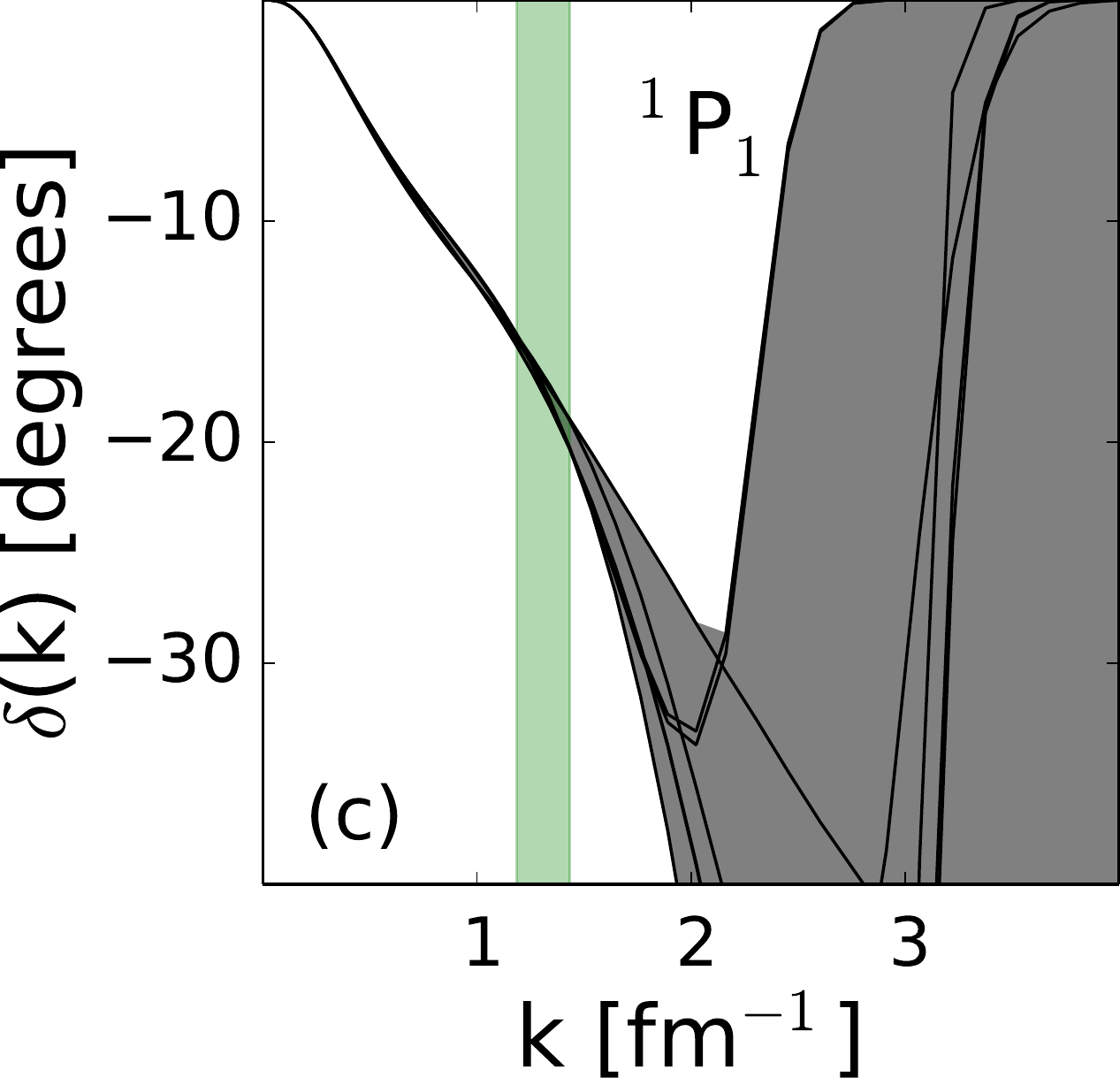}
   \vspace*{-.1in}
	\caption{ Phase shifts of various realistic potentials (see text) in the (a) \oneSzero,
	(b) \threeSone, and (c) \onePone\ partial waves.  The shaded regions show the range between
	the largest and smallest phase shifts. The vertical bands indicate the region
	where phase-shift equivalence between the potentials ends, as defined
	by Eqs.~\eqref{eq:epsilon} and \eqref{eq:vertband}.
	\label{fig:ps_modern}}
\end{figure*}

We have chosen a representative phenomenological potential and a set of \chieft\
potentials to evolve and examine in various partial waves.  
The phenomenological potential is \avpot\ (AV18), which employs basis operators in position representation and fits the coupling constants to elastic scattering data~\cite{av18orig,av18}.  
We use the N$^{3}$LO \chieft\  potential from Entem and Machleidt with a cutoff of 
500 MeV~\cite{EM500} and then  
five N$^{3}$LO \chieft\  potentials with various cutoffs from Epelbaum \textit{et al}.~\cite{EB}.  
These \chieft\  potentials have different regularization and phase shift fitting schemes,
which creates significant differences in the matrix elements of the potentials. 
As can be seen in Fig.~\ref{fig:ps_modern}, all of these potentials reproduce the same low-energy phase shifts.

\begin{figure}
	\includegraphics[width=.3 \textwidth]{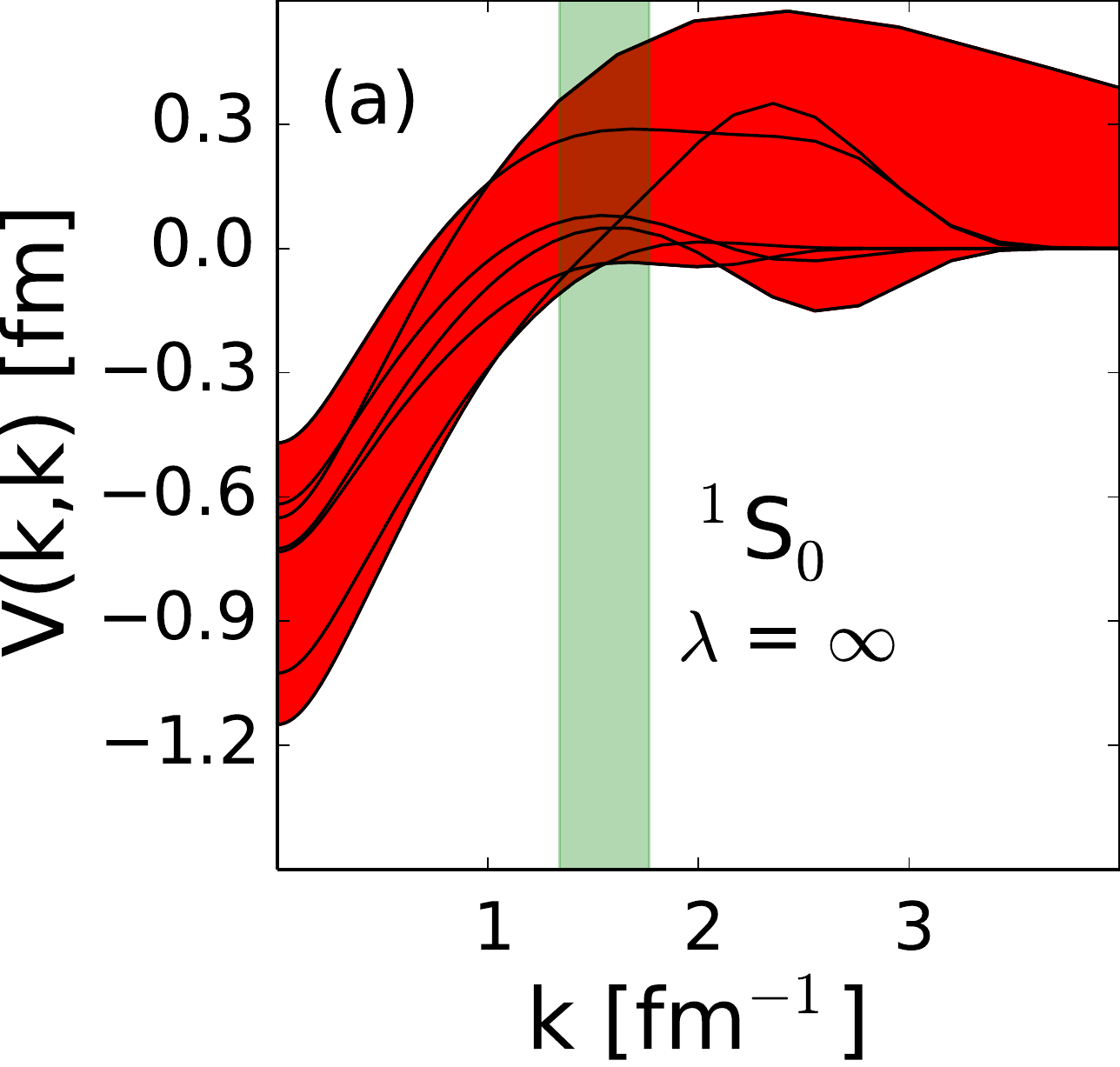}~~~%
	\includegraphics[width=.3 \textwidth]{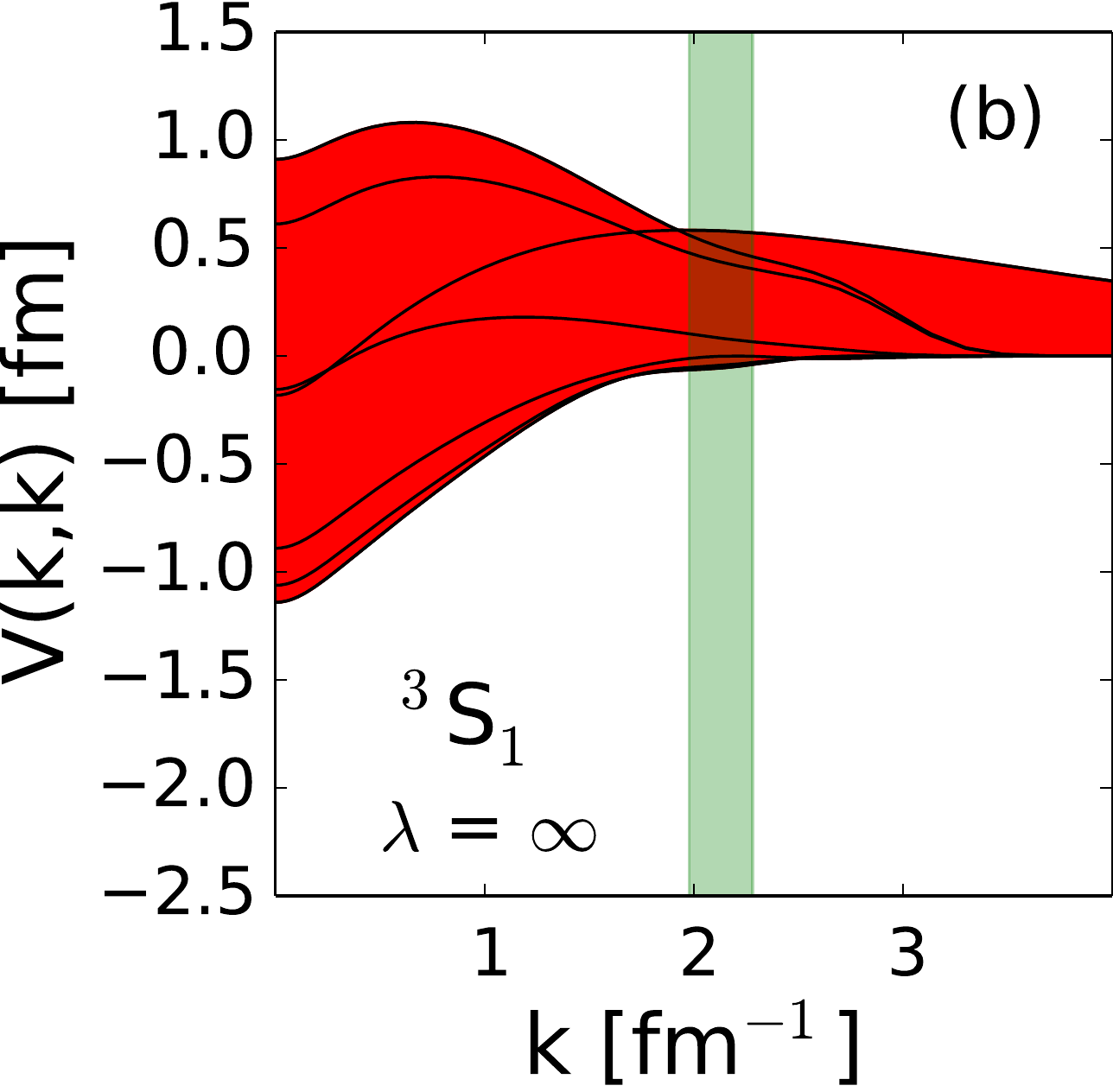}~~~%
	\includegraphics[width=.3 \textwidth]{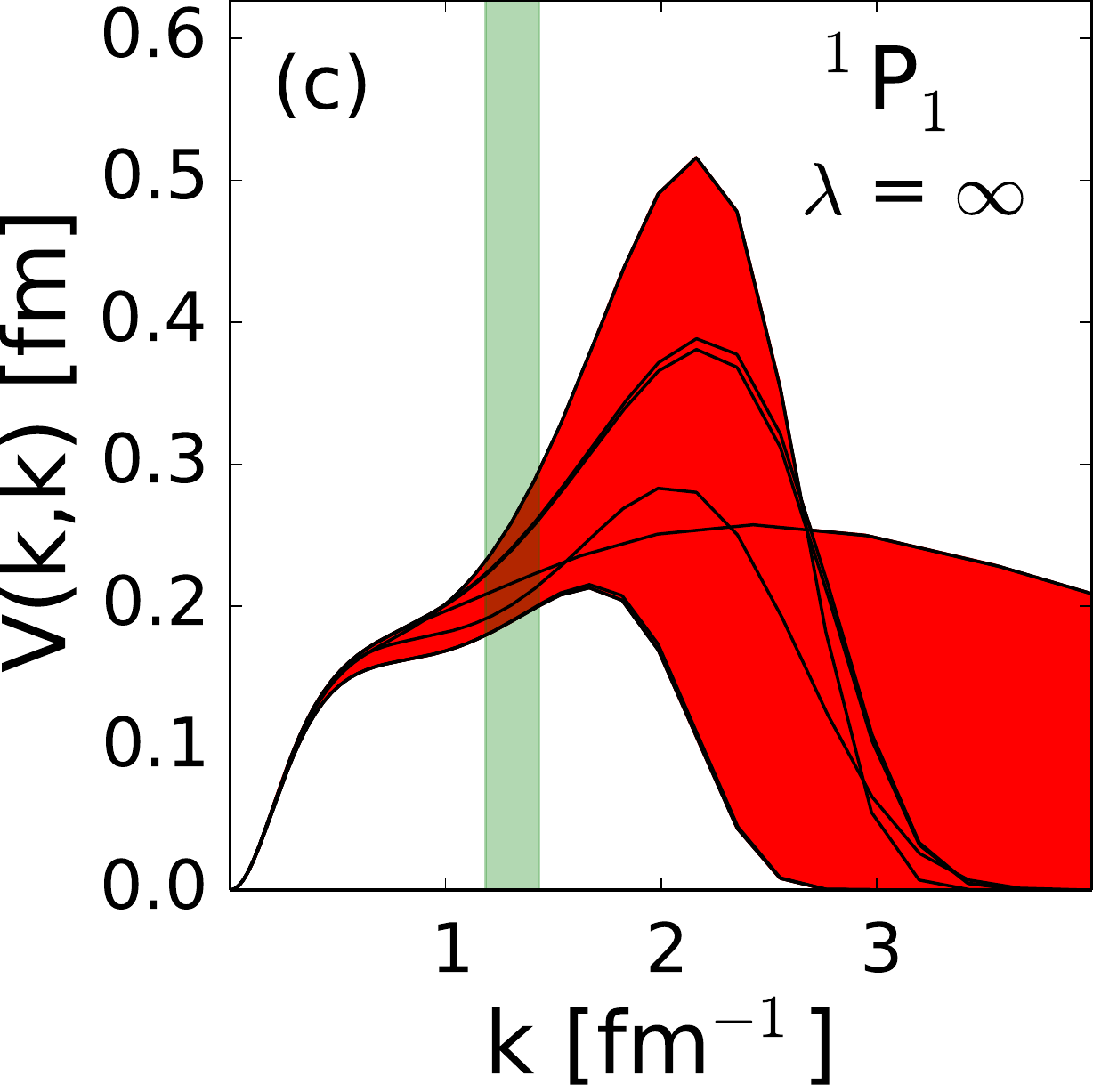}
   \vspace*{-.3in}
	\caption{  Diagonal matrix elements $V(k,k)$ of various unevolved realistic potentials (see text) in the (a) \oneSzero, (b) \threeSone, and (c) \onePone\ partial waves.
	The shaded regions show the range of values and the vertical bands are from
	Fig.~\ref{fig:ps_modern}. 
	\label{fig:diag_modern_inf}}
   \vspace{.1in}
	\includegraphics[width=.3 \textwidth]{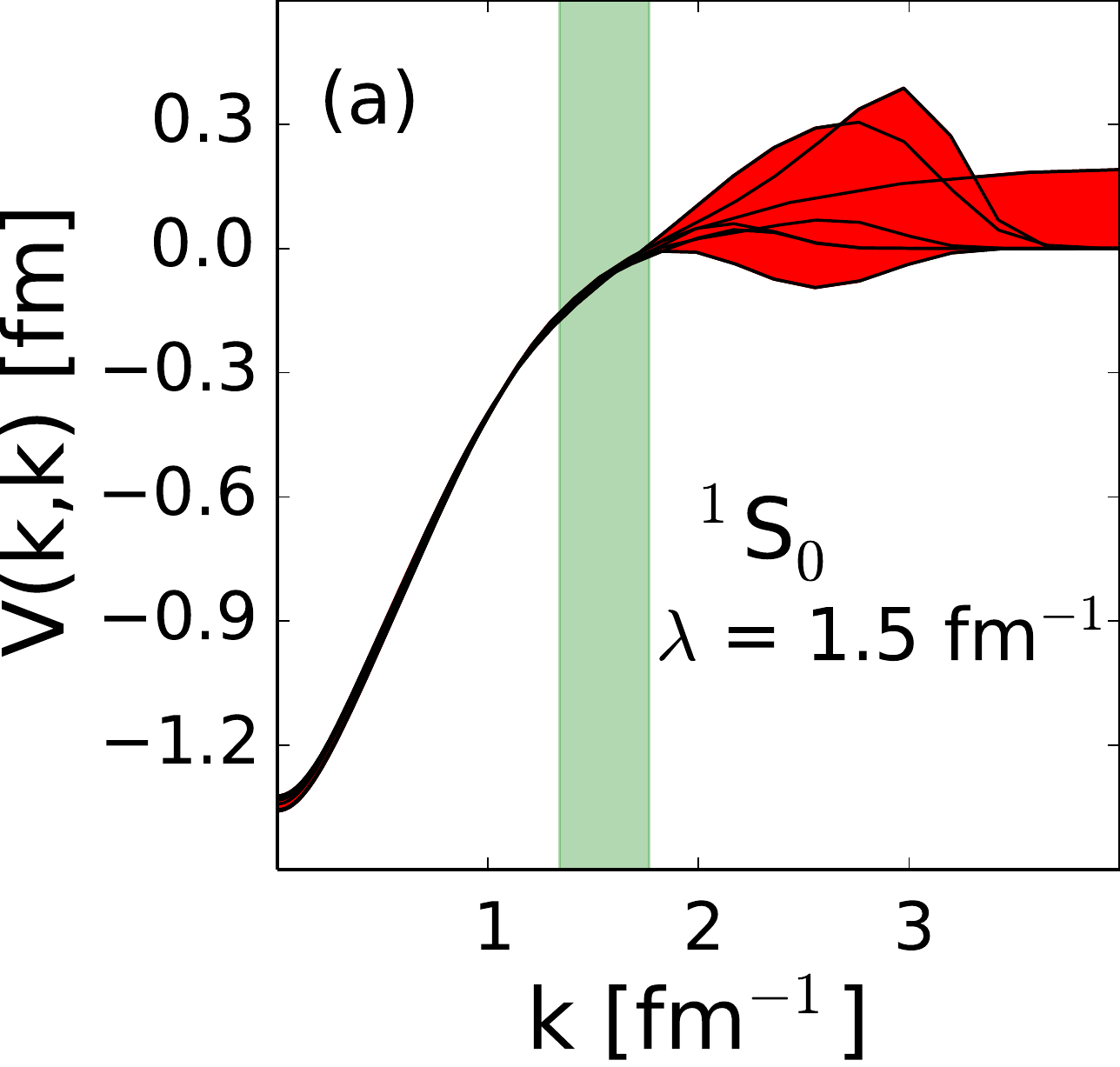}~~~%
	\includegraphics[width=.3 \textwidth]{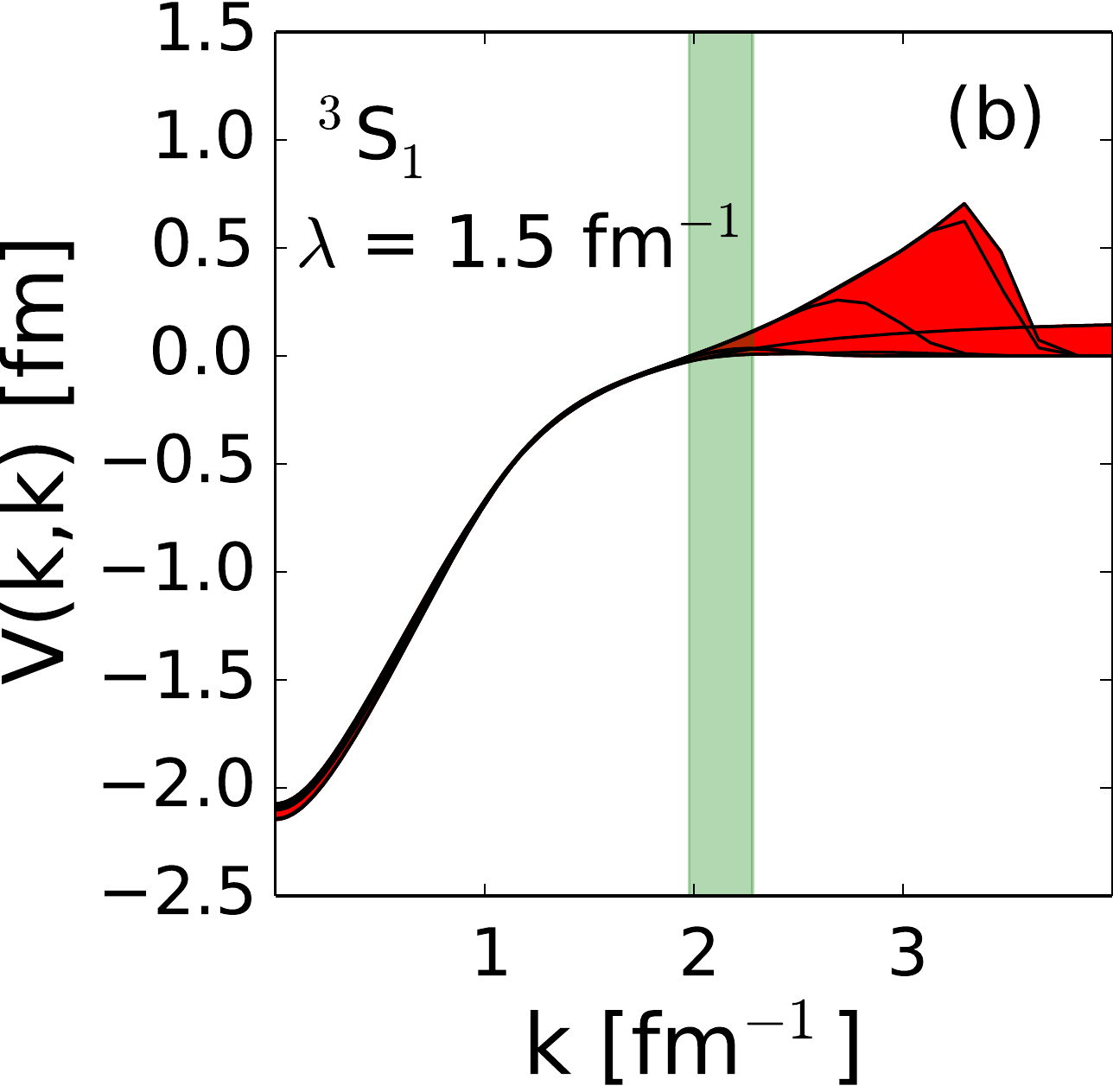}~~~%
	\includegraphics[width=.3 \textwidth]{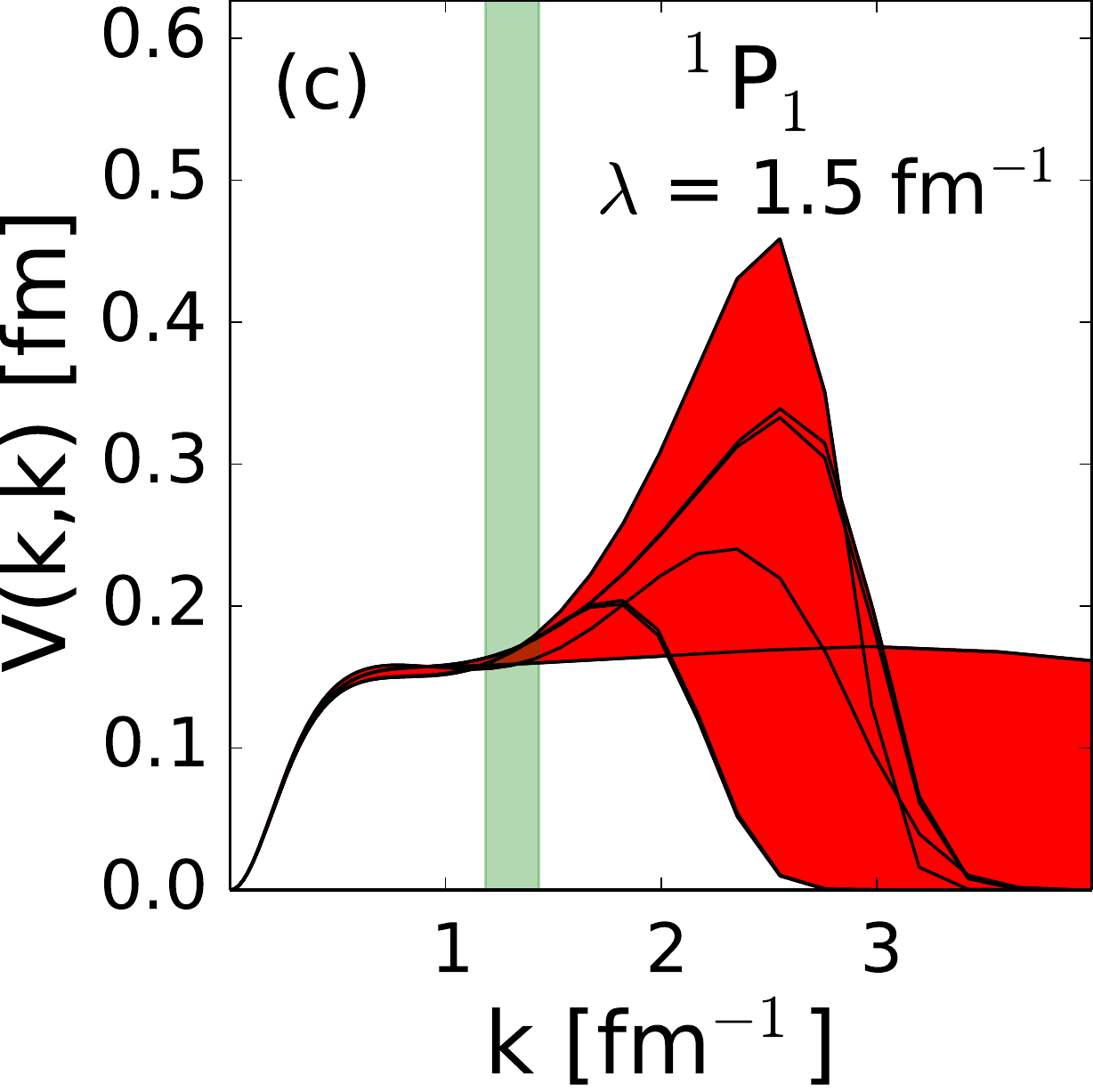}
   \vspace*{-.3in}
	\caption{  Diagonal matrix elements of various realistic potentials
	 in the (a) \oneSzero, (b) \threeSone, and (c) \onePone\  partial waves evolved 
	by the SRG to $\lambda$ = 1.5 fm$^{-1}$. 
	The shaded regions show the range of values and the vertical bands are from
	Fig.~\ref{fig:ps_modern}. 
	\label{fig:diag_modern_15}}
   \vspace*{.1in}
	\includegraphics[width= .9 \textwidth]{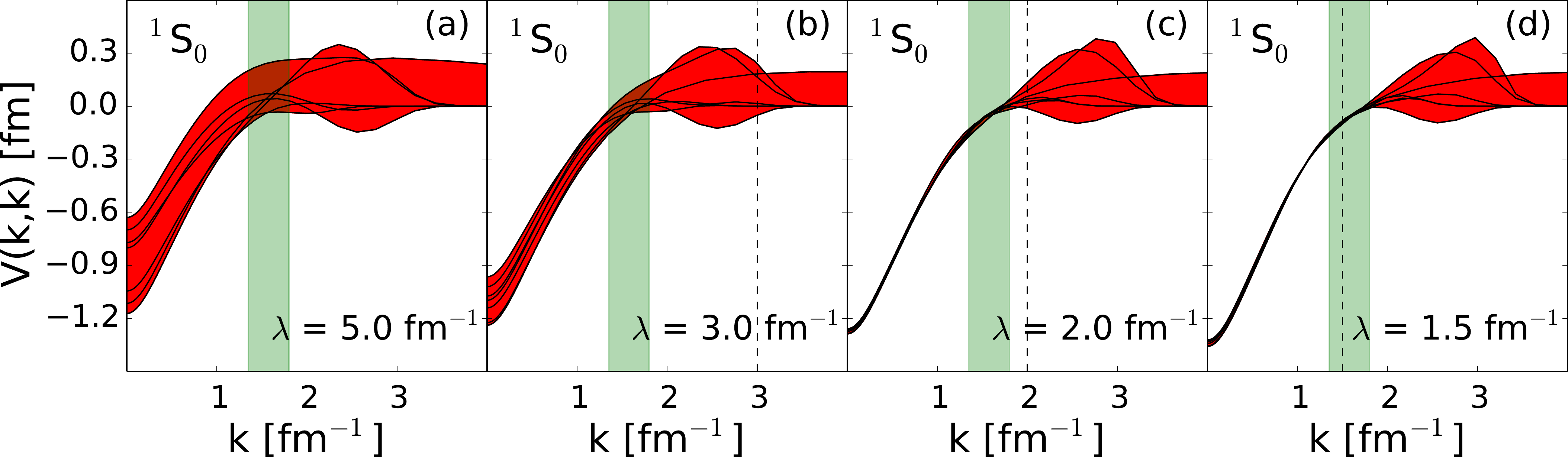}
   \vspace*{-.3in}
	\caption{ Diagonal matrix elements of various realistic potentials in
	the \oneSzero\ partial wave 
	evolved by the SRG to $\lambda$ = (a) 5.0 fm$^{-1}$, (b) 3.0 fm$^{-1}$, (c) 2.0 fm$^{-1}$, (d) 1.5 fm$^{-1}$ (marked by the vertical dashed line). 
	The shaded regions show the range of values and the vertical bands are from
	Fig.~\ref{fig:ps_modern}. 
	\label{fig:diag_modern_multi_cut}}
\end{figure}

From Fig.~\ref{fig:diag_modern_inf} one can see that the diagonals of the 
initial potentials
in momentum representation are quite different (the differences are particularly
evident in lower partial waves, so we focus on those).
In making these comparisons, we do not single out individual potentials but use a
shaded region to highlight the range of matrix element variation.
As advertised, after evolution the matrix elements collapse at low momentum
to a universal dependence on momentum (the result at fixed $\lambda = 1.5\infm$
is shown in Fig.~\ref{fig:diag_modern_15}).  
This feature is not restricted to the diagonal elements; low-energy off-diagonal matrix
elements of the potentials also evolve to universal values (see Fig.~\ref{fig:dsuniv} below).  
At higher momentum, the potential matrix elements deviate.

\begin{figure}
	\subfigure{\includegraphics[width=.3 \textwidth]{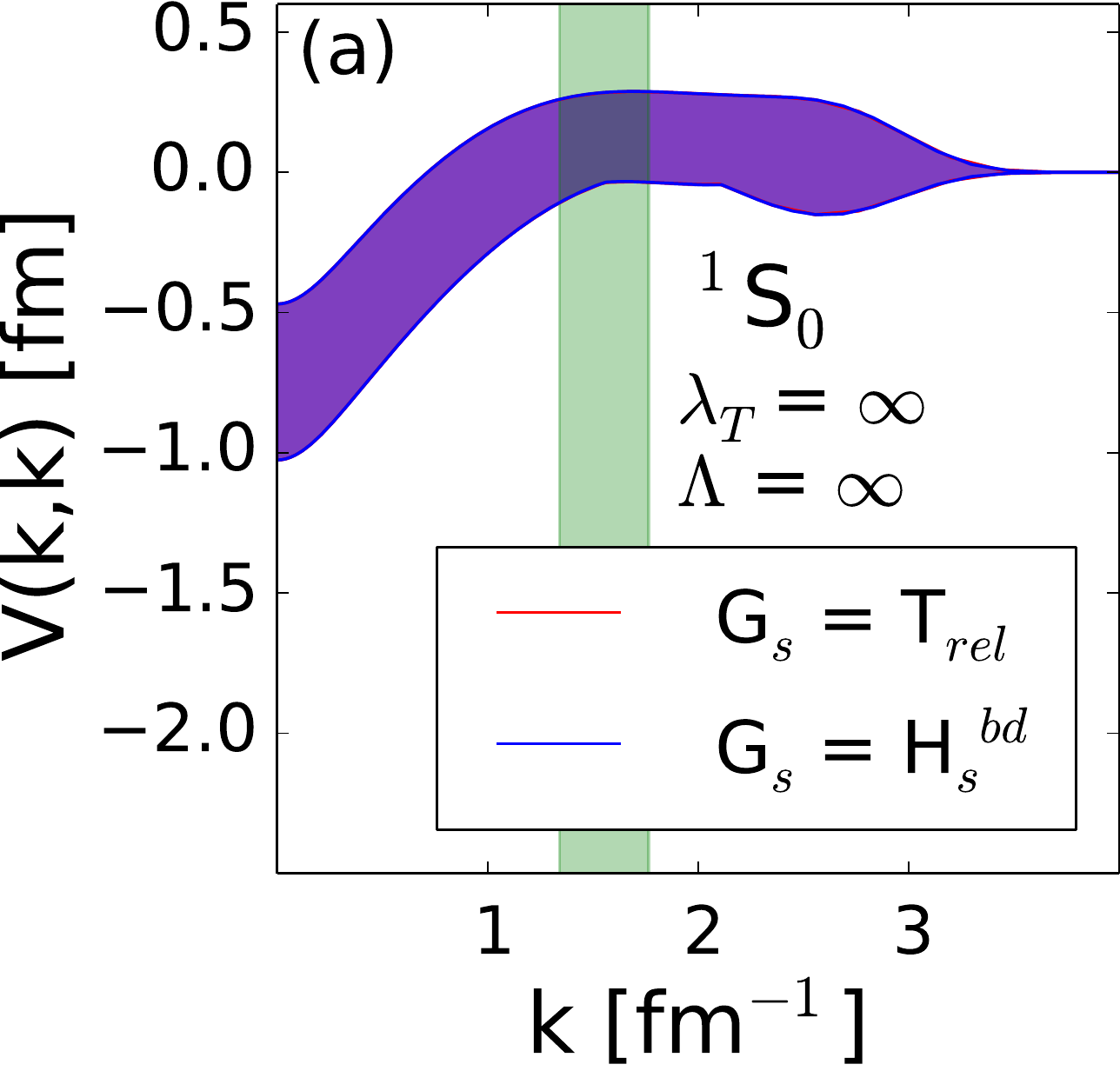}}
	\subfigure{\includegraphics[width=.3 \textwidth]{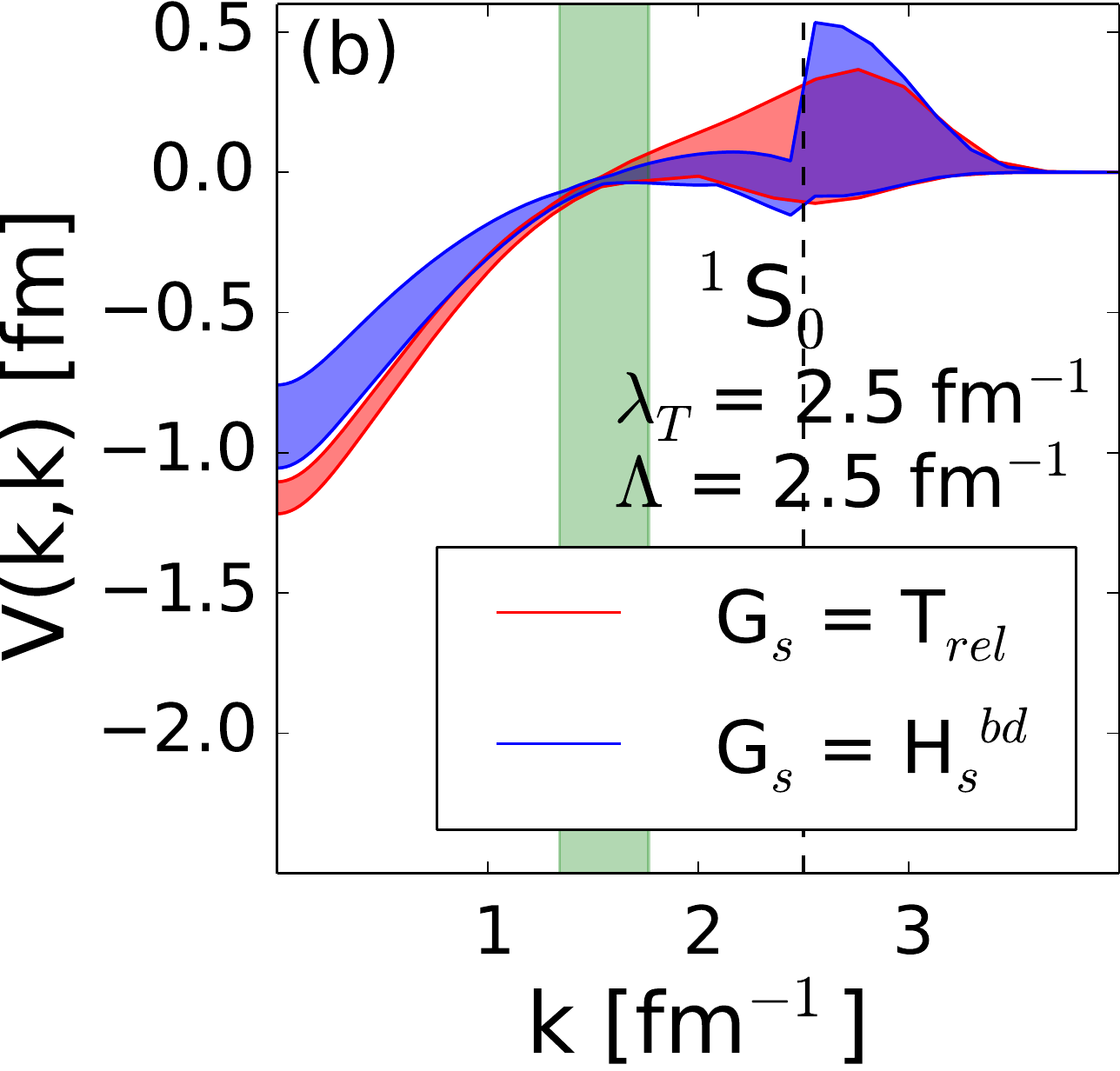}}
	\subfigure{\includegraphics[width=.3 \textwidth]{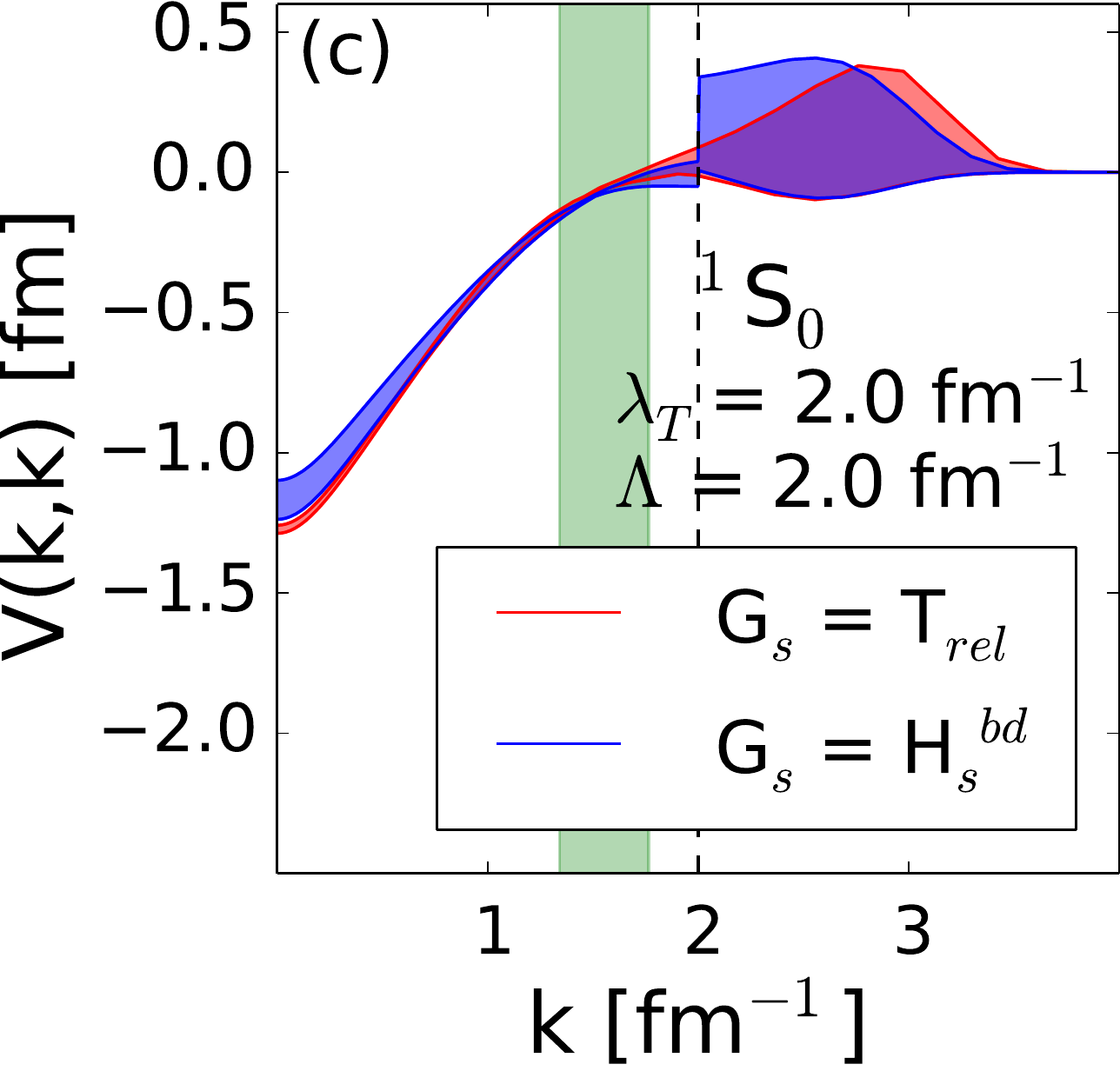}}
%
	\subfigure{\includegraphics[width=.3 \textwidth]{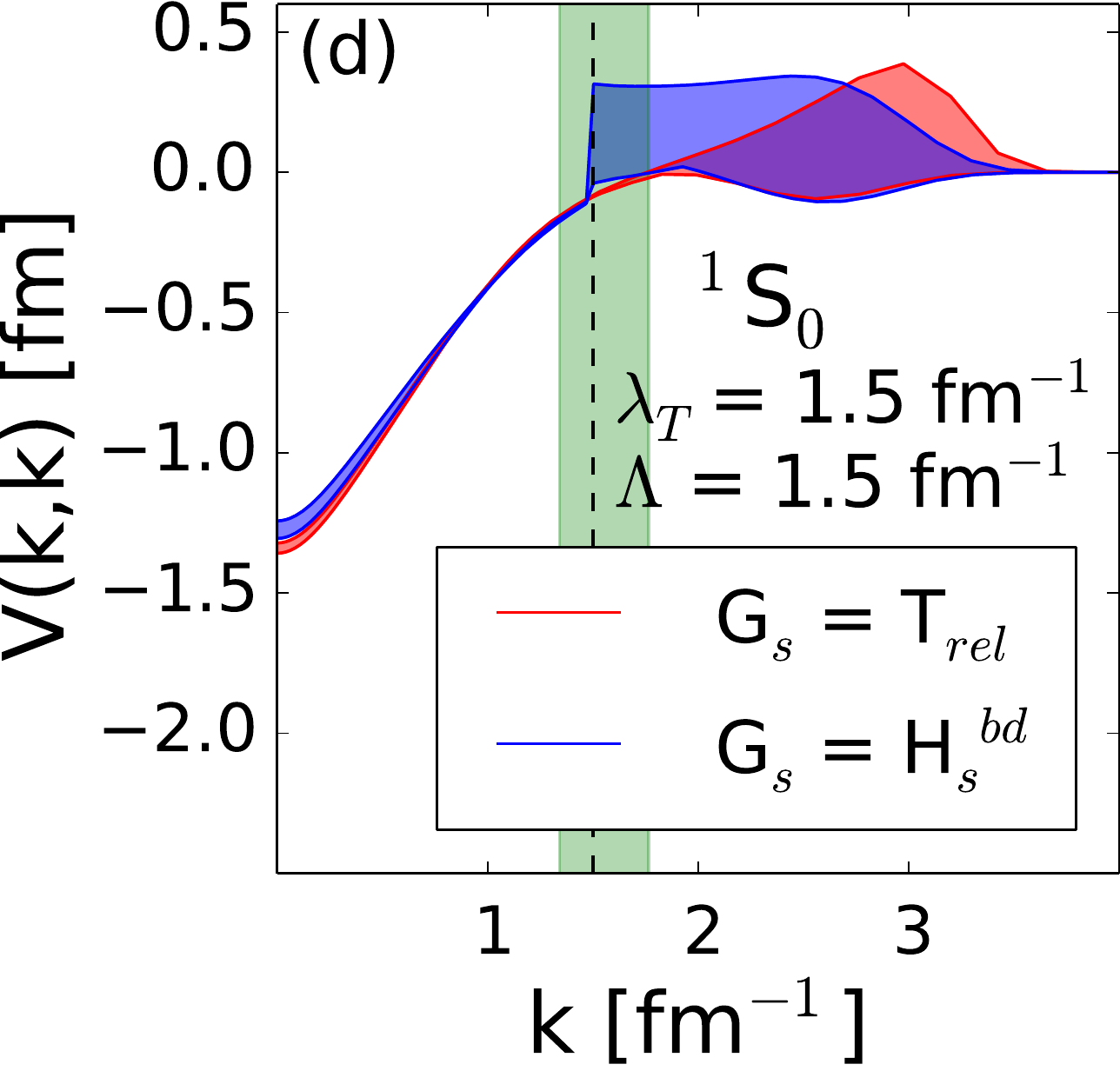}}
	\subfigure{\includegraphics[width=.3 \textwidth]{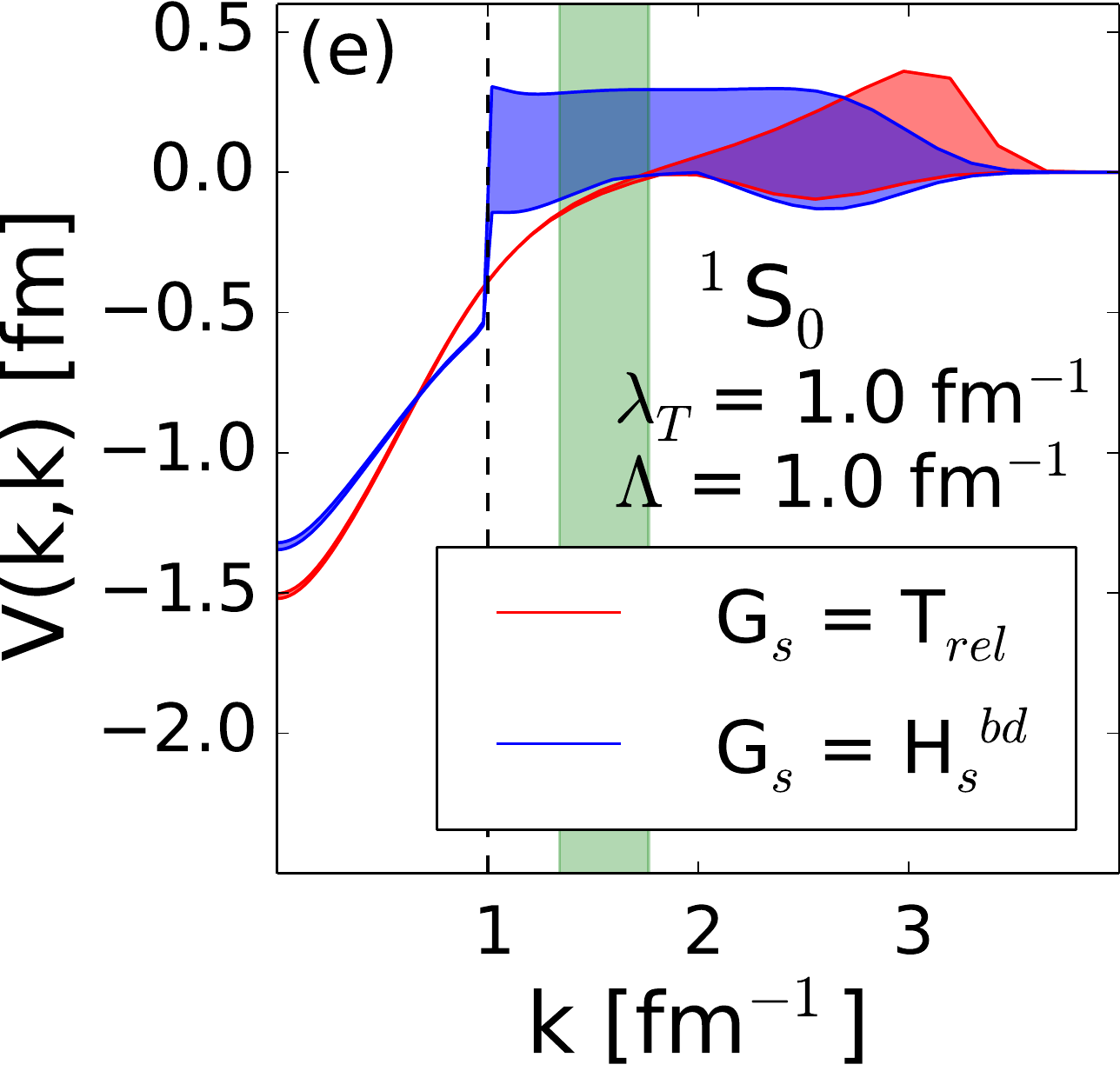}}
	\subfigure{\includegraphics[width=.3 \textwidth]{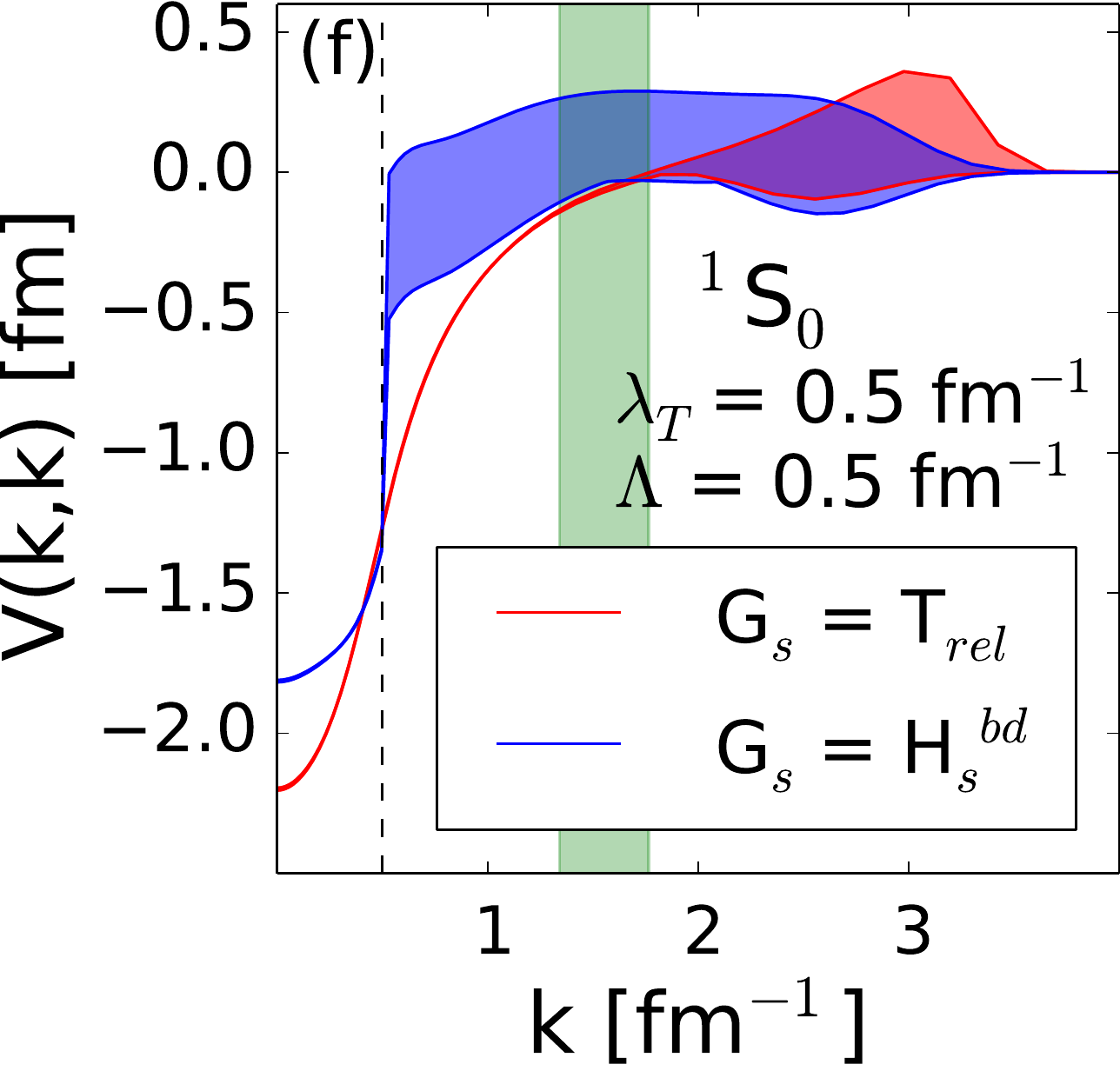}}
	\caption{ The spread of diagonal matrix elements of various \chieft\ potentials (see text) 
	in the \oneSzero\ 
	partial wave are shown as shaded regions for the unevolved potential and then after
	evolution to $\lambda_{T}$ = 2.5, 2.0, 1.5, 1.0, and 0.5 fm$^{-1}$ with the $T$ generator 
	(red or light gray).
	These are compared to the spread of the corresponding matrix elements for the $H_{s}^{bd}$ generator
	with $\Lambda$ = 2.5, 2.0, 1.5, 1.0, and 0.5 fm$^{-1}$fm$^{-1}$, all 
	evolved to $\lambda$ = 0.5 fm$^{-1}$ (blue or medium gray).
	The vertical bands are from Fig.~\ref{fig:ps_modern} and the vertical dashed lines mark
	$\lambda_{T}$ or $\Lambda$.  
	\label{fig:blockdiag}}
\end{figure}

Following Ref.~\cite{vlowkuniv}, we compare phase shift and matrix element deviations
to identify the correlations between phase-shift equivalence and matrix element
universality.
In Fig.~\ref{fig:ps_modern}, we have identified vertical bands within which the phase-shift equivalence among the various potentials ends and significant deviation begins.
While identifying an exact point marking this deviation will be somewhat arbitrary,
we can roughly choose a normalized width description that is consistent with 
visual assessments of the phase shift plots. 
In particular, for each partial wave, the vertical band represents the region characterized by:
\begin{equation}
   0.03 < \epsilon(k) < 0.1 \;,
   \label{eq:epsilon}
\end{equation}
where
\begin{eqnarray}
	\epsilon(k) &\equiv& \frac{\delta_{\rm high}(k)-\delta_{\rm low}(k)}{\Delta} 
	  \;. 
   \label{eq:vertband}
\end{eqnarray} 
The numerator is the range of phase shifts at a fixed $k$ while $\Delta$ is the range of phase 
shifts for the entire universality region. 
Our studies imply that the precise definition of $\epsilon$ is not important; as long as it consistently identifies
the regions where phase-shift equivalence ends it can be used to consistently compare to the regions where 
the universality of matrix elements end.

Comparing Figs.~\ref{fig:diag_modern_inf} and \ref{fig:diag_modern_15}, we see that
while diagonal matrix elements of the initial potentials differ significantly in the region
where phase-shift equivalence ends, this same region corresponds to where the matrix
elements have collapsed to universal values by $\lambda = 1.5\infm$.  
This suggests the hypothesis that \emph{a prerequisite for 
matrix element universality is phase-shift equivalence}.  Namely, if there are 
\emph{local} regions in energy
in which potentials are not phase-shift equivalent, then there is no universality in those regions
(this is tested further in Section~\ref{sec:ISSP}).  
Examining the diagonals of the potentials more closely, we observe that for the \oneSzero\
and \threeSone\ channels, the lowest matrix elements are not exactly the same.  
This may be a consequence of not evolving $\lambda$ further.
From the $T$ generator curves in Fig.~\ref{fig:blockdiag}, we can see that the slight width of the band 
decreases as we evolve chiral potentials to $\lambda = 0.5 \infm$.
Also, as we will see below, differences in the binding energy of the deuteron play 
an important role in the low-energy matrix elements of the \threeSone\ potential.

How low must $\lambda$ be before we see universality? 
Figure~\ref{fig:diag_modern_multi_cut} shows the diagonals of the \oneSzero\ potential evolved
to four different $\lambda$ values.  
The vertical bands correspond to the same region where phase-shift equivalence ends for the
\oneSzero\ channel as in Fig.~\ref{fig:ps_modern}, while the vertical dashed line shows the 
value of $\lambda$.  
We see in this partial wave (and in others not shown as well) that universality in the matrix
elements does not occur until $\lambda$ approaches the vertical band.
A natural hypothesis is that the matrix elements will not fully collapse to universal form 
until $\lambda$ reaches the region of phase-shift equivalence.  
There may be an intrinsic low-energy scale common to each of these
potentials that determines at what $\lambda$ universality in potential matrix elements will
appear.  A possibility is that this scale is a consequence of explicit treatment of pion physics in each of the modern realistic potentials.  
To test the latter explanation, a potential with phase-shift equivalence at much higher momenta 
and no explicit pion physics is required, which we consider in the next section.

As described earlier, the block-diagonalizing generator H$_{s}^{bd}$ will drive the potential
matrix elements to a different universal form than $T$.  
This is illustrated in Fig.~\ref{fig:blockdiag} with a set of \chieft\ potentials in
the \oneSzero\ channel.
When evolved to $\lambda \leq 2\infm$ with the $T$ generator, the universal form of diagonal potential matrix elements emerges over the full region of phase-shift equivalence.  
For the block diagonal generator with $\Lambda \leq 2\infm$, however, only diagonal matrix elements below $\Lambda$ become universal and with a different
flow than the matrix elements evolved with $T$.  
The universality is only up to $\Lambda$ because this SRG only decouples one block from
the other, so matrix elements at momenta above $\Lambda$ still couple to matrix elements 
in phase-shift-inequivalent regions and therefore do not collapse to a universal form. 
(Note that in the $\vlowk$ RG, the higher block is set to zero.) 
We will discuss only $G_s = T$ in the rest of this study but emphasize that the ideas about universality apply to both generators, although only in the low-momentum block for the 
H$_{s}^{bd}$ SRG.

The region of phase-shift equivalence for the realistic potentials is limited by the energies 
to which they can be fit to elastic scattering phase shifts.
The potentials are only fit up to the inelastic threshold, about 350 MeV, where contributions from pion production become non-negligible.  
Because of this, if we wish to investigate different regions of universality, we must use a method that can \lq{}fit\rq{} the phase shifts in a controlled range of energies.  
One of the simplest approaches is solving the inverse scattering problem with a separable potential, which we consider in the next section.


\section{Separable Inverse Scattering Potential} \label{sec:ISSP}

Instead of fitting coupling constants for predetermined operators to the phase shifts, 
an inverse scattering procedure constructs a potential directly from the phase shifts.
Separability is just a constraint to define a unique potential, chosen here due to its 
simplicity.
For instance, when solving the Lippmann-Schwinger equation, a separable potential 
reduces the problem of solving an integral equation to simply evaluating an integral.  
The three-body Faddeev equations also simplify for a separable potential, as one of the integrals 
over internal momenta becomes trivial. 
A key feature of the ISSP for this study is that the potential is entirely created from 
the phase shifts and binding energy of the deuteron; no \emph{explicit} pion exchange 
or other physics is imposed.  
This allows us to determine whether or not universality requires extra physics, 
such as explicit long-range pion terms or other phenomenological considerations.
We start with a brief summary of the inverse scattering separable potential for two
nucleons. 

\subsection{Formalism}

The form of a rank-$n$ separable potential is:
\begin{equation}
		V = \sum_{i,j=0}^{n-1} \ket{\nu_{i}} \Lambda_{ij} \bra{\nu_{j}} \;.
\end{equation}
For our purposes a rank-1 separable potential will be sufficient, but future studies
may benefit from a higher-rank potential.
A rank-1 potential in momentum representation takes the form:
	\begin{equation}
		V(k,k\rq{})=\sigma \nu(k) \nu (k\rq{}) \;,
		\label{eq:separable}
	\end{equation}
where $\sigma$ is simply $\pm 1$.	
Details of the rank-1 separable inverse scattering problem are well 
documented~\cite{BrownJackson,Kwong}; here, we simply state the main results, 
some limitations, and how to work around the limitations (see appendix~\ref{app:app_issp} for derivation).   
The solution to the separable inverse scattering problem 
is~\cite{BrownJackson}:
\begin{eqnarray}
		\sigma \nu^{2}(k) &=& - \frac{k^{2} - k_{b}^{2}}{k^{2}} \;
		   \frac{\sin(\delta(k))}{k} \; e^{-\Delta(k)} \;,\\
		\Delta(k) &=& \frac{1}{\pi} P \int_{0}^{\infty} \frac{dk\rq{} \delta(k\rq{})}{k\rq{} - k} \;,\\
		E_{b} &=& \frac{\hbar^{2} k_{b}^{2}}{2 m}  \;,
\end{eqnarray}
where $k_{b}$ is zero if there is no bound state and equal to the binding momentum 
for a single bound state with binding energy $E_b$
(for a rank-1 separable potential there can be at most one bound state, which is the case for the two-nucleon problem in nuclear physics).
	
Once $\nu(k)$ is determined, the entire potential is known from Eq.~\eqref{eq:separable}.  
The binding energy $E_{b}$ can be tuned independently of the phase shifts.  
A limitation of rank-1 separable potentials is that if the phase shift as a function of 
momentum crosses zero, then so too must the potential, and a rank-1 ISSP 
as defined thus far can never change signs if $\nu$ is real.
This point is clear from
Eq.~\eqref{eqzeroes}, which follows from the Lippmann-Schwinger equation for a 
separable potential (with standing wave boundary conditions):
	\begin{equation}\label{eqzeroes}
		\frac{1}{k} \tan(\delta_{l}(k)) = 
		-\frac{ V_{l}(k,k)}{1+\frac{2}{\pi}\mathcal{P}\int 
		    \frac{\textstyle dp \; p^{2} V_{l}(p,p)}{\textstyle p^{2}-k^{2}}} \;.
	\end{equation}
A zero-crossing in $\delta(k)$ corresponds to a zero-crossing on the right side
of this equation, which can only be achieved by the numerator crossing zero if 
the denominator remains finite.

Because some of the phase shifts for nucleon-nucleon partial waves exhibit zero crossings, 
we need an inverse scattering potential that allows this feature.  
We can still use the same rank-1 formalism, however, if we split the problem into two energy
regimes, above and below the zero crossing~\cite{Kwong}.  Then we can define:
	\begin{eqnarray}
		\delta_{<}(k) &\equiv& \delta(k) \theta(k_{0} - k) \;, \\
		\delta_{>}(k) &\equiv& \delta(k) \theta(k - k_{0}) \;, \\
		V(k,k\rq{}) &=&  V_{<}(k,k\rq{}) + V_{>}(k,k\rq{}) \;,
	\end{eqnarray}
and determine $V_{<}$ and $V_{>}$ separately using the rank-1 formalism 
with $\delta_{<}$ and $\delta_{>}$ as input, respectively.  We have confirmed
numerically that potentials created with this prescription
accurately reproduce the input phase shifts.

\begin{figure*}[htp!]
	{\includegraphics[width=.32\textwidth]{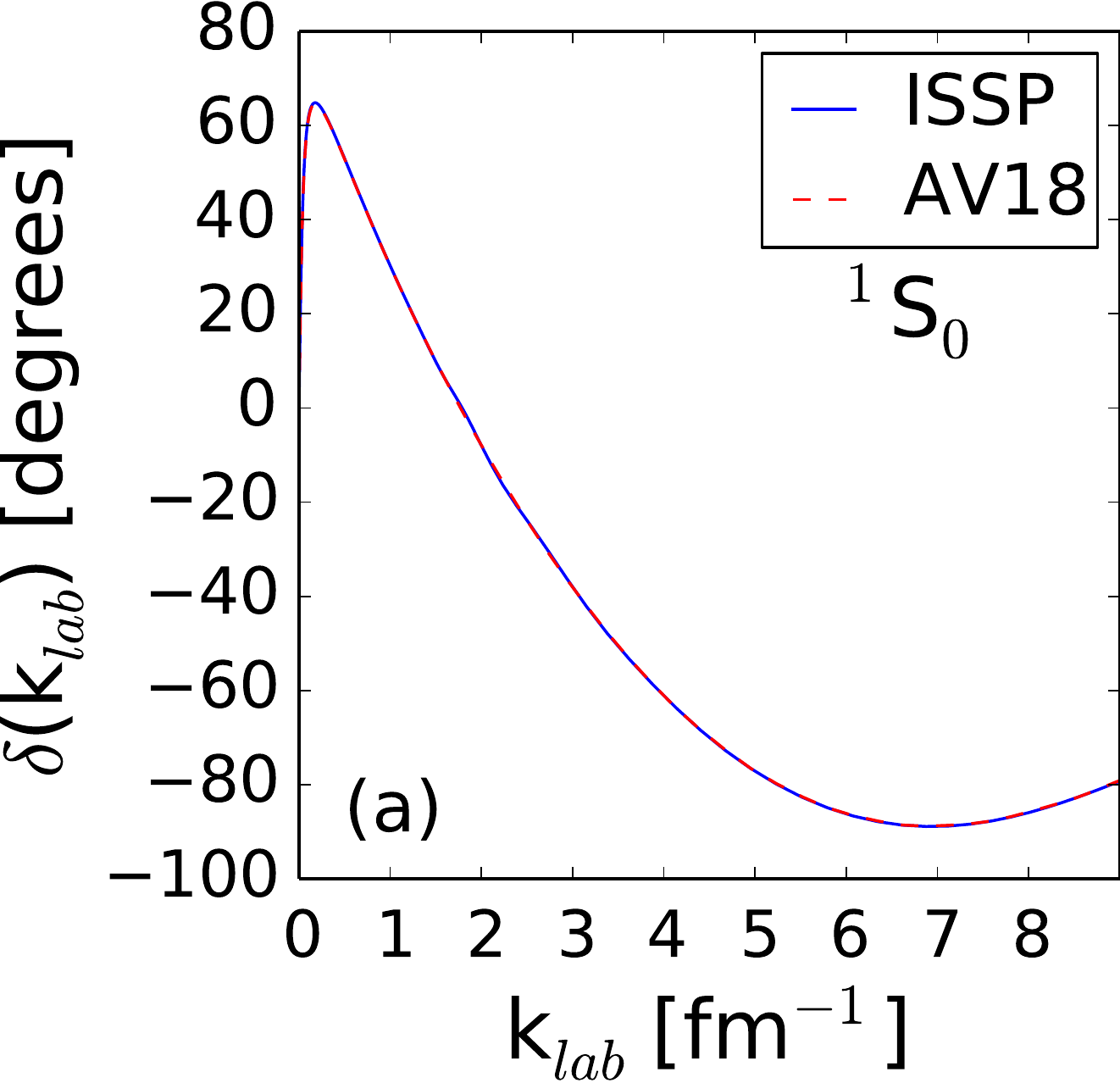}}
	\hfill
	{\includegraphics[width=.32\textwidth]{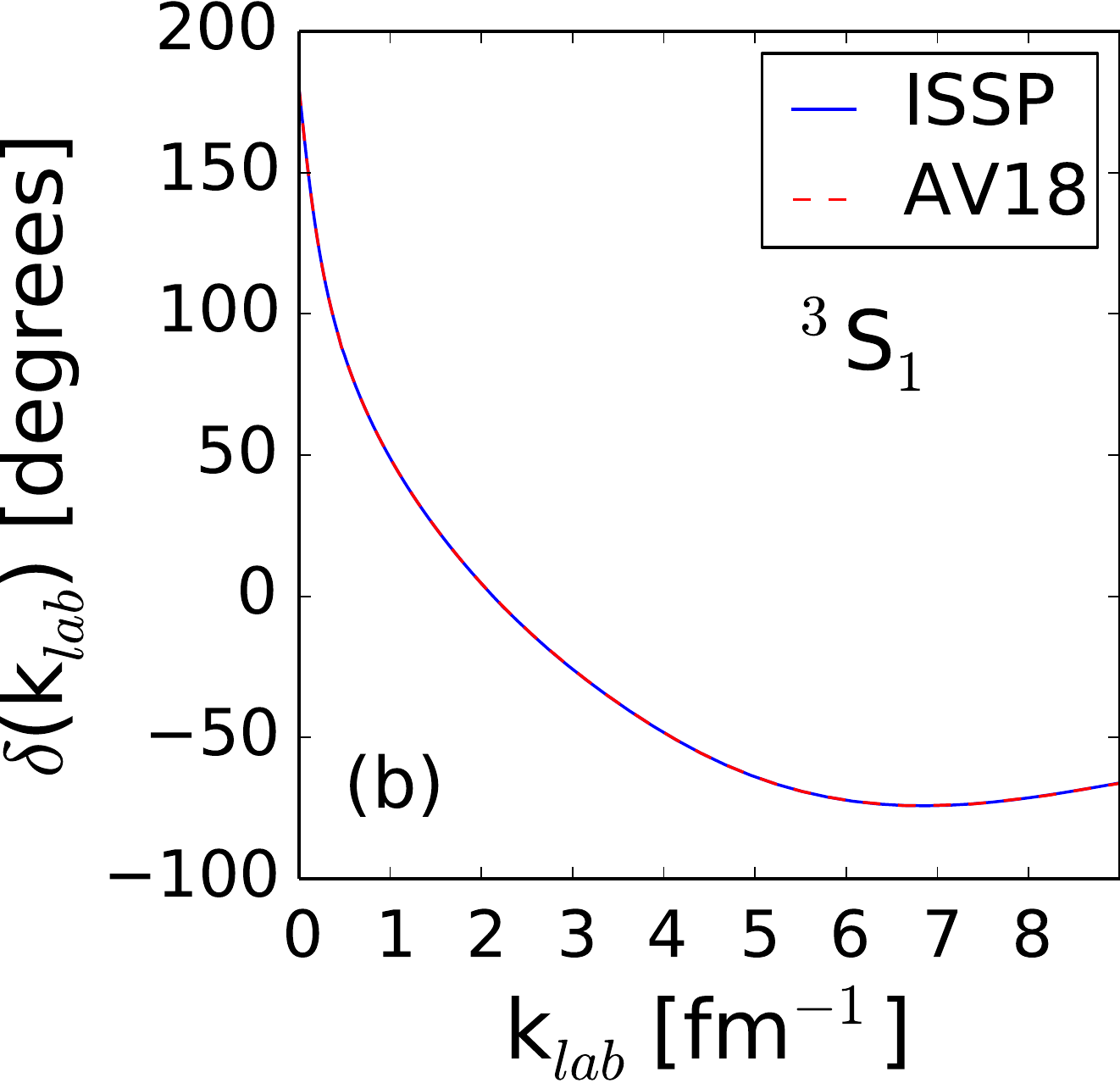}}
	\hfill
	{\includegraphics[width=.31\textwidth]{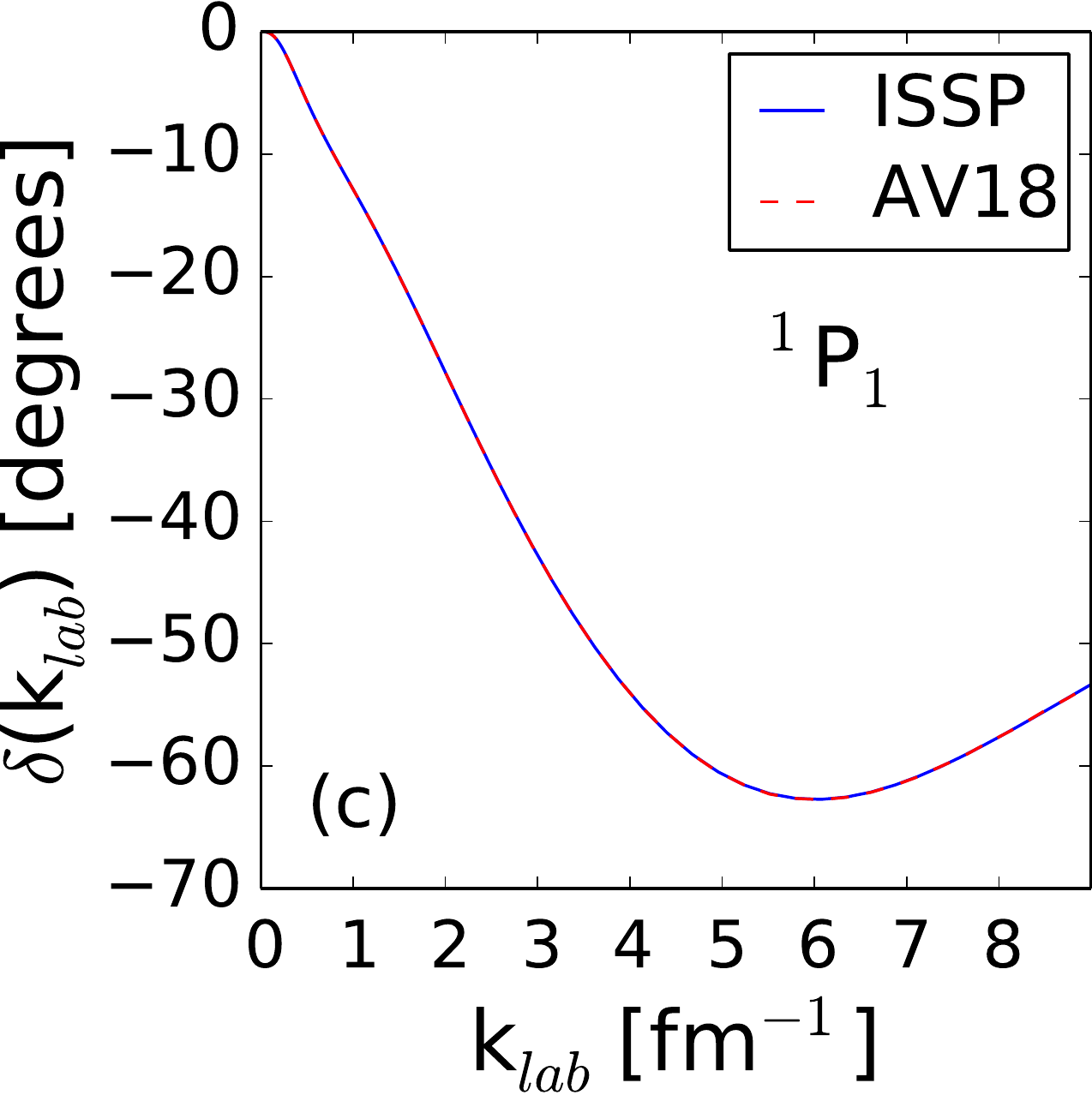}}
	\vspace*{-.1in}
	\caption{ Phase shifts using the AV18 potential and the ISSP up to high lab
	momentum $k_{\rm lab}$ in the (a) \oneSzero, (b) \threeSone, and
	(c) \onePone\ partial waves.
	\label{fig:ps_issp_av18_all}}
\end{figure*}

\begin{figure*}[htp!]
	{\includegraphics[width=.32\textwidth]{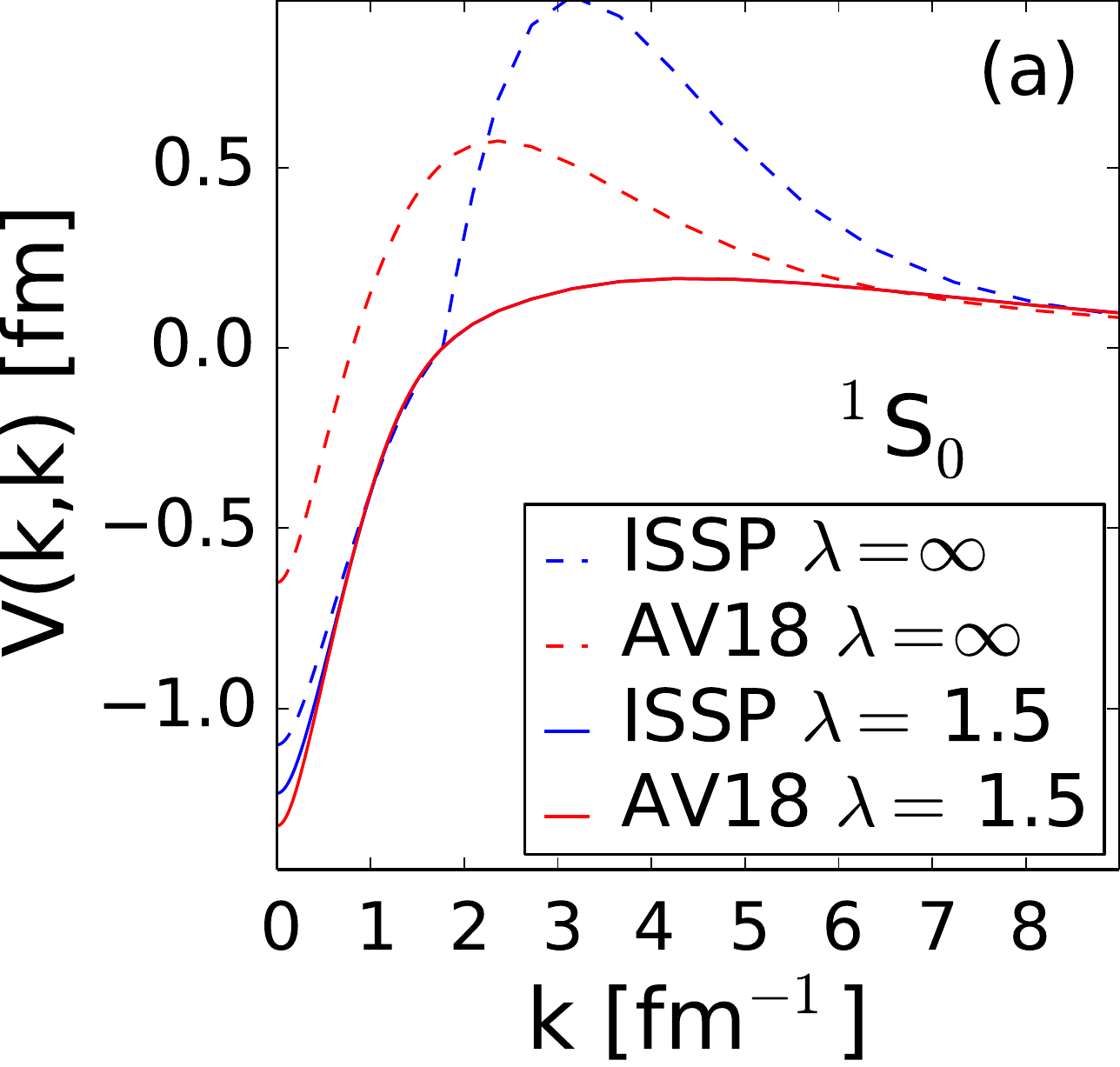}}
	\hfill
	{\includegraphics[width=.32\textwidth]{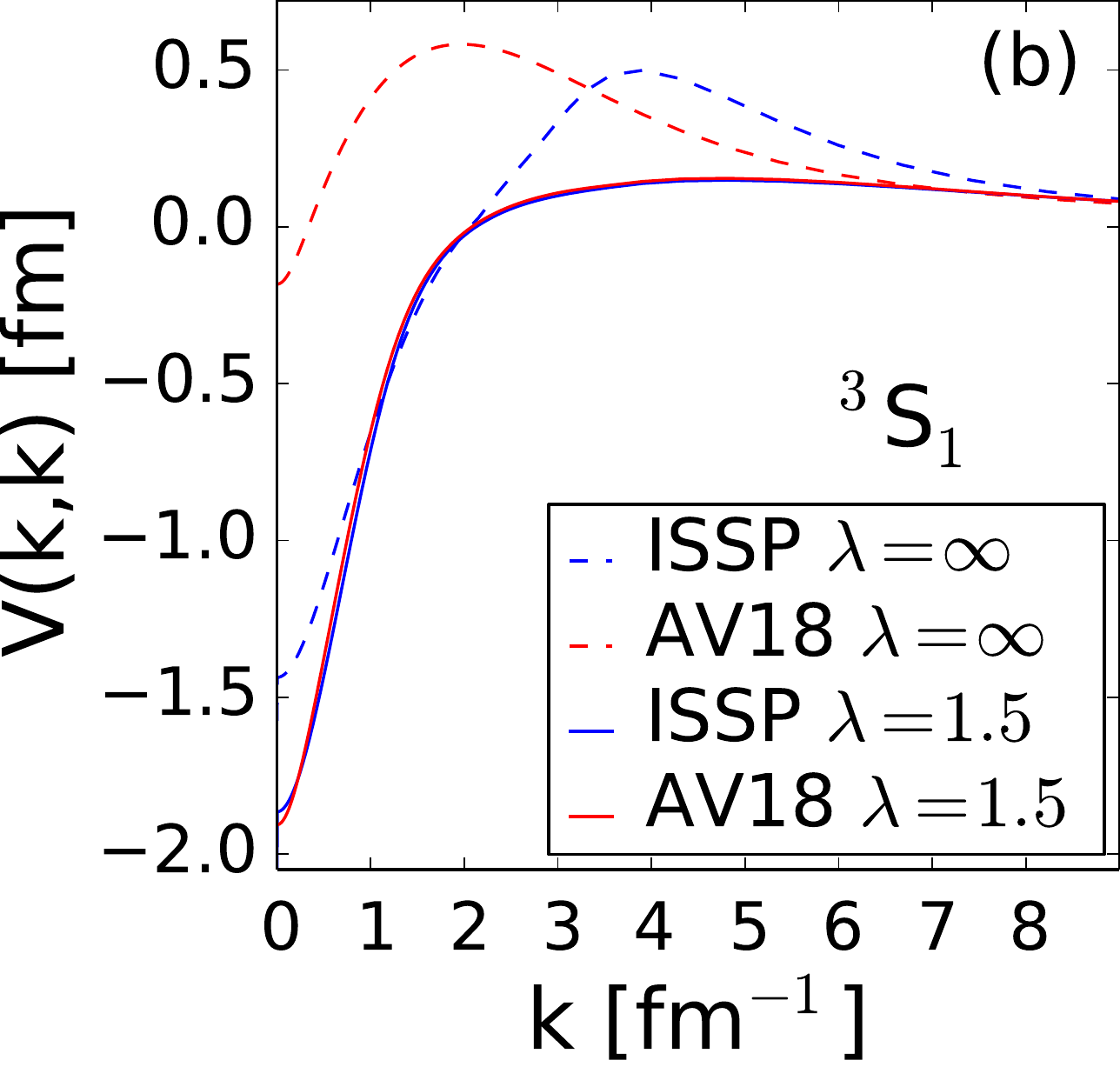}}
	\hfill
	{\includegraphics[width=.32\textwidth]{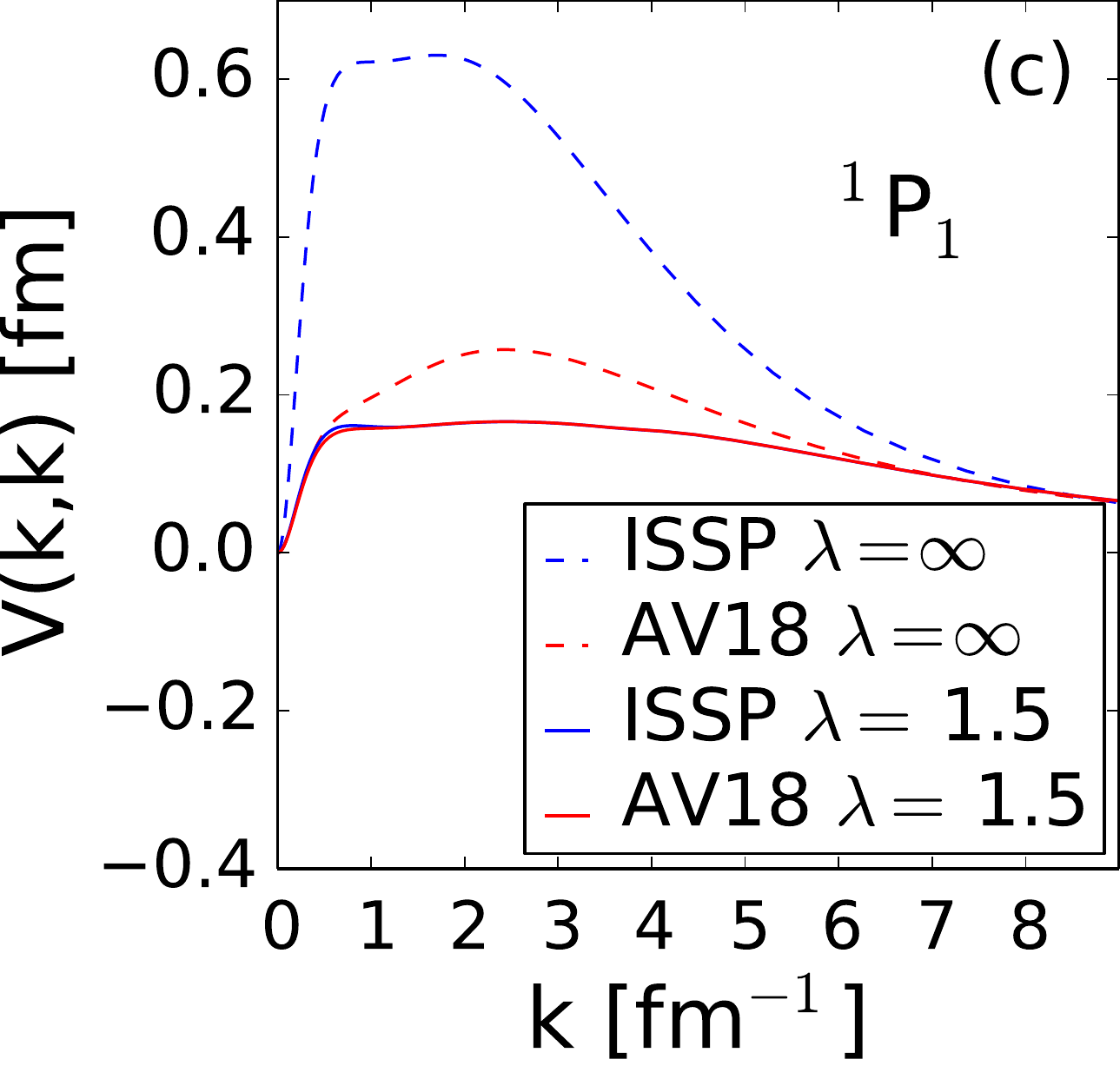}}
	\vspace*{-.1in}
	\caption{ Diagonal matrix elements of the AV18 potential and the ISSP up to high lab
	momentum $k_{\rm lab}$ in the (a) \oneSzero, (b) \threeSone, and
	(c) \onePone\ partial waves.  Cutoff $\lambda$ is in units of fm$^{-1}$.
	\label{fig:diagss_issp_av18_all}}
\end{figure*}



The method described thus far works directly for uncoupled channels, but for 
NN scattering we must also account for
coupled channels, where some further formalism is required.  
For this purpose, we use the Blatt-Beidenharn (BB) convention for phase shifts in the coupled channel for our calculations~\cite{BBps,Kwong}.  (In the plots we employ the more typically used 
Stapp-$\overline{N}$ convention for the phase shifts for visualization~\cite{Nbarps}.)  
The BB convention can be summarized as:
	\begin{eqnarray}
		\mathbf{S}(k) &=& \mathbf{U}^{\dag}(k) \mathbf{\widehat{\Delta}}(k) \mathbf{U}(k) \;,\\
		\mathbf{\widehat{\Delta}}(k)&=& \begin{pmatrix} e^{2 i \delta_{0}(k)} & 0 \\ 
								  0 & e^{2 i \delta_{1}(k)}
			 			     \end{pmatrix} \;,\\
		\mathbf{U}(k)&=& \begin{pmatrix} \cos(\epsilon(k)) & \sin(\epsilon(k)) \\ 
								  -\sin(\epsilon(k)) & \cos(\epsilon(k))
			 			     \end{pmatrix} \;.
	\end{eqnarray}
Here, S(k) is the scattering matrix which relates incoming and outgoing wavefunctions~\cite{qm2landau}, with k the momentum corresponding to
the interaction energy.
Then the inverse scattering potential can be written as:
\begin{equation}
		\mathbf{V}(k,k\rq{}) = 
		\mathbf{U}^{\dag}(k) \mathbf{\hat{V}}(k,k\rq{}) \mathbf{U}(k\rq{}) \;,
\end{equation}
where
\begin{equation}		
		\mathbf{\hat{V}}(k,k\rq{}) =  \begin{pmatrix} \hat{V}_{0}(k,k\rq{}) & 0 \\ 
								                        0 & \hat{V}_{1}(k,k\rq{})
			 			               \end{pmatrix} 
		\;.
\end{equation}
To proceed, one uses the inverse scattering method for uncoupled channels
to find $\hat{V}_{0}(k,k\rq{})$ from $\delta_{0}(k)$ and $\hat{V}_{1}(k,k\rq{})$ 
from $\delta_{1}(k)$.  
The complete potential is then found by a rotation by the mixing parameter, $\epsilon(k)$.  
With this complete separable inverse scattering formalism, we can now create a phase-equivalent potential at all energies in any given partial-wave channel.  

 
\subsection{Universality in separable inverse scattering potentials} \label{sec:universality}

We use phase shifts from \avpot\ to create the phase-equivalent ISSP.  
In Fig.~\ref{fig:ps_issp_av18_all}, we see that the elastic phase shifts are 
quantitatively reproduced well above the inelastic threshold.  
We choose \avpot\ specifically because it has phase shifts that extend to this high
energy, but any realistic potential could be used for starting phase shifts.  
(Note: for simplicity we treat the problem non-relativistically with only elastic scattering 
because we are interested in testing universality and low-energy effects, not to
have a realistic description of high-energy physics.)  
The ISSP\rq{}s from chiral potentials exhibit similar behavior, except that the internal 
cutoffs drive matrix elements and phase shifts to zero at high energies, which is
less useful for the present investigation.  
The accuracy of the ISSP in reproducing phase shifts can be further increased 
simply by using more grid points and increasing the maximum momentum if the phase 
shifts are nonzero above this momentum.  

Figure~\ref{fig:diagss_issp_av18_all} shows the diagonal matrix elements of \avpot\ 
and the ISSP for three different partial waves before and after SRG evolution.  
We observe that after SRG evolution to $\lambda=1.5$ fm$^{-1}$, universality in the 
diagonal matrix elements also extends to the full range of energies.  
In fact, the only discernible difference in the evolved potential diagonals is below 
the SRG cutoff.  
Above $\lambda$ the matrix elements in the region shown are completely collapsed to universal values.

\begin{figure}[th!]
	\begin{center}
	\includegraphics[width=3.2in]{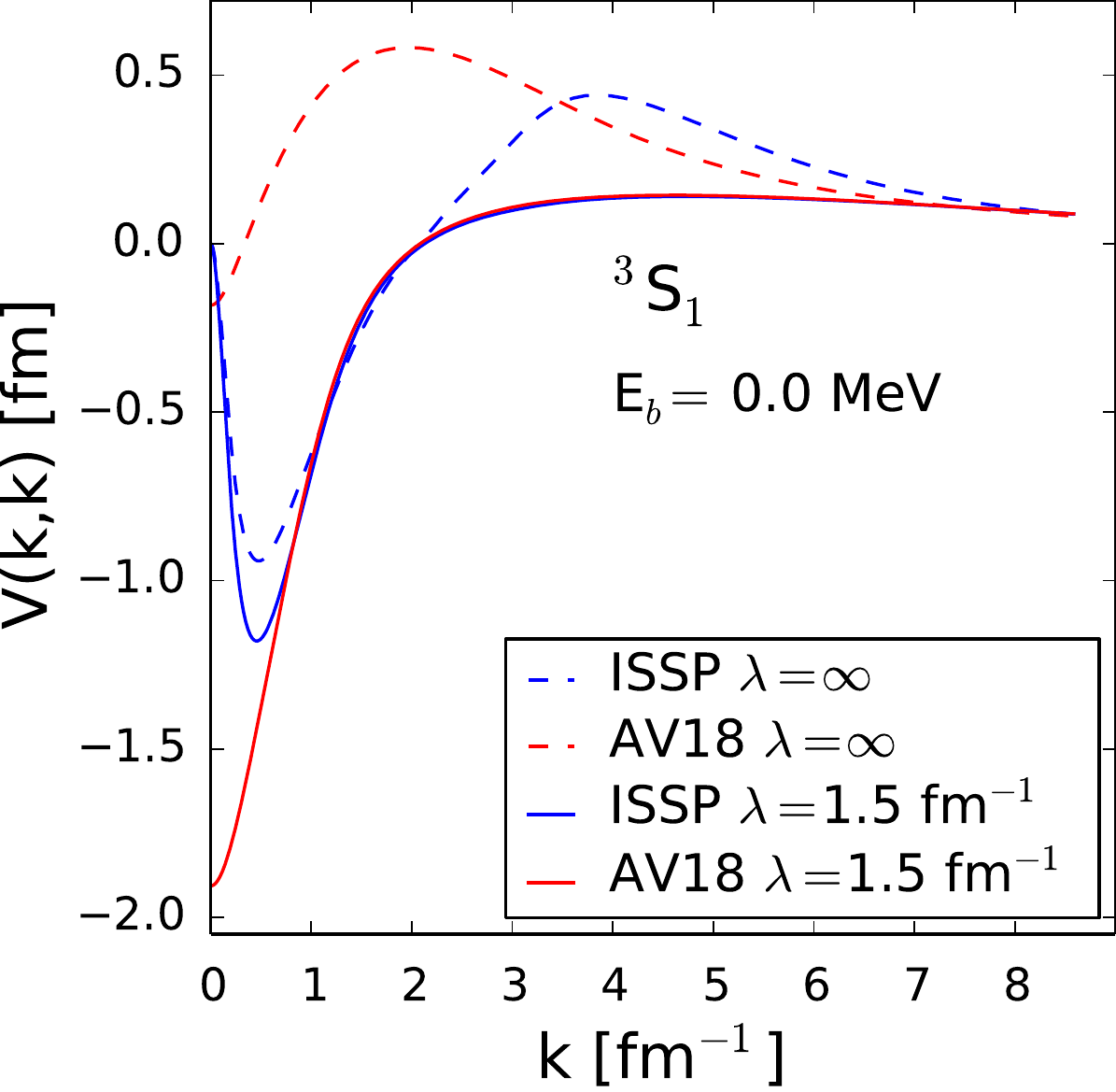}
	\caption{ Initial and evolved diagonal matrix elements in the $^{3}$S$_{1}$ channel
	for AV18 and an ISSP with a binding energy of 0\,MeV. 
	\label{fig:diagss_issp_av18_0_bd}}
	\end{center}
\end{figure}

\begin{figure}[th!]
	\begin{center}
	\includegraphics[width=3.2in]{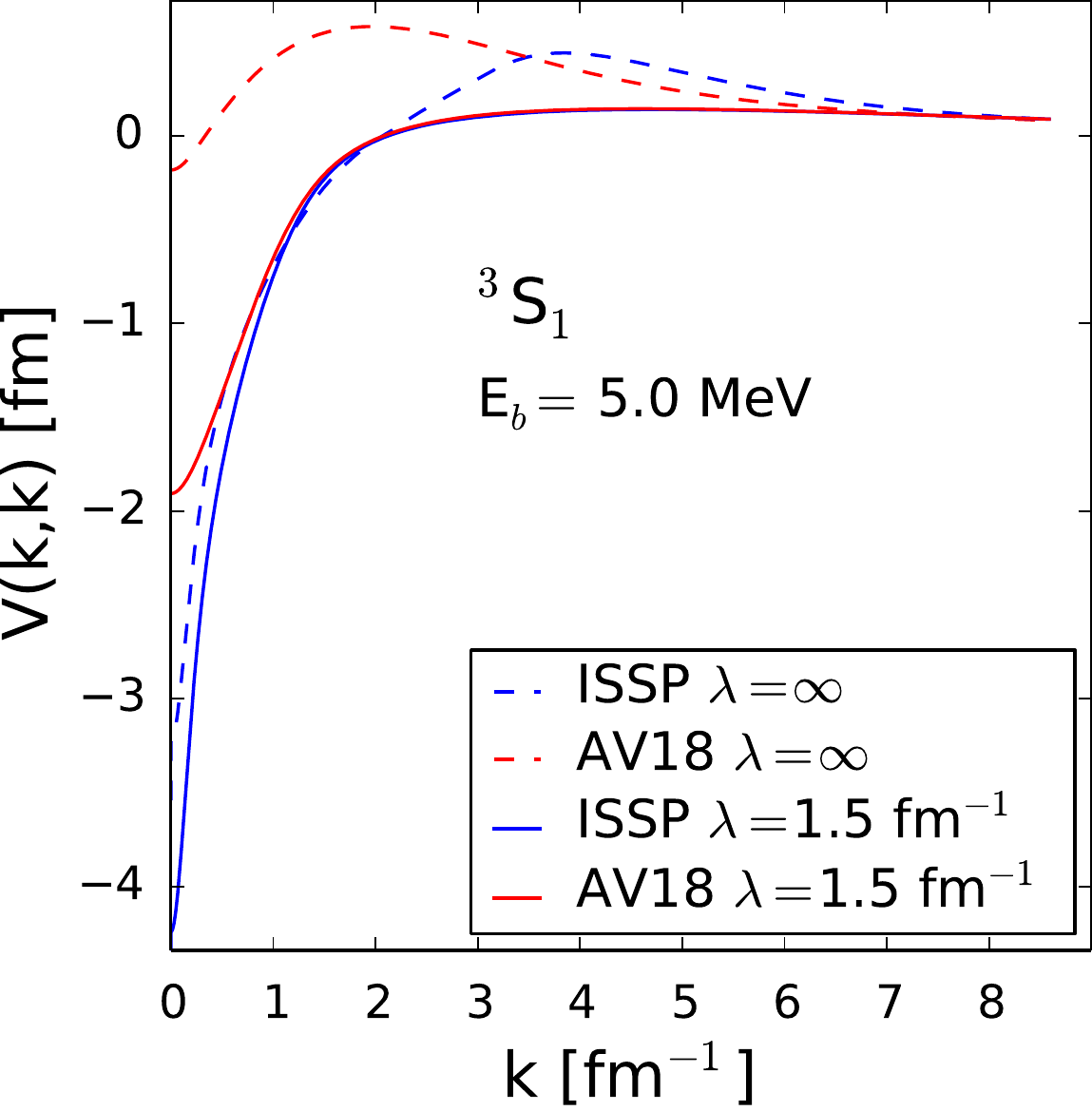}
	\caption{ Initial and evolved diagonal matrix elements in the $^{3}$S$_{1}$ channel
	for AV18 and an ISSP with a binding energy of $5$\,MeV. 
	\label{fig:diagss_issp_av18_5_bd}}
	\end{center}
\end{figure}

Because the binding energy in the ISSP formalism is independently tuned from the phase shifts,
we can investigate in the deuteron
$^{3}$S$_{1}$-$^{3}$D$_{1}$ coupled channel how universality
in potential matrix elements is affected by differences in the bound-state energy.
Figures~\ref{fig:diagss_issp_av18_0_bd} and \ref{fig:diagss_issp_av18_5_bd} show the effect of
phase-shift-equivalent potentials having the wrong binding energy.  
In Fig.~\ref{fig:diagss_issp_av18_0_bd}, the ISSP is created from the phase shifts of the 
\avpot\ potential in the deuteron channel, but with a binding energy of 0\,MeV
instead of 2.224\,MeV. 
It is evident that the effect on diagonal matrix elements is substantial.  
The low-energy matrix elements of the bare ISSP tend towards zero as the momentum decreases.  
As the potentials evolve, the diagonal matrix elements are driven to universal values except
that the ISSP is constrained by its binding energy to approach zero as the momentum approaches zero.

A similar effect can be seen in Fig.~\ref{fig:diagss_issp_av18_5_bd} where instead of 0\,MeV as input binding energy, the ISSP is created with input binding energy of $5$\,MeV.  
The ISSP reproduces this energy better than 100\,eV.  
This potential is overbound and its lowest momentum matrix elements are forced lower 
than if it had the physical deuteron binding energy.  
Again, the higher momentum matrix elements flow towards a universal form because of phase-shift equivalence.  
These plots show that phase-shift equivalence is not the only prerequisite for universality in the diagonal potential matrix elements, but a correct binding energy is also necessary.  
(That is, we need S-matrix equivalence for negative energies as well.)
This may account for the small deviations in the potentials at lowest momenta in 
Fig.~\ref{fig:diag_modern_inf}.  The $^{3}$D$_{1}$ partial wave plots of the corresponding ISSP potentials with different binding energies are indistinguishable.  
This effect only appears in the \threeSone\ potentials.  
It is possible that a virtual bound state in the \oneSzero\ partial wave has a similar effect 
on the evolved low-momentum potential matrix elements, but the ISSP cannot tune virtual bound states and residues in the same way it accommodates bound states, thus we do not investigate this point further.

\begin{figure*}[tp!]
	{\includegraphics[width=.44\textwidth]{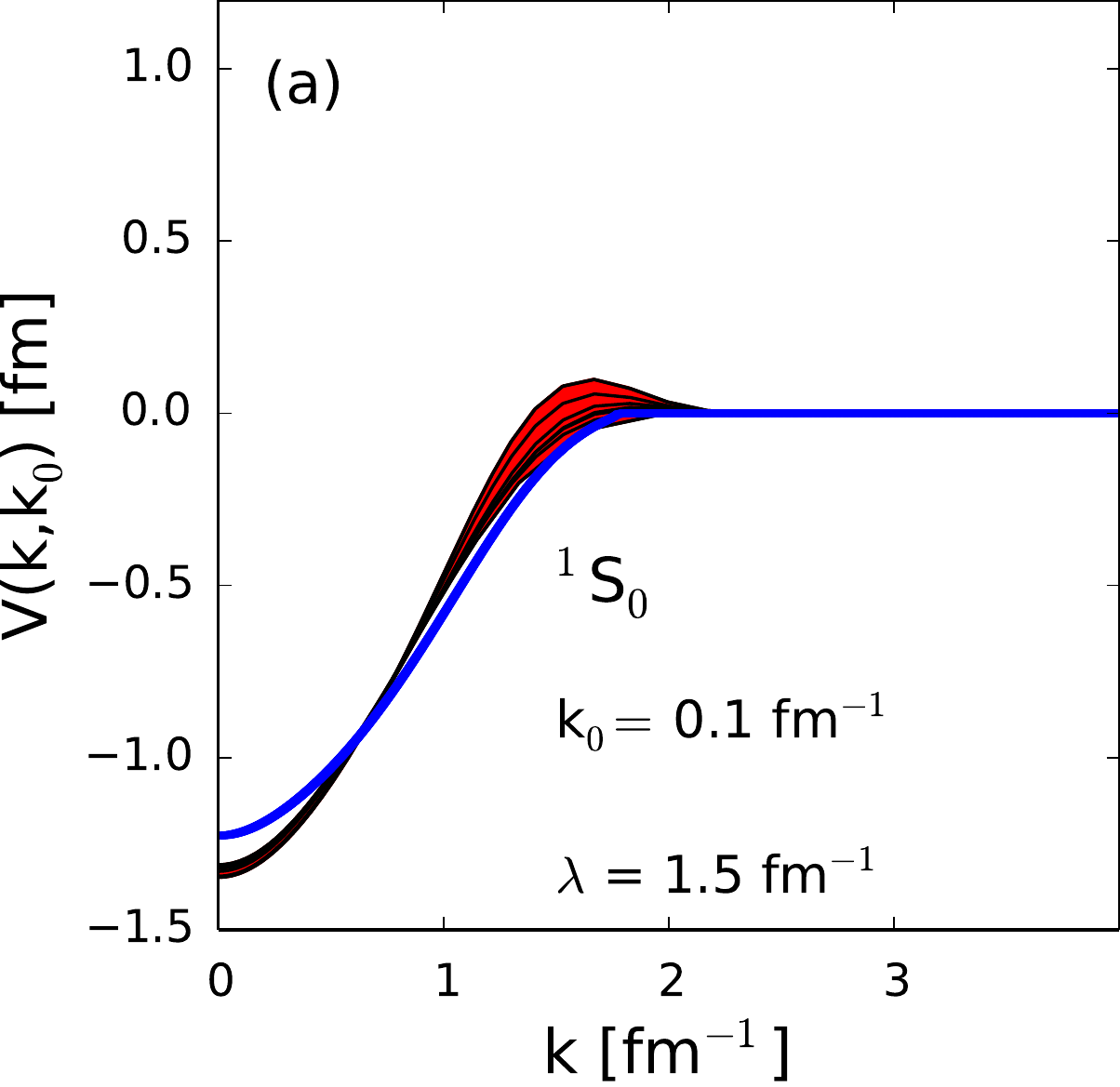}}
	\hspace*{.5in}
	{\includegraphics[width=.44\textwidth]{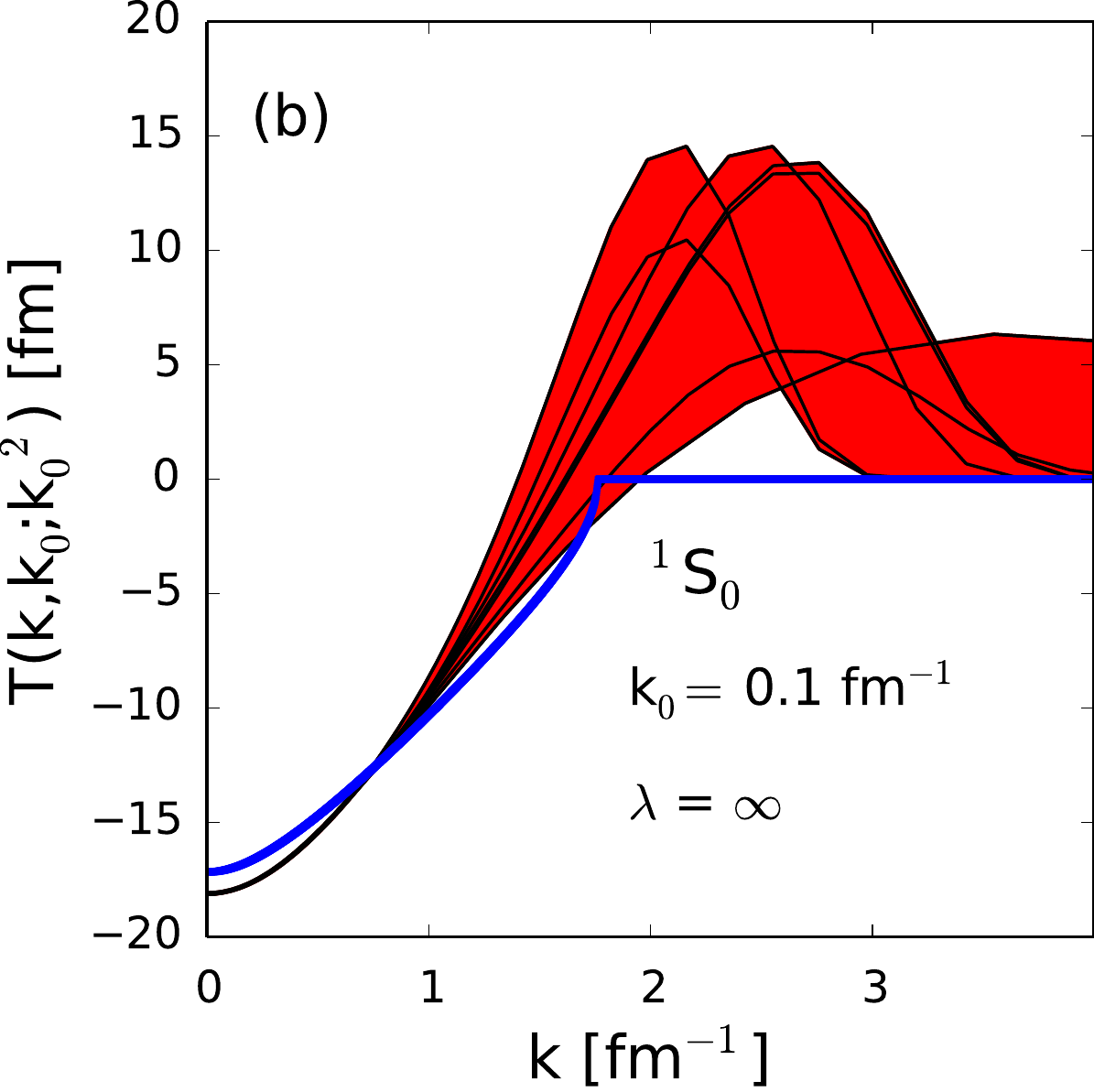}}
	\vspace*{-.1in}
	\caption{ (a) Off-diagonal SRG evolved potential matrix elements
	$V(k,k_0)$ with $k_0 = 0.1\infm$ and (b) unevolved half-on-shell T matrices
	$T(k,k_0;k_0^2)$.  
	In both figures, the thick line is from the ISSP while the bands are various realistic
	potentials. 
	\label{fig:isspno}}
\end{figure*}


Next we turn to off-diagonal matrix elements.
Figure~\ref{fig:isspno} shows the potential matrix elements $V(k_0,k)$ for $k_0 = 0.1\infm$
as a function of $k$ for the ISSP and all of the realistic potentials evolved to $\lambda=1.5\infm$.  
We can see that although these off-diagonal cuts for the modern potentials agree at 
$\lambda=1.5\infm$, the ISSP matrix elements do not.  
By using a diagonalizing SRG transformation (that is, $G_s = T$), the off-diagonal potential
matrix elements are exponentially suppressed.  
Because of this, it appears that the ISSP \emph{approaches} a universal form, but unlike the
realistic potentials, there is no finite $\lambda$ at which the ISSP collapses to universal 
form.  
Figure~\ref{fig:isspno} shows low-energy half-on-shell (HOS) T matrices from each of the unevolved realistic potentials and the ISSP.  
We observe that the realistic potentials, which will evolve to a universal form, have 
essentially the same low-momentum, low-energy HOS T-matrix elements, 
while the ISSP does not.  
This is consistent with carrying over to the SRG the suggestion from Ref.~\cite{vlowkuniv} that HOS T-matrix equivalence
is required for off-diagonal universality in $\vlowk$ RG-evolved matrix elements, 
much like phase-shift equivalence is required for universality of diagonals.  
We only show the \oneSzero\ partial waves, but the same pattern holds for all partial waves.  
Clearly, matching observables is not enough to produce fully universal potentials 
after evolution, and in the next section we will examine if matching observables and also including
the same explicit one-pion exchange potential will be enough for potentials to evolve to a low-energy
universal form.

\begin{figure*}[bhtp!]
	{\includegraphics[width=.44\textwidth]{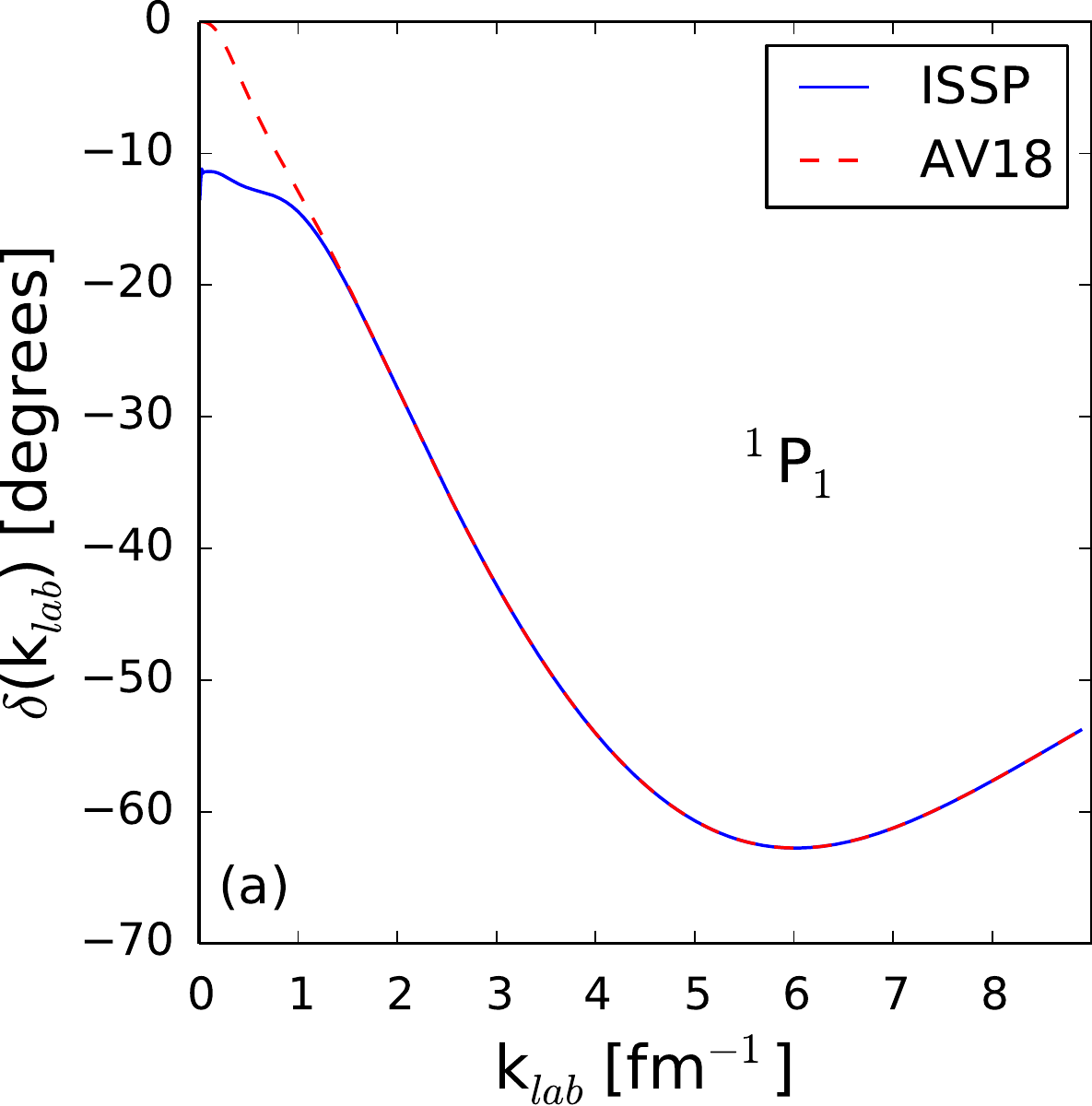}}
	\hspace*{.5in}
	{\includegraphics[width=.44\textwidth]{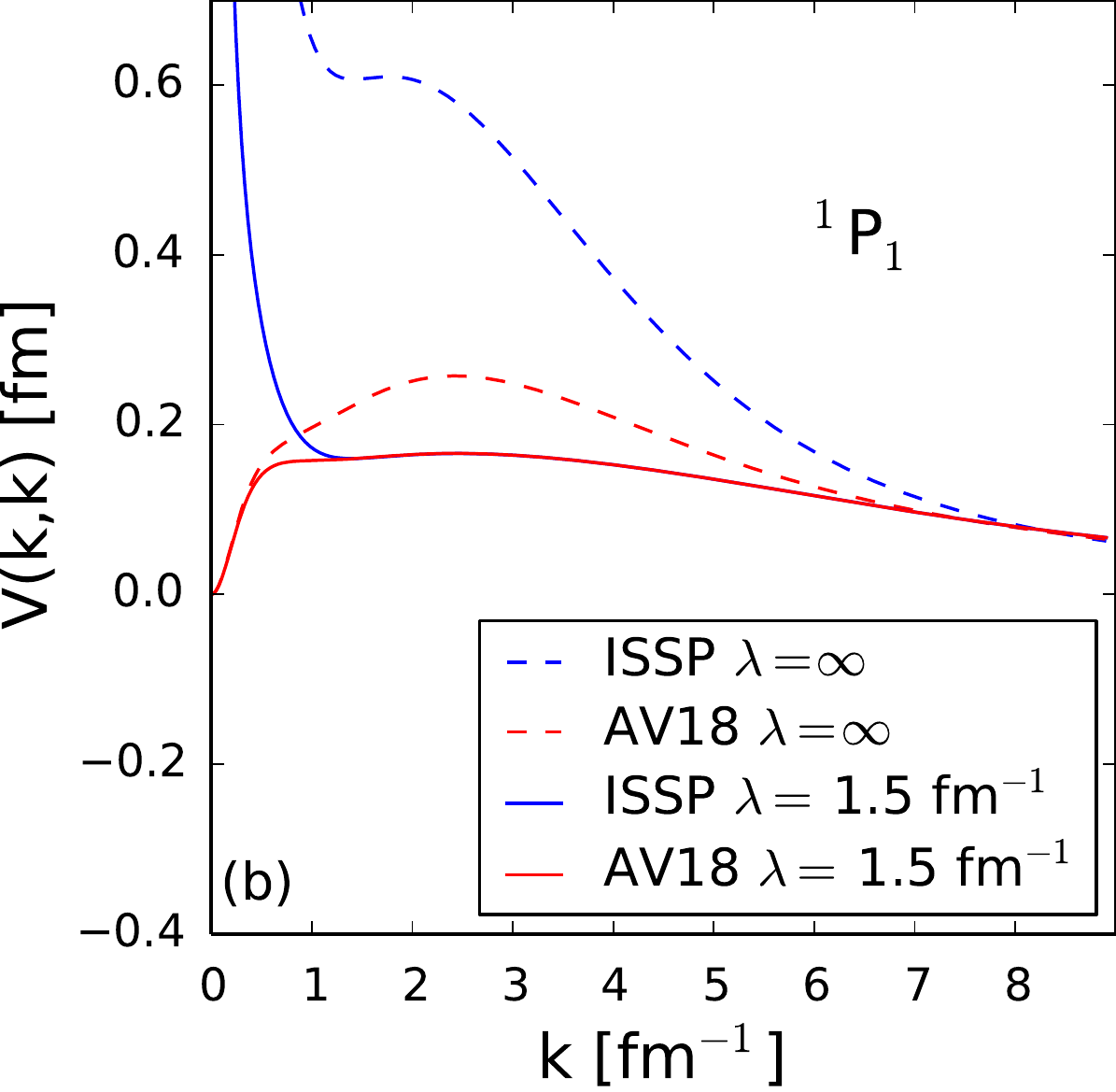}}
	\vspace*{-.1in}
	\caption{ Effects of low-energy phase-shift difference on universality. (a) phase shifts, (b) diagonals of potentials\label{fig:altered_low}}
\end{figure*}

\begin{figure*}[htp!]
	{\includegraphics[width=.44\textwidth]{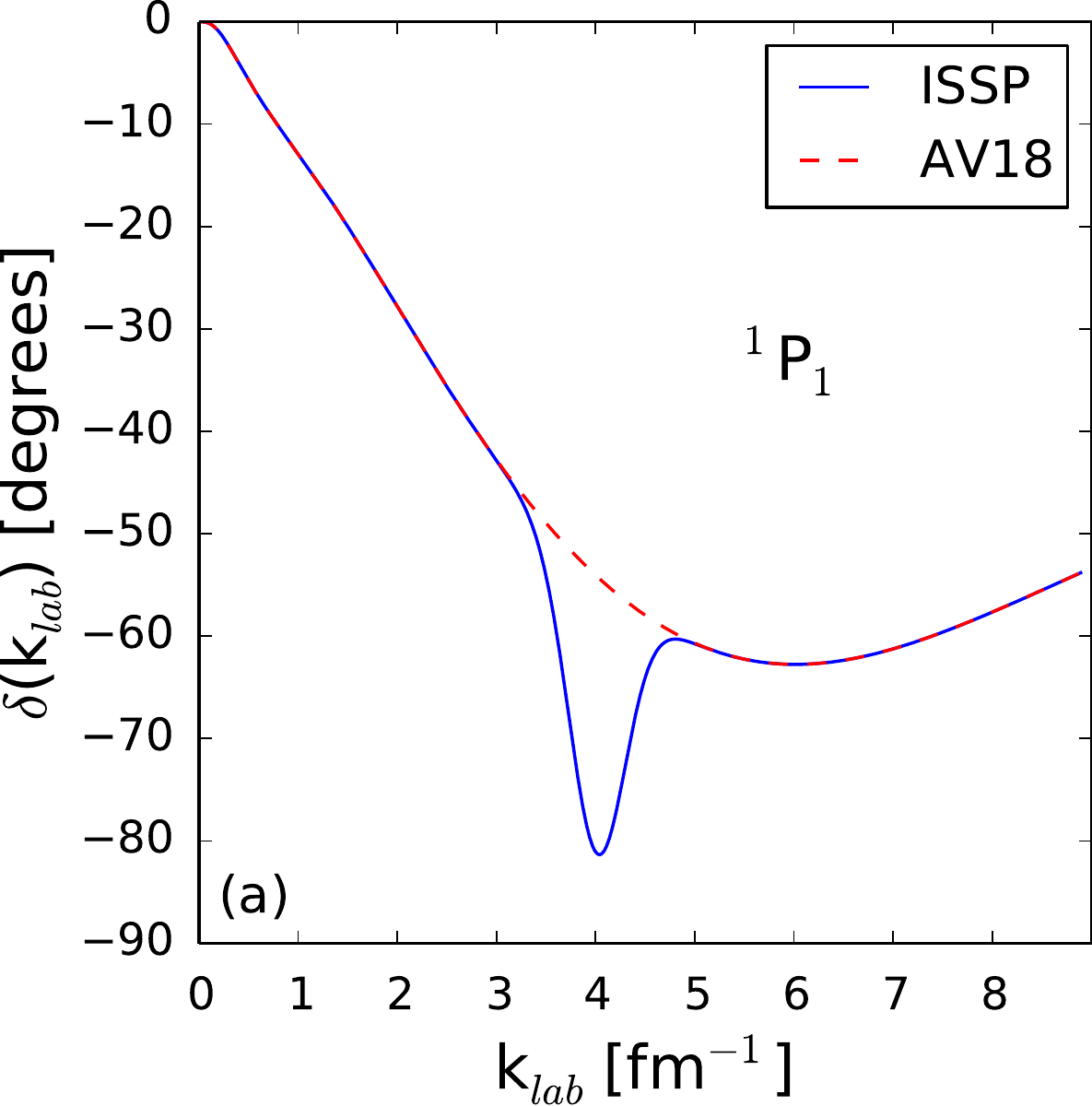}}
	\hspace*{.5in}
	{\includegraphics[width=.44\textwidth]{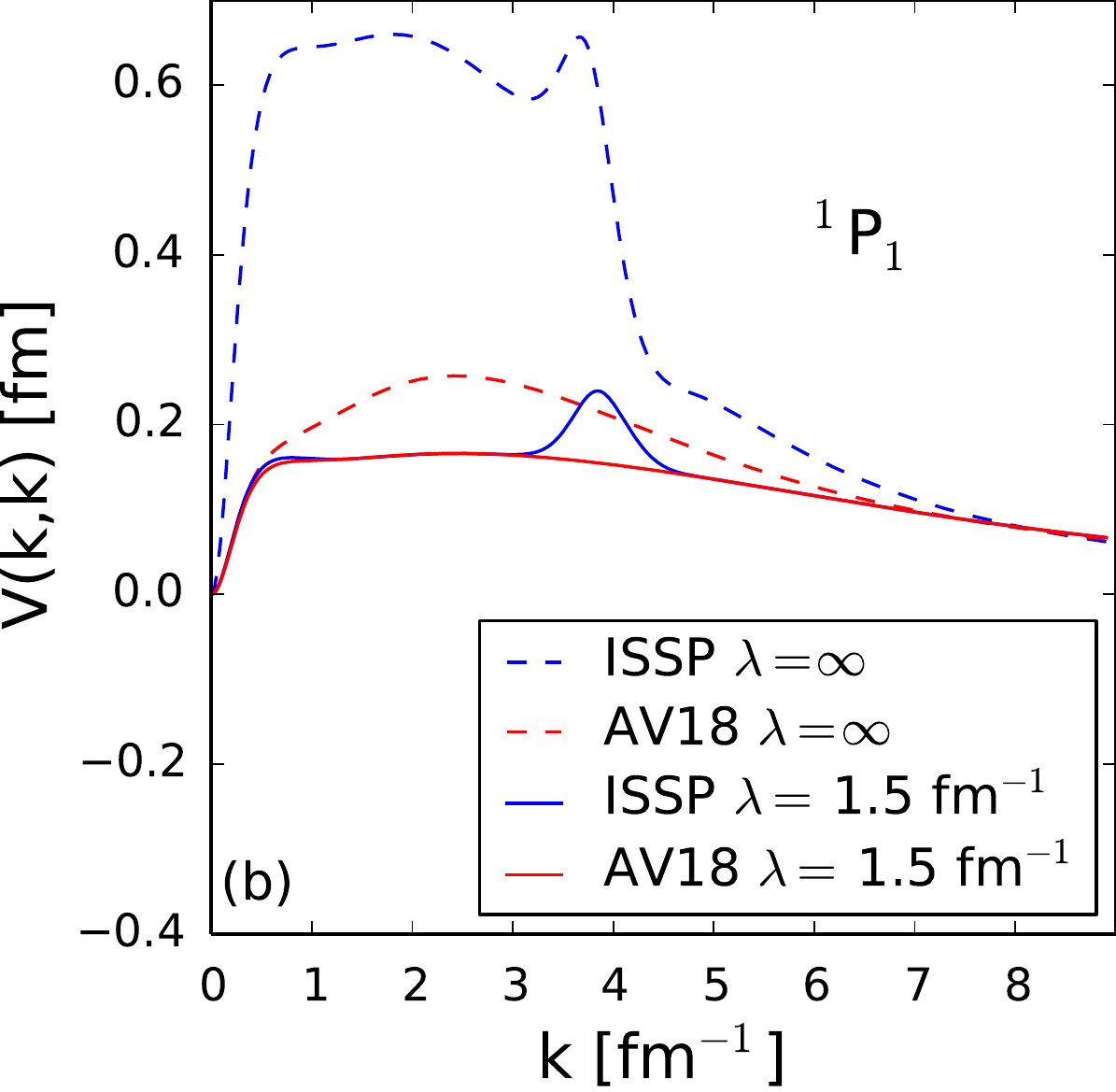}}
	\vspace*{-.1in}
	\caption{ Effects of intermediate-energy phase-shift difference on universality.  (a) phase shifts, (b) diagonals of potentials\label{fig:altered_bump}}
\end{figure*}



As a further test, we created ISSP\rq{}s using altered phase shifts in \emph{localized} regions
of energy to see if the flow to universal diagonal matrix elements is disturbed
only locally.
We use the \onePone\ channel for clarity.  
Figure~\ref{fig:altered_low}(a) shows the \onePone\ phase shifts for \avpot\ and 
for an ISSP that is phase-shift equivalent \emph{except} for a Gaussian bump that we
impose by hand at low energy.
In Fig.~\ref{fig:altered_low}(b) we see that the potentials evolve to the same
diagonal values everywhere but at low energy.
Another potential was constructed by creating low- and high-energy regions of phase-shift equivalence,
and imposing a Gaussian bump (around $k_{\rm lab}=4.0\infm$) to create a difference in the intermediate energy phase shifts, see Fig.~\ref{fig:altered_bump}(a).  
In Fig.~\ref{fig:altered_bump}(b) the evolution to common diagonal values again
works everywhere except near where the phase shifts disagree. 

We conclude from these figures (and other tests not shown) 
that the SRG evolved diagonal potential matrix elements are 
altered only in a region localized near the altered phase shifts.  
This suggests that an SRG softened potential is \emph{locally decoupled} such that 
the integral in the Lippmann-Schwinger (LS) equation for the on-shell T matrix can be truncated as:
\begin{eqnarray}
	T_{l}(k,k;k^2) = V_{l}(k,k)  
	   &+& \frac{2}{\pi} P\!\int_{k-\Lambda}^{k+\Lambda} dp\, p^{2} 
   \nonumber \\ & &  \null \times
	\frac{V_{l}(k,p)T_{l}(p,k;k^2)}{k^{2}-p^{2}}
	\;,
	\label{localLS}
\end{eqnarray}
where the lower limit of the integral is taken to be zero if $k - \Lambda < 0$.
In Eq.~\eqref{localLS}, $\Lambda$ represents the local decoupling scale, which we will set to 
SRG $\lambda$.
(In fact $\lambda$ appears to be a conservative upper bound for $\Lambda$ to quantitatively reproduce phase shifts.)

\begin{figure}
	\begin{center}
	\includegraphics[width=3.2in]{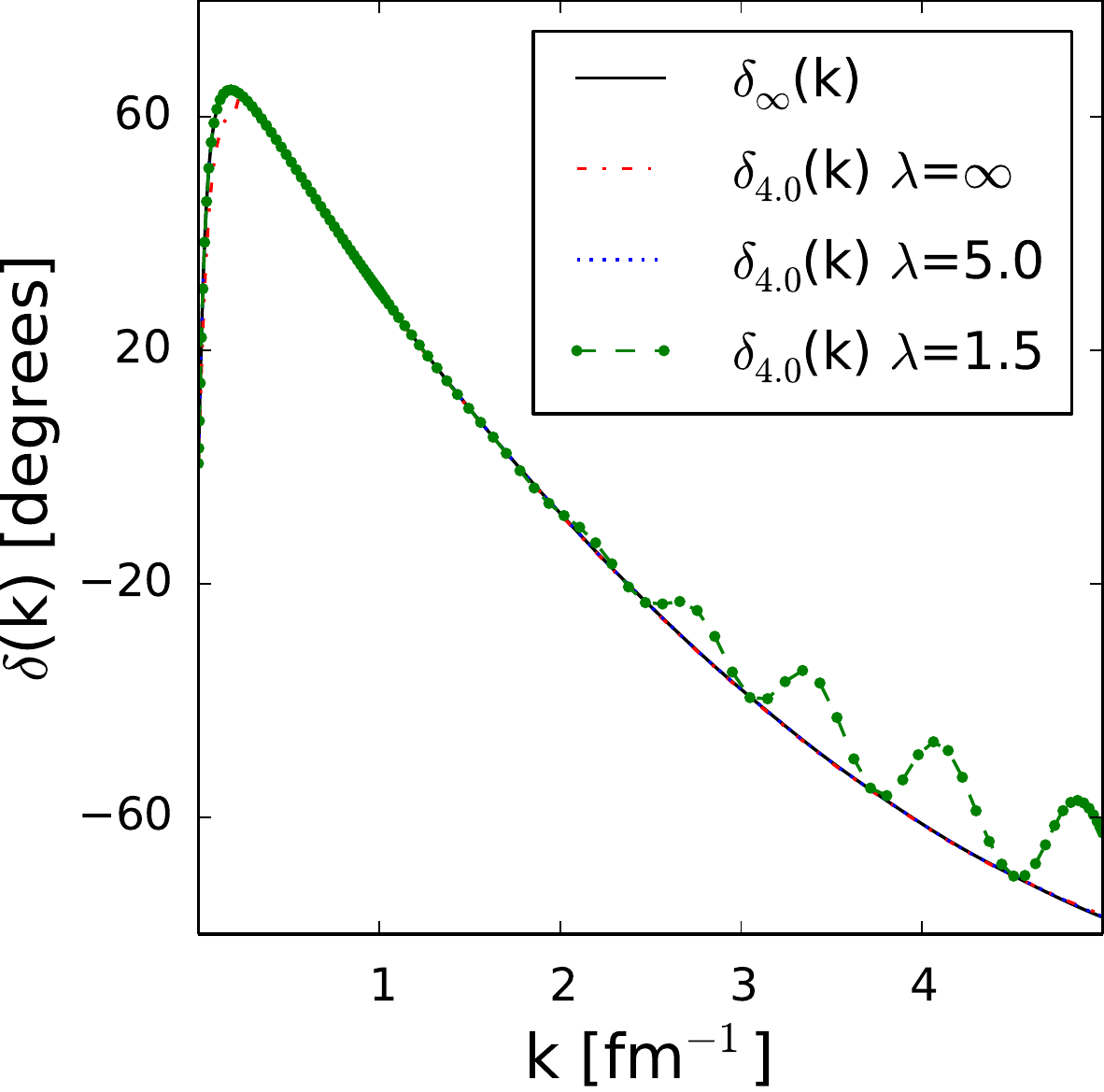}
	\caption{ Phase shifts of Argonne $v_{18}$ potential and truncated phase shifts of evolved potentials with $\Lambda$ = 4.0 fm$^{-1}$.  Cutoff $\lambda$ is in units of fm$^{-1}$.
	\label{fig:localps40}}
	\end{center}
\end{figure}

\begin{figure}
	\begin{center}
	\includegraphics[width=3.2in]{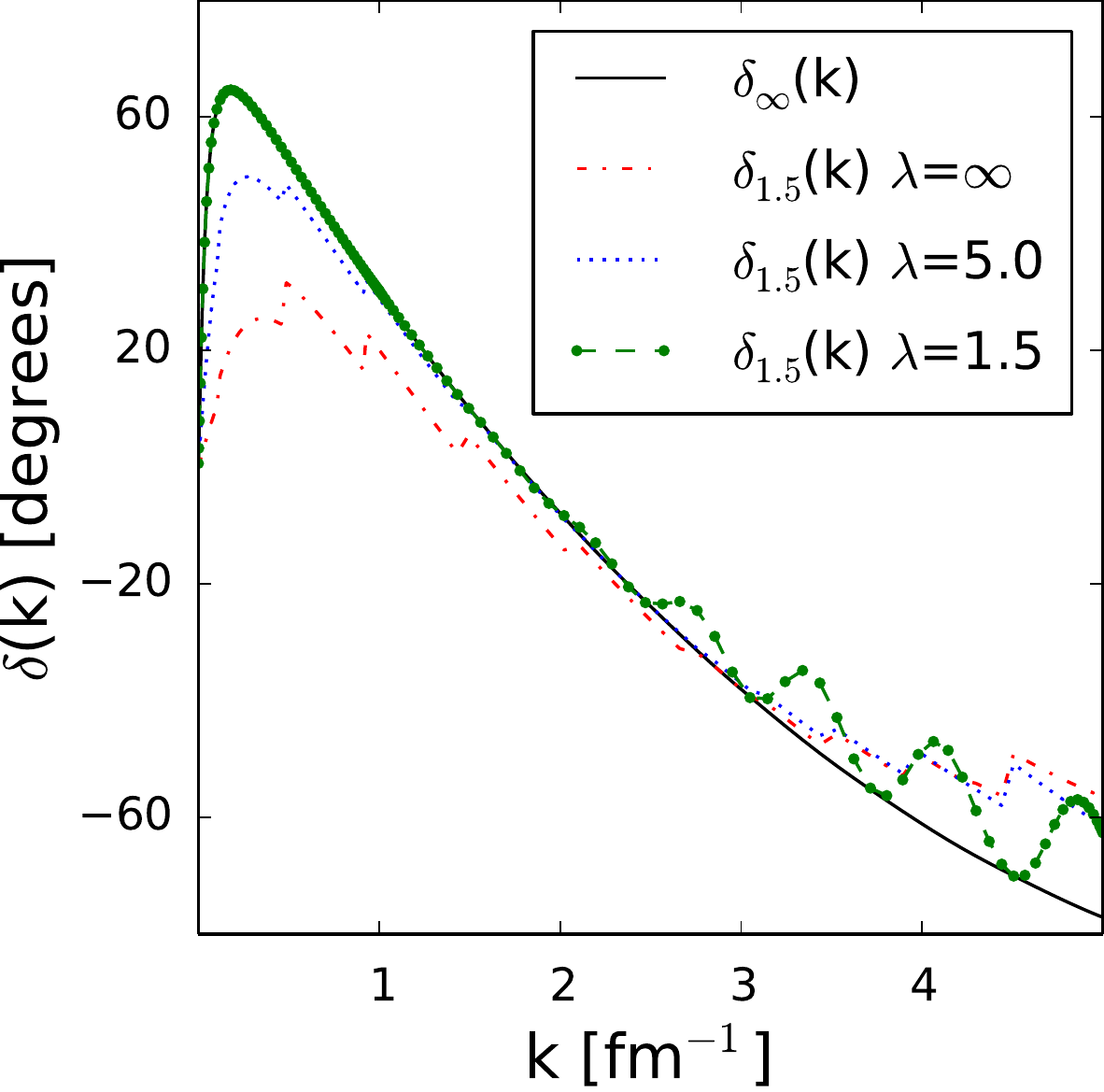}
	\caption{ Phase shifts of Argonne $v_{18}$ potential and truncated phase shifts of evolved potentials with $\Lambda$ = 1.5 fm$^{-1}$.  Cutoff $\lambda$ is in units of fm$^{-1}$.
    \label{fig:localps15}}
    \end{center}
\end{figure}

Figure~\ref{fig:localps40} shows phase shifts calculated from
Eq.~\eqref{localLS} with $\Lambda = 4.0\infm$ 
in the \oneSzero\ channel for the Argonne $v_{18}$ potential evolved to three different SRG 
$\lambda$'s.  These are compared to the actual phase shifts of the unevolved potential.  
We see that with this large value of $\Lambda$, the truncated phase shifts
for even the unevolved potential are largely reproduced and  
the low-momentum phase shifts from evolved potentials are indistinguishable from
the actual phase shifts.  
(The periodicity at high momentum for $\lambda = 1.5\infm$ is a numerical
grid artifact.) 
In Fig.~\ref{fig:localps15} we more severely truncate the integral in the LS equation 
to $\Lambda = 1.5\infm$.  
We see clearly that the potential evolved to $\lambda = 4.0\infm$ is not 
decoupled enough to 
reproduce the original phase shifts, but the potential evolved to $\lambda = 1.5\infm$ 
has phase shifts identical to the previous plot.  
This suggests that evolution with $T$ does locally decouple energy scales.


\begin{figure*}[htp!]
	{\includegraphics[width=.44\textwidth]{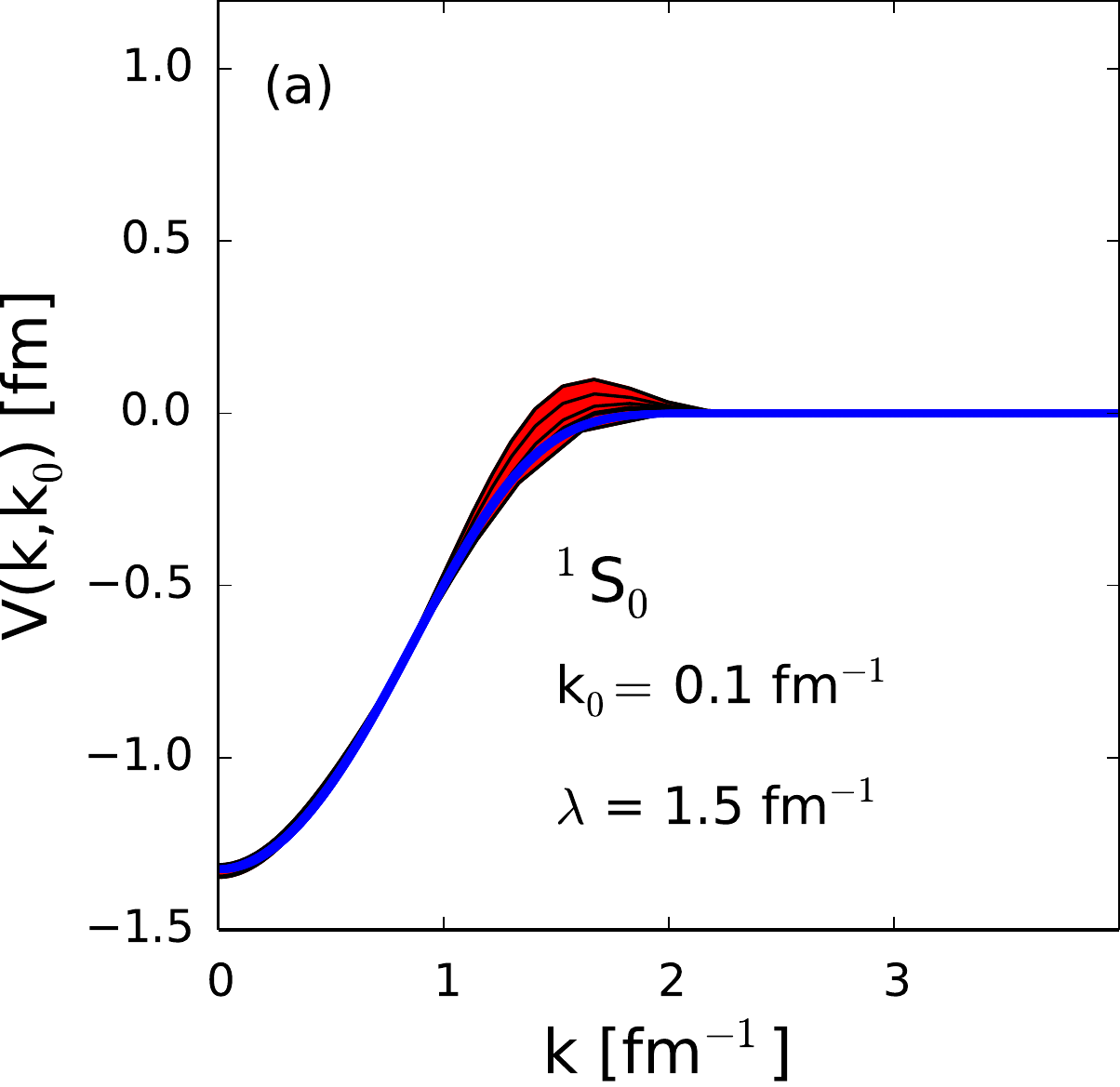}}
	\hspace*{.5in}
	{\includegraphics[width=.44\textwidth]{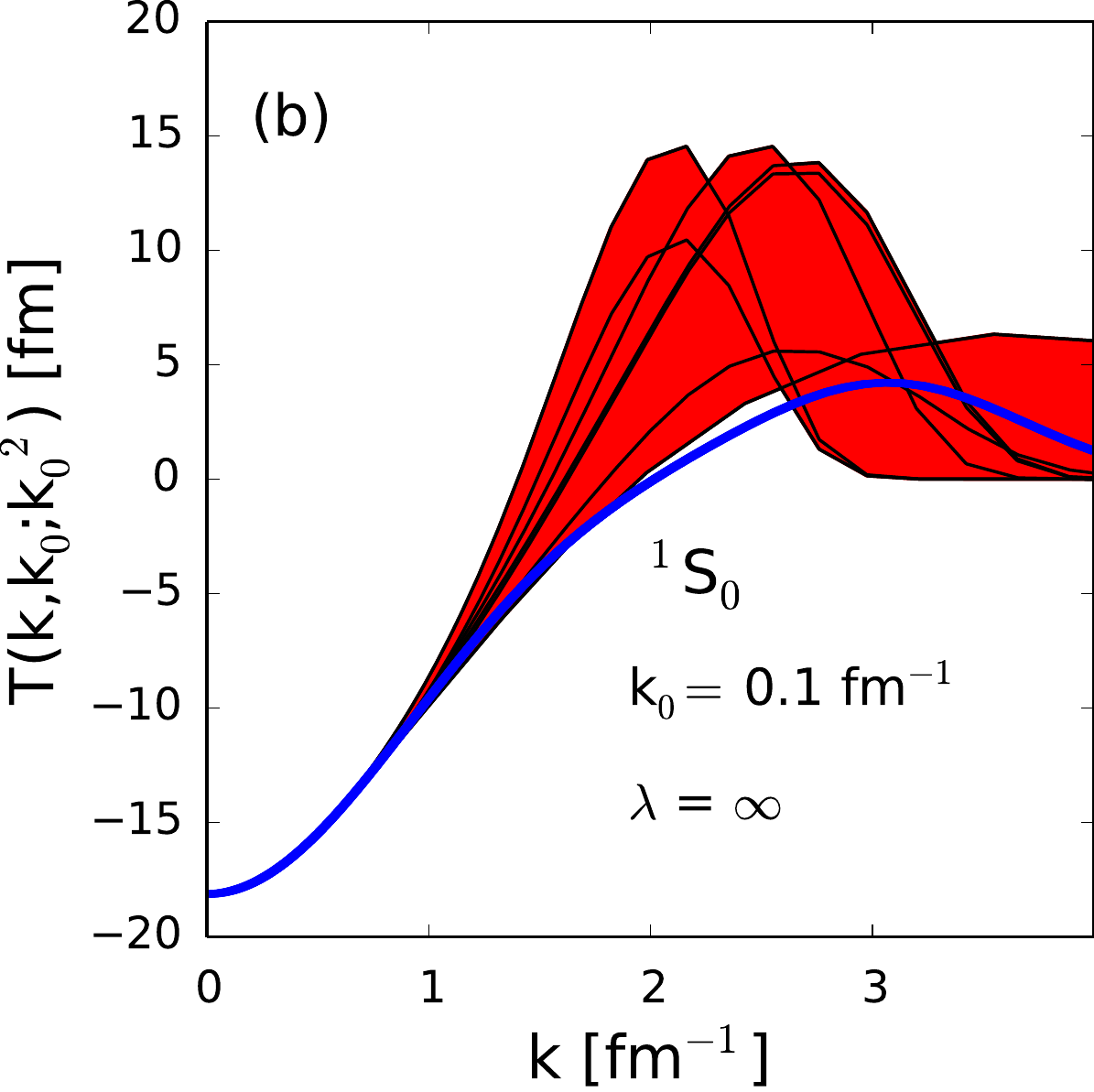}}
	\vspace*{-.1in}
	\caption{ (a) Off-diagonal SRG evolved potential matrix elements. (b) Unevolved half-on-shell T matrices.  In both figures, the thick line is the $\delta$-shell plus OPE potential 
	while the bands are from realistic modern potentials. 
	\label{fig:dsuniv}}
\end{figure*}

\begin{figure*}[htp!]
	{\includegraphics[width=.32\textwidth]{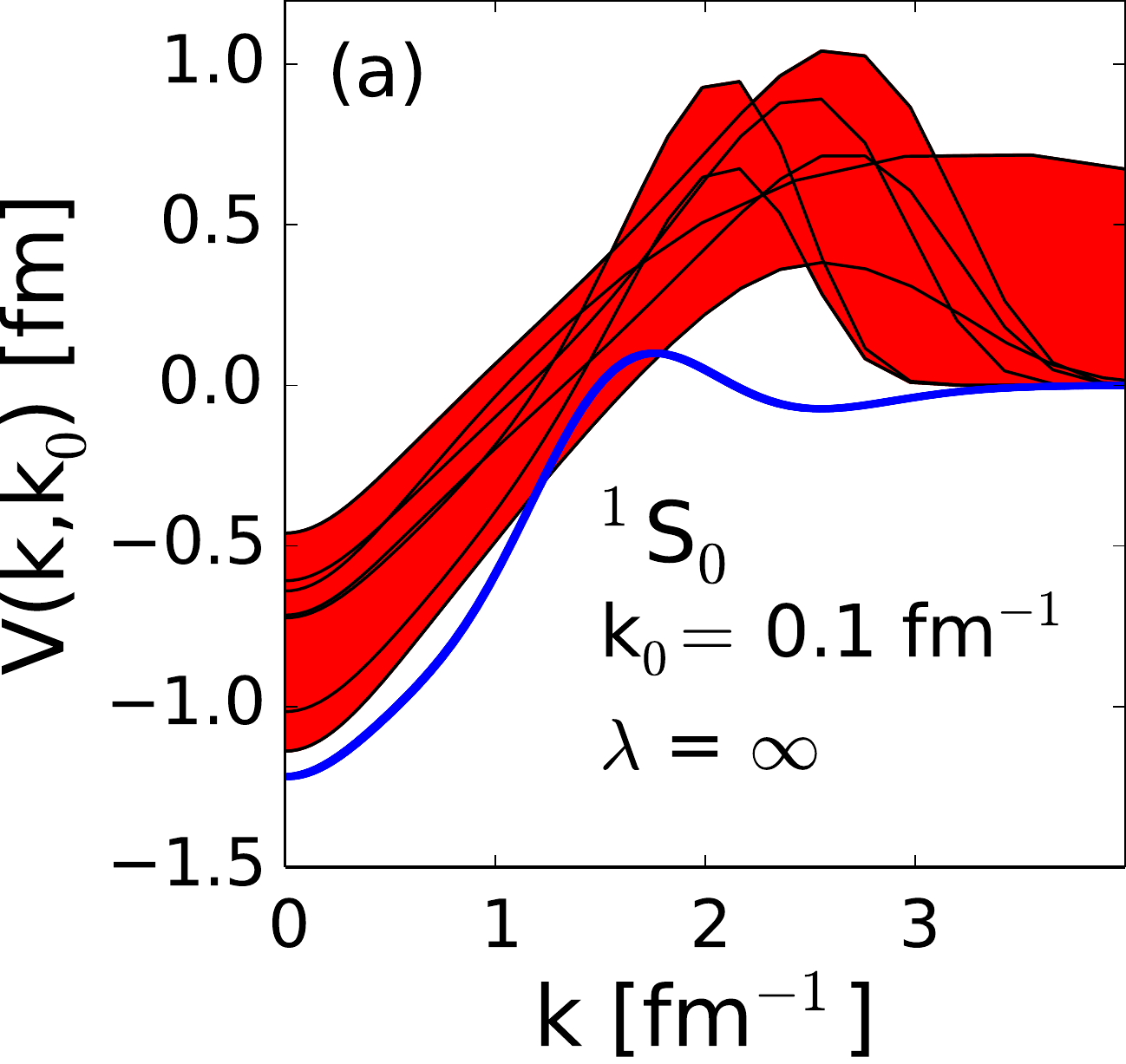}}
	\hfill
	{\includegraphics[width=.32\textwidth]{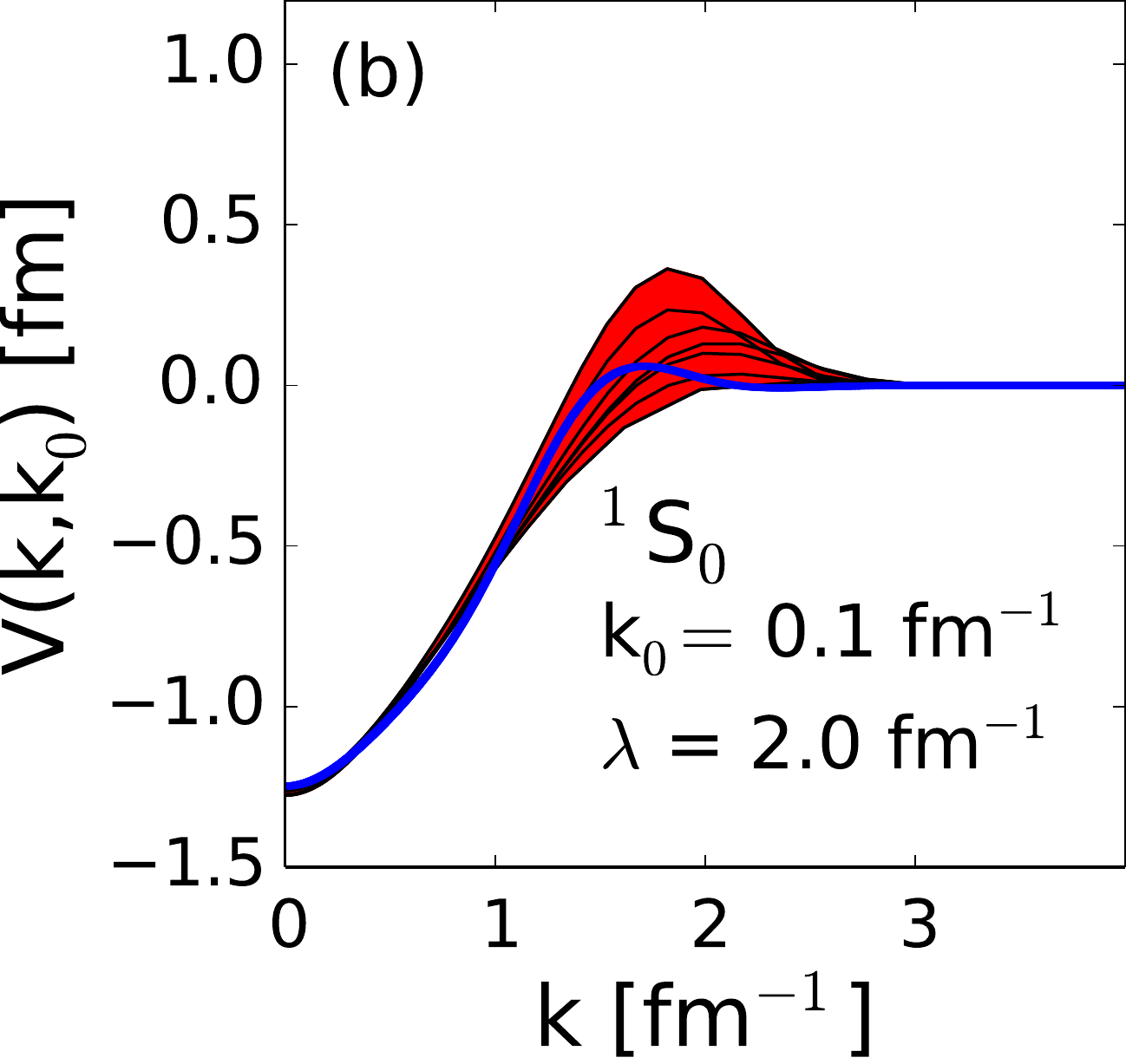}}
	\hfill
	{\includegraphics[width=.32\textwidth]{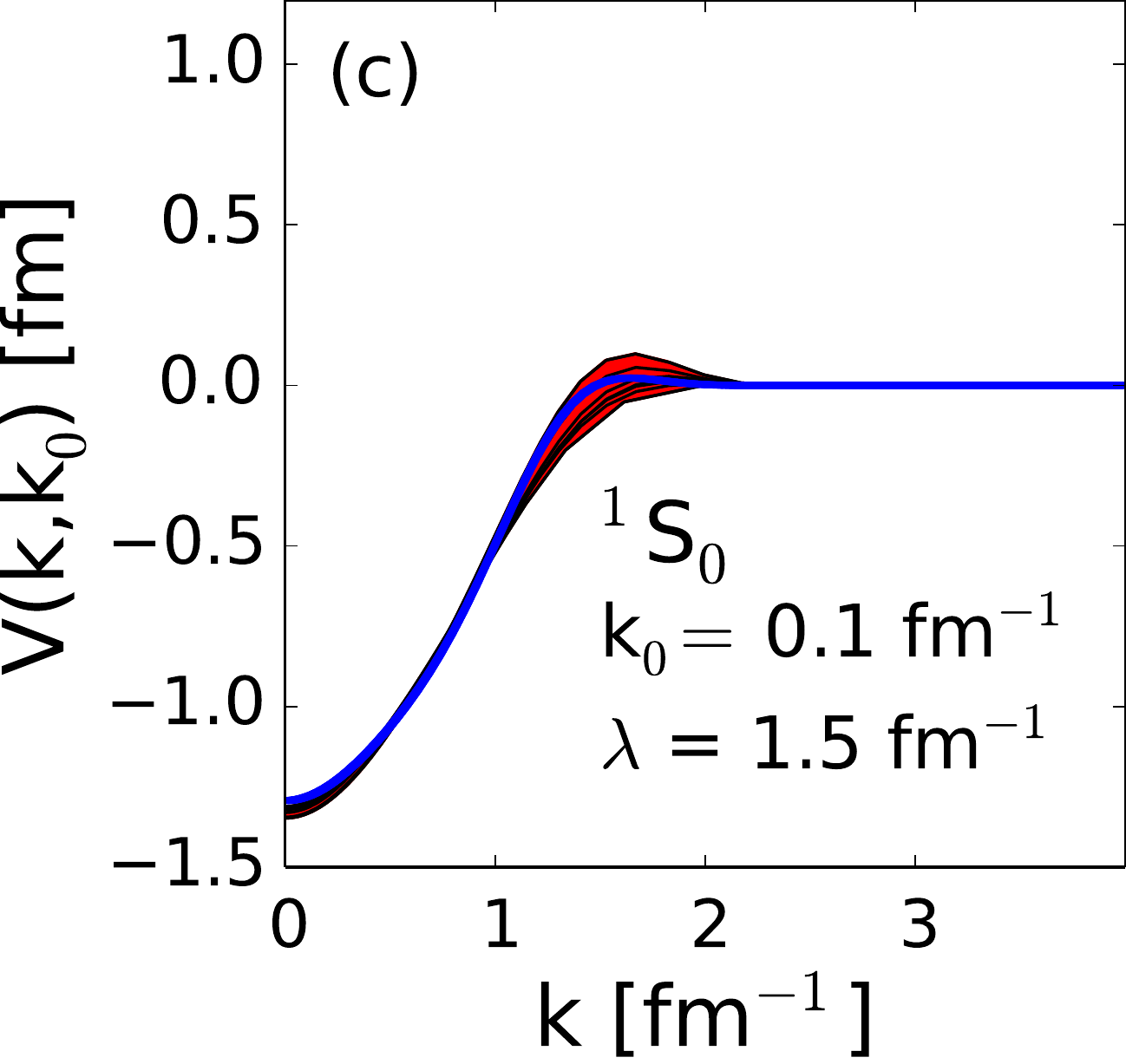}}
	\vspace*{-.1in}
	\caption{ Off-diagonal matrix elements of (a) chiral N$^3$LO potentials and the JISP16 potential (blue line)  
	in the \oneSzero\
  channel and the same potentials evolved by the SRG to (b) $\lambda = 2.0\infm$ and (c) $\lambda = 1.5\infm$.
	\label{fig:1s0_jisp_all}}
\end{figure*}




\section{OPE plus $\delta$-shell} \label{sec:OPE}

Here we further test the suggestion that explicit treatment of the longest-ranged physics is a 
requirement for potentials to evolve to a universal form~\cite{vlowkuniv}. 
In particular, we develop a simple test potential that is (approximately)
phase-shift equivalent in the same
momentum regions as the realistic potentials but also has the same explicit long-range forces.  
We use the model from Navarro P\'erez \textit{et al}.\ that combines the one-pion exchange (OPE) potential with a sum of $N$ $\delta$-shell potentials~\cite{Arriola,Perez:2013jpa} in each
partial wave:  
	\begin{equation}
		V_{l}(r)= V_{l}^{\rm OPE}(r) + \sum_{i=1}^{N} g^l_{i}\ \delta(r-r_{i}) \;.
	\end{equation}
The explicit form of the OPE potential is~\cite{OPE},
\bea
	V_{1\pi}(\bm{r}) &=& \frac{m_{\pi}^{3}}{12\pi} \left(\frac{g_{A}}{2 f_{\pi}}\right)^{2}
						 \bm{\tau}_{1} \cdot \bm{\tau}_{2}\ [T(r)\ S_{12} + 
						 Y(r)\ \bm{\sigma}_{1} \cdot \bm{\sigma}_{2}], \\
	T(r) &=& \frac{e^{-m_{\pi}r}}{m_{\pi}r} \left[ 1 + \frac{3}{m_{\pi}r} + \frac{3}{(m_{\pi}r)^{2}} \right] ,\\
	Y(r) &=& \frac{e^{-m_{\pi}r}}{m_{\pi}r}, \\
	S_{12} &=& 3 (\bm{\sigma_{1}} \cdot \bm{r})(\bm{\sigma_{2}} \cdot \bm{r}) - \bm{\sigma}_{1} \cdot \bm{\sigma}_{2}.
\eea
We choose the $\{r_{i}\}$ as short-range lengths (under $2$ fm), 
and fit the $\{g^l_{i}\}$ to match low-momentum phase shifts.  
For efficiency in momentum representation, we choose a different regulator than Ref.~\cite{Arriola,Perez:2013jpa}, instead regulating the
potential in momentum representation with a separable form factor:
	\begin{equation}
		f_{\rm reg}(k,k')= e^{-({k}/{\Lambda})^{4}}e^{-({k'}/{\Lambda})^{4}} \;,
	\end{equation}
for which we choose $\Lambda = 3$ fm$^{-1}$.
We now have a potential with explicit long-range pion terms and adjustable short range terms, 
which is phase-shift equivalent at low momentum to the realistic potentials.


\subsection{Universality in OPE plus $\delta$-shell}

We can see from Fig.~\ref{fig:dsuniv} that the OPE plus $\delta$-shell 
off-diagonal potential elements evolve to the same universal form as the modern realistic potentials.  
Also, Fig.~\ref{fig:dsuniv} shows the corresponding unevolved HOS T matrices.  
We see that the OPE plus $\delta$-shell potential has the same low-energy low-momentum 
HOS T matrix and shows a corresponding low-momentum universality in off-diagonal matrix 
elements.  
This behavior is not unique to the \oneSzero\ partial wave, but appears for all partial waves.  
This simple potential explicitly contains only the longest range OPE potential and has very simple short-range terms, but it collapses to the same universal low-momentum potential 
after SRG evolution.  
Combined with the ISSP results, this is strong evidence that the same explicit inclusion of
the longest-range contributions to the potential, which is reflected in
low-energy HOS T-matrix equivalence, is required for collapse to a universal form.  

\subsection{JISP potential}  \label{subsec:jisp}

In principle, a good test of our observations about universality is the JISP16 potential,
which is a realistic potential
constructed using the $J$-matrix version of inverse scattering theory~\cite{Shirokov:2003kk,Shirokov:2005bk}.
Because there is no explicit incorporation of a pion-exchange tail in the functional
form of the potential, we might expect the Hamiltonian to exhibit non-universal evolution
with the SRG for off-diagonal matrix elements.
In fact, the unevolved JISP potential is already soft and changes only slightly under
SRG evolution.  
But as shown in Fig.~\ref{fig:1s0_jisp_all} in the \oneSzero\ channel for a set of off-diagonal
matrix elements (and true for the diagonal and other partial waves), JISP16 is already close
to the universal form reached by the chiral N$^3$LO potentials.  There are still differences,
but they are small.  However the JISP HOS T matrix is also close to the others (perhaps as
the result of additional adjustments of the potential using the freedom of the inverse
scattering framework ~\cite{Shirokov:2005bk}), so there is no inconsistency with our general conclusions.


\section{Recap and moving into 3-body}  \label{sec:conclusion}

Modern realistic  two-nucleon potentials exhibit a
flow to universal potential matrix elements under the similarity
RG. High and low momenta are decoupled in this universal matrix,
allowing us to truncate the matrix and drastically simplify 
low-energy bound state and reaction calculations. Any initial interaction
that yields this universal matrix after SRG evolution is equally effective. This is of little
practical importance for the two-nucleon potential, but it could be extremely
useful if many-nucleon potentials display this same type of universality. 
Producing accurate realistic few-nucleon potentials is extremely difficult, and much effort is spent working on producing momentum representation matrix elements for additional terms in \chieft. 
Our results suggest that any convenient potential that includes long-range pion exchange 
interactions can be used to produce universal many-nucleon interactions
when evolved with an SRG transformation.
 
Our study of universality for two-body potentials
yields the following observations:
\begin{itemize}
\item Inverse scattering separable potentials, with no explicit consideration 
of long-range pion exchange, exhibit a universal collapse
of diagonal matrix elements  after evolution
in regions of phase-shift equivalence. 

\item
If an intermediate region of phase-shift inequivalence
is imposed, the collapse does not occur in this region, but still occurs in every region of phase-shift equivalence.
This implies that SRG softened potentials are actually \emph{locally}
decoupled in energy/momentum. 

\item
An incorrect binding energy has a strong effect on the lowest potential 
matrix elements and will prevent flow towards a universal form.  

\item
While phase-shift equivalence and correct binding energies (i.e., S-matrix equivalence) are apparently
requirements for universality in two-body potential matrix elements,  
the ISSP example shows that these are not sufficient to guarantee a potential 
that will flow to the same off-diagonal values as conventional realistic
potentials. 

\item
However, a potential that reproduces low-energy observables and 
contains explicit long-range (OPE) terms does flow to universal form, which
is consistent with observations made for $\vlowk$ evolution in
Ref.~\cite{vlowkuniv}.

\item
To the extent that low-energy HOS T-matrix equivalence indicates long-range
equivalence of potentials, it signals off-diagonal universality in evolved
potential matrix elements.

\smallskip

\item
For universality to appear, the SRG decoupling parameter must be sufficiently low that potential 
matrix elements in the low-momentum region of HOS T-matrix equivalence are decoupled 
from high-momentum matrix elements.

\end{itemize}

\noindent
These considerations address the onset of universality for the two-body part
of the inter-nucleon potential but
for a complete discussion we have to consider
the full many-body Hamiltonian.  It is well established that evolution of $\lambda$ induces many-body forces of increasing importance~\cite{3bsrgsimple,bognerfurnstahlschwenk,Furnstahl:2013oba} and the SRG transformations
will only be approximately unitary if they are omitted.
This entails 
a lower limit to the region of universality in practical applications.
The following chapters will detail the framework in which we search for universality in a simple 3-body problem, but before that, the next section will detail an important simplification of the fitting procedure.

\section{Phase-shift equivalence and eigenvalue equivalence} \label{sec:ps_eig_equiv}

In the 3-body problem, the same prescription of fitting $\delta$-shell potentials to phase shifts is made much more complicated, because 3-body scattering is much more complicated.
It is well known that a relationship exists between the phase shifts and the bound states of a system given certain boundary conditions (eg. Luscher's method~\cite{Luscher:1985dn}).
Using this as a guideline, we can confirm that fitting the low-energy eigenvalues (including positive eigenvalues) for a Hamiltonian matrix also fits the phase shifts in the same energy-regime.
Because the eigenvalues are trivial to find, many-body scattering formalism is unnecessary for our 3-body study.

The bound-state eigenvalues, in general, will be basis and mesh independent as long as transformations are accurate and numerical precision adequate.
This is not so for positive eigenvalues, however, which are highly dependent on the basis and mesh we choose to work with.
In momentum representation, for instance, the scattering eigenvalues of the Hamiltonian matrix are almost completely dominated by the  eigenvalues of the kinetic energy matrix at large momenta.
For this reason, it is convenient for numerical precision and visualization to use the difference of the eigenvalues of the Hamiltonian matrix minus the eigenvalues of the relative kinetic energy matrix, all divided by the mesh weights.
We will call this vector, $\widetilde{\xi}$.

\beqn
	\widetilde{\xi}_{i} = \frac{eig(H)_{i} - eig(T)_{i}}{w_{i}}.
\eeqn
$\widetilde{\xi}$ is clearly defined in uncoupled channels, but how to divide out the weights in coupled channels is a complication to this method.
Our simple three-body model will not have coupled channels, thus we do not develop this method further to take into account coupled channels.

\begin{figure}
	\includegraphics[width=.45 \textwidth]{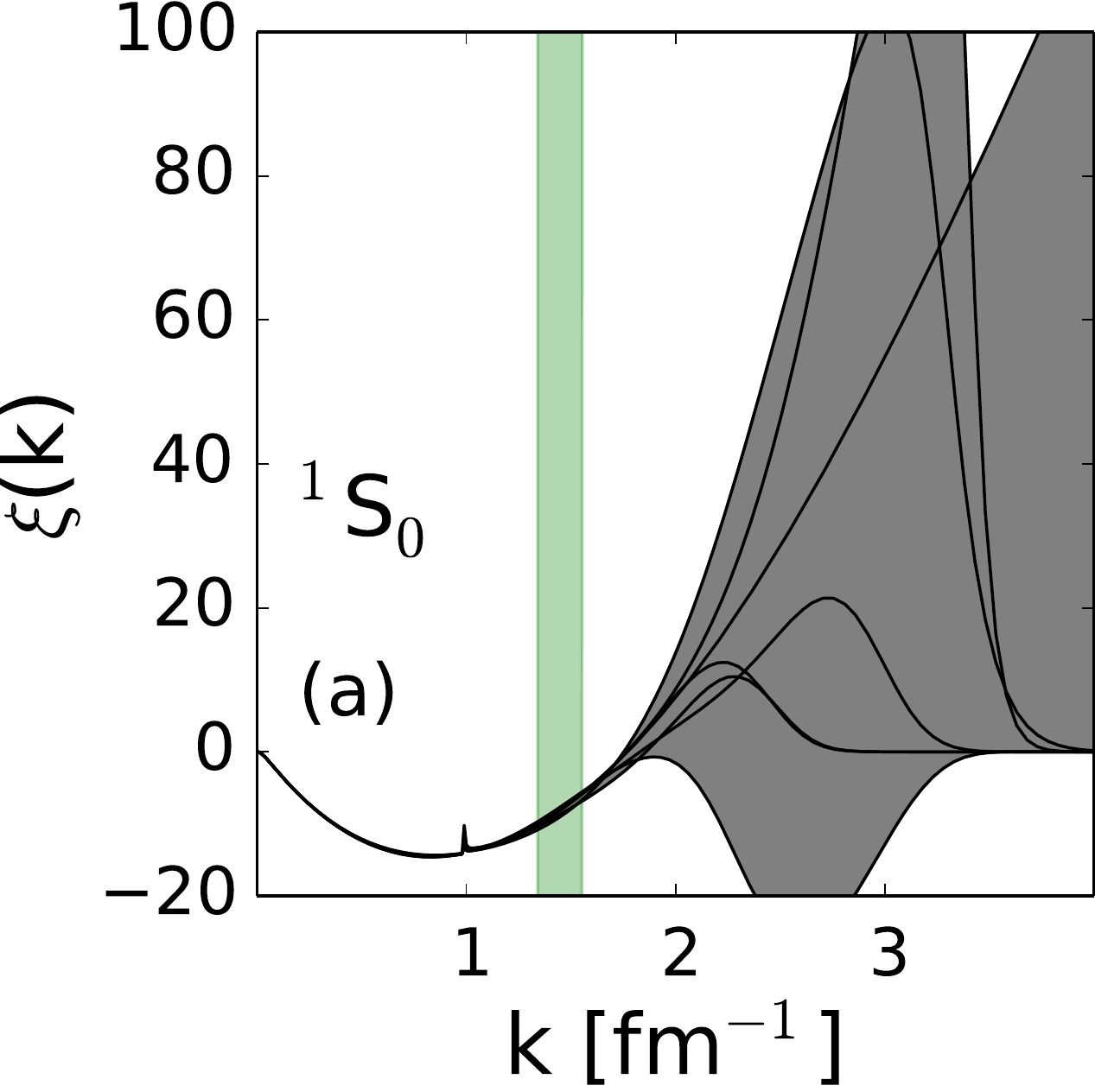}~~~%
	\includegraphics[width=.45 \textwidth]{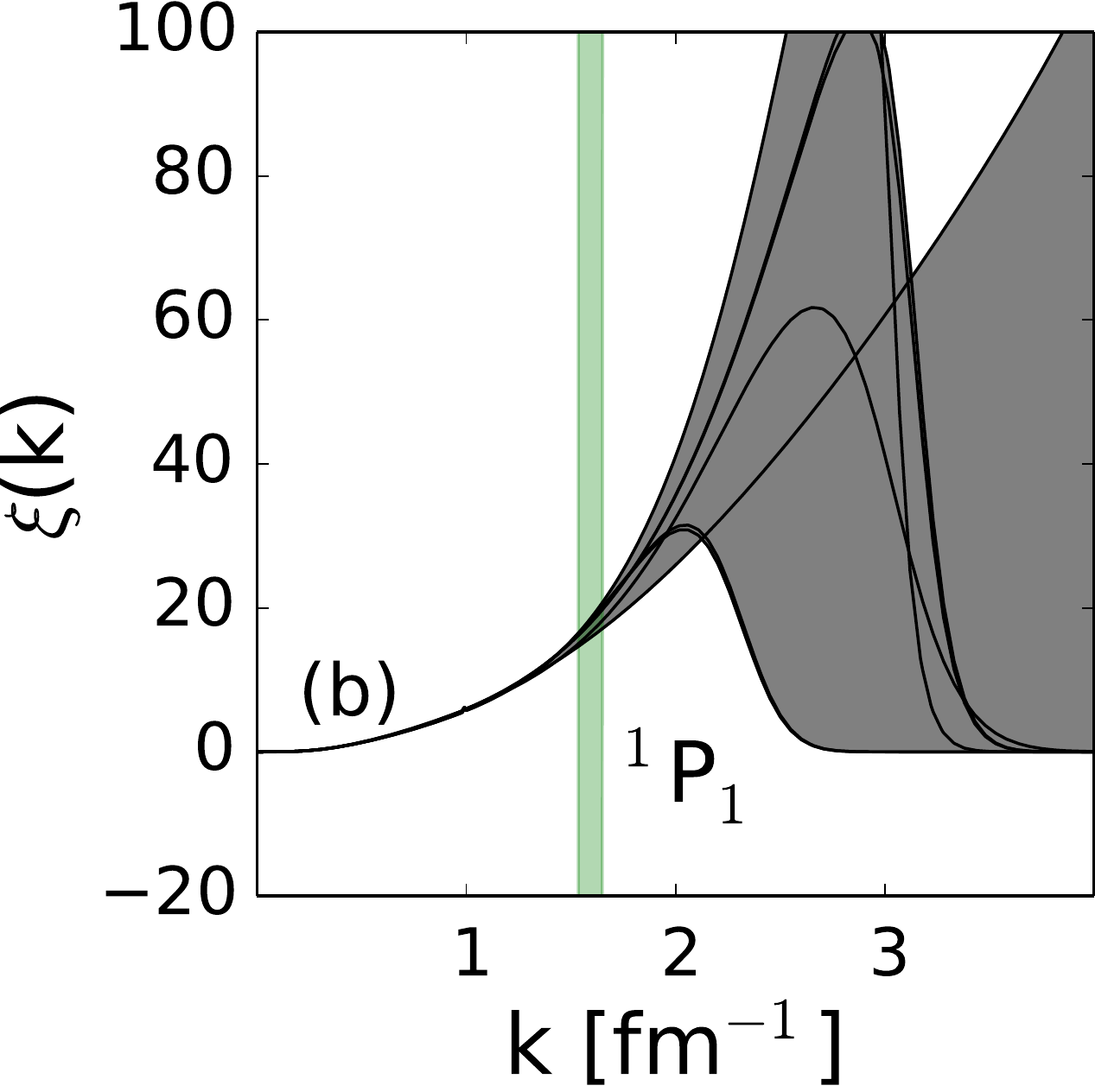}~~~%
   \vspace{-.3in}
	\caption{  Hamiltonian matrix eigenvalues minus relative kinetic energy matrix eigenvalues all divided by mesh weights ($\widetilde{\xi}(k)$) of various unevolved realistic potentials (see text) in the (a) \oneSzero, (b) \threeSone, and (c) \onePone\ partial waves.
	The shaded regions show the range of values and the vertical bands are calculated in the same manner as
	Fig.~\ref{fig:ps_modern}, but with denser meshes for better numerical precision. 
	\label{fig:eigminust_modern_inf}}
\end{figure}

Fig.~\ref{fig:eigminust_modern_inf} shows $\widetilde{\xi}$ calculated for the same chiral and phenomenological potentials as before.
Comparing to Fig.~\ref{fig:ps_modern}, the same qualitative behavior exists between the phase shifts and $\widetilde{\xi}$ (although $\widetilde{\xi}$ requires more mesh points for precision).
We see low-energy eigenvalue equivalence in the same regions that exhibit low-energy phase-shift equivalence, and equivalence in $\widetilde{\xi}$ from each potential matrix begins to break down at the same energy as the phase-shift equivalence.
Therefore, we can safely fit $\delta$-shells to low-energy eigenvalues rather than phase shifts.

\chapter{Harmonic Oscillator Basis}
\label{chapt:HOBasis}

Choosing a convenient basis is a critical step in solving low-energy nuclear problems.
Until now, we have chosen to work with a plane-wave basis with the momenta of particles serving as continuous degrees of freedom.
Plane waves are of great use in simple analytic problems and for uniform systems (e.g. nuclear matter \cite{Hebeler:2013ri}), but for finite bound-state calculations, it is far more efficient to expand in a different basis.
In nuclear few-body bound-state problems, it is most common to use the harmonic oscillator (HO) basis.
The basis functions of the HO basis are the eigenfunctions of the harmonic oscillator Schr\"{o}dinger equation.

The HO-basis has a number of useful features which simplify three- and many-body problems compared to the plane-wave basis.  
Most useful to our calculation is that the HO-basis is a discrete basis.  
This simplifies operator equations, which in plane-wave basis involve integrals over continuous momenta.  Instead of discretizing these integrals with momentum meshes and weights, and representing them as matrix multiplication, the HO-basis already represents all operators as discrete matrices.  
The HO-basis also simplifies bound states calculations; simply choose your favorite eigenvalue routine and input the Hamiltonian matrix; this is true even for many-body bound states which require Fadeev formalism in a plane-wave basis~\cite{Akerlund:2011cb}.  
An extremely difficult aspect of an A-body calculation is the presence of \emph{spectator} delta-functions that arise from the \emph{spectator} particle when embedding fewer-body potentials into the A-body space (we will discuss this further in section~\ref{ssec:HO1d3b}).  
In the HO-basis, the numerically difficult to handle Dirac-delta functions become Kronecker delta-functions, but the symmetrization process still connects off-diagonal matrix elements in the potential.  
The simplification of the many-body problem in a HO-basis are significant, because the basis size required is greatly reduced.
SRG evolution of realistic nuclear 3-body potentials in a plane-wave basis is cutting-edge research~\cite{Hebeler:2012pr,Wendt:2013bla}, and these potentials must still be transformed into HO-basis for use in many-body methods.

In this chapter, we start with the formalism for HO-basis expansions in Section~\ref{sec:HOformulation}.
We then examine some of the limitations of the finite basis transformation and the effects on potential matrix elements in a simple model and with realistic 2-body nuclear potentials in Section~\ref{sec:HOtrunc}.
Finally, we will show how the choice of basis affects generators and SRG flow of potential matrix elements in Section~\ref{sec:HOSRG}.


\section{Formulation}
\label{sec:HOformulation}

Before we jump into our 3-body calculation in 1-D, some of the specifics of HO-basis expansions are important to understand.  
We plan on observing matrix elements of evolved potentials, so a firm understanding of exactly how the matrix elements are generated is critical.
The following section will detail the HO-basis expansion first in 1-D for 2 particles, followed by 2 particles in a 3-D partial wave expansion, and lastly in 1-D for 3 particles.
Because we will examine three identical spin-zero bosons, we will build the formalism with that model in mind, but provide some generalization to fermions when appropriate.
For nonzero spin, both spatially symmetric and antisymmetric states combine with spin and isospin to create total states that are antisymmetric (symmetric) for fermions (bosons) under the interchange of particles, thus examining spin-zero bosons focuses on a subset of channels used in the more complicated nuclear many-body problem.


\subsection{HO basis for 1-dimension 2-body}
\label{ssec:HO1d2b}

One can use the HO-basis to expand the wave functions in eigenfunctions of the harmonic oscillator Hamiltonian.
The Schr\"{o}dinger equation is
\bea
	H_{HO} \ket{\phi_{n}} &=& E  \ket{\phi_{n}}, \label{eq:HOschro} \\
	H_{HO} &=& \frac{\hat{P}^{2}}{2m} + \frac{1}{2} m \Omega^{2} \hat{X}^{2}.
\eea
It is convenient to define the oscillator parameter (taking $m=1$) as
\beqn
	b = \sqrt{ \frac{1}{\Omega}}.
\eeqn
The solutions of~\ref{eq:HOschro} in momentum representation are, for integer $n$~\cite{Shankar},
\beqn
	\phi_{n}(p) = \sqrt{ \frac{b}{\sqrt{\pi} 2^{n} n! }} H_{n}(b p) e^{-\frac{1}{2} (b p)^2} \label{eq:HOwf}.
\eeqn
The $H_{n}(b k)$ appearing in Eq.~\eqref{eq:HOwf} are Hermite polynomials~\cite{Shankar}.
We see that normalizable wavefunctions can be expanded into the orthonormal set of HO-basis functions, typically with some truncation, $N_{max}$, as
\beqn
	\ket{\psi} = \sum_{i = 0}^{N_{max}} c_{i} \ket{\phi_{i}}.
\eeqn
This expansion is exact if $N_{max} = \infty$, but numerically, we must choose some finite $N_{max}$ which will create truncation errors.
We can transform potential matrix elements from plane-waves to HO-basis:
\bea
	V_{nm} &=& \bra{n} V \ket{m} , \nonumber \\
	V_{nm} &=& \int_{-\infty}^{\infty} dp\ dp'\ \braket{n}{p}\bra{p} V \ket{p'}\braket{p'}{m}. \label{eq:HOtrans}
\eea
When numerically generating the oscillator wave functions, it is best to use a recurrence relationship for Hermite polynomials,
\beqn
	H_{n} = 2 x H_{n-1} + 2 (n-1) H_{n-2}, \label{eq:HOrecurrence}
\eeqn
as multiplying and dividing factorials can quickly lead to numerical round-off errors.
It is important when expanding with large $N_{max}$ to utilize a finer-grained momentum mesh to accurately integrate over the many oscillations of the HO wave functions.

Once we have the HO-basis potential matrix elements, we can also simply construct the relative kinetic energy matrix using ladder operators.
We use the identity~\cite{Shankar},
\beqn
	\hat{P} =  i \sqrt{\frac{1}{2 b^{2}}}(a^{\dagger} - a ),
\eeqn
in the HO-basis and find the relative kinetic energy (see appendix~\ref{app:app_ho}),
\bea
	\bra{n} T^{(2)} \ket{m} = \frac{1}{4 b^{2}} ( (2n+1) \delta_{n,m} &+& \sqrt{(m+1)(m+2)} ~\delta_{n,m+2} \nonumber \\
	&+& \sqrt{(n+1)(n+2)} ~\delta_{n+2,m} ). \label{eq:HOT}
\eea
In~\eqref{eq:HOT}, the superscript $(2)$ means that this is the 2-body relative kinetic energy, and $\delta_{i,j}$ is a Kronecker delta.  
We see that the relative kinetic energy takes an infinite tri-diagonal form when the basis is not truncated.
As we will explore later, truncating $T$ in either momentum representation or in an HO-basis will have consequences when transforming between the two bases.

For indistinguishable particles, we must symmetrize (antisymmetrize) the Hamiltonian for bosons (fermions).
This is done quite simply for two particles in the HO-basis.
The 2-body problem can be split into purely even and odd states which are orthogonal to each other.
For spin-zero bosons, we keep the symmetric states (even-n) and omit all odd states (odd-n), and for spin-$\frac{1}{2}$ fermions we can mix the spatially symmetric even-n (antisymmetric odd-n) states with the antisymmetric singlet (symmetric triplet) spin states to create overall antisymmetric states.

The usual method for finding bound state energies will be to start with an analytic form of the potential in momentum representation, and then transform into HO-basis with Eq.~\eqref{eq:HOtrans}.  
Then we add the HO-basis relative kinetic energy matrix to form the full Hamiltonian matrix.
The Hamiltonian is diagonalized with any standard routine on a computer.
The 2-body problem can be solved numerically in momentum representation just as easily as in a HO-basis, but as we will see, adding a third particle complicates the problem in momentum representation, and thus the HO-basis greatly simplifies the calculation.



\subsection{HO basis for 3-dimension 2-body}
\label{ssec:HO3d2b}

Before adding a third particle, it is important to note that the HO-basis is used in realistic, 3-D calculations as well.
The 1-dimensional formalism can be extended to 3 dimensions.
We expand into partial waves which introduces a new parameter, $l$, the orbital angular momentum.
The solutions of the 3-D partial-wave harmonic oscillator Schr\"{o}dinger equation in momentum representation, with energy, $E = 2 n + l + \frac{3}{2}$, are~\cite{Furnstahl:2013vda}:
\beqn
	\psi_{nlm}(\mathbf{k}) = (-i)^{l} \sqrt{\frac{2n! b}{\Gamma(n+l+\frac{3}{2})}} \frac{(kb)^{l+1}}{k} e^{-\frac{(kb)^{2}}{2}} L_{n}^{l+\frac{1}{2}}\left( (kb)^{2} \right) Y_{lm}(\mathbf{\hat{k}}),
\eeqn
where $L_{n}^{l+\frac{1}{2}}(x)$ are the generalized Laguerre polynomials, $\Gamma(x)$ is the gamma function, and $Y_{lm}(\bf{\hat{k}})$ are the spherical harmonics.


\subsection{HO basis for 1-dimension 3-body}
\label{ssec:HO1d3b}

\begin{figure*}
	{\includegraphics[width=.8\textwidth]{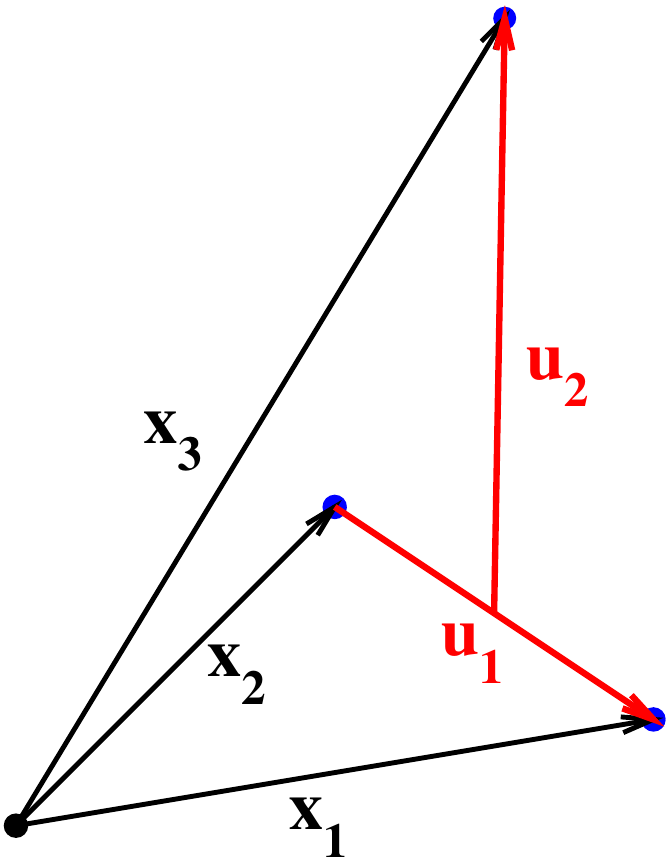}\caption{ Jacobi coordinates, $u_{1} = \sqrt{2}  r$ and $u_{1} = \sqrt{\frac{3}{2}} R$, and particle positions as vectors in 2-D.}\label{fig:ch4_jacobivec}}
	\hfill
\end{figure*}

The three-body problem introduces a great deal of complexity into solving bound- and scattering-state problems.  
Here we examine only bound states of our 1-D systems.
The first step in simplifying the problem, much like in two-body systems, is changing the degrees of freedom from the individual particles' positions or momenta to a reduced set of degrees of freedom by subtracting out the center-of-mass motion.
Taking the position of each particle as $x_{1},x_{2},x_{3}$ and assuming equal-mass particles, we can define a set of Jacobi coordinates $(r,R)_{i}$, where $i$ indicates the \emph{spectator} particle.
Unless otherwise noted, we will choose particle 3 as the spectator particle, and then we define:
\bea
	r &=& \frac{1}{\sqrt{2}}( x_{1} - x_{2} ) , \\
	R &=& \sqrt{\frac{2}{3}} \left( \frac{1}{2}( x_{1} + x_{2} ) - x_{3} \right).
\eea
We see that, up to a constant, $r$ corresponds to the separation of the \emph{pair} of particles 1 and 2, and that $R$ corresponds to the distance of particle three from the center of mass of particles 1 and 2 (if all masses are equal).
The choice of the constant is to ensure that the relative kinetic energy operator is symmetric in terms of the conjugate momenta $p$ and $q$, defined in terms of single particle momenta, $k_{i}$, as:
\bea
	p &=& \frac{1}{\sqrt{2}}( k_{1} - k_{2} ) , \\
	q &=& \sqrt{\frac{2}{3}} \left( \frac{1}{2}( k_{1} + k_{2} ) - k_{3} \right).
\eea
Thus the total relative kinetic energy operator is:
\beqn
	\hat{T}^{(3)} = \frac{\hat{p}^{2}}{2 m} + \frac{\hat{q}^{2}}{2 m},
\eeqn
where $m$ is the mass of each particle.
Figure~\ref{fig:ch4_jacobivec} shows the Jacobi coordinates for a 2-D arrangement of three particles.
We show two dimensions for better visualization of each vector.
From here we can transform from plane-wave basis in Jacobi momenta (basis states $\ket{p q} \equiv \ket{p} \ket{q}$) to HO-basis (basis states $\ket{\np \nq}  \equiv \ket{\np} \ket{\nq}$) by transforming each momentum separately, namely:
\bea
	\braket{p q}{ \np \nq} &=&  \phi_{\np}(p) \phi_{\nq}(q) \\
	\bra{\np \nq} V \ket{\npp \nqp} &=& \int dp\ dq\ dp'\ dq'\ \braket{ \np \nq}{p q} \bra{p q} V \ket{p' q'} \braket{p' q'}{ \npp \nqp}.
\eea
It is important to make our $N_{max}$ truncation on \emph{total} oscillator number, rather than on each $\np$ and $\nq$ independently.  
This is equivalent to making a truncation in total energy.
The truncation is important later to ensure that transformations between bases with different spectator particles are unitary, which is required for the symmetrization process.

Now we can emphasize the difficulty with embedding 2-body potentials into a 3-body plane-wave basis, and the simplification of HO-basis.
The 3-body Hamiltonian has three different 2-body interactions; one for each unique pair of particles.
If we choose to work in Jacobi momenta with particle 3 as the spectator, the interaction between particles 1 and 2 has a simple form:
\beqn
	V^{(2,12)}(p,q;p',q') = V^{(2)}(p,p') \delta(q-q').
\eeqn
Here, the superscript, $(2,12)$ refers to the 2-body pair interaction between particles 1 and 2 embedded into the 3-body basis.

Thus we see the \emph{spectator delta-function} for the non-interacting \emph{spectator} particle.
In this basis, however, the other two pair interactions are more complicated.
There is a linear transformation between sets of Jacobi momenta with different spectator particles, so we can write each Jacobi momentum with a given spectator index as a function of both Jacobi momenta with a different spectator index;
\bea
	p_i ( p_j q_j ) = \alpha p_j + \beta q_j , \\
	q_i ( p_j q_j ) = \gamma p_j + \delta q_j .
\eea
Here, the subscript on momenta refers to the spectator particle, and thus which set of Jacobi momenta $p$ or $q$ belongs to.

The values of the constants $\{\alpha, \beta,\gamma,\delta\}$ are calculated given the definition of Jacobi momenta.
Now, the other two pair-wise interactions are written in their own Jacobi momenta, which are taken as functions of the first set of Jacobi momenta:
\bea
	V^{(3,23)}(p,q;p',q') = V^{(2)}\left(p_1(p , q),p_1'(p , q)\right) \; \delta \left(q_1(p , q)-q_1'(p , q)\right), \\
	V^{(3,31)}(p,q;p',q') = V^{(2)}\left(p_2(p , q),p_2'(p , q)\right) \; \delta \left(q_2(p , q)-q_2'(p , q)\right).
\eea
This often requires interpolation when solving numerically, as the Jacobi momenta for different spectators cannot be all set on the same momentum mesh and thus must be integrated out~\cite{Hebeler:2012pr,Akerlund:2011cb}.
There are other subtleties in constructing the potential matrix involving consistent domains for momentum integrals~\cite{Hebeler:2012pr}, and after the potential is constructed, one must use Fadeev methods; finding poles in the 3-body multi-channel $T$-matrix; to find the bound state~\cite{Hebeler:2012pr}.

In the HO-basis, the delta-functions are not Dirac delta-functions but rather Kronecker delta functions.
Thus, the 2-body potential between particles 1 and 2 can be embedded into a 3-body basis with spectator particle 3 as:
\beqn
	V^{(2,12)}_{\np \nq \npp \nqp} = V^{(2)}_{\np \npp} \; \delta_{\nq \nqp}.
\eeqn
Because we are interested in spinless bosons for our 3-body study, we must symmetrize our basis.
We follow the same strategy as Jurgenson in Ref.~\cite{Jurgenson:2008jp}.
First, we symmetrize the 2-body states.
In the 2-body system, the symmetrizer is,
\beqn
	S^{(2)} = \frac{1}{2}( 1 + P_{1 2}),
\eeqn
where $P_{1 2}$ is the permutation operator between particles 1 and 2, and
\beqn
	P_{1 2} \ket{\np} = (-1)^{\np} \ket{\np}.
\eeqn
Now, the symmetrized states are the eigenstates of the symmetrizer with eigenvalue one.
We can clearly see that n-even states fulfill this requirement, thus to symmetrize the 2-body sector, we omit n-odd states.
We restrict our basis states  to $\ket{N_{p} \nq} \equiv \ket{N_{p}} \ket{\nq}$, where $N_{p}$ is an even integer.
We can then apply the 3-body symmetrizer,
\beqn
	S^{(3)} = \frac{1}{6} (1 + P_{1 2} + P_{2 3} + P_{1 2} P_{2 3} + P_{2 3} P_{1 2} + P_{1 2} P_{2 3} P_{1 2} ),
\eeqn
upon our partially-symmetrized basis states.
Because $\ket{N_{p}}$ is an eigenstate of $P_{1 2}$ with eigenvalue one, the symmetrizer reduces to
\beqn
	S^{(3)} = \frac{1}{3} (1 + 2 P_{2 3} ).
\eeqn
Refer to appendix \ref{app:app_ho} to calculate $\bra{N_{p} \nq} P_{2 3} \ket{N'_{p} \nqp}$, after which the HO-basis matrix elements of the 3-body symmetrizer, $\bra{N_{p} \nq} S^{(3)} \ket{N'_{p} \nqp}$, follow directly.

It is convenient to rearrange our basis once more, because our truncation is in total $N_{3} = N_{p} + \nq$.
We order states by energy and a degeneracy index, rather than the two oscillator parameters.
Doing this reorganization reduces the number of matrix elements by omitting many that have value zero in our total-n truncation scheme, and it also organizes the matrix elements into energy blocks.
This is not a transformation, but just a reorganization of matrix elements and a shrinking of the matrix by removing zero-valued elements.
The new basis can be written as $\ket{N_{3},i}$, where $i$ is simply a label for the degenerate states at each energy.
For our matrices, we use $i = \nq$.
Finding the set of eigenvectors with eigenvalue one of the 3-body symmetrizer in HO-basis will give us the coefficients of fractional parentage,
\beqn
	\braket{N_{3}, i}{N'_{3}, i'}_{S}.
\eeqn
The whole process can be set up numerically as a series of matrix multiplications to first transform from Jacobi plane-wave basis, $\ket{p q}$, then to the energy-truncated, partially-symmetric HO-basis, $\ket{N_{3},i}$, and finally to the fully-symmetric energy-truncated basis, $\ket{N_{3},i}_{S}$.
The embedded 2-body potential in the fully-symmetric basis is identical under any interchange of particles, thus the problem of separately treating each pair interaction is solved.  
One simply must multiply by a combinatoric factor, ${3 \choose 2}$, for the total embedded 2-body potential.
Any explicit 3-body potential will receive the same treatment, transforming from momentum representation to a partially symmetric HO-basis, to the full symmetric basis, but explicit 3-body potentials have no combinatoric factor. 

We must also build the relative kinetic energy operator in HO-basis with ladder operators in both the $n_{p}$ space and the $n_{q}$ space (see appendix~\ref{app:app_ho} for full calculation).
This matrix reduces to:
\beqn
	T^{(3)}_{\np \nq \npp \nqp} = T^{(2)}_{\np \npp} \delta_{\nq \nqp} + T^{(2)}_{\nq \nqp} \delta_{\np \npp}.
\eeqn
We create this matrix in the partially-symmetrized $\ket{N_{3},i}$ basis, then symmetrize by transforming into the symmetric basis, just like for the potential.

Once the embedded 2-body potentials, the explicit 3-body potential, and the relative kinetic energy are all in the symmetric HO-basis, eigenvalues and eigenvectors can be found by a simple matrix diagonalization routine.  
To extend the process to fermions, one needs only to use the odd $\np$ states, and then find the eigenvectors of the antisymmetrizer with eigenvalue one.  
The antisymmetrizer is constructed simply by multiplying each permutation operator in the symmetrizer by $(-1)$.
For fermions, one can attach a spin-like variable to emulate spin-$\frac{1}{2}$ particles, which complicates the antisymmetrization process.
For instance, in the 2-body sector the spin singlet ($S = 0$) is odd and must be paired with even $\np$, and the triplet with odd $\np$.
Thus, the potential would appear band-diagonal in spin-space.
The 3-body antisymmetrization operator becomes more complicated, but the process to find antisymmetric states remains the same.


\section{Basis transformations, truncations, and cutoffs}
\label{sec:HOtrunc}

We can now examine some subtleties in using HO-basis expansions for few-body calculations.  
As mentioned before, choosing a finite truncation, $N_{max}$, is required to numerically solve problems, but also creates truncation errors.
Some of these errors can be negated by tuning the HO parameter and keeping basis size sufficiently large, but certain operators are only accurately transformed numerically with an infinite basis (e.g. singular operators) .

\subsection{Cutoffs imposed by HO-basis expansion}
\begin{figure*}[htp!]
	{\includegraphics[width=.45\textwidth]{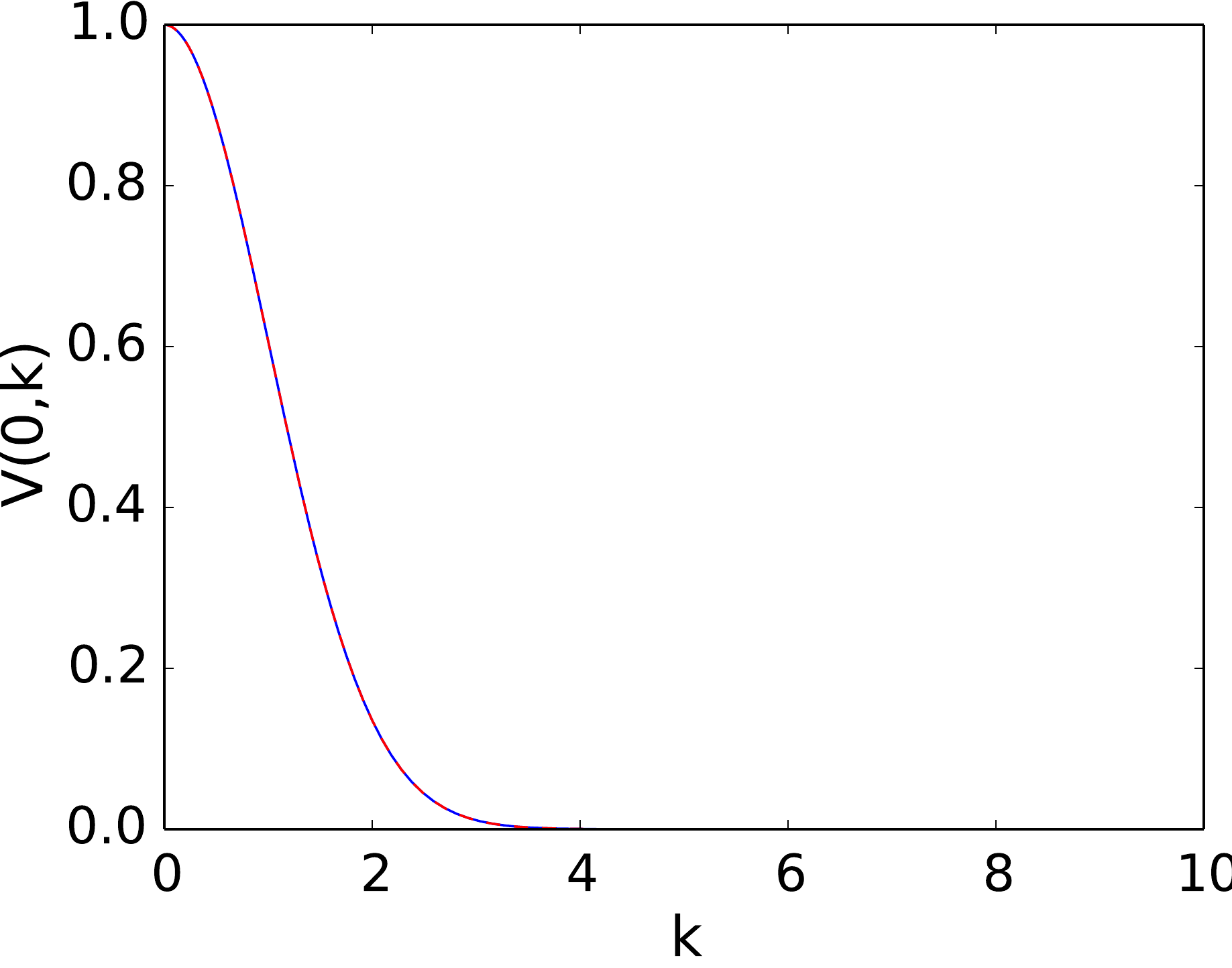}}
	{\includegraphics[width=.45\textwidth]{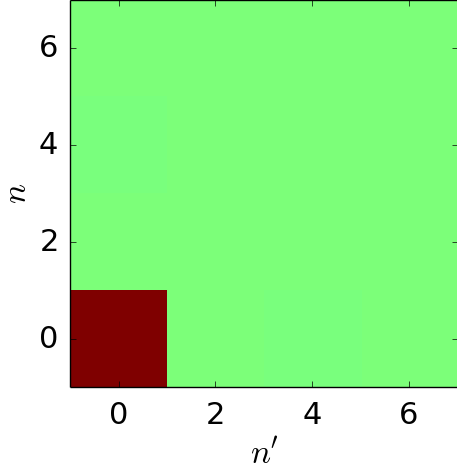}}\caption{ (Left) Off-diagonal matrix elements of our test potential matrix elements in momentum representation (solid blue line) and plotted again after transformation to HO-basis ($b=1$, $\nmax = 6$) and back (dashed red line). (Right) HO-basis potential matrix elements ($b=1$, $\nmax = 6$).  To best show occupancy, we plot $V_{n,n'} / V_{\rm max}$, the matrix elements divided by the max value.}
	\hfill
	\label{fig:ch4_trunc_1}
\end{figure*}

Because of the popularity of the HO-basis in realistic nuclear calculations, much work has been done to understand the basis truncation errors.
As proposed in Refs.~\cite{Konig:2014hma,More:2013rma}, using a finite HO-basis imposes both an infrared (IR) and ultraviolet (UV) cutoff.
We can think of these cutoffs as a length cutoff and momentum cutoff.
For 1-D these are,
\bea
	L_{HO} &=& b \sqrt{2 N_{max} + 5} \;,\\
	\Lambda_{HO} &=& \frac{1}{b} \sqrt{2 N_{max} + 5} \;.
\eea
Or, for 3-D S-waves,
\bea
	L_{HO} &=& b \sqrt{4 N_{max} + 7} \;,\\
	\Lambda_{HO} &=& \frac{1}{b} \sqrt{4 N_{max} + 7} \;.
\eea
As long as all non-negligible potential matrix elements are at separations below $L_{HO}$ in coordinate representation and momenta below $\Lambda_{HO}$ in momentum representation, then we can transform between plane-wave and HO-basis with high precision.
This gives us a method by which we can calculate the requirements on $b$ and $N_{max}$ for accuracy in our basis transformations.
If our basis size is too small, or $b$ is not tuned properly, then the HO-basis truncation will truncate non-negligible matrix elements, and thus transformation back into plane-waves will produce oscillatory errors in the matrix elements.
These effects are shown in Figs.~(\ref{fig:ch4_trunc_1}-\ref{fig:ch4_trunc_2c}).
We can see an indication of an inaccurate basis transformation in the matrix elements of the potential in HO-basis if the matrix elements along the edges are nonzero.
This implies that we have truncated nonzero matrix elements and thus transformation to momentum representation will produce errors in the potential matrix elements.
The transformation back into plane waves is missing the high-$n$ basis functions, and thus we see oscillations in the matrix elements.

To illustrate this point, we choose a very simple potential matrix,
\bea
	V(p,p') = e^{-\frac{p^{2}}{2}} e^{-\frac{p'^{2}}{2}}, \\
	V(r,r') = e^{-\frac{r^{2}}{2}} e^{-\frac{r'^{2}}{2}}.
\eea
%
We choose this form because of the ease of Fourier transforming, the presence of a cutoff momentum, and the symmetry in momentum and coordinate representation, which allows us to illustrate the requirements both on $\Lambdaho$ and $\Lho$.
Observing Fig.~\ref{fig:ch4_trunc_1}, we see the off-diagonal matrix elements of the potential in momentum representation and a contour plot of the potential in HO-basis.
We choose $b=1.0$ because of the symmetry between momentum and coordinate representation, and $\nmax=6$ for convenience (for this simple case, it is clear that only one oscillator is necessary).
This potential is essentially an outer product of the $n=0$ HO wave function, thus only $n=0$ is required, but taking $b$ close to, but not equal to $1.0$ continuously adds small values to higher-$n$ matrix elements.

This choice of $b$ and $\nmax$ allows for accurate transformation back and forth between HO-basis and plane-waves.
This is no surprise, as calculation of the HO-induced cutoffs yields $\Lambdaho = \Lho = \sqrt{17}$.
Observing the off-diagonal slice of the potential, this is in the area where potential matrix elements are negligible.
Because of the symmetry between momentum and coordinate representation for this potential, and $b=1$, the same is true for coordinate representation.

\begin{figure*}[htp!]
	{\includegraphics[width=.45\textwidth]{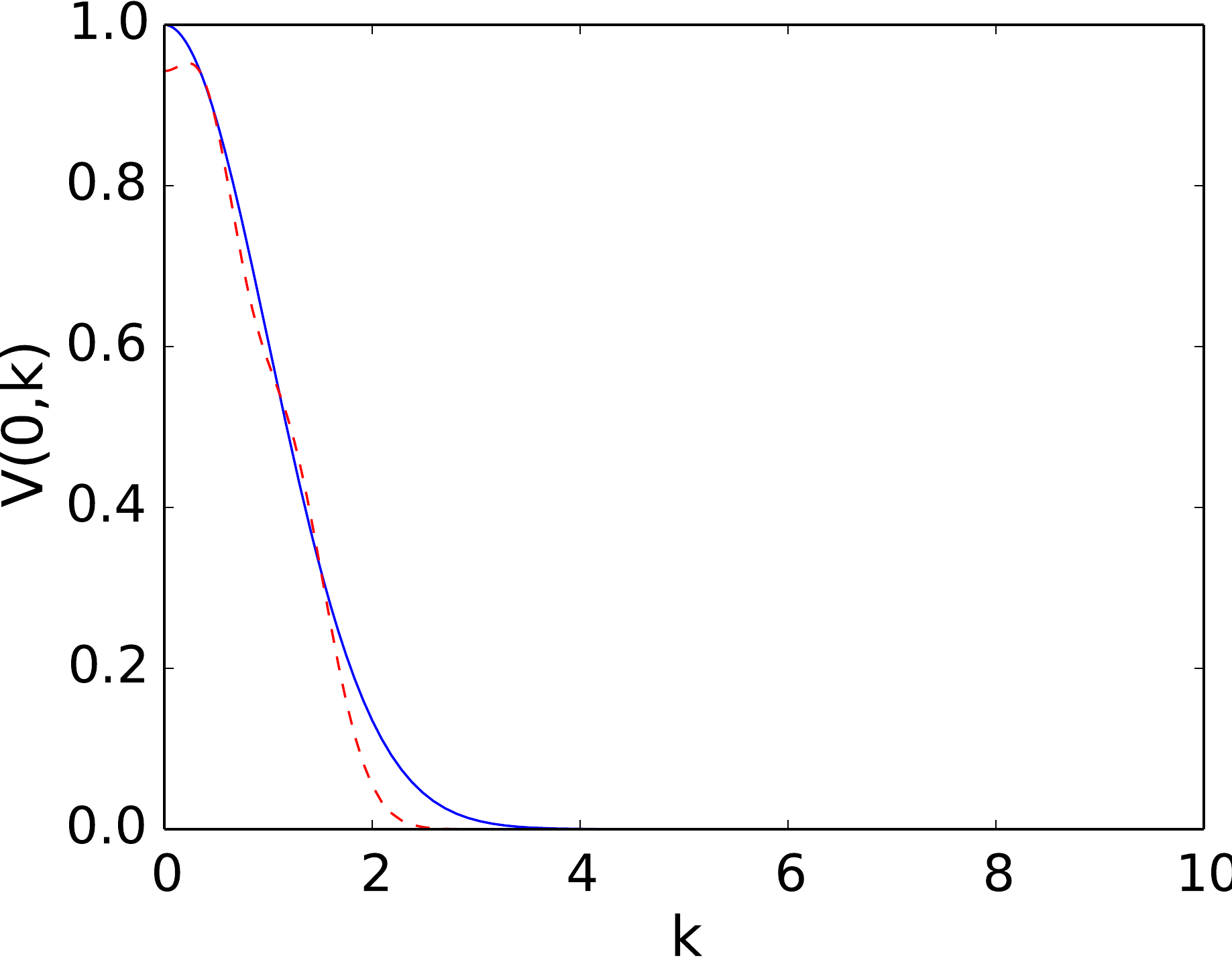}}
	{\includegraphics[width=.45\textwidth]{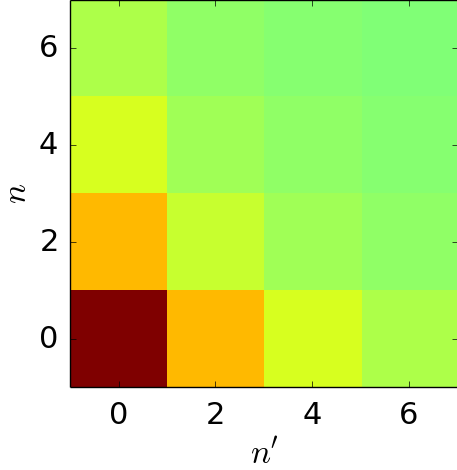}}
	\caption{Same as Fig.~\ref{fig:ch4_trunc_1}, but for $b=2$, $\nmax = 6$}
	\hfill
	\label{fig:ch4_trunc_2a}
\end{figure*}

Now, if instead we transform to an HO-basis with $b=2.0$, and the same $\nmax$, we can calculate the HO-induced cutoffs and find $\Lambdaho = \frac{\sqrt{17}}{2}$ and $\Lho = 2 \sqrt{17}$.
The IR error actually improves, but the induced momentum cutoff truncates non-negligible matrix elements, thus there will be some error in transforming from HO-basis to plane-wave basis and visa versa (actually, there is no error transforming from coordinate representation to HO-basis, but going back, or any transformation to and from momentum representation will not be accurate).
We can see a contour plot of the new HO-basis potential matrix elements and a cut of the momentum representation matrix elements of the original potential and twice transformed potential (transformed from momentum representation to HO, and then back) in Fig.~\ref{fig:ch4_trunc_2a}.

We can clearly see that the HO-basis matrix elements are non-negligible at the edge of the basis truncation (and will see later that indeed we have truncated nonzero elements).
The effect of this truncation is clear when observing the cut of momentum representation potential matrix elements; the cutoff forces higher momentum matrix elements to zero and oscillatory errors appear at momenta below the cutoff.

\begin{figure*}[htp!]
	{\includegraphics[width=.45\textwidth]{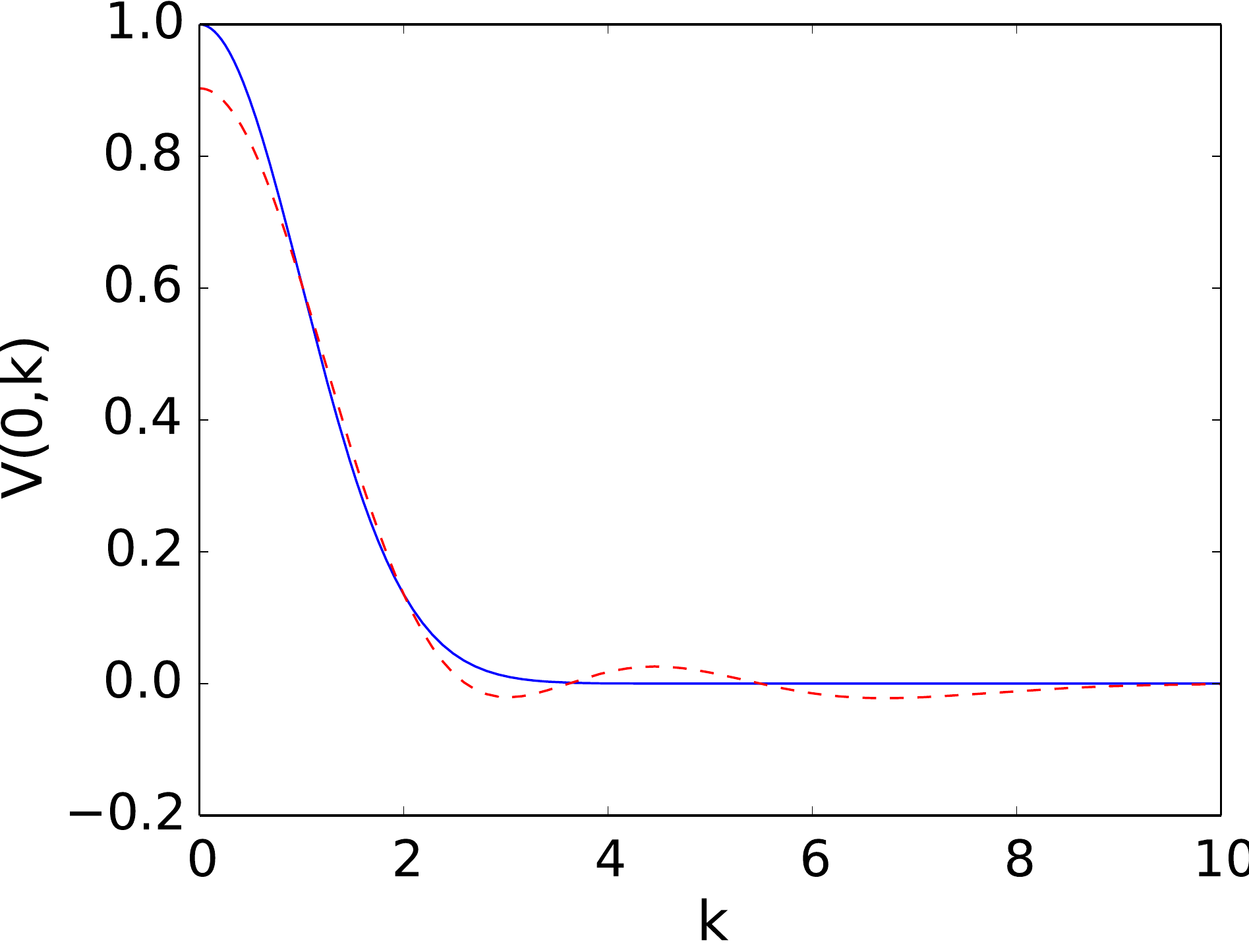}}
	{\includegraphics[width=.45\textwidth]{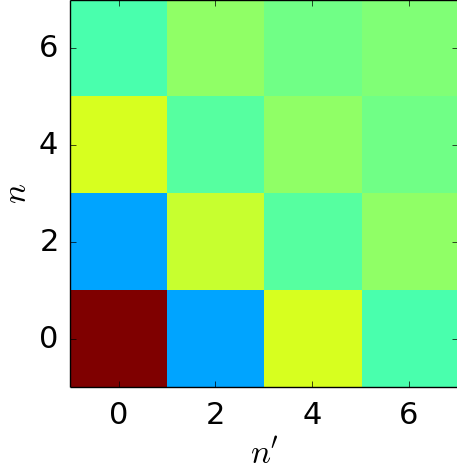}}
	\caption{Same as Fig.~\ref{fig:ch4_trunc_1}, but for $b=0.5$, $\nmax = 6$}
	\hfill
	\label{fig:ch4_trunc_2b}
\end{figure*}

The message is similar if we choose $b=0.5$ as well.
Now, the UV cutoff is higher, thus transformation from momentum representation to HO-basis will actually be more accurate, but transformation back, or to/from position representation will have truncation errors.
We can see this in Fig.~\ref{fig:ch4_trunc_2b}.

The truncation introduces IR errors, and sure enough, observation of the cut of momentum representation matrix elements reveals the characteristic oscillation of a long-range truncation.
A plot in coordinate representation would look very much the same as Fig.~\ref{fig:ch4_trunc_2a}.
Once again, the HO-basis matrix elements are nonzero at the edges, and thus there will be an incomplete transformation.

Often, the choice of $b$ is made to minimize UV or IR errors, or the error on an observable, so to increase accuracy of a calculation we must increase $\nmax$.
For our test potential, we saw that $b=1.0$ is ideal for basis transformations, but if we want to get the same UV cutoff and accuracy for $b=2.0$, we must quadruple $\nmax$.

\begin{figure*}[htp!]\label{fig:ch4_trunc_2c}
	{\includegraphics[width=.45\textwidth]{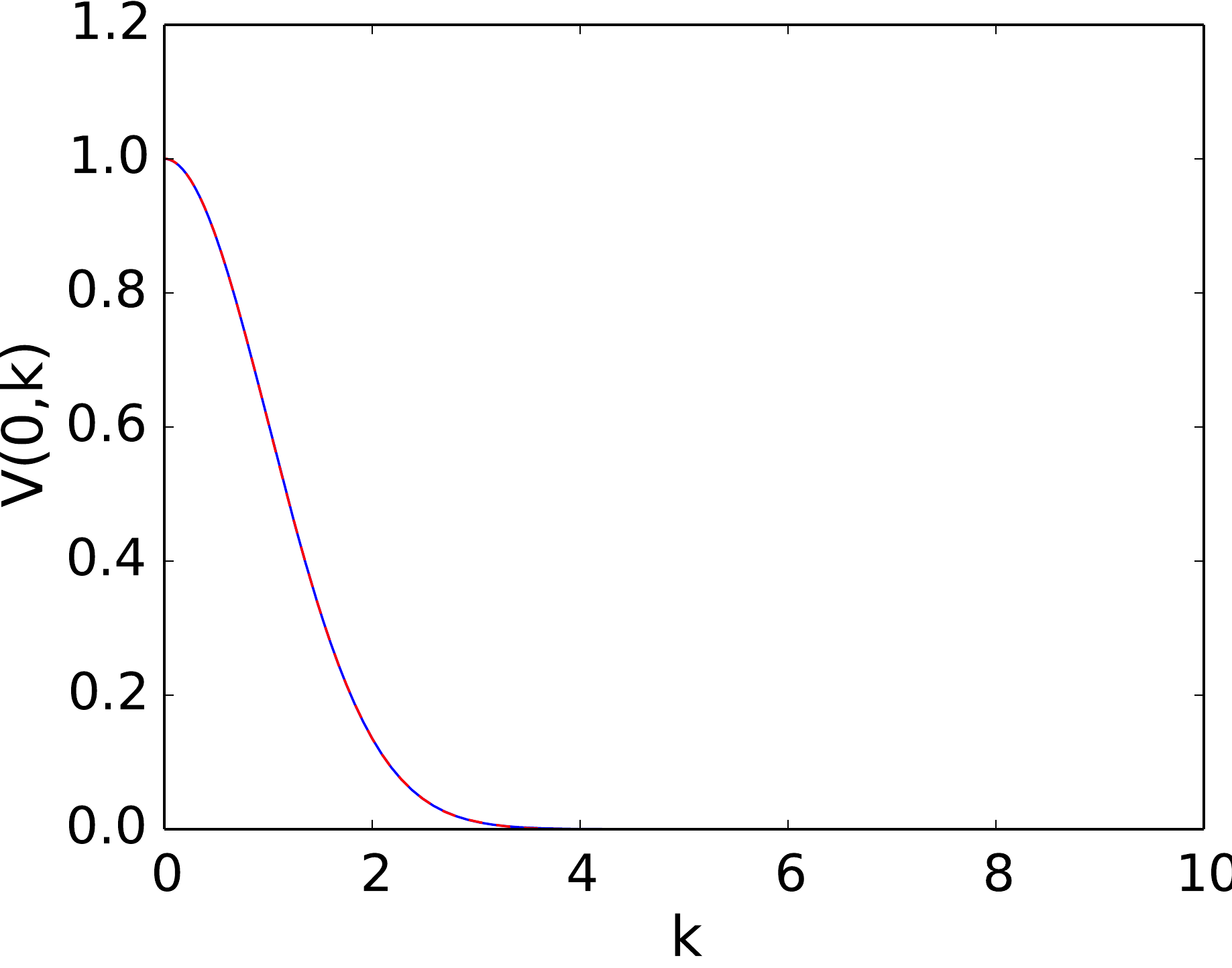}}
	{\includegraphics[width=.45\textwidth]{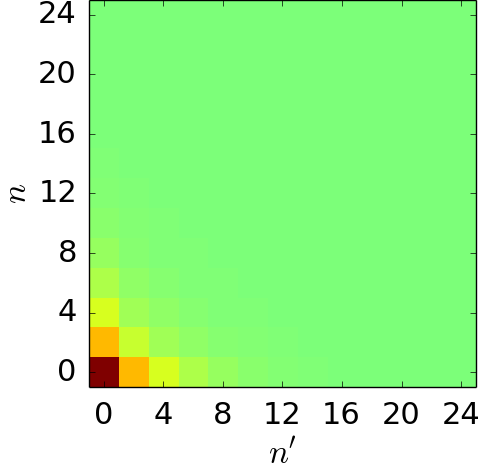}}
	\caption{Same as Fig.~\ref{fig:ch4_trunc_1}, but for $b=2$, $\nmax = 24$}
	\hfill
	\label{fig:ch4_trunc_2c}
\end{figure*}

Figure~\ref{fig:ch4_trunc_2c} shows the HO-basis potential matrix elements and a cut of the momentum representation matrix elements of the original potential and the doubly transformed potential.
We can clearly see that our new $\nmax$ truncation is large enough such that only zero-valued matrix elements are omitted.
Increasing accuracy in HO-basis transformations by increasing $\nmax$ is a costly game, as the scaling of the cutoff is proportional to $N_{\rm max}^{2}$.

In practice, when observing potential matrix elements, we can look for the same indications of inaccurate basis transformations due to truncation.
If we observe unexplained oscillations in momentum representation potential matrix elements, or nonzero valued matrix elements at the edges of our HO-basis potential matrix, then we know that our truncation has induced errors in the matrix elements.

Our goal is to observe universality in the matrix elements, thus oscillatory errors will make it impossible to follow the same strategy as before, so we require very accurate basis transformations.
Often, such accuracy in basis transformations is unnecessary for calculating certain observables (e.g. ground state energy of a 2-body system), but accuracy in all non-zero matrix elements and likely all many-body observables requires larger HO-bases.


For our calculation, our choice of $\nmax$ will be large enough such that there are no transformation errors identified as before, and $b$ will be such that the 2-body ground state binding energy is accurate to several digits (which will mean low IR errors).
This will allow us to choose a preliminary basis size to accurately transform between plane-wave and HO-basis for \emph{unevolved} potentials, but as we will see in Section ~\ref{sec:HOSRG}, accurate SRG evolution of the potentials requires more careful consideration of oscillator parameter and basis size.


\subsection{Transforming singular functions}
\label{ssec:transsingfunc}

Numerical transformation of singular functions, specifically the Dirac $\delta$-function between plane-wave and HO-basis is one major limitation of choosing a finite $N_{max}$ truncation.
Our $\delta$-shell fitting procedure is introduced using Dirac $\delta$-functions in coordinate representation, and the relative kinetic energy operator introduces a $\delta$-function in momentum representation.
The fitting procedure requires only narrowly peaked functions which will be fit to data, thus we can instead use regulated $\delta$-functions and completely bypass basis-transformation complications, but altering the $\delta$-function in the kinetic energy imposes unavoidable errors.

To better understand the limitations of singular functions, we examine the $\delta$-functions in coordinate representation.  
One may assume that a simple solution would be transforming to HO-basis analytically and then truncating the space.
We see this is simply done:
\bea
	\Delta_{n m}(a) &=& \int dr\ dr'\ \phi_{n}(r) \delta(r - a) \delta(r - r') \phi_{m}(r'), \\
	\Delta_{n m}(a) &=& \phi_{n}(a) \phi_{m}(a).
\eea
The trouble lies in the normalization factor in the oscillator wave-functions.
If we wish to keep errors negligible, then we should choose a sufficiently large $N_{max}$ such that the truncated matrix elements are negligible.
Using the closed form for Hermite polynomials with argument zero (called Hermite numbers)~\cite{NIST:DLMF,Olver:2010:NHMF},
\bea
	\Delta_{n m}(0) &=& \phi_{n}(0) \phi_{m}(0), \\
	\Delta_{n m}(0) &=& \frac{(n-1)!!}{\sqrt{n!}}\frac{(m-1)!!}{\sqrt{m!}} ,
\eea
where $n!!$ denotes the double factorial.
Utilizing an identity of the double factorial~\cite{NIST:DLMF,Olver:2010:NHMF}~\cite{NIST:DLMF,Olver:2010:NHMF}, this can be expressed for non-zero, even $n$ and $m$ as
\beqn
	\Delta_{n m}(0) = \left( \prod_{i=1}^{n} \sqrt{\frac{2 i - 1}{ 2 i}} \right) \left( \prod_{j=1}^{m} \sqrt{\frac{2 j - 1}{ 2 j}} \right).
\eeqn
This is \emph{extremely} slow to converge and thus any finite basis HO-basis we hope to use will have truncated non-negligible matrix elements.

Instead, we choose to use regulated $\delta$-functions for our fitting procedure.
We do not require explicit $\delta$-functions for our fitting procedure, simply narrow-peaked functions in coordinate representation, and by regulating, we ensure that we avoid the complications of non-zero matrix elements at the edge of our basis truncation.
We regulate the coordinate representation $\delta$-functions by first Fourier-transforming to momentum representation and then imposing a soft momentum cut-off with a regulating function; just like before in Section~\ref{sec:OPE}.
Then, we transform this regulated $\delta$-function to HO-basis.
If the regulating function is not too sharp, and the cut-off is sufficiently small, then this regulated $\delta$-function can be transformed between HO-basis and plane-wave basis with high accuracy.
Because the fitting routine we use does not rely on the functions specifically being $\delta$-functions, this regulated $\delta$-function works just the same in creating the 2-body potential.


Alternatively, the relative kinetic energy operator in a plane-wave basis is a $\delta$-function in momentum representation,
\beqn
	\bra{p} T \ket{p'} = \frac{p^{2}}{2m}\delta(p - p').
\eeqn
We can not accurately replace this operator with a regulated $\delta$-function, so we cannot accurately transform the relative kinetic energy matrices between plane-wave and HO-basis.
Regulating (or simply imposing a momentum cutoff) causes errors in all of the high-n, HO-basis kinetic energy matrix elements which will make calculations for observables impossible.
Instead, we use the analytic expression in each basis.
Because of basis truncation, the operator represented in one basis will not be identical to the representation in another basis.
Because the kinetic energy is tri-diagonal in HO-basis, this method isolates the truncation error to the two matrix elements above our truncation which mix the omitted and retained matrix elements.
Using this method, we are able to accurately reproduce observables in the 2- and 3-body sectors.
Because of this restriction, we can only transform potentials between bases and not the full Hamiltonian.
To transform a Hamiltonian in one basis to another, we subtract the relative kinetic energy matrix calculated in the original basis, then transform the potential and add back the relative kinetic energy matrix calculated in the new basis.
In the next section, we will see that differences in $\Trel$ in each basis will create differences in the SRG evolution of potentials.


\section{SRG differences in plane-wave and HO basis}
\label{sec:HOSRG}

Because of the subtleties in transforming certain operators from momentum representation to HO-basis, certain generators of SRG flow become impractical when evolving in HO-basis.
For instance the Wegner generator, $G(s)=H_{d}(s)$, is basis-dependent, as the diagonal potential matrix elements in momentum representation transformed to HO-basis is no longer a diagonal matrix.
If we were to use the Wegner operator, at each step in the SRG evolution routine, we would have to transform the potential to the plane-wave basis, then transform only the diagonal elements back into HO-basis to form the generator, making the evolution computationally cumbersome.
The best candidates for evolution in HO-basis are well defined entirely in HO-basis at every $\lambda$.
For this study, we will use $\Trel$, as it is completely defined in HO-basis and also is not $\lambda$ dependent, although we have discussed that the basis truncation makes $\Trel$ different in different bases.

We have seen earlier that evolving potentials with $\Trel$ as a generator in momentum representation softens the potential.
This has the effect of reducing the required $N_{max}$ for accurate transformation to HO-basis. 
One can truncate the decoupled high-momentum matrix elements of a SRG-softened potential, effectively reducing momentum cutoff, $\Lambda_{\rm pot}$, above which matrix elements are zero.
This means that $\Lambda_{HO}$ can be smaller, and $N_{max}$ and $b$ can be tuned such that a smaller basis has no additional truncation error.

Because of the basis truncations and finite mesh sizes, the $\Trel$ matrix in plane-wave and HO-basis is no longer identically the same operator, and we should expect to see differences in evolved potential matrix elements at some point in the SRG evolution.
In momentum representation, $\Lambda_{\rm pot}$ is effectively reduced by SRG evolution, but what about in HO-basis?
A very simple argument can be used to at least suggest that $\Lambda_{\rm pot}$ in momentum representation cannot \emph{increase} with SRG flow with generator $\Trel$ (or any generator diagonal in the basis).
We can define a corresponding $N_{pot} \leq N_{max}$ in HO-basis corresponding to some value of $n$ above which all matrix elements are identically zero, but at which the potential is nonzero.
The same argument shows that under SRG flow with generator $\Trel$ in HO-basis, there are terms which act to increase $N_{pot}$.

Take four matrices, $D_{i j},T_{i j},M_{i j}$ and $V_{i j}$ corresponding to a diagonal matrix, tri-diagonal matrix, and two different truncated matrices respectively.
The truncated matrices are constructed so that we can define some $n$ such that
\beqn
	V_{i j},M_{i j} \equiv 0 \text{ if } i \text{ or } j > n.
\eeqn
Then, looking at what happens when $V_{i j}$ is multiplied by $D_{i j},T_{i j},$ or $M_{i j}$
\bea
	(DV)_{i j} &=& d_{i} V_{i j}, \\
	(TV)_{i j} &=& a_{i} V_{i j} + b_{i} V_{i j-1} + c_{i} V_{i-1 j}, \text{ if } i,j > 1, \\
	(M V)_{i j} &=& \sum_{l} M_{i l} V_{l j}.
\eea
The values of coefficients $a_{i}$, $b_{i}$, $c_{i}$ and $d_{i}$ are generally not important.
We see then
\bea
	(DV)_{(n+1) j} &=& V_{(n+1) j} \equiv 0, \\
	(TV)_{(n+1) j} &=& a_{i} V_{(n+1) j} + b_{i} V_{(n+1) (j-1)} + c_{i} V_{n j} = c_{i} V_{n j}, \text{ if } i,j > 1, \\
	(M V)_{(n+1) j} &=& \sum_{l} M_{(n+1) l} V_{l j} \equiv 0.
\eea
Therefore, the only one of these matrix multiplications that increases the value of $n$ is $TV$.
The same is true for reversed-order multiplications.
If we examine the RHS of the flow Eq.~\eqref{eq:srgdiffeq}, we see that it is comprised entirely of these three operations and their reverse orders.
Particularly, the RHS of the flow equation in momentum representation has no tri-diagonal terms, thus elements with index above $n$ will never change from zero under SRG flow.
If we extend this from a discrete matrix back to a continuous potential, then, $\Lambda_{\rm pot}$ will not grow as we evolve in momentum representation with $\Trel$.
Alternatively, because $\Trel$ in HO-basis is tri-diagonal, SRG flow can increase $N_{pot}$ at each step.
Indeed, when we examine the evolution of potentials in HO-basis, we see that they spread to higher n-values than required for unevolved potentials.
We shall see that SRG evolution in the HO-basis can push the truncation requirements higher than we observed for unevolved potentials in Section~\ref{sec:HOtrunc}.
It is also important to remember that the SRG is a continuous set of \emph{unitary} transformations, so even if the HO-basis truncation is too small and will create errors in matrix elements, it will still reproduce the same eigenvalues as the unevolved potential up to numerical errors in the algorithm.

Another source of differences between evolution in each basis is the finite momentum mesh size.  
For larger $\nmax$, we must choose a momentum grid with enough points to accurately integrate over the many oscillations of the oscillator wave function.
As the potential is evolved in momentum representation, however, the high-momentum matrix elements become sharply peaked on the diagonal.
For this reason, more points must be included in our momentum mesh for an accurate transformation to HO-basis.
This manifests as IR-errors, as the momentum difference between mesh points in momentum representation is related to a long-range cutoff in coordinate representation.

\begin{figure*}[htp!]
	{\includegraphics[width=.45\textwidth]{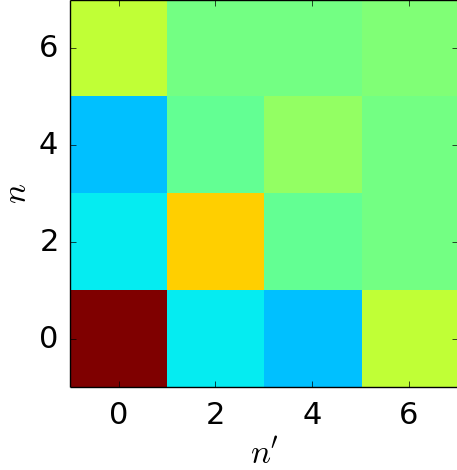}}
	\hfill
	{\includegraphics[width=.45\textwidth]{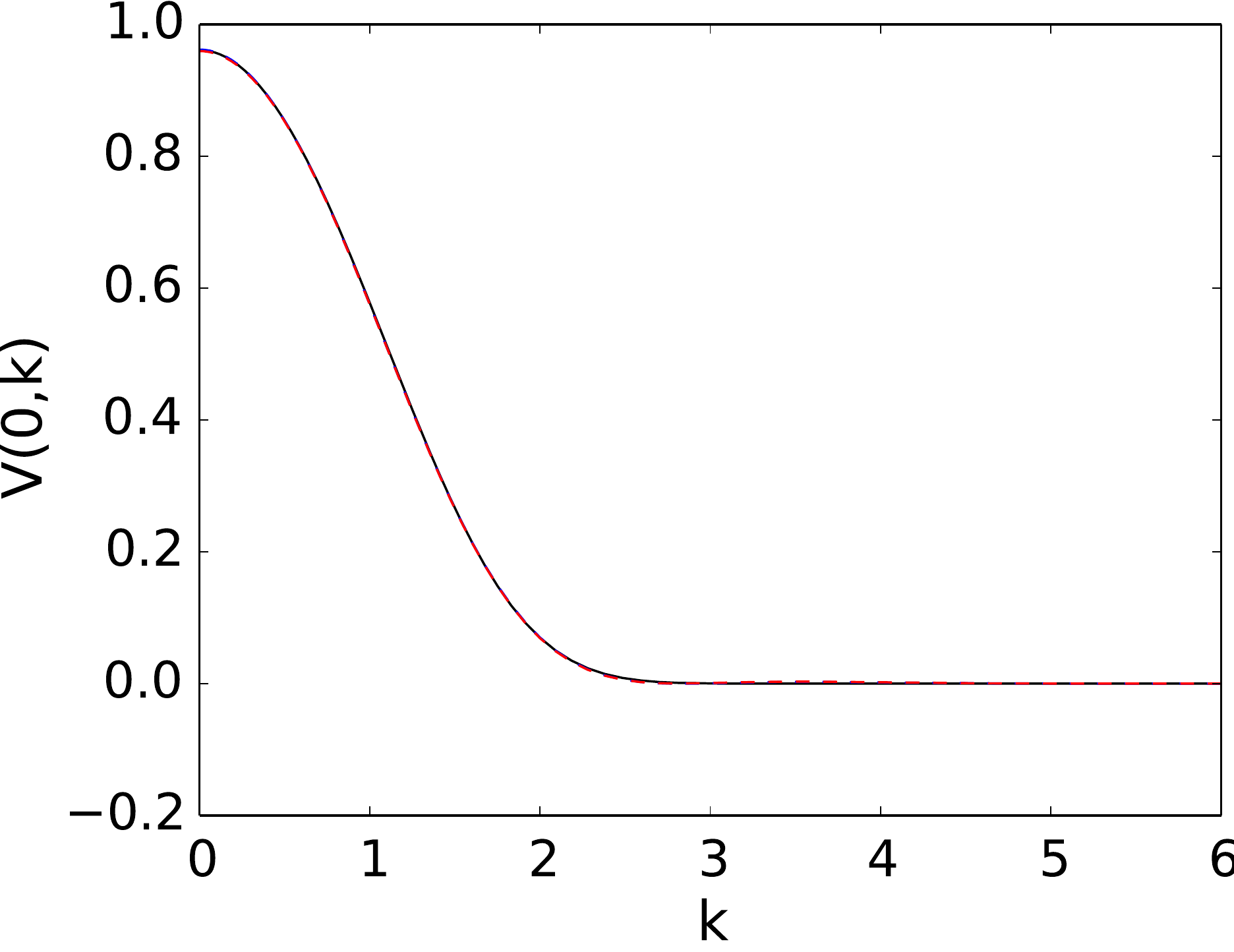}}
	\caption{ (Left) HO-basis ($b=1.0$, $\nmax = 6$) matrix elements of our test potential evolved to $\lambda = 2.0$ in HO-basis and scaled to 10 times their absolute maximum value.  (Right) Vertical slice of potential matrix elements evolved to $\lambda = 2.0$ in momentum representation.  The solid black line is calculated completely in momentum representation, the dashed red line is evolved in HO-basis, then transformed to momentum representation, and the dashed blue line is evolved in momentum representation then evolved to HO-basis and back.}
	\hfill
	\label{fig:ch4_srg_20}
\end{figure*}

\begin{figure*}[htp!]
	{\includegraphics[width=.45\textwidth]{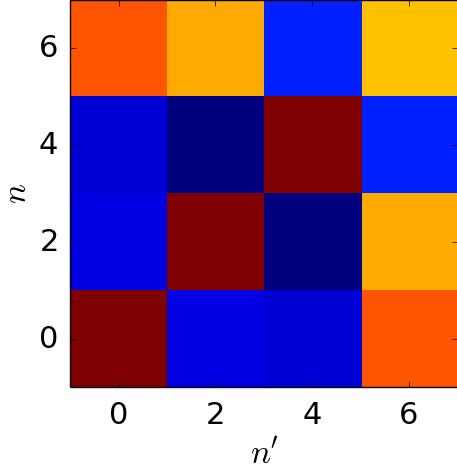}}
	\hfill
	{\includegraphics[width=.45\textwidth]{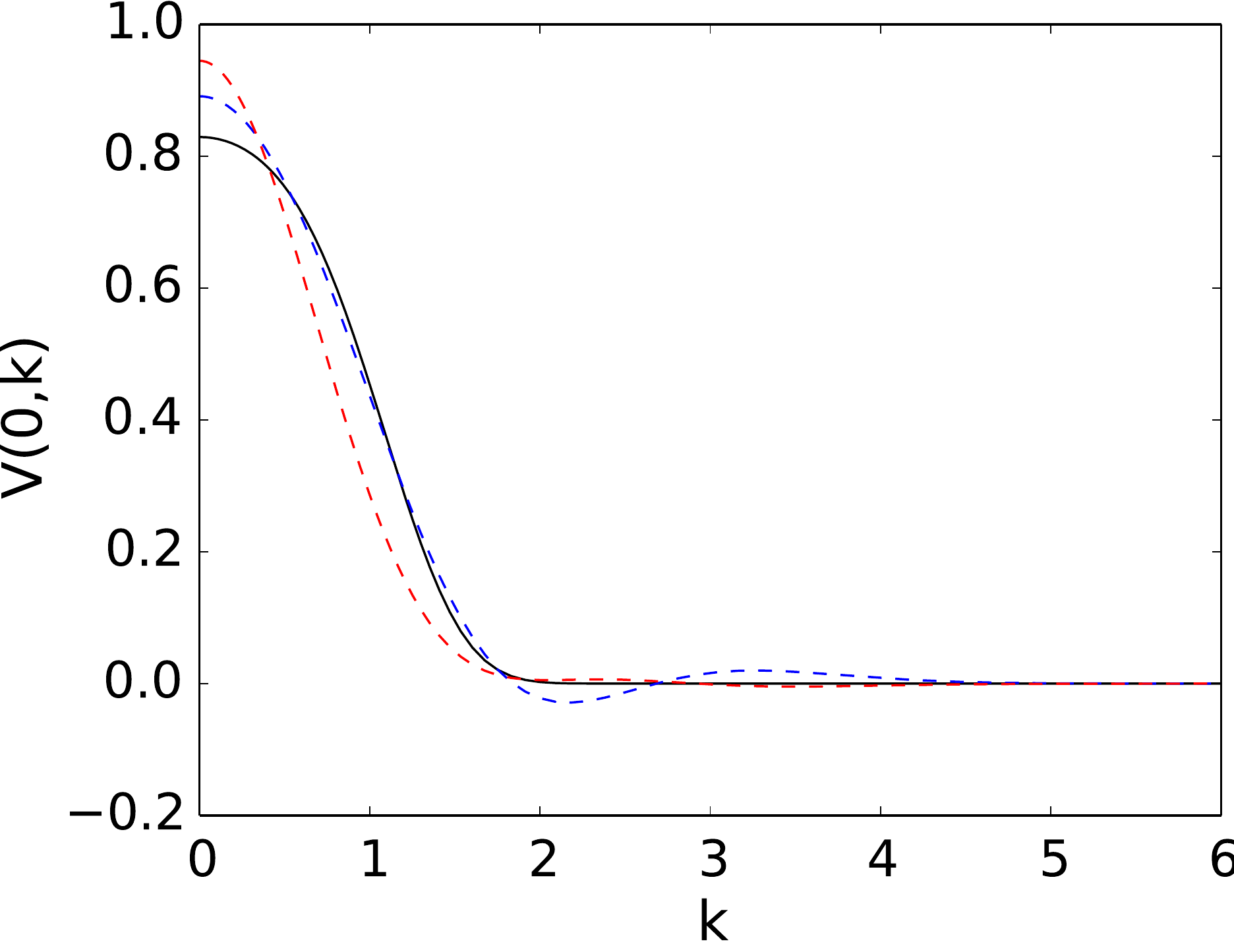}}
	\caption{Same as Fig.~\ref{fig:ch4_srg_20}, but for $\lambda = 0.9$}
	\hfill
	\label{fig:ch4_srg_9}
\end{figure*}

Figure~\ref{fig:ch4_srg_20} shows a contour plot of HO-basis potential matrix elements and a vertical slice of momentum representation matrix elements for our simple test potential evolved to $\lambda = 2.0$.
We can see that the HO-basis matrix elements along the edge of our truncation are small, but nonzero.
Looking at the momentum representation matrix elements, we see very small differences emerging at some momenta.
For this test potential, SRG evolution in each basis is the same until $\lambda = 2.0$.
If we evolve further to $\lambda = 0.9$, shown in Fig.~\ref{fig:ch4_srg_9}, we see that evolution in each basis becomes different.
At $\lambda = 0.9$, the HO-basis matrix elements have increased along the edge of the basis space.
The errors in momentum representation potential matrix elements have also increased.
The dashed blue line, which is produced by evolving in momentum representation then transforming to HO-basis and back, shows that there are errors from the basis transformation.
The dashed red line, which is produced by evolving in HO-basis and then transforming to momentum representation, differs from both the black and dashed blue line, showing that the evolution is different in each basis.

Unlike the unevolved imposed cutoffs HO-basis, the new requirements on basis truncation and oscillator parameter for accurate SRG evolution are not analytically understood, but a larger basis size allows for accurate SRG evolution to lower $\lambda$.
For this reason, some choose to transform between HO-basis with different $b$ and $N_{max}$ dynamically during their evolution~\cite{PhysRevC.90.024325}.
For our study, we will simply start with a sufficiently large basis and tuned oscillator parameter such that we do not observe the oscillatory errors and nonzero edges in the 2-body evolved potential matrix elements.


\section{HO-basis Recap}
\label{sec:HOrecap}

In this chapter, we have explored the many sources of errors when transforming between finite bases.
Transforming to HO-basis adds a cutoff in both momentum representation (UV cutoff, $\Lambda_{HO}$) and coordinate representation (IR cutoff, $L_{HO}$).
In the 2-body sector, if these cutoffs are large enough such that only zero-value matrix elements are omitted, then there will be no truncation errors in transformation between HO-basis and plane-wave basis.

Because singular functions cannot be accurately transformed between bases, we choose to use the analytic expression for the kinetic energy in each basis.
The kinetic energy matrix in a finite HO-basis is not the same operator as in a finite momentum-representation plane-wave basis, because of the truncation.
For this reason, SRG evolution performed in each basis will have differences that increase as $\lambda$ decreases, even if the unevolved potentials can be transformed between the basis without truncation error.
If we increase basis size, we can reduce the SRG differences in each basis, thus we can use the 2-body sector to define an $\nmax$ large enough such that SRG errors are negligible for the $\lambda$ we evolve to.

\chapter{Toward 3-Body Universality in a ``1-D Laboratory''}
\label{chapt:3BUniv}

With the formalism in place, we can begin the 3-body calculation in 1-D.
We will introduce a potential utilized in previous studies~\cite{Akerlund:2011cb,Jurgenson:2008jp} to confirm binding energies, then introduce a better potential for use in our study.
We will show that SRG evolution errors appear in the 3-body sector at much higher $\lambda$ than expected from the 2-body sector analysis of the previous chapter.
We can identify these errors in the potential matrix elements as truncation errors originating from the spectator $\delta$-functions.
Although the potentials reproduce 3-body observables, each step of the SRG evolution compounds the truncation errors, causing increasing errors in the evolved potential matrix elements as $\lambda$ decreases.
These errors make our study of universality in 3-body matrix elements impossible, but understanding them brings us closer to precision Hamiltonians for many-body nuclear problems.

\section{Model Potentials}

First, we will consider as a starting point the potential proposed in Ref.~\cite{Akerlund:2011cb,Jurgenson:2008jp}.
\beqn
	V_{2}(r) = \sum_{i=1}^{2} \frac{c_{i}}{\sigma_{i} \sqrt{\pi}}\ e^{-\frac{r^{2}}{\sigma_{i}^{2}}},
\eeqn
where $r$ is the separation between particles.
This potential is easily transformed into momentum representation.
We must be consistent in our 2- and 3- body bases, so we will use the Jacobi-coordinate, $l = r / \sqrt{2}$ and conjugate momentum, $p$, instead of the usual 2-body center of mass momentum.
Then, we see
\bea
	\tilde{V}_{2}(p,p') &=& \int_{0}^{\infty} \frac{dl}{2 \pi}\ e^{-\i l (p-p')}\ V_{2}(\sqrt(2) l) \\
	\tilde{V}_{2}(p,p') &=& \sum_{i=1}^{2} \frac{c_{i}}{2 \pi \sqrt{2}}\ e^{-\frac{\sigma_{i}^{2} (p-p'))^{2}}{8}}.
\eea
We can confirm the binding energies of Ref.~\cite{Akerlund:2011cb} for different choices of $\sigma_{i}$ and $c_{i}$, using $b=0.5 fm$ and $\nmax = 120$.
These results also employ a 3-body potential of the form,
\beqn
	V(p,q;p',q')= c_{3}\ e^{-(\frac{p^{2}+q^{2}}{4})^{4}}e^{-(\frac{p'^{2}+q'^{2}}{4})^{4}}.\label{eqn:3bpot}
\eeqn
\begin{center}
\begin{table}[h]
\resizebox{1.0\textwidth}{!}{\begin{minipage}{\textwidth}
\begin{center}
\begin{tabular}{|r | r | r | r | r|}
	\hline
	$ \alpha_{v2} $ & $c_{3}$ & $E_{2}$ & $E_{3}$ \\ \hline
	1 & -0.1 & -0.920 & -3.225 \\ \hline
	1 & -0.05 & -0.920 & -2.885 \\ \hline
	1 & 0.0 & -0.920 & -2.567 \\ \hline
	1 & 0.05 & -0.920 & -2.279 \\ \hline
	1 & 0.1 & -0.920 & -2.027 \\ \hline
	2 & -0.1 & -0.474 & -2.571 \\ \hline
	2 & -0.05 & -0.474 & -2.133 \\ \hline
	2 & 0.0 & -0.474 & -1.708 \\ \hline
	2 & 0.05 & -0.474 & -1.307 \\ \hline
	2 & 0.1 & -0.474 & -0.951 \\ \hline
\end{tabular}
\caption{3-body ground-state energies in fm$^{-2}$ for different parameters chosen in Ref. [30]. $\alpha_{v2}=1$ means that $c_{1} = 12.$ fm$^{-1}$, $c_{2} = -12.$ fm$^{-1}$, $\sigma_{1} = 0.2$ fm, and $\sigma_{2} = 0.8$ fm.  $\alpha_{v2}=2$ means that $c_{1} = 0.$ fm$^{-1}$, $c_{2} = -2.$ fm$^{-1}$, and $\sigma_{2} = 0.8.$ fm.  } 
\label{tab:constvals}
\end{center}
\end{minipage} }
\end{table}
\end{center}
%
This potential has one very nice benefit of a clear separation of scales; each Gaussian has a well defined length-scale, $\sigma$.
With a choice of $\sigma_{i}$, we can set up a ``long-range'' and ``short-range'' piece of the potential and carry out our $\delta$-shell fits.
The potential, however is not ideal for our calculation for a number of reasons.
The values of $\sigma_{i}$ and $c_{i}$ chosen in Ref.~\cite{Akerlund:2011cb,Jurgenson:2008jp} require a very large number of points for accurate matrix elements when transforming back and forth between momentum representation and HO-basis, or SRG evolution (due to the diagonal never diminishing at high energies).
Secondly, a Gaussian is not a good analogy for the long-ranged nature of the OPE potential.
Lastly, as is shown in Ref~\cite{Glazek:2008pg}, having a large ground state energy causes trouble if we need to evolve very low in $\lambda$.

Therefore, we will choose a better 2-body potential for studying universality in matrix elements.
We still utilize a Gaussian for the short-ranged piece, but will use an exponential for the long-ranged piece.
\beqn
	V_{2}(r) = \frac{c_{1}}{\sigma_{1} \sqrt{\pi}}\ e^{-\frac{r^{2}}{\sigma_{1}^{2}}} + c_{2}\ e^{- \frac{r}{\sigma_{2}}}.
\eeqn
We choose the constants such that there is some separation, $r$, above which the potential is almost entirely the long-ranged piece, and that the ground state binding energy is smaller than the $\lambda$'s we wish to evolve to.
For instance, for the values in Table~\ref{tab:constvals}, at $r=1.0$, the contribution from the long-range exponential is $-10.8$ fm$^{-2}$, while the contribution from the short-range piece is only $1.9\times10^{-9}$ fm$^{-2}$.
Also, we must make sure that the short-range piece is non-negligible in the calculation, such that it acts as a hard core and is strongly repulsive at short-range.

In Jacobi-momentum representation, the potential is:
\beqn
	\tilde{V}_{2}(p,p') = \frac{c_{1}}{2 \pi \sqrt{2}}\ e^{-\frac{\sigma_{1}^{2} (p-p'))^{2}}{8}} + \frac{c_{2} \sqrt{2}}{2 \pi} \frac{\sigma_{2}}{\frac{\sigma_{2}^{2} (p-p')^{2}}{2} + 1}.
\eeqn
We then, analogous to realistic chiral potentials, multiply by a separable regulating function:
\beqn
	\tilde{V}_{2}(p,p') \to \tilde{V}_{2}(p,p') e^{-(\frac{p}{\Lambda})^{2n}} e^{-(\frac{p'}{\Lambda})^{2n}}.
\eeqn

The following section will detail universality in the 2-body sector for our model potential, followed by our findings in the 3-body sector.

\section{2-body Sector}

For our study, we choose the values shown in Table~\ref{tab:constvals} to construct our potential.
\begin{center}
\begin{table}[h]
\resizebox{1.0\textwidth}{!}{\begin{minipage}{\textwidth}
\begin{center}
\begin{tabular}{|c | c | c | c | c| c|}
	\hline
	$c_{1}$ & $c_{2}$ & $\sigma_{1}$ & $\sigma_{1}$ & $\Lambda$ & $n$ \\ \hline
	48 fm$^{-1}$ &  -24 fm$^{-2}$ & 0.2 fm & 1.25 fm & 3.0 fm$^{-1}$ & 4 \\
	\hline
\end{tabular}
\caption{Parameters for 2-body potential}
\label{tab:constvals}
\end{center}
\end{minipage} }
\end{table}
\end{center}
This potential has a single bound state with energy, $E_{bd} = 0.402$ fm$^{-1}$.
The momentum regulator defines a well defined minimum for the UV cutoff from our HO-basis truncation.
We choose $b$ and $\nmax$ such that the truncation cutoff is above $\Lambda = 4.0$ fm$^{-1}$, as the regulator at this value is negligible.
We can choose a value for our coordinate-space truncation, $L$, such that the long-range part of the potential is negligible.  
$L = 10$ fm is a safe estimate based on the Fourier transformation of the regulator.
With these values in mind, we chose $b=2.0$ fm and $\nmax = 120$.
The truncation cutoffs from this (roughly $\Lambda \sim 7.8$ fm$^{-1}$ and $\L \sim 31.3$ fm) are more than enough for accuracy in transforming between bases, allowing accuracy in evolution to lower values of SRG $\lambda$.

\begin{figure}
	\includegraphics[width=\textwidth]{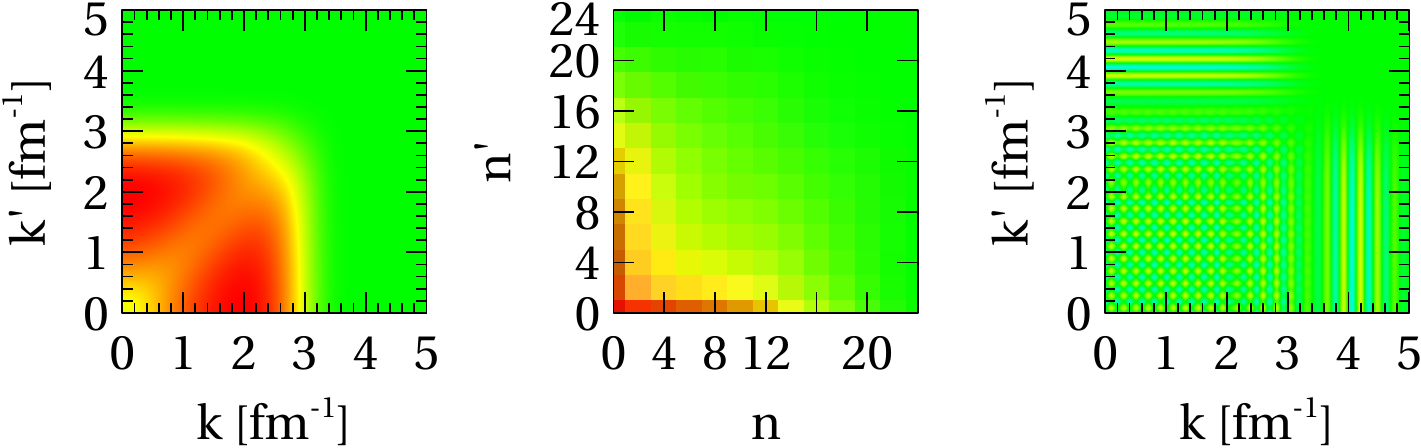}~~~%
	\caption{  Left: Matrix elements $V(p,p')$ of our model potential. Center: HO-basis matrix elements of the same potential transformed into HO-basis.  Right: Difference of matrix elements between the original potential and the twice-transformed potential, multiplied by 10$^7$. 
	\label{fig:orig_ho_diff}}
\end{figure}
%
Fig.~\ref{fig:orig_ho_diff} shows from left to right: the model potential in momentum representation, in HO-basis, and the difference between the original and twice-transformed potential in momentum representation.
We can see very clearly that above 4 fm$^{-1}$, momentum representation matrix elements of our potential are negligible, thus there should be no error in transforming our potential to HO basis.
The cutoffs imposed at this momentum are $e^{- (\frac{4}{3})^8} = 4.6 * 10^{-5}$, thus we can safely say above 4 fm$^{-1}$, matrix elements are negligible with several digits of precision.
Observing the HO-basis matrix elements, we can see that even an $\nmax$ of 24 would be large enough to transform the potential between HO-basis and plane-wave bases.
Finally, to explicitly show the difference in matrix elements of the original potential and the twice transformed potential in momentum-representation, the third picture shows these matrix elements multiplied by 10$^{7}$ such that they can be seen in the plot.
This error in the 7th decimal is much more precise than we would expect from a typical realistic calculation, but must be kept low when looking for other, more substantial errors in calculation.
We start with negligible errors in our basis-transformation and more points than necessary so that we can clearly identify errors that will arise from the SRG-evolution.

The next step in our procedure is to generate a potential that will evolve to the same universal form in low-energy matrix elements as our model potential.
Much like before, we construct this potential as the sum of the long-ranged exponential present in the model potential, but replace the short-ranged Gaussian with a sum of regulated $\delta$-shells.
Each shell has a different coupling which we fit to the lowest eigenvalues of the model potential.
This potential conforms to the rules for universality we observed in Chapter~\ref{chapt:2BUniv} as it has the same explicit long-range potential and is low-energy eigenvalue equivalent to the model potential.
We also create an eigenvalue-equivalent potential that does not have the explicit long-range piece of the potential (thus it will not evolve to a universal low-energy form) by fitting only $\delta$-shells without the long-range piece of the potential to the low-energy eigenvalues of the model potential.
%
\begin{figure}
	\includegraphics[width=.3 \textwidth]{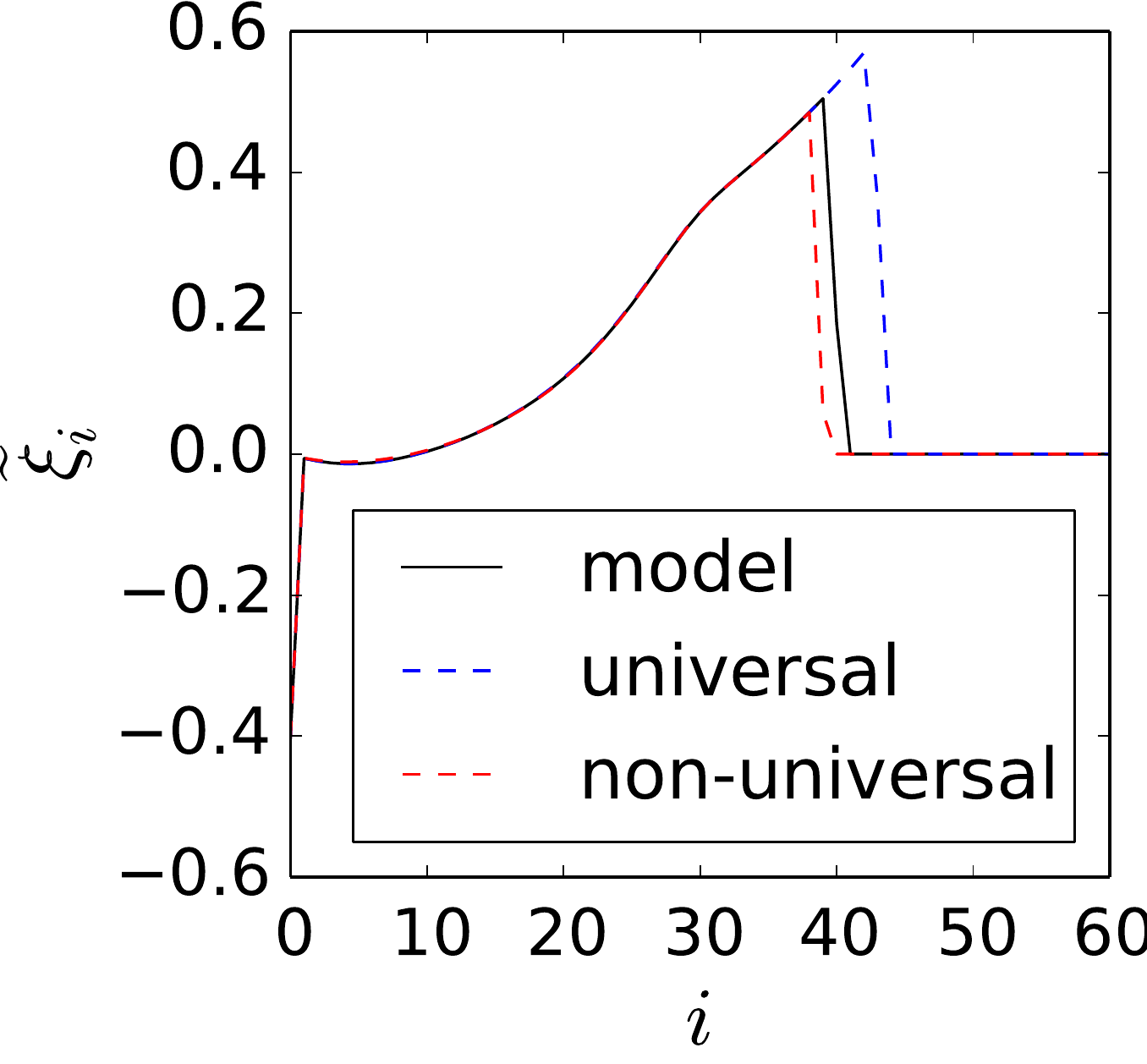}~~~%
	\includegraphics[width=.3 \textwidth]{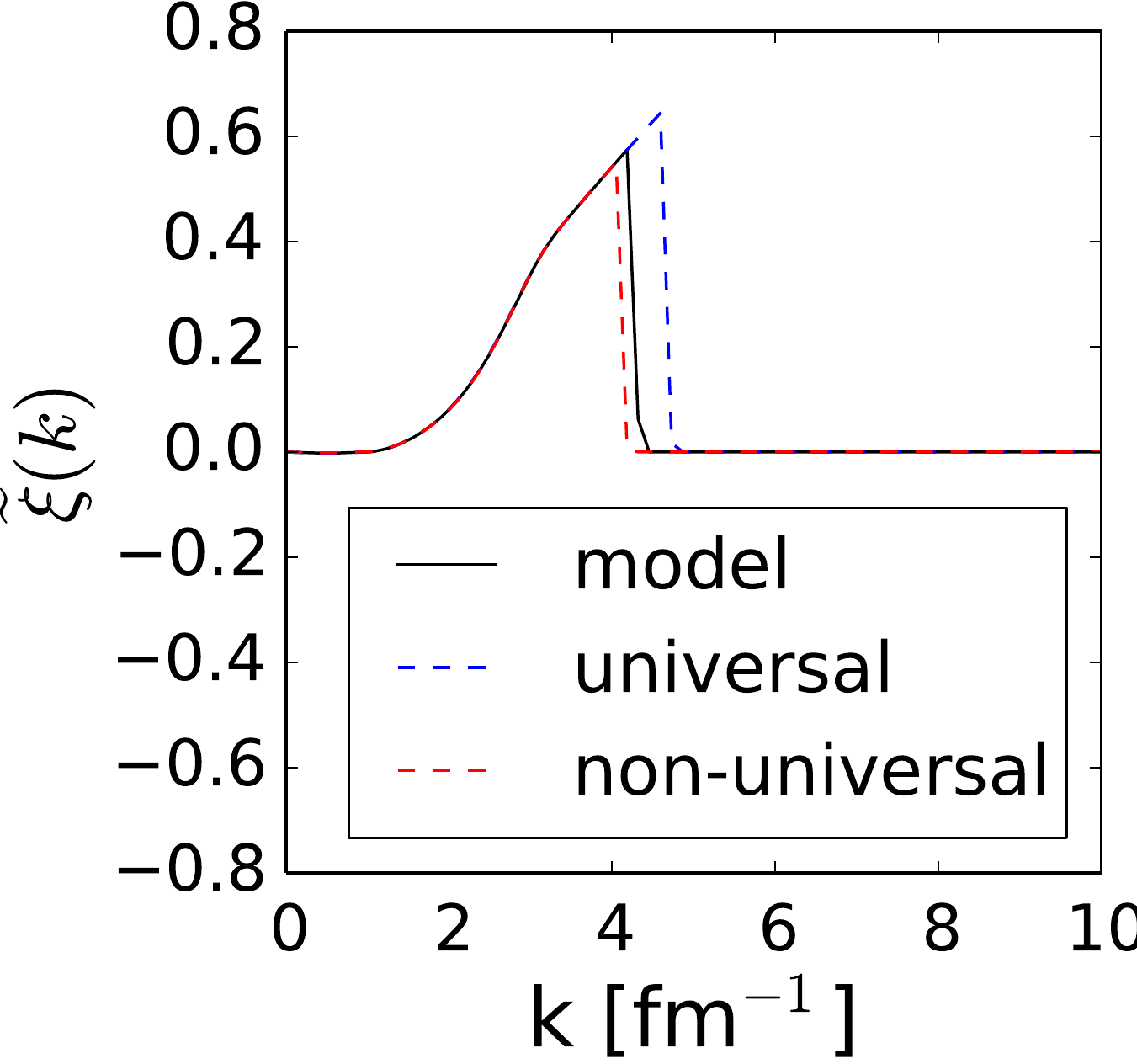}~~~%
	\includegraphics[width=.3 \textwidth]{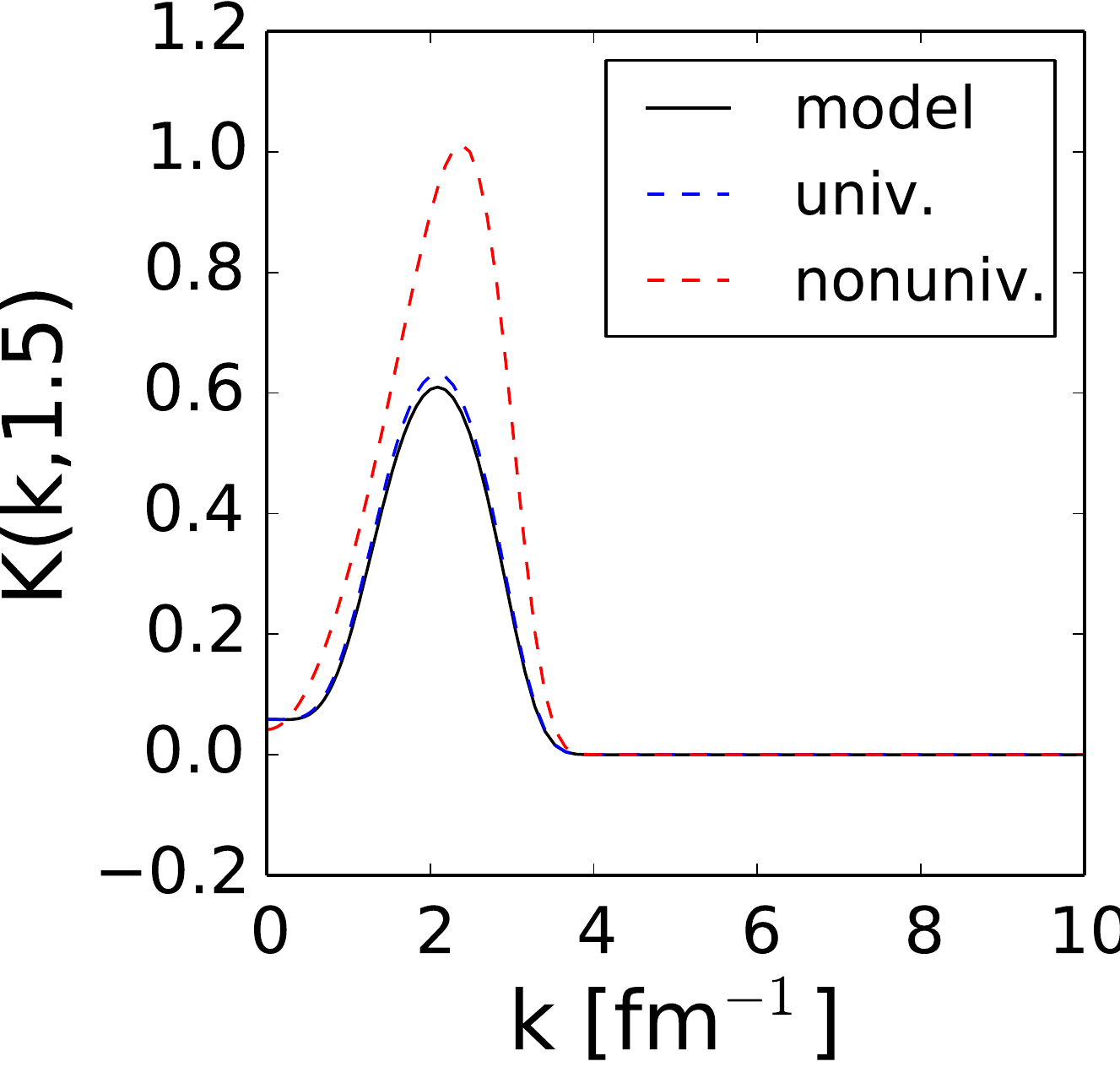}~~~%
   \vspace{-.3in}
	\caption{  (Left) $\widetilde{\xi}_{i}$ for model, universal, and non-universal potential matrices in HO-basis; $n = 2 i$. (Middle) $\widetilde{\xi}$(k) for model, universal, and non-universal potential matrices in momentum representation. (Right) Half-on-shell K matrix elements, K(k,k$_{0}$), for k$_{0} =$ 1.5 fm$^{-1}$.  The value of $\widetilde{\xi}$ is grid-dependent, thus the different scales for the first two plots.
	\label{fig:eigminust_hosk_model}}
\end{figure}

Figure~\ref{fig:eigminust_hosk_model} shows the eigenvalues in HO-basis and momentum representation and a vertical slice of the half-on-shell K matrix elements of the three potentials.
Notice that the eigenvalues are very similar for all three potentials, but the low-energy HOS K-matrix elements for the non-universal potential are quite different from the model and universal potentials.
At higher energies, the HOS K-matrix elements are all different, as expected.
The HOS K-matrix equivalence between the model and universal potentials seems to end approximately above 1.5 fm$^{-1}$.
While we are still concerned with 2-body potentials, we have the luxury of being able to plot momentum representation eigenvalues and HOS K matrix elements, so we have done so to double check that the procedure is the same as before.
We can see that the model and universal potentials exhibit the criteria to evolve to a universal form, and the nonuniversal, yet low-energy eigenvalue-equivalent potential does not.

Now, we can evolve the potential matrix elements in momentum-representation and HO-basis.
%
\begin{figure}
	\includegraphics[width=.45 \textwidth]{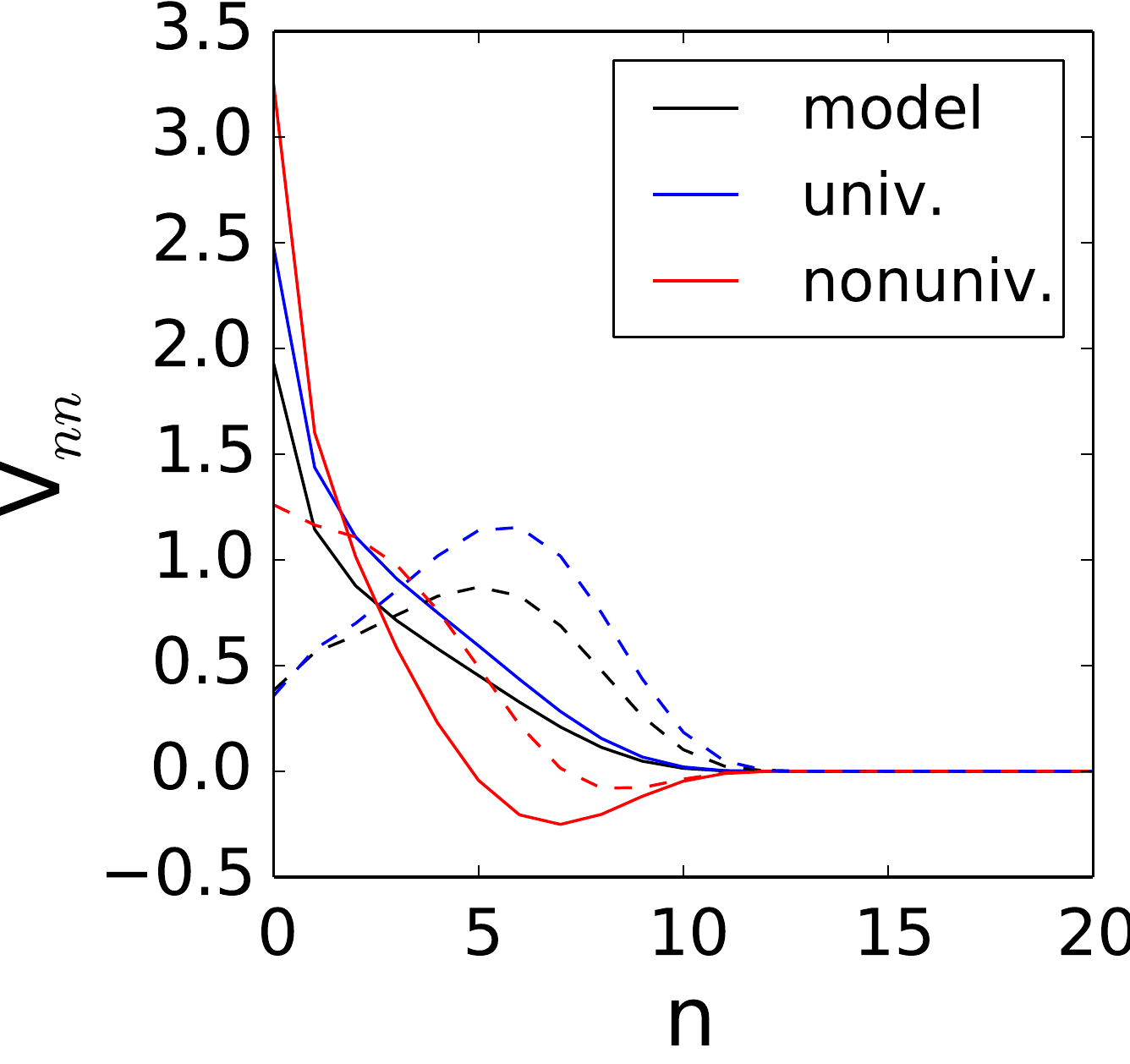}~~~%
	\includegraphics[width=.45 \textwidth]{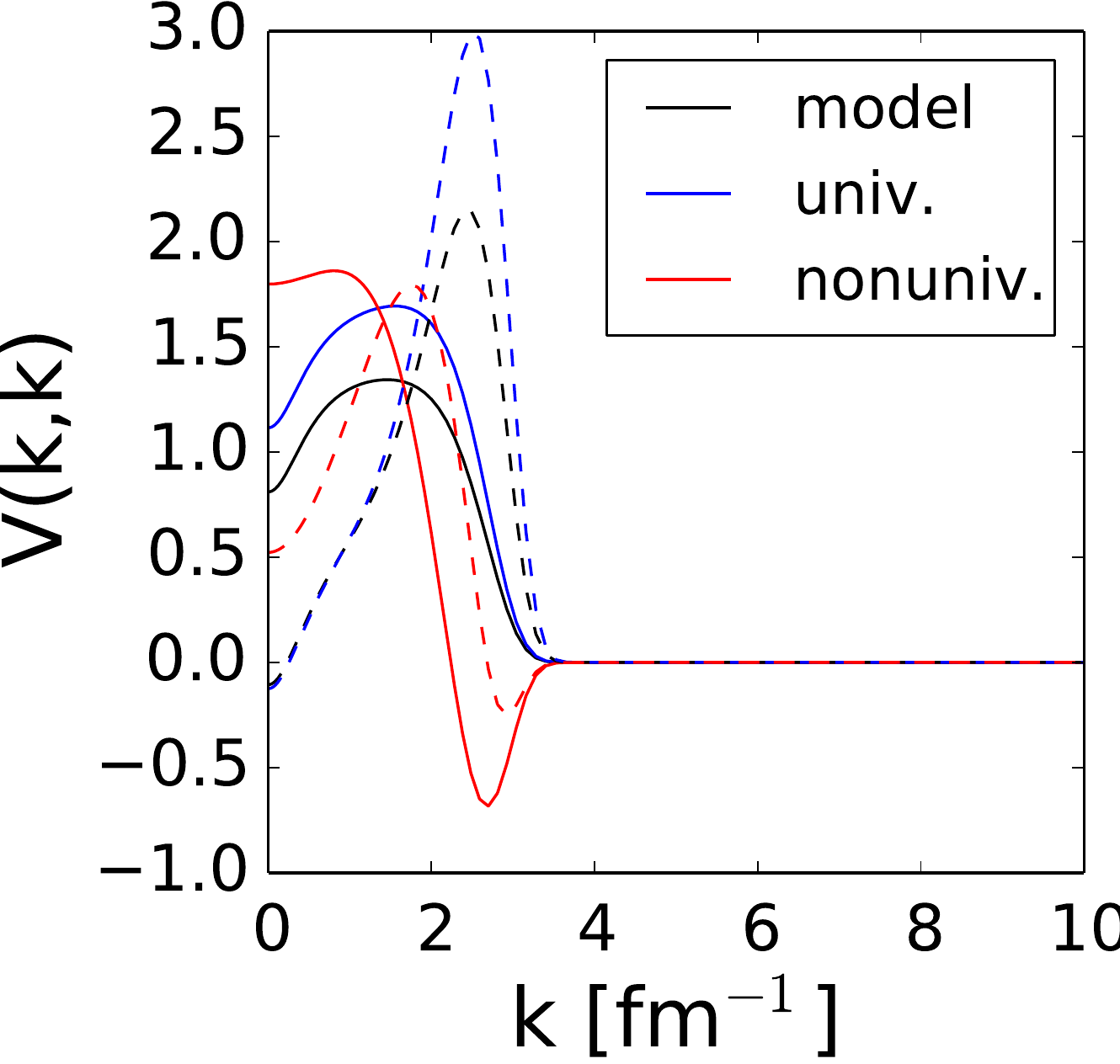}~~~%
   \vspace{-.3in}
	\caption{  (Left) HO-basis diagonal matrix elements of our three potentials, unevolved (solid) and evolved in HO-basis to $\lambda$ = 2.0 fm$^{-1}$ (dashed).  (Right)  Momentum representation diagonal matrix elements of our three potentials, unevolved (solid) evolved in momentum representation to $\lambda$ = 2.0 fm$^{-1}$ (dashed).
	\label{fig:pot_evo_20_diag}}
\end{figure}
\begin{figure}
	\includegraphics[width=.45 \textwidth]{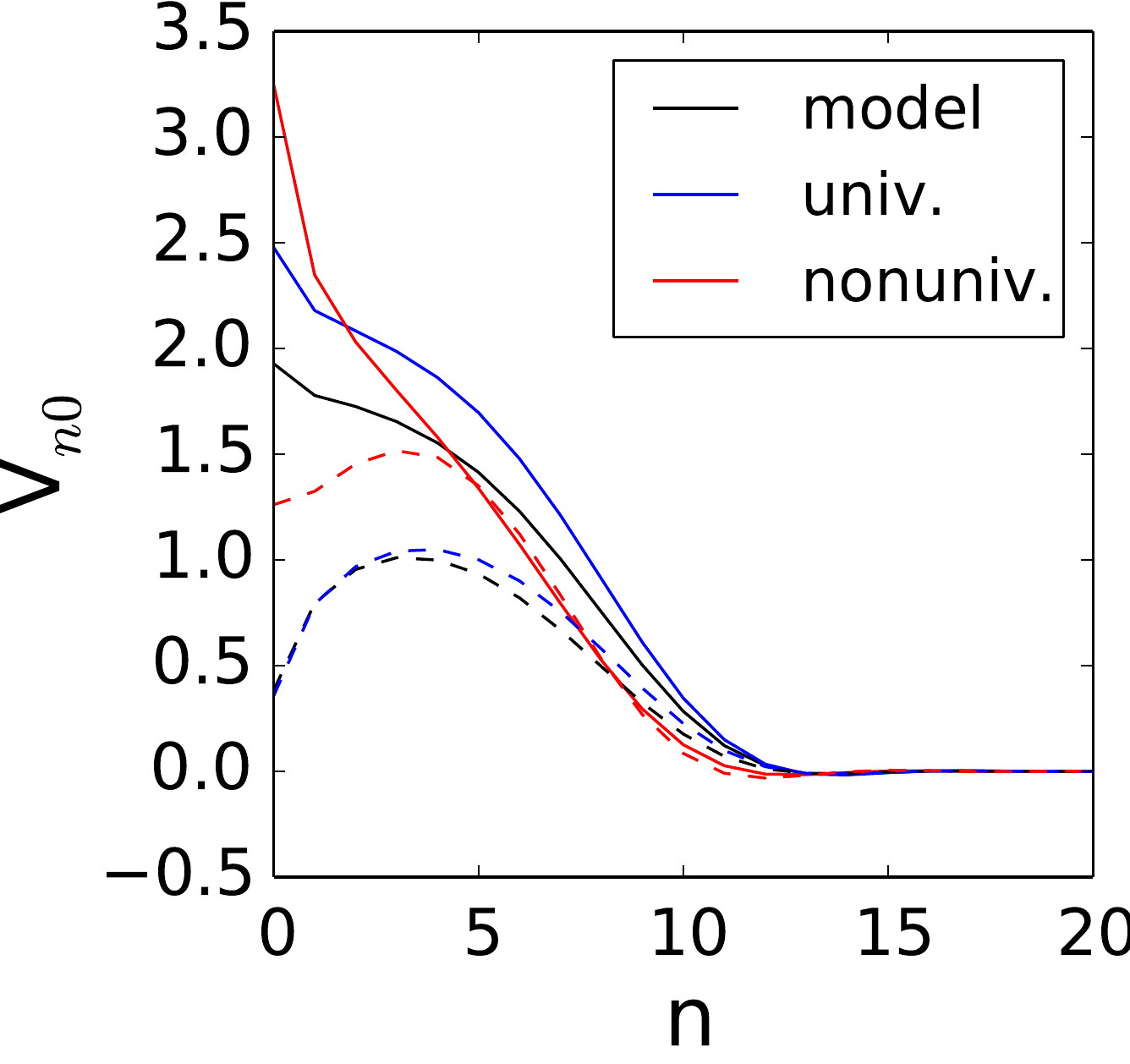}~~~%
	\includegraphics[width=.45 \textwidth]{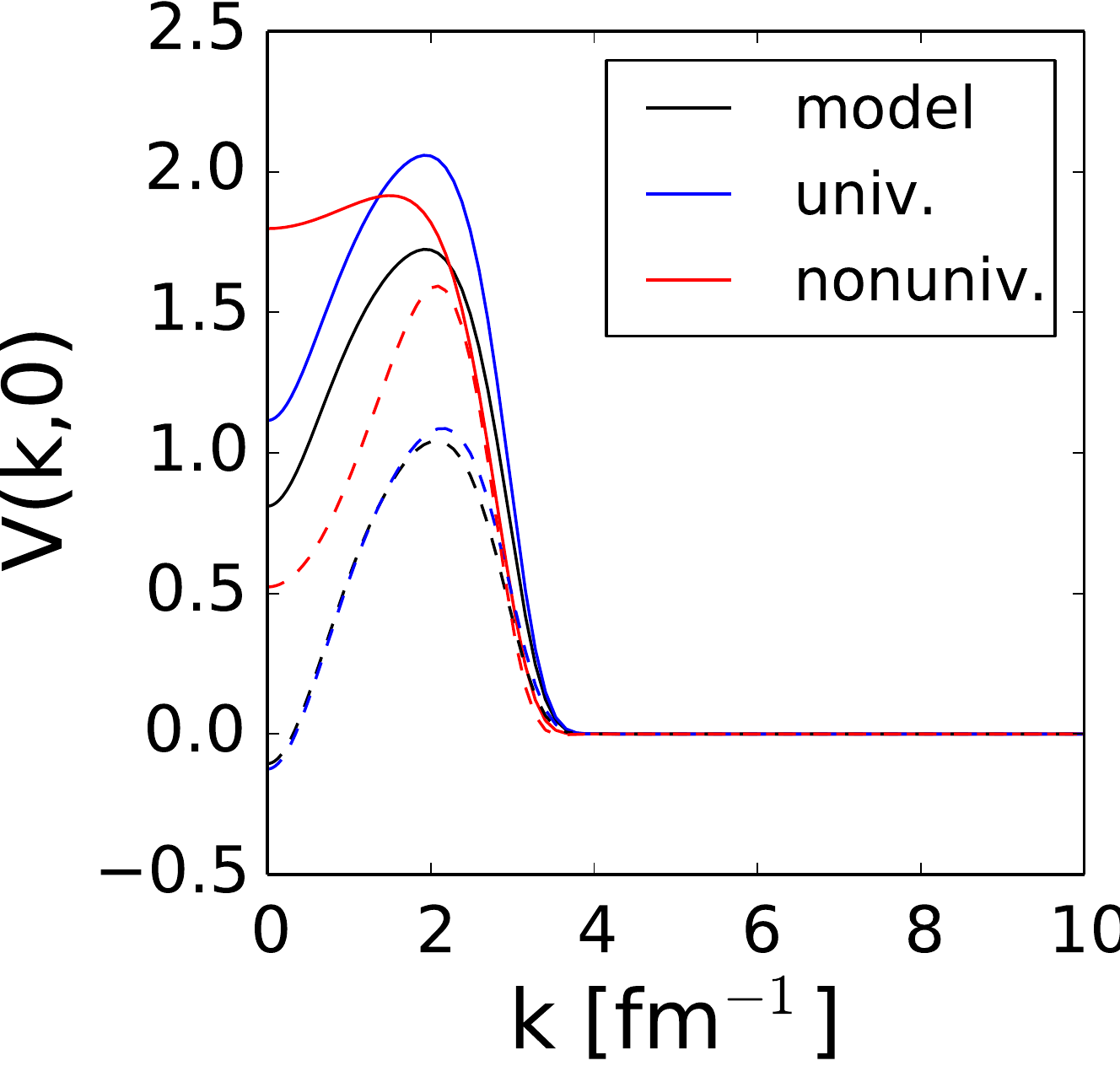}~~~%
   \vspace{-.3in}
	\caption{  (Left) HO-basis off-diagonal matrix elements of our three potentials, unevolved (solid) and evolved in HO-basis to $\lambda$ = 2.0 fm$^{-1}$ (dashed).  (Right)  Momentum representation off-diagonal matrix elements of our three potentials, unevolved (solid) evolved in momentum representation to $\lambda$ = 2.0 fm$^{-1}$ (dashed).
	\label{fig:pot_evo_20_offdiag}}
\end{figure}
\begin{figure}
	\includegraphics[width=.45 \textwidth]{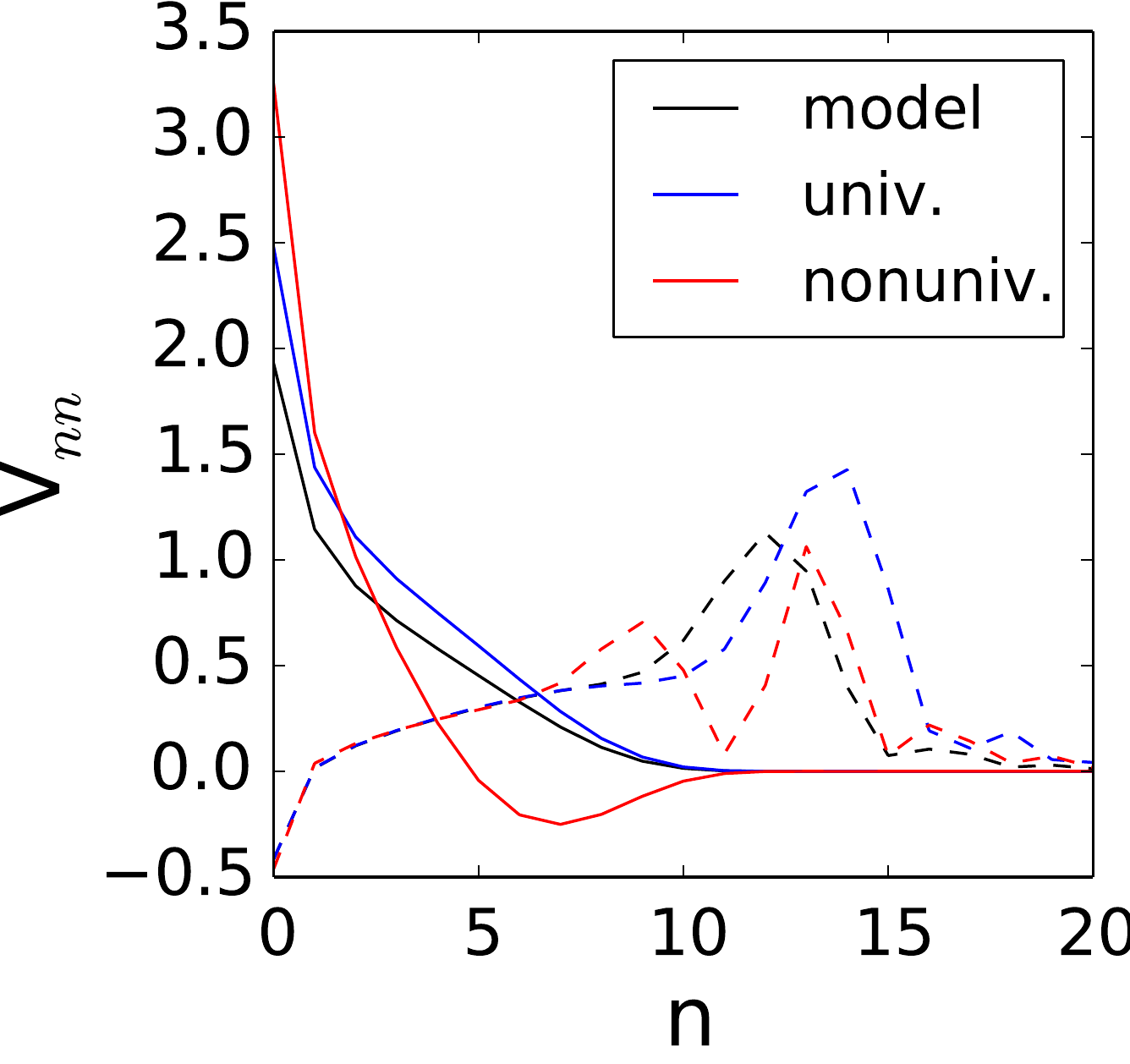}~~~%
	\includegraphics[width=.45 \textwidth]{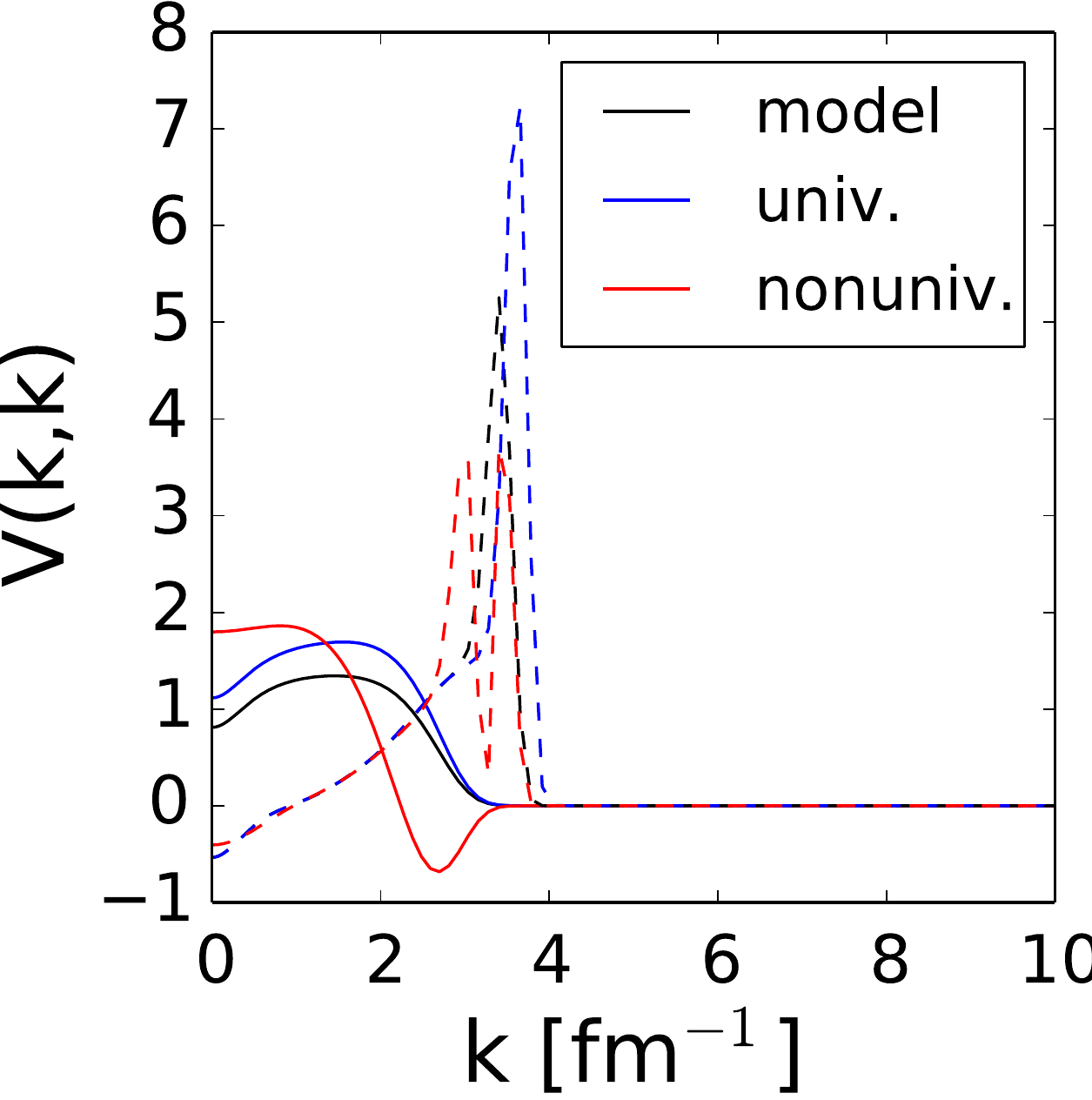}~~~%
   \vspace{-.3in}
	\caption{  (Left) HO-basis diagonal matrix elements of our three potentials, unevolved (solid) and evolved in HO-basis to $\lambda$ = 1.0 fm$^{-1}$ (dashed).  (Right)  Momentum representation diagonal matrix elements of our three potentials, unevolved (solid) evolved in momentum representation to $\lambda$ = 1.0 fm$^{-1}$ (dashed).
	\label{fig:pot_evo_10_diag}}
\end{figure}
\begin{figure}
	\includegraphics[width=.45 \textwidth]{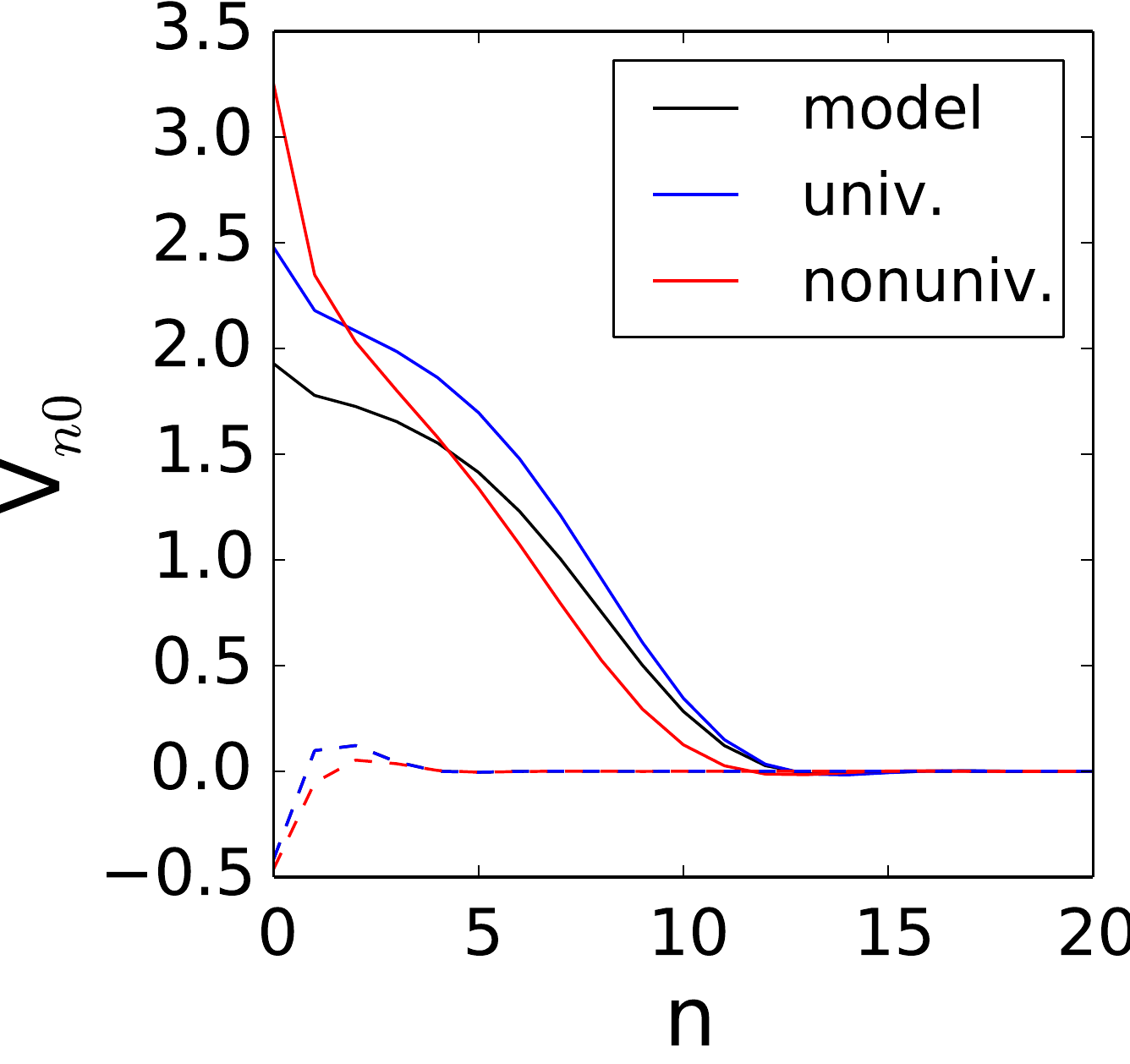}~~~%
	\includegraphics[width=.45 \textwidth]{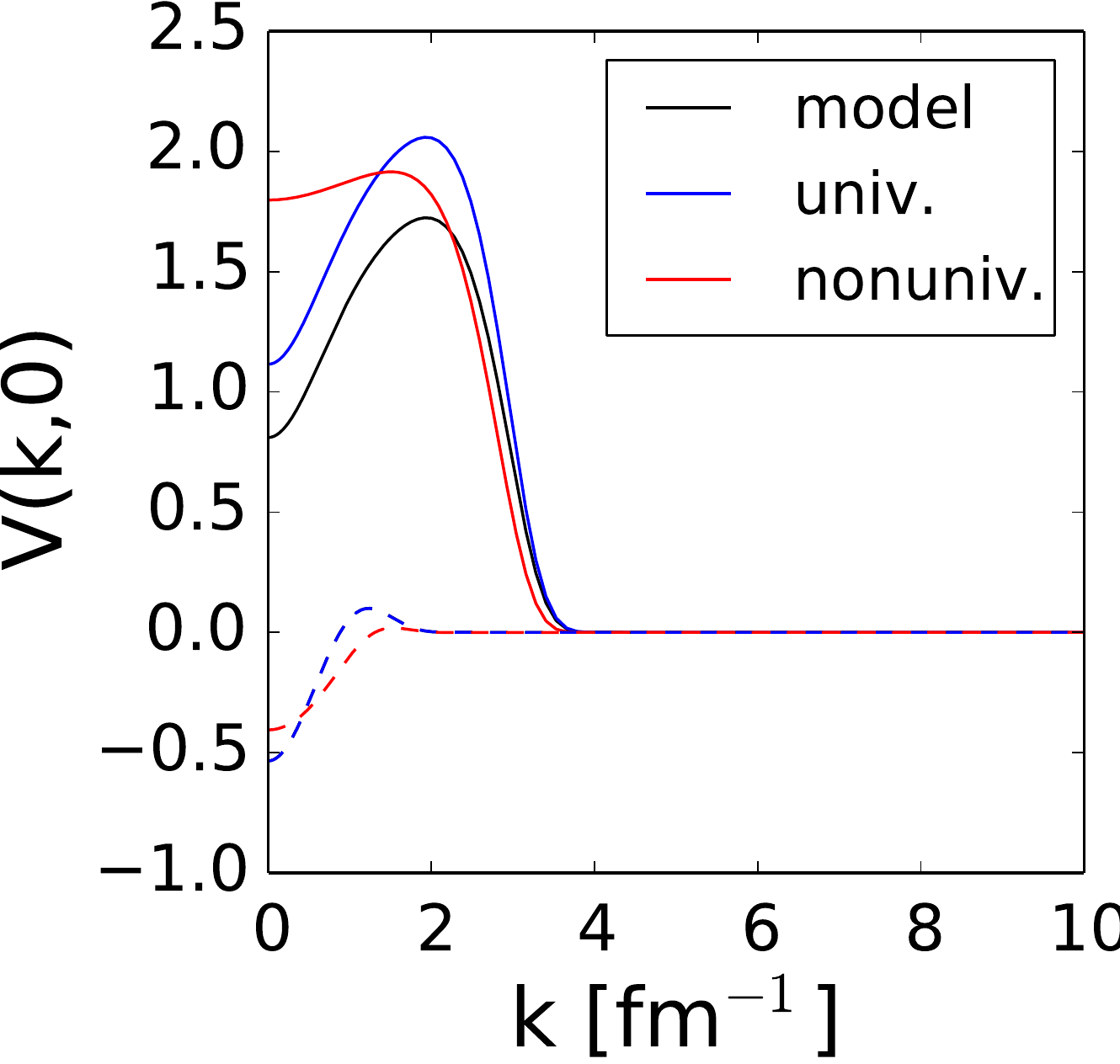}~~~%
   \vspace{-.3in}
	\caption{  (Left) HO-basis off-diagonal matrix elements of our three potentials, unevolved (solid) and evolved in HO-basis to $\lambda$ = 1.0 fm$^{-1}$ (dashed).  (Right)  Momentum representation off-diagonal matrix elements of our three potentials, unevolved (solid) evolved in momentum representation to $\lambda$ = 1.0 fm$^{-1}$ (dashed).
	\label{fig:pot_evo_10_offdiag}}
\end{figure}
Observing Figs.~\ref{fig:pot_evo_20_diag}-\ref{fig:pot_evo_10_offdiag}, we see the emergence of universal matrix elements.
These figures show potential matrix elements of the three potentials in (left) HO-basis and (right) momentum-representation both unevolved (solid lines) and evolved (dashed lines).
The potential designed to be non-universal, as planned, does not exhibit a universal low-energy form.
We can see this in Figs.~\ref{fig:pot_evo_20_diag}~\ref{fig:pot_evo_20_offdiag}.
Notice that below $1.5$ fm$^{-1}$, the momentum-representation matrix elements evolved to $\lambda$ = 2.0 fm$^{-1}$ of the model and universal potentials (black-dashed and blue-dashed lines respectively) fall on top of each other, while the non-universal potential matrix elements (red-dashed) are very different.
Viewing the HO-basis matrix elements at this $\lambda$ is inconclusive, as only the first two matrix elements in HO-basis are the same between the universal potentials, while the rest differ.

Evolving further to $\lambda$ = 1.0 fm$^{-1}$ (Figs.~\ref{fig:pot_evo_10_diag},\ref{fig:pot_evo_10_offdiag}), we see that more matrix elements align in HO-basis, but also that the high-n diagonal matrix elements are no longer zero.
This is also true for the (not-shown) near-diagonal elements.
This means that transformation between bases will not be as accurate as with the unevolved potentials.
As $\lambda$ decreases further, oscillator errors will appear in the high-momentum matrix elements of the potentials in momentum representation.

We can see that momentum representation is ideal for viewing matrix elements and searching for a low-momentum universal form.
A momentum scale appears around 1.5 fm$^{-1}$, below which matrix elements are universal, and it is easier to identify this universal form at $\lambda >1.0$ fm$^{-1}$ in momentum representation than in HO-basis.
It may be possible to identify universality in HO-basis matrix elements, but as we evolve to lower $\lambda$, we accumulate SRG errors in HO-basis and potentially run closer to the binding momentum which creates pathologies in the evolution with $\Trel$.
In the 3-body problem, we cannot evolve in momentum representation, nor can we evolve to as low a $\lambda$.
Our strategy for the 3-body problem will be to evolve in HO-basis and observe matrix elements in HO-basis and then those transformed into momentum representation.
We have chosen our HO-basis parameters to have much more precision than required in the 2-body sector, such that we can isolate imprecisions that appear in the 3-body problem.

\section{3-body HO-basis evolution}






After embedding the 2-body potentials of the previous section, we can calculate bound state energies.
The model 2-body potential has a 3-body ground state energy of 0.991 fm$^{-1}$, the universal potential a ground state energy of 0.989 fm$^{-1}$, and the non-universal potential 0.961 fm$^{-1}$.  
Each potential has a number of three-body excited bound states.
None of these ground state energies need be identical; as has been shown in Ref.~\cite{Jurgenson:2010wy}; different universal 2-body potentials may yield different 3-body ground state energies.
Realistic 2-body potentials do not predict exactly the same triton binding energy, and the predictions do not reproduce the experimental value without a 3-body force~\cite{bognerfurnstahlschwenk}.
The predictions should not be off by an order-of-magnitude, however, as this could be evidence of strong many-body forces that break the many-body hierarchy.

When evolving in HO-basis, it is standard (for both realistic and other model potentials) to build the entire 3-body Hamiltonian and then plug it into the SRG flow equation.
The method of Refs.~\cite{3bsrgsimple,Hebeler:2012pr,Bogner:2006pc} to subtract out the disconnected diagrams in the 3-body evolution is naturally defined in a 3-body plane-wave basis, but much more difficult, if possible, in HO-basis.
After evolution, we can isolate the 3-body piece by evolving the 2-body potential in a 2-body basis, embedding it into a 3-body basis, and then subtracting it from the total 3-body potential (embedded 2-body plus explicit 3-body) evolved in the 3-body basis.
This evolution is unitary and will preserve the binding energies at any $\lambda$ up to numerical errors in the differential equation solver.

\begin{figure}
	\includegraphics[width=.9 \textwidth]{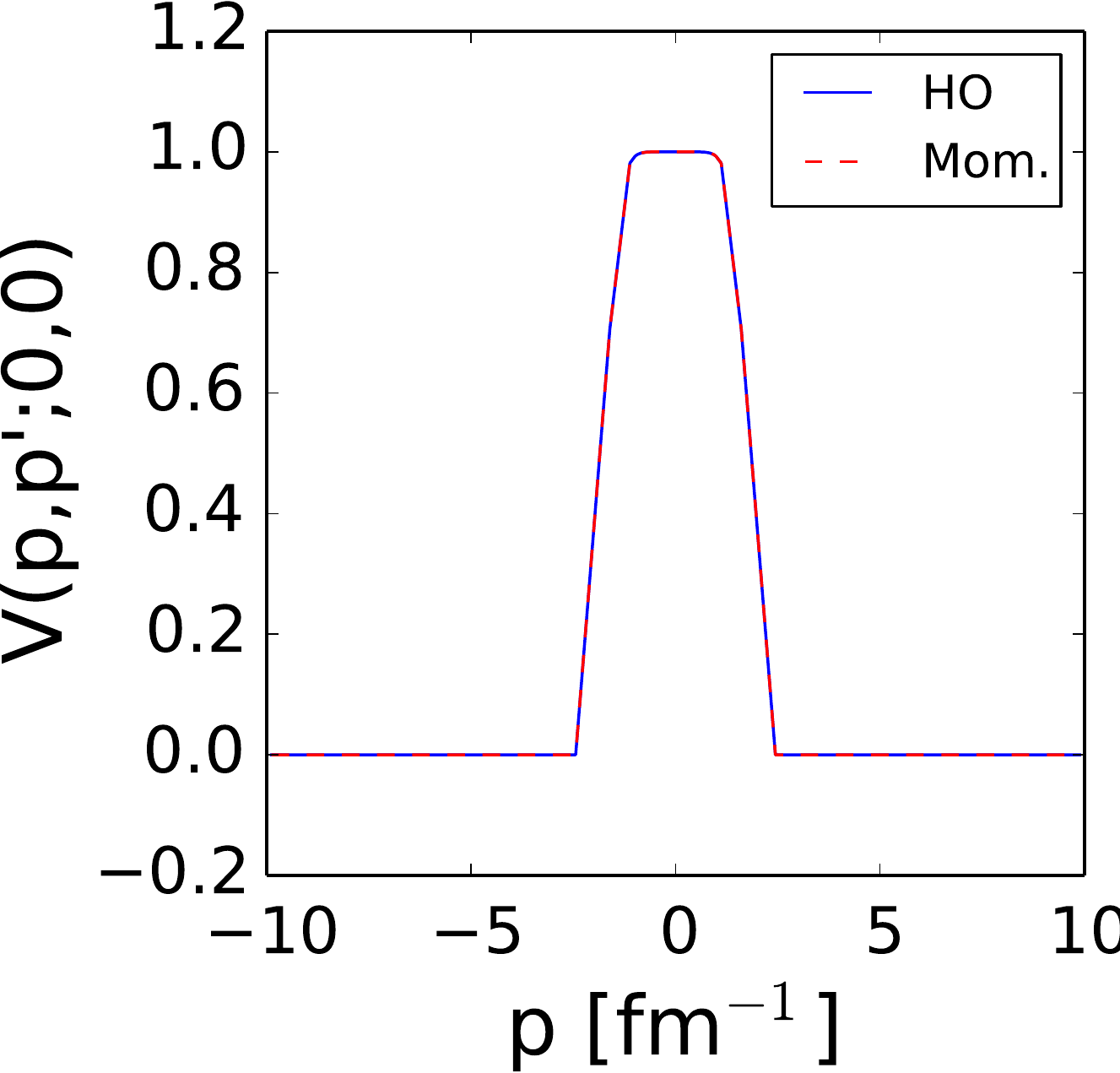}%
   \vspace{-.3in}
	\caption{  Momentum representation matrix elements diagonal in pair momentum with q=q'=0 of the explicit 3-body potential, V(p,p';q,q').  The dashed red line is in momentum representation where the potential is initially defined.  The solid blue line is the potential transformed into HO-basis and combined with the 2-body embedded potential, which is then removed and then the matrix is transformed back to momentum representation.  The maximum difference in any matrix element is 7.61$\times$10$^{-8}$.  The 3-body potential used is from Eq.~\eqref{eqn:3bpot} with c$_{3}=1.0$.
	\label{fig:pot3_check}}
\end{figure}

Once the evolved 3-body potential (which can contain both explicit and induced pieces) is isolated, transforming back to momentum representation is straightforward.
When viewing potential matrix elements of the evolved 3-body potential in momentum representation, oscillations indicative of basis transformation errors appear at $\lambda$ values much higher than expected.
In fact, the oscillations appear even at $\lambda$ values so high that no discernible changes occur in the 2-body evolution (or in an explicit 3-body potential we could include).
Our method accurately transforms unevolved 3-body potentials (Eq.~\eqref{eqn:3bpot}) from HO-basis to momentum representation, thus it is clear that the evolution causes the oscillations in momentum representation of evolved 3-body potentials (see Fig.~\ref{fig:pot3_check}).
%

%
\begin{figure}
	\includegraphics[width=.45 \textwidth]{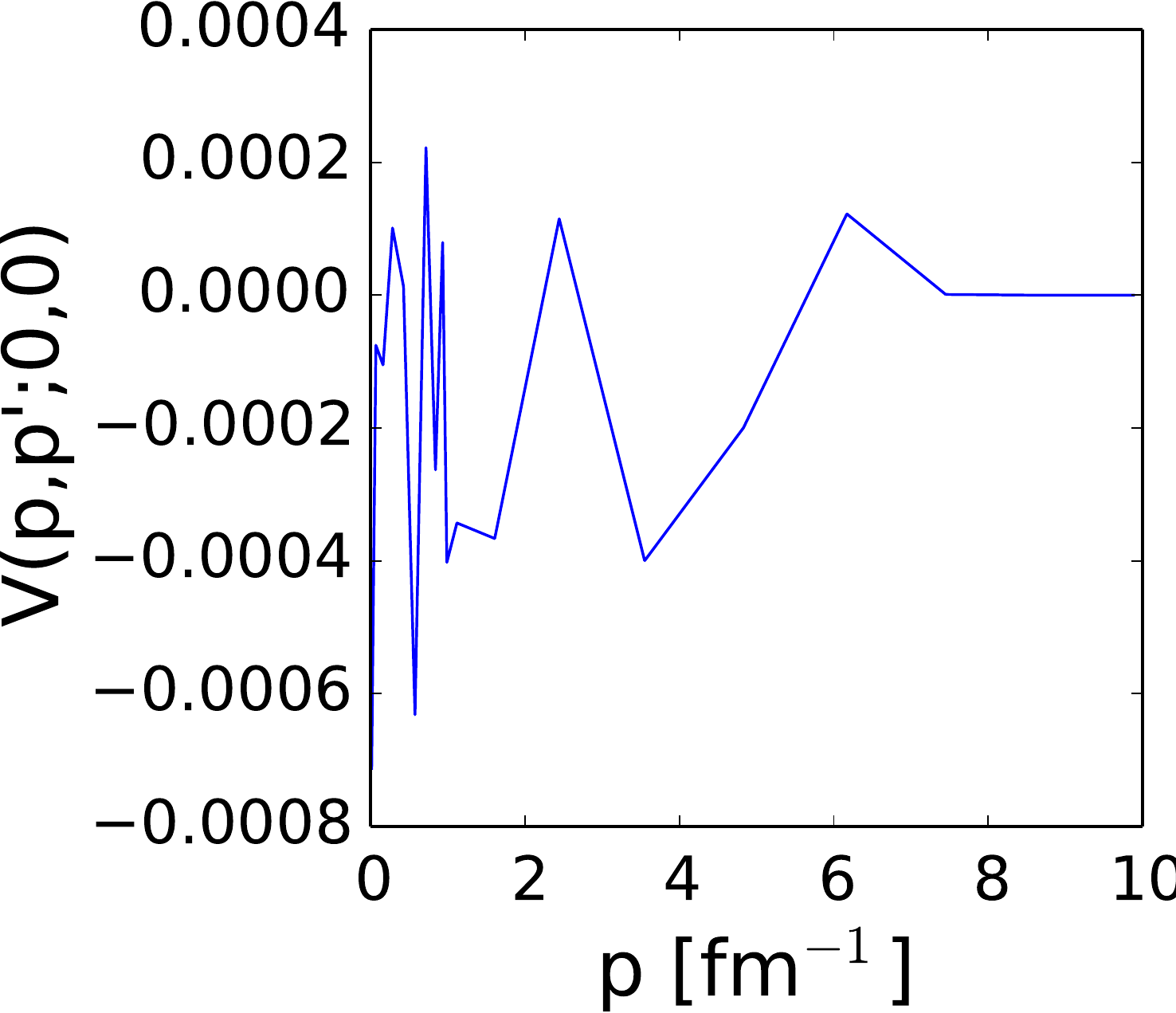}%
	\includegraphics[width=.45 \textwidth]{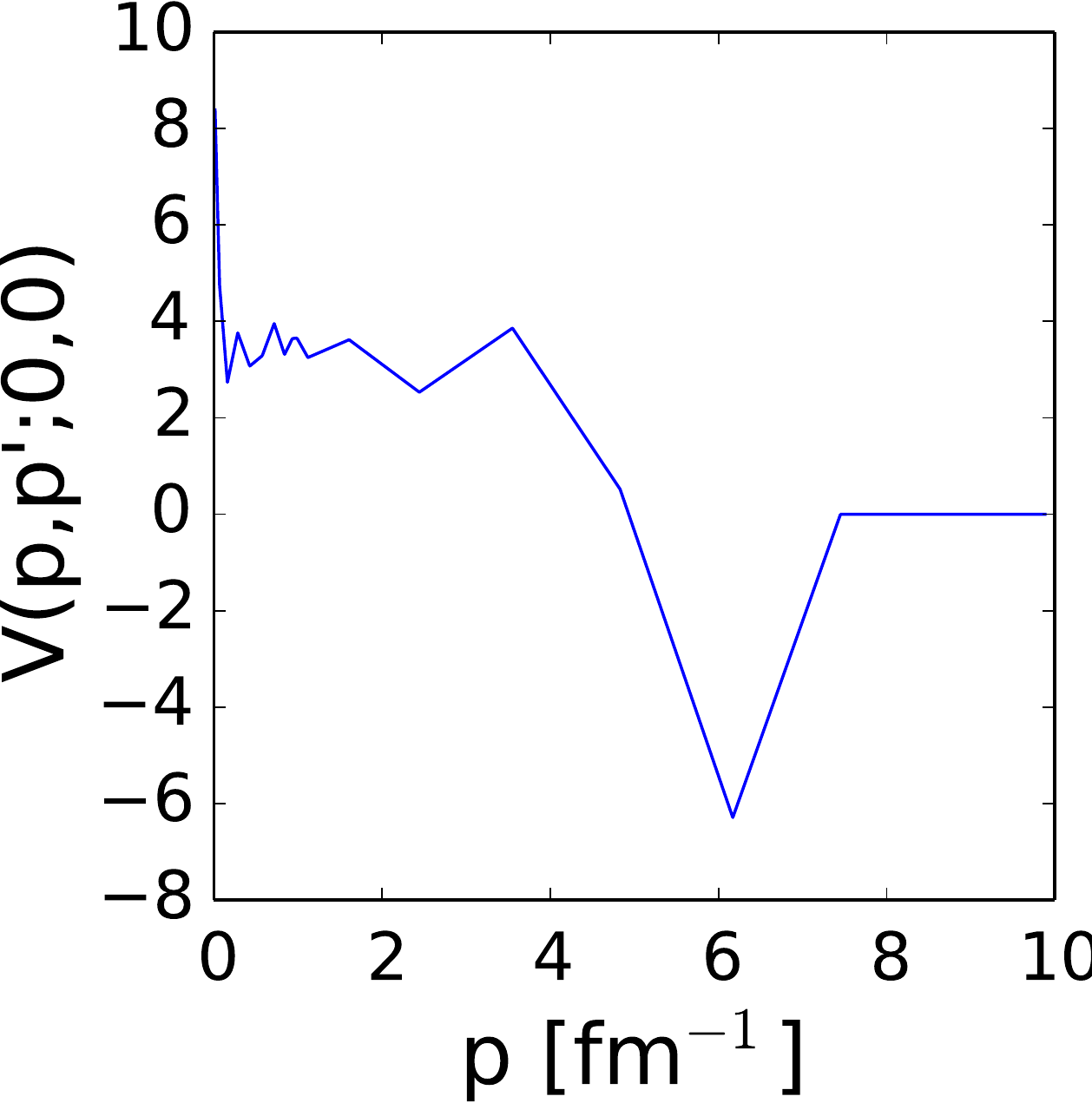}%
   \vspace{-.3in}
	\caption{  Momentum representation matrix elements diagonal in pair momentum with q=q'=0 of the induced 3-body potential, V(p,p';q,q'), (left) evolved to $\lambda = 14$ fm$^{-1}$ and (right) evolved to $\lambda = 1.5$ fm$^{-1}$.  Notice the magnitude of the oscillation in these low-momentum matrix elements increases as $\lambda$ decreases.  The oscillations appear due to truncation of nonzero high-n HO-basis matrix elements, and thus appear as high-frequency oscillations.
	\label{fig:pot3_induced_osc}}
\end{figure}
Fig.~\ref{fig:pot3_induced_osc} shows momentum representation matrix elements for the induced 3-body potential.  
We select matrix elements with zero spectator momentum and see oscillatory behavior in the low pair-momentum elements which increases in magnitude as $\lambda$ is decreased.
This oscillation is similar to what we observe for a basis truncation error in the 2-body sector; appearing as high-frequency oscillations due to omitted high-n oscillator wave-functions.
Furthermore, the oscillatory behavior appears at much higher $\lambda$ than we expect, based on the evolution in the 2-body sector (or lack thereof).
For this reason, the oscillatory behavior most likely is not representative of the induced 3-body potential, but rather a truncation error on the matrix elements of the full, evolved 3-body potential.

%
\begin{figure}
	\includegraphics[width=.3 \textwidth]{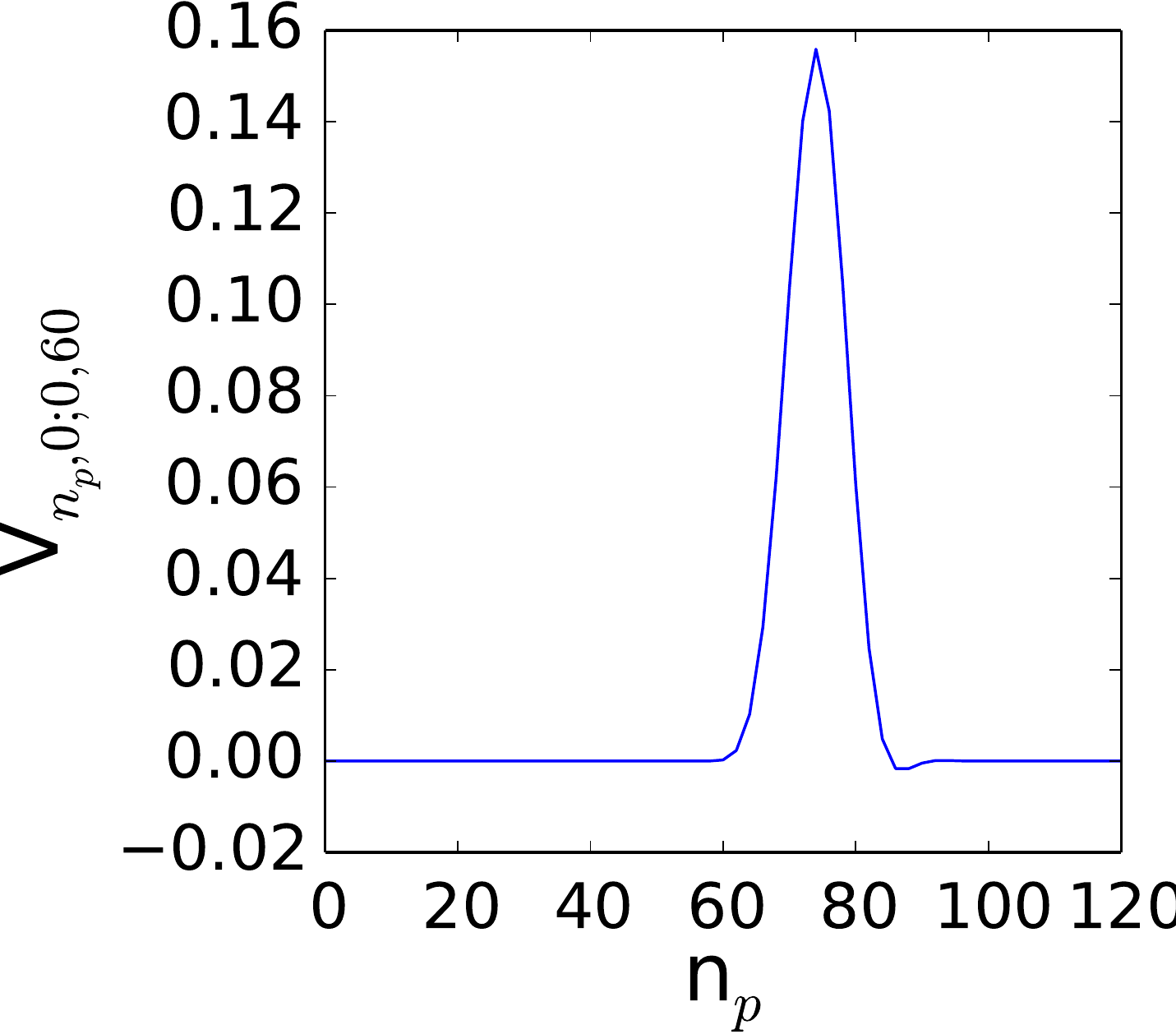}%
	\includegraphics[width=.3 \textwidth]{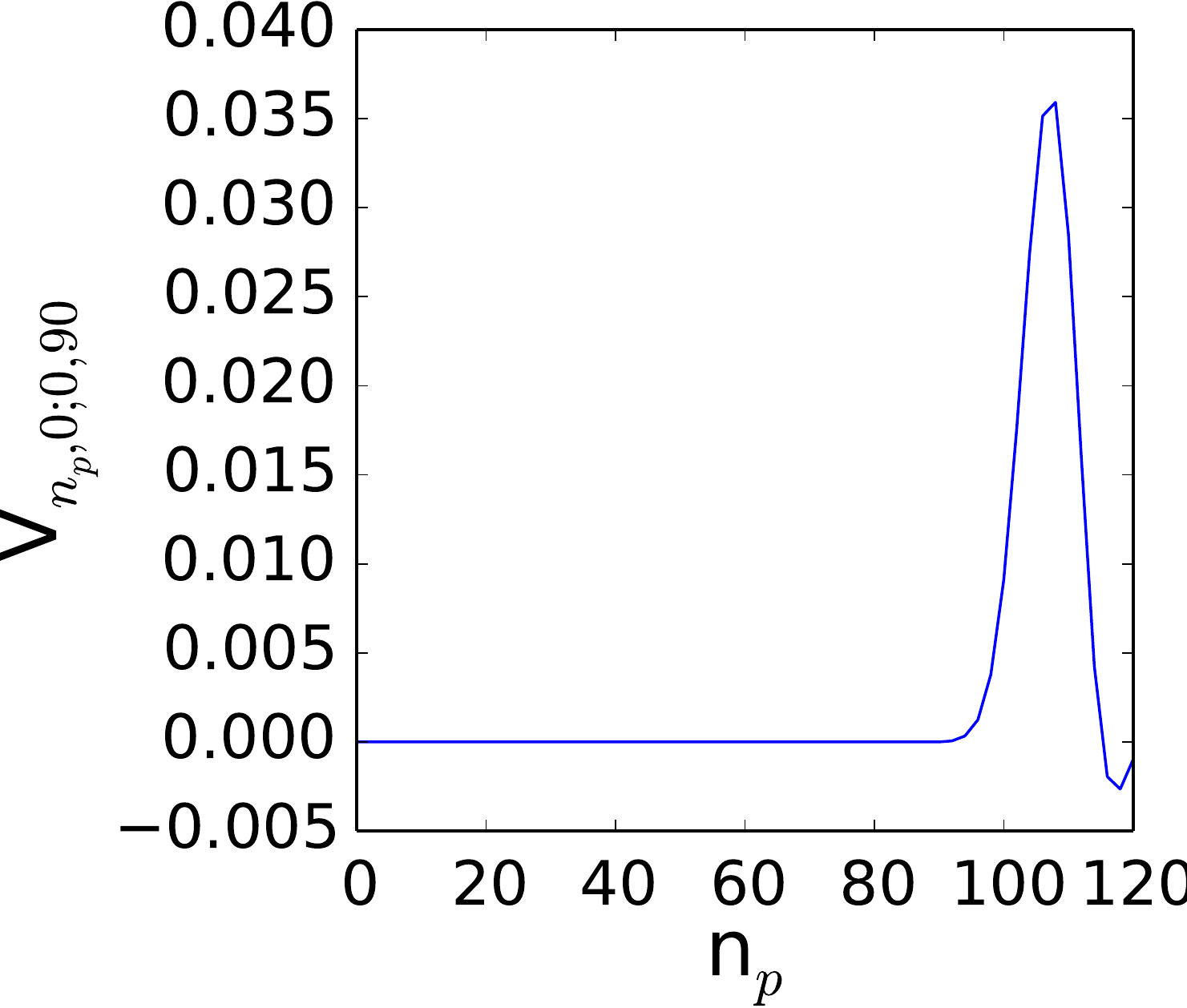}%
	\includegraphics[width=.3 \textwidth]{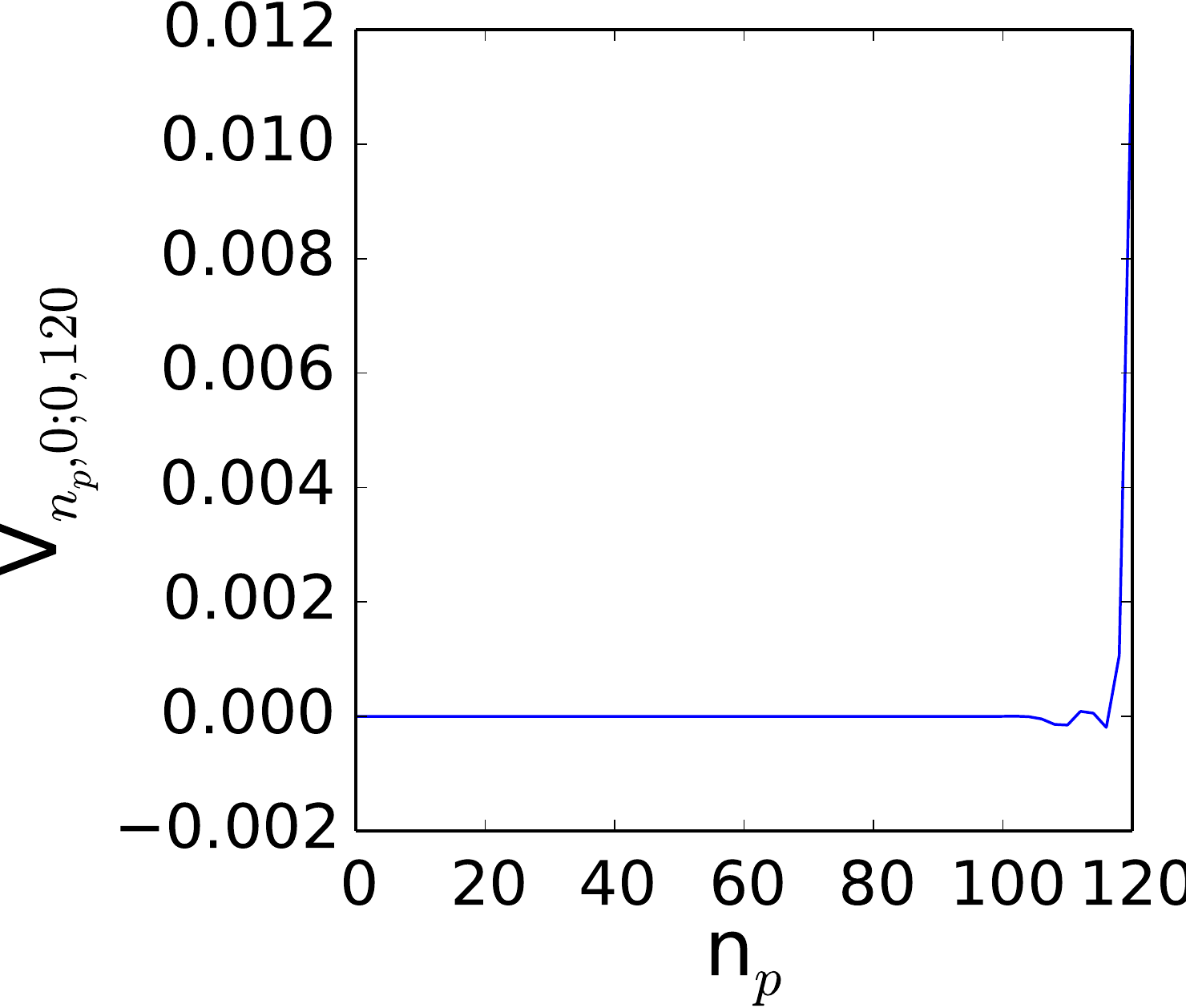}%
   \vspace{-.3in}
	\caption{  HO-basis matrix elements of the unevolved embedded 2-body potential as a function of $n_{p}$ with $n'_{p} = 0, n_{q} = 0$ and (left) $n'_{q} = 60$,  (middle) $n'_{q} = 90$, (right) $n'_{q} = 120$.  Notice the shape and decreasing magnitude of each sample of matrix elements as $n'_{q}$ increases (the vertical scale is different in each plot to better resolve shapes).}
	\label{fig:pot3_ho_faroffdiag}
\end{figure}
For further evidence that the momentum-representation oscillatory matrix elements of the induced 3-body potential are a sign of basis-truncation errors in the matrix elements, we can examine the HO-basis matrix elements of the embedded, unevolved 2-body potential.
Fig.~\ref{fig:pot3_ho_faroffdiag} shows a selection of off-diagonal potential matrix elements of the unevolved embedded 2-body potential.
We can see that the matrix elements for $n'_{q} = 60$ and $90$ are roughly localized just above the point where $n_{p} + n_{q} = n'_{p} + n'_{q}$.
Therefore, when we look at the extreme off-diagonal matrix elements ($n'_{q} = 120$), we can see that they are increasing up to the basis truncation ($\nmax = 120$), and using the shape when $n'_{q} = 60$ and $90$ as a guide, we presume that a higher truncation would produce nonzero matrix elements at $n_{p}$ values slightly greater than 120.
Much like the oscillatory truncation errors that appear in the 2-body sector, these truncated 3-body matrix elements produce oscillatory errors when evolving in the 3-body sector.
Unlike the 2-body errors, these truncated matrix elements are unavoidable in the total $n$ truncation scheme we use.

A truncation on each individual $n$, specifically $n_{p}$ much greater than $n_{q}$, might solve the problem of truncating non-negligible matrix elements from the embedded 2-body potential, but the total $n$ truncation is required for a unitary symmetrizer.
If the symmetrizer is not unitary, there will be errors in the symmetrized basis.
Furthermore, a much more inhibiting problem with a non-unitary symmetrizer is that the eigenvalues will no longer be either zero or one, but somewhere in between.
This means that at best one must choose only nearly symmetric states (eigenvalue close to 1), and throw out barely symmetric states (eigenvalue close to 0), and at worst states may have eigenvalues in between such that the process cannot differentiate symmetric states from the rest at all.
Lastly, some of the matrix elements we would omit in the symmetrizer with a truncation on each $n$ would be the very same far off-diagonal matrix elements that are causing the error in SRG evolution, thus the scheme with the least problems is a total-$n$ truncation.

At the heart of the problem is the 2-body potential between one of the pair particles and the spectator.
The momentum entering is a combination of the pair and spectator momentum, and thus because the spectator momentum (or oscillator number) can be anything, the embedded potential will have coupling between low and high pair momentum (or oscillator number) matrix elements.
In the unevolved potential, these far off-diagonal matrix elements play a small role, thus we can find accurate binding energies, but as the potential evolves in HO-basis, the errors in matrix elements accumulate with each step in the ordinary-differential-equation solver.

%
\begin{figure}
	\includegraphics[width=.45 \textwidth]{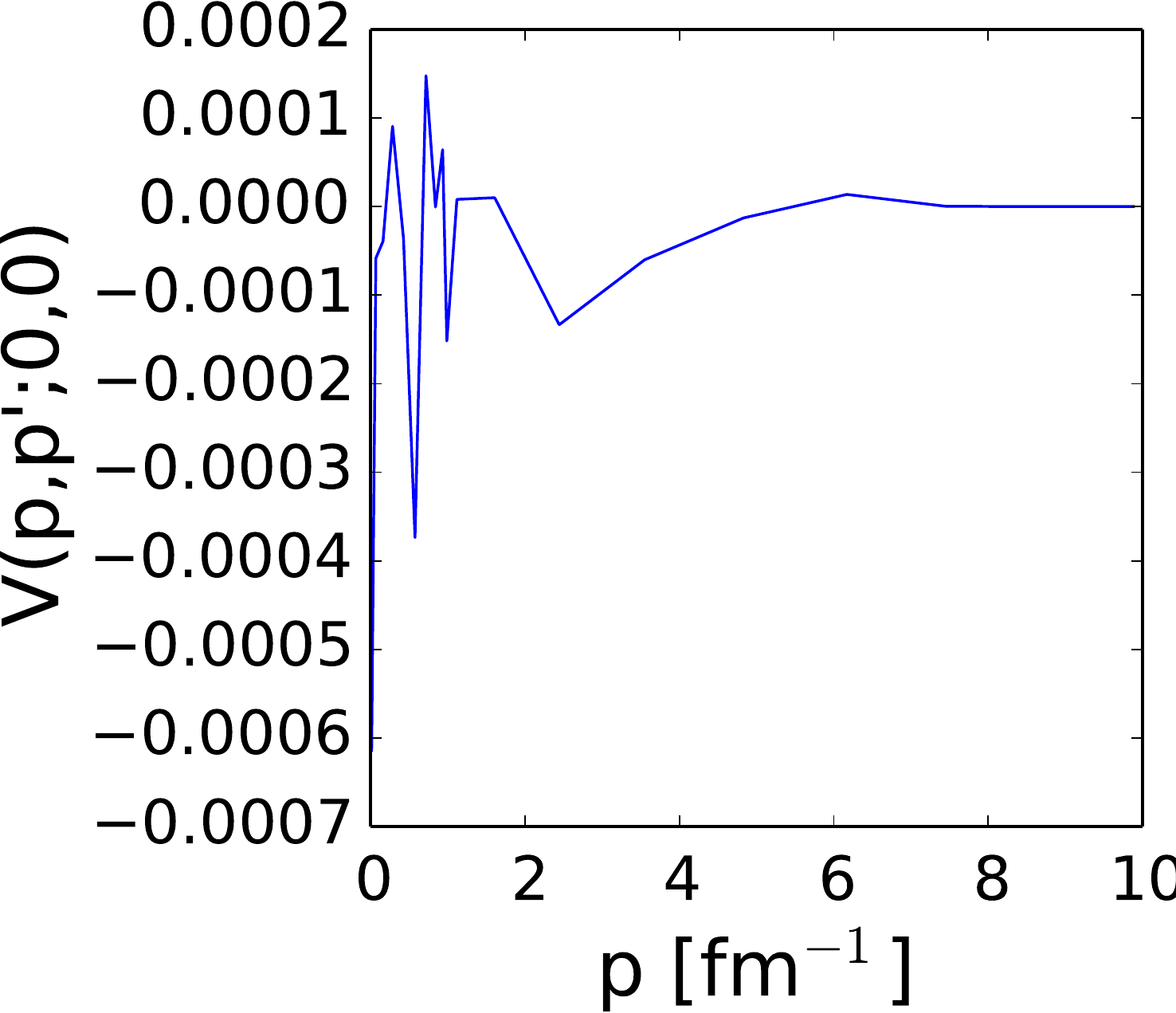}%
	\includegraphics[width=.45 \textwidth]{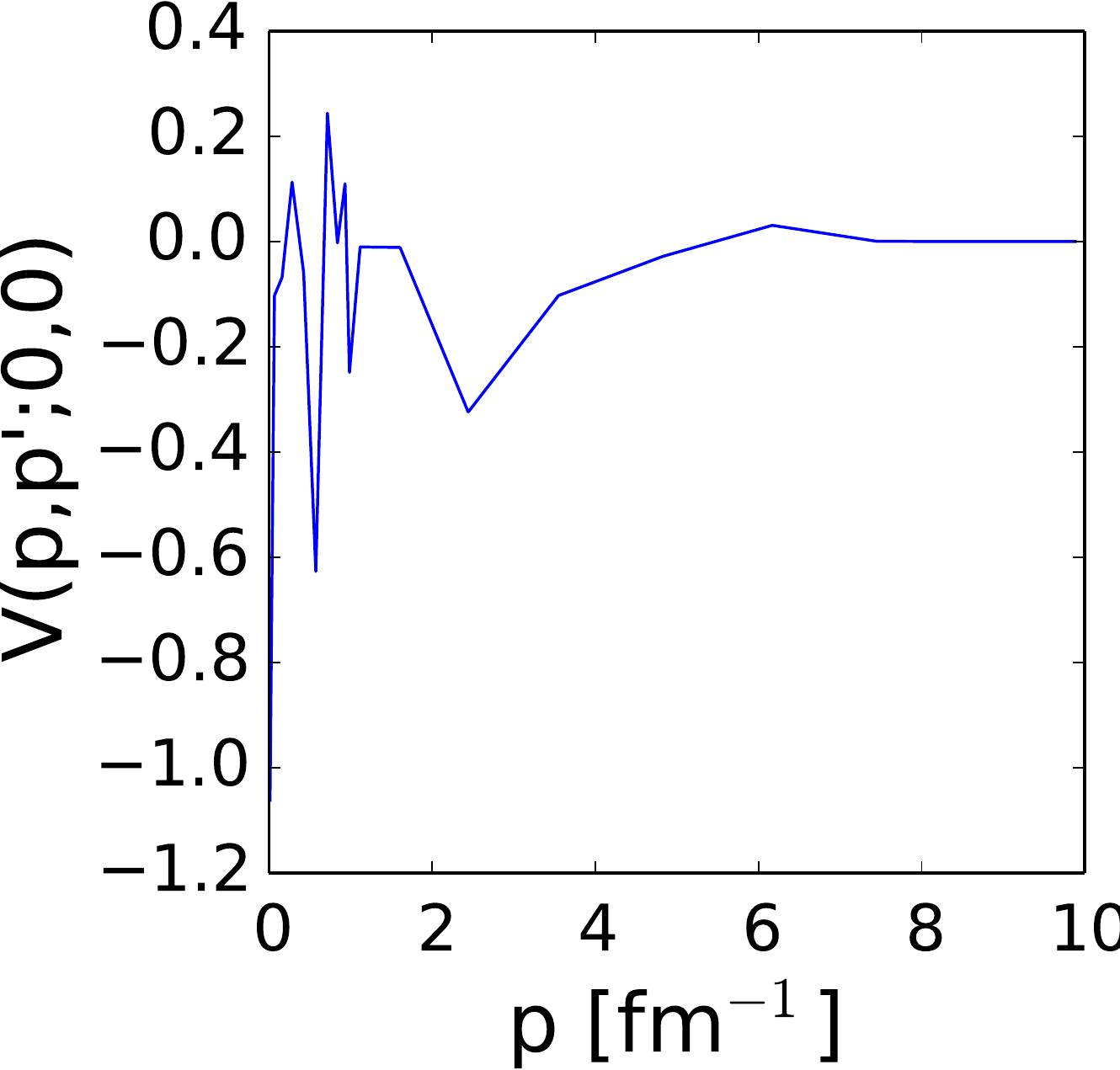}%
   \vspace{-.3in}
	\caption{  Error in momentum representation matrix elements of the embedded 2-body potential evolved with only the quadratic in $\Trel$ term ($q=q'=0$ and $p=p'$). (left) evolved to $\lambda = 14$ fm$^{-1}$ and (right) evolved to $\lambda = 1.5$ fm$^{-1}$.  Notice the magnitude of the oscillation in these low-momentum matrix elements increases as $\lambda$ decreases.
	\label{fig:pot3_zeroerror_osc}}
\end{figure}
To further clarify these errors as an error in the HO-basis SRG evolution and not just as a part of a pathological induced 3-body potential, we can examine the 3-body evolution equation only using the first term, quadratic in $T$.
This term does not induce any 3-body potential, and without an explicit 3-body potential it is identical to the first term in the 2-body evolution equations for each 2-body potential.
We can see each 2-body evolution equation with just the leading-order term looks like:
\beqn
	\frac{d V_{i j}}{ds} = \left[ [ T_{i j} , V_{i j}] , T_{i j}\right]. \label{eq:leadingordersrg}
\eeqn
The 3-body evolution equation with just the leading-order term,
\beqn
	\sum_{i \neq j}\frac{d V_{i j}}{ds} = \sum_{i \neq j} \left[ [ T_{rel} , V_{i j}] , T_{rel}\right],
\eeqn
can be simplified using:
\bea
	[ T_{rel} , V_{i j}] &=& [ T_{i j} , V_{i j}] \\
	\left[ [ T_{i j} , V_{i j}] , T_{rel}\right] &=& \left[ [ T_{i j} , V_{i j}] , T_{i j}\right],
\eea
which are easily worked out in either HO-basis or momentum representation.
We then see that the 3-body leading-order evolution equation is simply a sum of each of the three 2-body terms.
Therefore, embedding the 2-body potential into the energy-truncated symmetric basis and evolving with the 3-body leading-order equation should be identical to evolving the 2-body potential with the leading-order term in the 2-body basis, then embedding into the energy-truncated symmetric basis.
Observing Fig.~\ref{fig:pot3_zeroerror_osc} we can clearly see the same oscillatory errors appearing in the momentum representation matrix elements which get worse as $\lambda$ decreases.
Therefore it is clear that there is an error in the HO-basis 3-body evolution that increases as $\lambda$ decreases.

It is important to note that this is an error in the matrix elements of the 3-body evolved potential.  
It will not appear as an error in the eigenvalues or any 3-body observables calculated in HO-basis, as the SRG is a unitary transformation.
If the HO-basis potential was transformed to momentum representation, and a calculation done in momentum representation, errors in calculated observables would likely appear.
As adding additional bodies increasingly adds importance to off-diagonal potential matrix elements through additional momentum integrations~\cite{3b_bible}, we expect to see the errors in potential matrix elements translate to many-body errors in observables, increasing with particle number.
Most apparent, however is that our prescription for observing momentum representation matrix elements in a search for universal form after evolution is impossible with the oscillatory errors arising from the evolution.

The errors in HO-basis evolution do, however, decrease with a larger basis size.  
We see in Fig.~\ref{fig:pot3_ho_faroffdiag} that the far-off diagonal matrix elements decrease in magnitude as total N increases.
For an accurate HO-basis evolution, we see that $\nmax$ must be taken to be much larger than is necessary to reproduce 3-body observables, and for a desired accuracy at a given $\lambda$, it may be possible to simply make $\nmax$ large enough.
It may also be possible to bypass the problem using the evolution method of~\cite{Bogner:2006pc} in a plane-wave basis, where it is well defined.
It may also be possible to subtract out the disconnected diagrams in HO-basis and evolve in an analogous way.


\section{3-body Recap}

We began this chapter confirming ground-state energies for a previously used 1-D potential for spinless bosons~\cite{Akerlund:2011cb,Jurgenson:2008jp}.
We introduced a 2-body potential more beneficial to our study because its long-range interaction is not as confined as a Gaussian, it requires smaller basis size for accurate calculations, and it is easier to fit with our $\delta$-shell procedure.
We confirmed that this potential follows the same rules for universality in evolved potential matrix elements as the realistic potentials of Chapter~\ref{chapt:2BUniv}.
This potential evolves accurately to $\lambda = 2.0$ fm$^{-1}$, but continuing the evolution to $\lambda = 1.0$ fm$^{-1}$ shows signs of truncation errors.

Continuing into the 3-body sector, we see that oscillatory evolution errors appear at much higher $\lambda$ than in the 2-body sector.
Small errors appear at $\lambda = 14$ fm$^{-1}$, and evolving further to $\lambda = 1.5$ fm$^{-1}$ increases the errors to be of comparable magnitude to the unevolved potential.
We identify the source of these errors as far-off diagonal matrix elements arising from the spectator particle.
To be sure that these oscillations are SRG evolution errors, we show the difference in matrix elements evolved with the term quadratic in $T$ of Eq.~\eqref{eq:leadingordersrg}, which should be the same if calculated in the 2-body sector then embedded into 3-body or calculated in the 3-body sector.
Fixing this SRG error in 3-body SRG evolution is a critical step in more precise Hamiltonians for many-body calculations.
\chapter{Conclusion}
\label{chapt:conclusion}

\section{Recap}

The modern precision 2- and 3-nucleon potentials are the product of decades of discovery and refinement, but their success in few-body systems is not yet mirrored in many-body calculations.
Even the soft \chieft\ potentials possess too strong a coupling between relevant low-energy and high-energy potential matrix elements, thus decoupling these matrix elements through SRG evolution is a critical step for solving many-body problems.
In an effort to improve 2-body potentials, we can use universality in evolved potential matrix elements as a guideline to simplify required aspects of the potential.
We have seen in chapter~\ref{chapt:2BUniv} that for matrix elements of different 2-body potentials to evolve to the same universal form in a low-energy region, their unevolved HOS T-matrix must be equivalent in the same energy region, and they must have the same binding energies (if any).
Potentials possessing the same explicit long-range terms exhibit the HOS T-matrix equivalence required for universal low-energy potential matrix elements.
One may also utilize this in reverse; understanding the effective range of an operator based on the effect in the HOS T-matrix elements.

The necessity of an explicit 3-body potential for accurate 3- and many-body calculations means that it is important to extend our understanding of universality into the 3-body sector.
The complexity of adding an explicit 3-body interaction that complements the 2-body potential has limited our methods for solving few- and many-body nuclear problems.
For this reason, evolution of potentials with explicit and induced 3-body forces for use in many-body calculations has been restricted to HO-basis with limited maximum oscillator number.
There is far less understanding of evolution precision in 3-body spaces than in 2-body, and explicitly observing differences in momentum and HO-evolution is important.

We have observed in chapter~\ref{chapt:3BUniv} that truncation errors imposed by the HO-basis are manageable in the 2-body sector; one has the ability to set momentum cutoffs and mesh sizes sufficiently high to achieve desired precision.
In 2-body potential matrix elements, it is possible to only truncate negligible matrix elements.
We observe that in the 3-body sector, the spectator particle adds strength to far off-diagonal matrix elements in the embedded 2-body potential.
The strength of these far off-diagonal elements decreases with our truncation of oscillator number, but at too slow a rate, which inhibits computation.
Errors in the 3-body observables from truncation of these non-zero matrix elements in the 3-body sector can be overcome with sufficiently large mesh size, however due to the nature of HO-basis SRG evolution with the $T$ generator, the errors compound and grow as $\lambda$ decreases.
We observe this as oscillatory errors in momentum-representation and differences in the quadratic-in-$T$ term evolution of 2-body potential matrix elements in the 2-body space and 3-body space.

More precise evolution of embedded 2-body potentials in HO-basis is required before universality in evolved potential matrix elements in the 3-body sector can be studied in a way analogous to chapter~\ref{chapt:2BUniv}.
Furthermore, these errors in potential matrix elements arising from HO-basis SRG evolution do not affect the 3-body observables, but they could be a source of error in many-body calculations.
As an increasingly important focus of low-energy nuclear physics is estimating theoretical error bars, a better understanding of the HO-basis evolution errors is important, and we have identified an apparent source of this evolution error.


\section{Looking Forward}

The natural next step in this study will benefit greatly from 3-body evolution in momentum representation.
Once precision evolution in a momentum basis (plane-wave Jacobi~\cite{Hebeler:2012pr} or otherwise~\cite{Wendt:2013bla}) is accomplished, it will be straight-forward to compare with HO-evolution and identify the extent of the differences.
Some many-body methods will still require the computational efficiency of HO-basis, but if the errors of HO-basis evolution are indeed large, a momentum-basis evolution followed by transformation to HO-basis would create better input Hamiltonians for many-body calculations.
Furthermore, we saw that observing universality in evolved matrix elements is much more straight-forward in momentum representation, thus a study of 3-body universality would be much clearer if precise momentum representation evolution was possible.
Should 4-body forces be recognized as important contributions to the input Hamiltonians, HO-basis evolution will certainly be much simpler to achieve than momentum representation evolution, and the lessons learned about the differences in the two evolution methods in the 3-body sector may provide valuable insight into the errors in 4-body evolution, where direct examination of the differences may not yet be possible.

Because of the difficulty of generating explicit 3-body potentials, far fewer of them exist than 2-body potentials.
In principle, each 2-body potential may have an infinite set of complimentary 3-body potentials which together accurately reproduce observables in the 3-body sector.
Understanding universality in the 3-body potentials may serve as an important simplification to generating them and may identify features required to improve accuracy of few-body input Hamiltonians in many-body calculations.


%
%
\bibliographystyle{unsrt}
\bibliography{mybib}

\begin{thebibliography}{10}

\bibitem{Kanada:1967uf}
H.~Kanada, S.~Otsuki, K.~Sakai, and M.~Yasuno.
\newblock {Analysis of proton proton scattering at high-energy}.
\newblock {\em Prog.Theor.Phys.}, 38:1134--1145, 1967.

\bibitem{av18orig}
R.~B. Wiringa, V.~G.~J. Stoks, and R.~Schiavilla.
\newblock Accurate nucleon-nucleon potential with charge-independence breaking.
\newblock {\em Phys. Rev. C}, 51:38--51, Jan 1995.

\bibitem{av18}
S.~Veerasamy and W.~N. Polyzou.
\newblock Momentum-space argonne v18 interaction.
\newblock {\em Phys. Rev. C}, 84:034003, Sep 2011.

\bibitem{Weinberg:1990rz}
Steven Weinberg.
\newblock {Nuclear forces from chiral Lagrangians}.
\newblock {\em Phys.Lett.}, B251:288--292, 1990.

\bibitem{vlowkuniv}
S.~K. Bogner, T.~T.~S. Kuo, and A.~Schwenk.
\newblock {Model independent low momentum nucleon interaction from phase shift
  equivalence}.
\newblock {\em Phys. Rept.}, 386:1--27, 2003.

\bibitem{bognerfurnstahlschwenk}
S.~K. Bogner, R.~J. Furnstahl, and A.~Schwenk.
\newblock {From low-momentum interactions to nuclear structure}.
\newblock {\em Prog. Part. Nucl. Phys.}, 65:94--147, 2010.

\bibitem{Bogner:2001jn}
S.K. Bogner, A.~Schwenk, T.T.S. Kuo, and G.E. Brown.
\newblock {Renormalization group equation for low momentum effective nuclear
  interactions}.
\newblock 2001. 
\newblock {arXiv: nucl-th/0111042}

\bibitem{Bogner:2001gq}
S.~K. Bogner, T.~T.~S. Kuo, A.~Schwenk, D.~R. Entem, and R.~Machleidt.
\newblock Towards a unique low momentum nucleon nucleon interaction.
\newblock {\em Phys. Lett. B}, 576:265--272, 2003.

\bibitem{GlazekWilson}
Stanis\l{}aw~D. G\l{}azek and Kenneth~G. Wilson.
\newblock Renormalization of hamiltonians.
\newblock {\em Phys. Rev. D}, 48:5863--5872, Dec 1993.

\bibitem{Wegner}
F.~Wegner.
\newblock Flow-equations for hamiltonions.
\newblock {\em Ann. Phys. (Leipzig)}, 3:77--91, 1994.

\bibitem{Bogner:2006pc}
S.~K. Bogner, R.~J. Furnstahl, and R.~J. Perry.
\newblock Similarity renormalization group for nucleon-nucleon interactions.
\newblock {\em Phys. Rev. C}, 75:061001, 2007.

\bibitem{Roth:2011vt}
R.~Roth, S.~Binder, K.~Vobig, A.~Calci, J.~Langhammer, and P.~Navratil.
\newblock {Ab Initio Calculations of Medium-Mass Nuclei with Normal-Ordered
  Chiral NN+3N Interactions}.
\newblock {\em Phys. Rev. Lett.}, 109:052501, 2012.

\bibitem{Furnstahl:2013oba}
R.~J. Furnstahl and K.~Hebeler.
\newblock {New applications of renormalization group methods in nuclear
  physics}.
\newblock {\em Rep. Prog. Phys.}, 76:126301, 2013.

\bibitem{decouplingSRG}
E.~D. Jurgenson, S.~K. Bogner, R.~J. Furnstahl, and R.~J. Perry.
\newblock {Decoupling in the Similarity Renormalization Group for
  Nucleon-Nucleon Forces}.
\newblock {\em Phys. Rev. C}, 78:014003, 2008.

\bibitem{SRGgeneral}
R.~J. Furnstahl.
\newblock {The Renormalization Group in Nuclear Physics}.
\newblock {\em Nucl. Phys. Proc. Suppl.}, 228:139--175, 2012.

\bibitem{Hebeler:2012pr}
K.~Hebeler.
\newblock {Momentum space evolution of chiral three-nucleon forces}.
\newblock {\em Phys. Rev. C}, 85:021002, 2012.

\bibitem{Wendt:2013bla}
Kyle~A. Wendt.
\newblock {Similarity Renormalization Group Evolution of Three-Nucleon Forces
  in a Hyperspherical Momentum Representation}.
\newblock {\em Phys. Rev. C}, 87:061001, 2013.

\bibitem{Hammer:2012id}
Hans-Werner Hammer, Andreas Nogga, and Achim Schwenk.
\newblock {Three-body forces: From cold atoms to nuclei}.
\newblock {\em Rev. Mod. Phys.}, 85:197, 2013.

\bibitem{Wilson:1974mb}
Kenneth~G. Wilson.
\newblock {The Renormalization Group: Critical Phenomena and the Kondo
  Problem}.
\newblock {\em Rev.Mod.Phys.}, 47:773, 1975.

\bibitem{Timoteo:2011tt}
V.~S. Timoteo, S.~Szpigel, and E.~Ruiz~Arriola.
\newblock {Symmetries of the Similarity Renormalization Group for Nuclear
  Forces}.
\newblock {\em Phys. Rev. C}, 86:034002, 2012.

\bibitem{Arriola:2013nja}
E.~Ruiz~Arriola, V.~S. Timoteo, and S.~Szpigel.
\newblock {Nuclear Symmetries of the similarity renormalization group for
  nuclear forces}.
\newblock {\em PoS}, CD12:106, 2013.

\bibitem{GlazekPerry}
Stanislaw~D. Glazek and Robert~J. Perry.
\newblock {The impact of bound states on similarity renormalization group
  transformations}.
\newblock {\em Phys. Rev. D}, 78:045011, 2008.

\bibitem{BlockDiag}
E.~Anderson, S.~K. Bogner, R.~J. Furnstahl, E.~D. Jurgenson, R.~J. Perry, and
  A.~Schwenk.
\newblock {Block Diagonalization using SRG Flow Equations}.
\newblock {\em Phys. Rev. C}, 77:037001, 2008.

\bibitem{Wendt:2011qj}
K.~A. Wendt, R.~J. Furnstahl, and R.~J. Perry.
\newblock {Decoupling of Spurious Deep Bound States with the Similarity
  Renormalization Group}.
\newblock {\em Phys. Rev. C}, 83:034005, 2011.

\bibitem{3b_evidence}
Steve Pieper.
\newblock private communication.

\bibitem{Binder:2013xaa}
Sven Binder, Joachim Langhammer, Angelo Calci, and Robert Roth.
\newblock {Ab Initio Path to Heavy Nuclei}.
\newblock {\em Phys.Lett.}, B736:119--123, 2014.

\bibitem{computational}
Rubin~H. Landau, Manuel Jos\'{e}~P\'{a}ez P\'{a}ez, and Cristian~C. Bordeianu.
\newblock {\em A Survey of Computational Physics}.
\newblock Princeton University Press, New Jersey, 2008.

\bibitem{Arriola}
R.~Navarro~P\'erez, J.~E. Amaro, and E.~Ruiz~Arriola.
\newblock {Coarse graining Nuclear Interactions}.
\newblock {\em Prog. Part. Nucl. Phys.}, 67:359--364, 2012.

\bibitem{Perez:2013jpa}
R.~Navarro~P\'erez, J.~E. Amaro, and E.~Ruiz~Arriola.
\newblock Coarse-grained potential analysis of neutron-proton and proton-proton
  scattering below the pion production threshold.
\newblock {\em Phys. Rev. C}, 88:064002, Dec 2013.

\bibitem{Akerlund:2011cb}
O.~Akerlund, E.J. Lindgren, J.~Bergsten, B.~Grevholm, P.~Lerner, et~al.
\newblock {The Similarity Renormalization Group for Three-Body Interactions in
  One Dimension}.
\newblock {\em Eur.Phys.J.}, A47:122, 2011.

\bibitem{Jurgenson:2008jp}
E.D. Jurgenson and R.J. Furnstahl.
\newblock {Similarity Renormalization Group Evolution of Many-Body Forces in a
  One-Dimensional Model}.
\newblock {\em Nucl.Phys.}, A818:152--173, 2009.

\bibitem{Szpigel:2000xj}
Sergio Szpigel and Robert~J. Perry.
\newblock {The Similarity renormalization group}.
\newblock 2000.

\bibitem{Kehrein:2006}
S.~Kehrein.
\newblock {\em The Flow Equation Approach to Many-Particle Systems}.
\newblock Springer, Berlin, 2006.

\bibitem{3bsrgsimple}
S.K. Bogner, R.J. Furnstahl, and R.J. Perry.
\newblock {Three-Body Forces Produced by a Similarity Renormalization Group
  Transformation in a Simple Model}.
\newblock {\em Annals Phys.}, 323:1478--1501, 2008.

\bibitem{EM500}
D.~R. Entem and R.~Machleidt.
\newblock {Accurate charge dependent nucleon nucleon potential at fourth order
  of chiral perturbation theory}.
\newblock {\em Phys. Rev. C}, 68:041001, 2003.

\bibitem{EB}
E.~Epelbaum, W.~Glockle, and Ulf-G. Meissner.
\newblock {The Two-nucleon system at next-to-next-to-next-to-leading order}.
\newblock {\em Nucl. Phys. A}, 747:362--424, 2005.

\bibitem{BrownJackson}
G.~E. Brown and A.~D. Jackson.
\newblock {\em The nucleon-nucleon interaction}.
\newblock North-Holland Pub. Co., Amsterdam, 1976.

\bibitem{Kwong}
N.~H. Kwong and H.~S. K\"ohler.
\newblock Separable $\mathrm{NN}$ potentials from inverse scattering for
  nuclear matter studies.
\newblock {\em Phys. Rev. C}, 55:1650--1664, Apr 1997.

\bibitem{BBps}
John~M. Blatt and L.~C. Biedenharn.
\newblock Neutron-proton scattering with spin-orbit coupling. i. general
  expressions.
\newblock {\em Phys. Rev.}, 86:399--404, May 1952.

\bibitem{Nbarps}
H.~P. Stapp, T.~J. Ypsilantis, and N.~Metropolis.
\newblock Phase-shift analysis of 310-mev proton-proton scattering experiments.
\newblock {\em Phys. Rev.}, 105:302--310, Jan 1957.

\bibitem{qm2landau}
Rubin~H. Landau.
\newblock {\em Quantum Mechanics II}.
\newblock John Wiley \& Sons, Inc., New York, 1996.

\bibitem{OPE}
A.~Nogga, R.~G.~E. Timmermans, and U.~van Kolck.
\newblock {Renormalization of one-pion exchange and power counting}.
\newblock {\em Phys. Rev. C}, 72:054006, 2005.

\bibitem{Shirokov:2003kk}
A.~M. Shirokov, A.~I. Mazur, S.~A. Zaytsev, J.~P. Vary, and T.~A. Weber.
\newblock {Nucleon nucleon interaction in the J matrix inverse scattering
  approach and few nucleon systems}.
\newblock {\em Phys. Rev. C}, 70:044005, 2004.

\bibitem{Shirokov:2005bk}
A.~M. Shirokov, J.~P. Vary, A.~I. Mazur, and T.~A. Weber.
\newblock {Realistic Nuclear Hamiltonian: 'Ab exitu' approach}.
\newblock {\em Phys. Lett. B}, 644:33--37, 2007.

\bibitem{Luscher:1985dn}
M.~Luscher.
\newblock {Volume Dependence of the Energy Spectrum in Massive Quantum Field
  Theories. 1. Stable Particle States}.
\newblock {\em Commun.Math.Phys.}, 104:177, 1986.

\bibitem{Hebeler:2013ri}
K.~Hebeler and R.J. Furnstahl.
\newblock {Neutron matter based on consistently evolved chiral three-nucleon
  interactions}.
\newblock {\em Phys.Rev.}, C87(3):031302, 2013.

\bibitem{Shankar}
R.~Shankar.
\newblock {\em Principles of Quantum Mechanics}.
\newblock Plenum Press, New York, 1994.

\bibitem{Furnstahl:2013vda}
R.J. Furnstahl, T.~Papenbrock, and S.N. More.
\newblock {Systematic expansion for infrared oscillator basis extrapolations}.
\newblock {\em Phys.Rev.}, C89:044301, 2014.

\bibitem{Konig:2014hma}
S.~König, S.K. Bogner, R.J. Furnstahl, S.N. More, and T.~Papenbrock.
\newblock {Ultraviolet extrapolations in finite oscillator bases}.
\newblock 2014.

\bibitem{More:2013rma}
S.N. More, A.~Ekström, R.J. Furnstahl, G.~Hagen, and T.~Papenbrock.
\newblock {Universal properties of infrared oscillator basis extrapolations}.
\newblock {\em Phys.Rev.}, C87(4):044326, 2013.

\bibitem{NIST:DLMF}
{NIST Digital Library of Mathematical Functions}.
\newblock http://dlmf.nist.gov/, Release 1.0.9 of 2014-08-29.
\newblock Online companion to \cite{Olver:2010:NHMF}.

\bibitem{Olver:2010:NHMF}
F.~W.~J. Olver, D.~W. Lozier, R.~F. Boisvert, and C.~W. Clark, editors.
\newblock {\em {NIST Handbook of Mathematical Functions}}.
\newblock Cambridge University Press, New York, NY, 2010.
\newblock Print companion to \cite{NIST:DLMF}.

\bibitem{PhysRevC.90.024325}
Robert Roth, Angelo Calci, Joachim Langhammer, and Sven Binder.
\newblock Evolved chiral $nn+3n$ hamiltonians for ab initio nuclear structure
  calculations.
\newblock {\em Phys. Rev. C}, 90:024325, Aug 2014.

\bibitem{Glazek:2008pg}
Stanislaw~D. Glazek and Robert~J. Perry.
\newblock {The impact of bound states on similarity renormalization group
  transformations}.
\newblock {\em Phys.Rev.}, D78:045011, 2008.

\bibitem{Jurgenson:2010wy}
E.D. Jurgenson, P.~Navratil, and R.J. Furnstahl.
\newblock {Evolving Nuclear Many-Body Forces with the Similarity
  Renormalization Group}.
\newblock {\em Phys.Rev.}, C83:034301, 2011.

\bibitem{3b_bible}
Walter Gl\"{o}ckle.
\newblock {\em The Quantum Mechanical Few-Body Problem}.
\newblock Springer-Verlag, Berlin, 1983.

\end{thebibliography}

\appendix
\chapter{ISSP Derivation}
\label{app:app_issp}

The following is the derivation of the ISSP from Ref.~\cite{BrownJackson}.
Starting from the Lippmann-Schwinger equation (LSE) in partial-wave momentum representation (omitting $l$-indecies),
\beqn
	\bra{k} K(E) \ket{k'} = \bra{k} V \ket{k'} + \frac{2}{\pi} P \int dp\ p^{2} \frac{\bra{k} V \ket{p} \bra{p} K(E) \ket{k'}}{E-p^{2}},
\eeqn
we can input a separable potential,
\beqn
	\bra{k} V \ket{k'} = \sigma \nu(k) \nu(k').
\eeqn
This simplifies the LSE to,
\bea
	\bra{k} K(E) \ket{k'} &=& \frac{\sigma \nu(k) \nu(k')}{D(\sqrt{E})}, \\
	D(x) &=& 1 - \frac{2}{\pi} P \int dp\ p^{2} \frac{\nu^{2}(p)}{x^{2}-p^{2}}. \label{eq:sep_k}
\eea
We can define,
\beqn
	D^{+}(x) = \lim_{\epsilon \to 0} \left( D(x + i \epsilon) \right).
\eeqn
Evaluating this limit, we find,
\beqn
	D^{+}(x) = D(x) + i \sigma \nu^{2}(x) x.\label{eq:imtakepart}
\eeqn
Combining this with equations~\ref{eq:sep_k} and~\ref{eqzeroes}, we see,
\beqn
	\tan(-\delta(k)) = \frac{Im[ D^{+}(x) ]}{Re[ D^{+}(x) ]}.
\eeqn
It follows to write,
\bea
	D^{+}(x) = |D^{+}(x)| e^{-i \delta(k)} , \\
	ln(D^{+}(x)) = ln|D^{+}(x)| - i \delta(k).
\eea
This final equations shows that the imaginary part of $ln(D^{+}(x))$ is determined by the phase shifts.

The next step requires making a contour integral.
Given a function, $f(z)$ analytic in a closed contour, C, containing point, z, we shall use Cauchy's theorem to write,
\beqn
	f(z) = \frac{1}{2 \pi i} \oint dx \frac{f(x)}{x-z}.
\eeqn
If we choose the contour including the real axis and a semicircle in the upper-half plane (UHP), and the function vanishes as $|z| \to \infty$ in the UHP, then adding $i \epsilon$ to z to guarantee we remain in the UHP this becomes,
\beqn
	f(z + i \epsilon) = \frac{1}{2 \pi i} \int_{-\infty}^{\infty} dx \frac{f(x)}{x-z-i \epsilon}.
\eeqn
Taking the real part, we see
\beqn
	Re [f(z + i \epsilon)] = \frac{1}{\pi} \int_{-\infty}^{\infty} dx \frac{Im [f(x)]}{x-z}.\label{eq:D_int}
\eeqn

We know that the potentials we consider fall off as $k \to \infty$, thus we can omit the semicircle part of the contour.
We also know that a bound state will produce a simple pole in the K-matrix, which is caused by a zero in $D^{+}(\sqrt{E})$.  We can modify $D^{+}(x)$ in a way that will remove the zeros in the UHP for our contour integration, and also preserve the asymptotic behaviour of the function.
Introducing $\kappa_{B} = \sqrt{E_{B}}$, this modification is
\beqn
	\hat{D}^{+}(k) = \frac{k+ i \kappa_{B}}{k - i \kappa_{B}} D^{+}(k).
\eeqn
Now we can relate the real part of $ln[D^{+}(k)]$ to an integral over the imaginary part, which is simply the phase-shifts.
We start by noticing that
\bea
	\rm{Re}[ ln (\hat{D}^{+}(k))] &=& ln (D^{+}(k)), \\
	\rm{Im}[ ln (\hat{D}^{+}(k))] &=& -i \delta(k) + ln (\frac{k+ i \kappa_{B}}{k - i \kappa_{B}}).
\eea
Now, using~\ref{eq:D_int}, we see
\bea
	ln |D^{+}(k))| &=& -\Delta(k) + \frac{1}{\pi i} P \int_{-\infty}^{\infty} \frac{dk'}{k'-k} ln(\frac{k+ i \kappa_{B}}{k - i \kappa_{B}}),\label{eq:log_issp} \\
	\Delta(k) &=&  \frac{1}{\pi} P \int_{-\infty}^{\infty} \frac{dk' \delta(k)}{k'-k}.
\eea
The second term (integral) in~\ref{eq:log_issp} is evaluated as $ln(\frac{k^2+\kappa^{2}_{B}}{k^2})$.
It follows, then that
\bea
	D^{+}(k) = \frac{k^2+\kappa^{2}_{B}}{k^2} e^{-\Delta(k) - i \delta(k)}.
\eea
Now, we simply equate the imaginary part of this equation, and equation~\ref{eq:imtakepart}, to find
\bea
	\sigma \nu^{2}(k) = - \frac{k^2+\kappa^{2}_{B}}{k^2} \frac{\sin(k)}{k} e^{-\Delta(k)}.
\eea

\chapter{Further Harmonic Oscillator Basis Derivation}
\label{app:app_ho}

In this appendix we show the derivation of $\Trel$ in HO-basis for 2- and 3-bodies and the derivation of the transition bracket.

To calculate $T^{(2)}$, the 2-body relative kinetic energy operator, we utilize ladder operators~\cite{Shankar}:
\bea
	\bra{n} T^{(2)} \ket{m} &=& \bra{n} \frac{\hat{P}^{2}}{2m} \ket{m} \\
	\hat{P} &=&  i \sqrt{\frac{m}{2 b^{2}}}(a^{\dagger} - a ) \\
	\bra{n} T^{(2)} \ket{m} &=& \bra{n} \frac{i \frac{m}{2 b^{2}} (a^{\dagger} - a )^{2} }{2m} \ket{m} \\
	\bra{n} T^{(2)} \ket{m}&=& \bra{n} - \frac{1}{4 b^{2}} (a^{\dagger} a^{\dagger} - a^{\dagger} a -a a^{\dagger} + a a) \ket{m} 
\eea
\bea
	\bra{n} T^{(2)} \ket{m}= - \frac{1}{4 b^{2}} [&&\sqrt{m+1}\sqrt{m+2} \braket{n}{m+2} \nonumber \\
	       						&&- (\sqrt{m}\sqrt{m}  + \sqrt{m+1}\sqrt{m+1}) \braket{n}{m} \nonumber \\
	      						&&+ \sqrt{n+1}\sqrt{n+2} \braket{n+2}{m} ]			
\eea
\bea
	\bra{n} T^{(2)} \ket{m}= - \frac{1}{4 b^{2}} [&&\sqrt{m+1}\sqrt{m+2} \delta_{n,m+2} \nonumber \\
	       						&&- (2n+1) \delta_{n,m} \nonumber \\
	      						&&+ \sqrt{n+1}\sqrt{n+2} \delta_{n+2,m} ]	.	
\eea

To calculate $T^{(3)}$ we take advantage of the fact that $\ket{n_{p}}$ and $\ket{n_{q}}$ are in orthogonal spaces, thus:
\bea
	\bra{n_{p} n_{q}} T^{(2)} \ket{m_{p} m_{q}} &=& \bra{n_{p} n_{q}} \frac{\hat{P}^{2}}{2m} + \frac{\hat{Q}^{2}}{2m} \ket{m_{p} m_{q}} \\
	\bra{n_{p} n_{q}} T^{(2)} \ket{m_{p} m_{q}} &=& \bra{n_{p}} \bra{n_{q}} \frac{\hat{P}^{2}}{2m} + \frac{\hat{Q}^{2}}{2m} \ket{m_{p}} \ket{m_{q}}
\eea
\bea
	\bra{n_{p} n_{q}} T^{(2)} \ket{m_{p} m_{q}} &=& \bra{n_{p}} \frac{\hat{P}^{2}}{2m} \ket{m_{p}} \braket{n_{q}}{m_{q}}  + \bra{n_{q}} \frac{\hat{Q}^{2}}{2m} \ket{m_{q}} \braket{n_{p}}{m_{p}} \\
	\bra{n_{p} n_{q}} T^{(2)} \ket{m_{p} m_{q}} &=& \bra{n_{p}} T^{(2)} \ket{m_{p}} \delta_{n_{q},m_{q}}  + \bra{n_{q}} T^{(2)} \ket{m_{q}} \delta_{n_{p},m_{p}} .
\eea

Lastly, we must calculate matrix elements of the transition bracket, $\bra{n_{p} \nq} P_{2 3} \ket{n'_{p} \nqp} \equiv \braket{n_{p} \nq}{n'_{p} \nqp}_{3}$.
The Jacobi momentum with particles 2 and 3 exchanged is readily calculated from Jacobi momentum definitions.
After some algebra,
\bea
	p' &=& \frac{1}{2} p + \frac{\sqrt{3}}{2} q , \\
	q' &=& \frac{\sqrt{3}}{2} p - \frac{1}{2} q .
\eea
We can define ladder operators for each oscillator basis state, $\eta^{\dagger n{i}}_{i} \ket{0} \equiv \frac{1}{\sqrt{n_{i} !}}\ket{n_{i}}$, which transform exactly the same as momenta:
\bea
	\eta'_{p} &=& \frac{1}{2} \eta_{p} + \frac{\sqrt{3}}{2} \eta_{q} , \\
	\eta'_{q} &=& \frac{\sqrt{3}}{2} \eta_{p} - \frac{1}{2} \eta_{q} .
\eea
Using these definitions, 
\bea
	\hspace{-.1\textwidth}\braket{n'_{p} \nqp}{n_{p} \nq}_{3} &=& \frac{1}{\sqrt{ n'_{p}! n'_{q}! n_{p}! n_{q}!}}\bra{0} \eta'^{n'_{p}}_{p} \eta'^{n'_{q}}_{q} \eta^{\dagger n_{p}}_{p} \eta^{\dagger n_{q}}_{q} \ket{0} \\
	= &&\hspace{-.05\textwidth}\frac{1}{\sqrt{ n'_{p}! n'_{q}! n_{p}! n_{q}!}}\bra{0} [\frac{1}{2} \eta_{p} + \frac{\sqrt{3}}{2} \eta_{q}]^{n'_{p}} \nonumber \\
	&&\hspace{.22 \textwidth}
	\times[\frac{\sqrt{3}}{2} \eta_{p} - \frac{1}{2} \eta_{q}]^{n'_{q}}
	\eta^{\dagger n_{p}}_{p} \eta^{\dagger n_{q}}_{q} \ket{0} \\
	= &&\hspace{-.05\textwidth}\frac{1}{\sqrt{ n'_{p}! n'_{q}! n_{p}! n_{q}!}}\bra{0}
	 \sum_{k=0}^{n'_{p}} \binom{n'_{p}}{k}[\frac{1}{2} \eta_{p}]^{(n'_{p} - k)} 
	 	 [\frac{\sqrt{3}}{2} \eta_{q}]^{k} \nonumber \\
	 	&& \hspace{.22 \textwidth}
	 \times \sum_{j=0}^{n'_{q}} \binom{n'_{q}}{j}[\frac{\sqrt{3}}{2} \eta_{p}]^{(n'_{q} - j)} 
	 	 [- \frac{1}{2} \eta_{q}]^{j} \nonumber \\
	 	&& \hspace{.22 \textwidth}
	 \times \eta^{\dagger n_{p}}_{p} \eta^{\dagger n_{q}}_{q} \ket{0}\\
	= &&\hspace{-.05\textwidth}\frac{1}{\sqrt{ n'_{p}! n'_{q}! n_{p}! n_{q}!}}
		\sum_{k=0}^{n'_{p}} \sum_{j=0}^{n'_{q}} \binom{n'_{p}}{k} \binom{n'_{q}}{j} \nonumber \\
		&& \hspace{.22 \textwidth}
			\times[\frac{1}{2}]^{(n'_{p} - k + j)} 
	 		[\frac{\sqrt{3}}{2}]^{(n'_{q} - j + k)} (-1)^{j} \nonumber \\
	 	&& \hspace{.22 \textwidth}
	 		\times n_{p}! n_{q}! \delta_{n'_{p} - k + n'_{q} - j , n_{p}} \delta_{k+j,n_{q}}  \\
	= &&\hspace{-.05\textwidth}\sqrt{\frac{n_{p}! n_{q}!}{n'_{p}! n'_{q}! }}
		\sum_{k=0}^{n'_{p}} \binom{n'_{p}}{k} \binom{n'_{q}}{n_{q} - k} [\frac{1}{2}]^{(n'_{p} + n_{q} - 2k)}
			[\frac{\sqrt{3}}{2}]^{(n'_{q} - n_{q} + 2k)} \nonumber \\
		&& \hspace{.22 \textwidth}
			\times \theta(n'_{q}-n_{q}+k)\theta(n_{q}-k) \nonumber \\
	 	&& \hspace{.22 \textwidth}
	 		\times (-1)^{n_{q}-k} \delta_{n'_{p} + n'_{q}, n_{p} + n_{q}}.
\eea
The third line utilizes the binomial theorem and the fourth line balances the ladder operators.

From this we can see why we need a truncation in total $\np+\nq$.  
If the truncation is made individually on each, we can examine the lowering operators in the previous equations.
They enter in powers, $[\frac{1}{2} \eta_{p} + \frac{\sqrt{3}}{2} \eta_{q}]^{n'_{p}} [\frac{\sqrt{3}}{2} \eta_{p} - \frac{1}{2} \eta_{q}]^{n'_{q}}$.
We see that each lowering operator can enter $\npp + \nqp$ times, connecting our originally truncated states to states with oscillator number above our truncation.
This makes our symmetrizer non-unitary, which causes eigenvalues to no longer be distinctly 1 or 0.
Rearranging the states and truncating on total oscillator number, however, solves this problem.
\chapter{Table of Abbreviations}
\label{app:app_abbr}

\begin{center}
\begin{table}[h]
\resizebox{1.0\textwidth}{!}{\begin{minipage}{\textwidth}
\begin{center}
\begin{tabular}{|l | l |}
	\hline
	AV18  &  \avpot \\ \hline
	HO & harmonic oscillator \\ \hline
	HOS & half on shell \\ \hline
	IR & infrared \\ \hline
	ISSP & inverse scattering separable potential \\ \hline
	LS & Lippmann-Schwinger \\ \hline
	OPE & one-pion exchange \\ \hline
	QCD & quantum chromodynamics \\ \hline
	SRG & similarity renormalization group \\ \hline
	UV & ultraviolet \\ \hline
	\chieft & chiral effective field theory \\ \hline
\end{tabular}
\caption{Table of abbreviations}
\label{tab:constvals}
\end{center}
\end{minipage} }
\end{table}
\end{center}

%
%

\end{document}